\documentstyle[preprint,aps,rmp]{revtex}
\newcommand{\bm}{\bbox}
\newcommand{\qq}{\begin{eqnarray}}
\newcommand{\qqq}{\end{eqnarray}}
\newcommand{\un}{\underline}
\begin{document}
\vskip -3.0cm
\title{Particles and Fields in Fluid Turbulence}
\author{G. Falkovich}
\address{Physics of Complex Systems, Weizmann Institute of Science,
Rehovot 76100, Israel}
\author{K. Gaw\c{e}dzki}
\address{CNRS, IHES, 91940 Bures-sur-Yvette and 
ENS-Lyon, 46 Allée d'Italie, 69364 Lyon, France}
\author{M. Vergassola}
\address{CNRS, UMR 6529 Observatoire de la C\^{o}te d'Azur, BP 4229, 06304 
Nice,  France}
\maketitle\vskip -1.0truecm
\begin{abstract}
The understanding of fluid turbulence has considerably progressed in
recent years. The application of the methods of statistical mechanics 
to the description of the motion of fluid particles, i.e. to
the Lagrangian dynamics, has led to a new quantitative theory 
of intermittency in turbulent transport. The first analytical
description of anomalous scaling laws in turbulence has been
obtained. The underlying physical mechanism reveals the role of
statistical integrals of motion in non-equilibrium systems.  For
turbulent transport, the statistical conservation laws are hidden in
the evolution of groups of fluid particles and arise from the
competition between the expansion of a group and the change of its
geometry.  By breaking the scale-invariance symmetry, the statistically 
conserved quantities lead to the observed anomalous scaling of 
transported fields. Lagrangian methods also shed new light on some 
practical issues, such as mixing and turbulent magnetic dynamo.
\end{abstract}
\tableofcontents

\vskip 0.5cm

{\quad\qquad "Well," said Pooh, 
" we keep looking for Home and not finding it, so I thought\hfill } 

{\quad\qquad  that if we looked for this Pit, 
we'd be sure not to find it, which would be a Good\hfill }

{\quad\qquad  Thing, because then we might find something that we 
{\it weren't} looking for, which\hfill }

{\quad\qquad might be just what we {\it were} looking for,
really". A. Milne, Tigger is unbounced.\hfill }

\section{Introduction}

Turbulence is the last great unsolved problem of classical physics
which has evaded physical understanding and systematic description for
many de\-ca\-des. Turbulence is a state of a physical system with many
degrees of freedom strongly deviating from equilibrium.  The first
obstacle to its understanding stems from the large number of degrees
of freedom actively involved in the problem.  The scale of injection,
where turbulence is excited, usually differs dramatically from the
scale of damping, where dissipation takes place. Nonlinear
interactions strongly couple the degrees of freedom by transferring
excitations from the injection to the damping scale throughout the
so-called inertial range of scales.  The ensuing complicated and
irregular dynamics calls for a statistical description.  The main
physical problem is to understand to what extent the statistics in the
inertial interval is universal, i.e. independent of the conditions of
excitation and dissipation. In such general formulation, the issue
goes far beyond fluid mechanics, even though the main examples and
experimental data are provided by turbulence in continuous media. From
the standpoint of theoretical physics, turbulence is a non-equilibrium
field-theoretical problem with many strongly interacting degrees of
freedom. The second deeply rooted obstacle to its understanding is
that far from equilibrium we do not possess any general guiding rule,
like the Gibbs principle in equilibrium statistical physics.  Indeed,
to describe the single-time statistics of equilibrium systems, the
only thing we need is the knowledge of dynamic integrals of motion.
Then, our probability distribution in phase space is uniform over the
surfaces of constant integrals of motion.  Dynamically conserved
quantities play an important role in turbulence description, too, as
they flow throughout the inertial range in a cascade-like
process. However, the conserved quantity alone does not allow one to
describe the whole statistics but only a single correlation function
which corresponds to its flux. The major problem is to obtain the rest
of the statistics.

In every case, the starting point is to identify the dynamical
integral of motion that cascades through the inertial interval.  Let
us consider the forced $3d$ Navier-Stokes equation
\begin{equation} \label{NS}
\partial_t {\bm v}({\bm r},t)+{\bm v}({\bm
r},t)\cdot{\bm \nabla} {\bm v}({\bm r},t)-\nu{\nabla}^2{\bm v}({\bm r},t)
=-{\bm \nabla}p({\bm r},t)+f({\bm r},t),
\end{equation} supplemented by the incompressibility condition
${\bm \nabla}\cdot{\bm v}=0$. An example of injection mechanism is a
random large-scale forcing ${\bm f}({\bm r},t)$ with correlation
length $L$.  The relevant integral of motion, conserved in the absence
of injection and dissipation, is kinetic energy $\int v^2 d{\bm r}/2$
and the quantity which cascades throughout the inertial interval is
energy density in wavenumber space.  The energy flux-constancy
relation was derived in Kolmogorov (1941) and it involves the
third-order moment of the longitudinal velocity increments:
\begin{equation} \Big\langle\left[\left({\bm v}({\bm r},t)-{\bm v}({\bm
0},t)\right)\cdot {\bm r}/r\right]^3\Big\rangle\,\equiv
\Big\langle(\Delta_r v)^3\Big\rangle=\,-{4\over 5}\,
\bar\epsilon_v\, r\ .  \label{45}
\end{equation} The separation $r$ is supposed to lie in the inertial 
interval, ranging from the injection scale $L$ down to the viscous
dissipation scale.  The major physical assumption made to derive the
so-called $4/5$ law is that the mean energy dissipation rate
$\bar\epsilon_v=\nu\langle ({\bm \nabla} {\bm v})^2 \rangle$ has a
nonzero limit as the viscosity $\nu$ tends to zero.  This clearly
points to the non-equilibrium flux nature of turbulence. The
assumption of finite dissipation gives probably the first example of
what is called ``anomaly'' in modern field-theoretical language: A
symmetry of the inviscid equation (here, time-reversal invariance) is
broken by the presence of the viscous term, even though the latter
might have been expected to become negligible in the limit of
vanishing viscosity. Note that the $4/5$ law (\ref{45}) implies that
the third-order moment is universal, that is, it depends on the
injection and the dissipation only via the mean energy injection rate,
coinciding with $\bar\epsilon_v$ in the stationary state. To obtain
the rest of the statistics, a natural first step made by Kolmogorov
himself was to assume the statistics in the inertial range be scale
invariant. The scale invariance amounts to assuming that the
probability distribution function (PDF) of the rescaled velocity
differences $r^{-h}\Delta_r v$ can be made $r$-independent for an
appropriate $h$. The $n$-th order moment of the longitudinal velocity
increments $\langle (\Delta_r v)^n\rangle$ (structure functions) would
then depend on the separation as a power law $r^{\sigma_n}$ with the
``normal scaling'' behavior $\sigma_n=hn$. The rescaling exponent may
be determined by the flux law, e.g.  $h=1/3$ for $3d$ Navier-Stokes
turbulence. In the original Kolmogorov theory, the scale invariance
was in fact following from the postulate of complete universality: the
dependence on the injection and the dissipation is carried entirely by
$\bar\epsilon_v$ not only for the third-order moment but for the whole
statistics of the velocity increments.  The velocity difference PDF
could then involve only the dimensionless combination $(\bar\epsilon_v
r)^{-1/3} \Delta_r v$ and would be scale invariant.  There are cases,
like weakly nonlinear wave turbulence (Zakharov {\em et al.}, 1992),
where both scale-invariance and complete universality are assured by
the fact that the statistics in the inertial range is close to
Gaussian. That does not hold for strongly nonlinear systems.  Already
in 1942, L.~D.~Landau pointed out that all the velocity structure
functions (except the third one) are averages of nonlinear functions
of the flux. They are therefore sensitive to its fluctuations, which
depend on the specific injection mechanisms. Consequently, the
velocity statistics in the inertial range may have nonuniversal
features.

Experiments do not support scale invariance either. The structure
functions are in fact found experimentally to have a power-law
dependence on the separation $r$. However, the PDF of the velocity
differences at various separations cannot be collapsed one onto
another by simple rescaling and the scaling exponent $\sigma_n$ of the
structure functions is a nonlinear concave function of the order
$n$. As the separation decreases in the inertial range, the PDF
becomes more and more non-Gaussian, with a sharpening central peak and
a tail that becomes longer and longer.  In other words, the smaller
the separations considered, the higher the probability of very weak
and strong fluctuations.  This manifests itself as a sequence of
strong fluctuations alternating with quiescent periods, which is
indeed observed in turbulence signals and is known as the phenomenon
of intermittency.  The violation of the dimensional predictions for
the scaling laws is referred to as ``anomalous scaling'' for it
reflects, again, a symmetry breaking. The Euler equation is
scale-invariant and the scales of injection and dissipation are
supposed to be very large and small (formally, the limits to infinity
and zero should be taken). However, the dynamics of turbulence is such
that the limits are singular and scale invariance is broken. The
presence of a finite injection scale $L$, irrespective of its large
value, is felt throughout the inertial range precisely via the
anomalies $\langle(\Delta_r v)^n\rangle\propto
(\bar\epsilon_v\,r)^{n/3}\left(L/r\right)^{n/3-\sigma_n}$.

The non-Gaussianity of the statistics, the anomalous scaling and the
intermittency of the field occur as a rule rather than exception in
the context of fluid dynamics. The same phenomenology is observed in
many other physical systems. An incomplete list includes compressible
Navier-Stokes turbulence, Burgers' turbulence, scalar and magnetic
fields.  Examples of scalar fields are provided by the temperature of
a fluid, the humidity in the atmosphere, the concentration of chemical
or biological species.  The advection-diffusion equation governing the
transport of a nonreacting scalar field by an incompressible velocity
is:
\begin{equation}
\label{scalare}
\partial_t\theta({\bm r},t)+{\bm v}({\bm r},t)\cdot{\bm
\nabla}\theta({\bm r},t)-\kappa{\nabla}^2\theta({\bm r},t)=\varphi({\bm r},t),
\end{equation}
where $\varphi$ describes the sources. For scalar dynamics, the space
integral of any function of $\theta$ is conserved in the absence of
sources and diffusion. In their presence, the corresponding relation
for the flux of $\theta^2$ was derived in Yaglom (1949):
\begin{equation}
\label{Yag}
\Big\langle\Big[\left({\bm v}({\bm r},t)-{\bm v}({\bm
0},t)\right)\cdot {\bm r}/r\Big]\Big[\theta({\bm r},t)-\theta({\bm
0},t)\Big]^2\Big\rangle=-{4\over 3}\,\bar\epsilon\, r.
\end{equation}
The major physical assumption is again that the mean scalar
dissipation rate
$\bar\epsilon=\kappa\langle\left(\nabla\theta\right)^2\rangle$ remains
finite even in the limit where the molecular diffusivity $\kappa$
vanishes. Consider the particular case when the advecting velocity
${\bm v}$ satisfies the $3d$ Navier-Stokes equation. Assuming again
scale-invariance, the flux relations (\ref{45}) and (\ref{Yag}) would
imply that the scaling exponent of both the velocity and the scalar
field is $1/3$. As it was expected for the velocity, the scalar
structure functions $S_n(r)=\langle\left[\theta({\bm r},t)-\theta({\bm
0},t)\right]^n\rangle$ would then depend on the separation as power
laws $r^{\zeta_n}$ with $\zeta_n=n/3$. Experiments indicate that scale
invariance is violated for a scalar field as well, that is
$\zeta_n\not=n/3$.  More importantly, the intermittency of the scalar
is much stronger than that of the velocity, in particular,
$n/3-\zeta_n$ is substantially larger than $n/3-\sigma_n$. It was a
major intuition of R.H.~Kraichnan to realize that the passive scalar
could then be intermittent even in the absence of any intermittency of
the advecting velocity.

The main ambition of the modern theory of turbulence is to explain the
physical mechanisms of intermittency and anomalous scaling in
different physical systems, and to understand what is really universal
in the inertial-interval statistics. It is quite clear that strongly
non-equilibrium systems generally do not enjoy the same degree of
universality as those in equilibrium. In the absence of a unified
approach to non-equilibrium situations, one tries to solve problems on
a case-by-case basis, with the hope to learn if any universal guiding
principle may be recognized. It is in solving the particular problems
of passive scalar and magnetic fields that an important step in
general understanding of turbulence has been recently made.  The
language most suitable for the description of the systems turned out
to be the Lagrangian statistical formalism, i.e. the description of
the motion of fluid particles.  This line of analysis, pioneered by
L.F.~Richardson and G.I.~Taylor in the twenties and later developed by
R.H.~Kraichnan and others, has been particularly effective here.  The
results differ from case to case.  Some fields are non-Gaussian but
scale invariance is not broken, while others have turned out to be
amenable to the first ever analytical description of anomalous scaling
laws.  The anomalous exponents have been found to be universal, but
not the constants appearing in the prefactors of generic correlation
functions. This has provided a quantitative clarification of Landau's
previously mentioned remark and of the aspects of turbulence
statistics that may still be expected to be universal.  More
importantly, the anomalous scaling has been traced to the existence of
statistical integrals of motion.  The mechanism is quite robust and
relevant for transport by generic turbulent flows. The nature of those
integrals of motion strongly differs from that of the dynamic
conservation laws that determine equilibrium statistics.  For any
finite number of fluid particles, the conserved quantities are
functions of the interparticle separations that are statistically
preserved as the particles are transported by the random flow.  For
example, at scales where the velocity field is spatially smooth, the
average distance $R$ between two particles generally grows
exponentially, while the ensemble average $\langle R^{-d}\rangle$ is
asymptotically time-independent in a statistically isotropic
$d$-dimensional random flow.  The integrals of motion change with the
number of particles and generally depend nontrivially on the geometry
of their configurations.  In the connection between the advected
fields and the particles, the order of the correlation functions is
equal to the number of particles and the anomalous scaling issue may
be recast as a problem in statistical geometry. The nonlinear behavior
of the scaling exponents with the order is then due to the dependence
of the integrals of motion on the number of particles. The existence
of statistical conservation laws signals that the Lagrangian dynamics
keeps trace of the particle initial configuration throughout the
evolution. This memory is what makes the correlation functions at any
small scale sensitive to the presence of a finite injection length
$L$.  We believe that, more generally, the notion of statistical
integrals of motion is a key to understand the universal part of the
steady-state statistics for systems far from equilibrium.

The aim of this review is a description of fluid turbulence from the
Lagrangian viewpoint. Classical literature on Lagrangian dynamics
mostly concentrated on turbulent diffusion and pair dispersion,
i.e. the distance traveled by one particle or the separation between
two particles as a function of time.  By contrast, in that general
picture that has emerged recently, the evolution of the
multiparticle-configuration geometry takes center stage.  The main
body of the review will present these novel aspects of Lagrangian
dynamics and their consequences for the advected fields.  We shall
adhere to the following plan.  The knowledge accumulated on one and
two particle dynamics has been extensively covered in literature
(Pope, 1994; Majda and Kramer, 1999).  The objective of the first
three Sections in Chapter~II is to point out a few fundamental issues,
with particular attention to the basic differences between the cases
of spatially smooth and nonsmooth velocity fields.  We then proceed to
the multiparticle statistics and the analysis of hidden statistical
conservation laws that cause the breakdown of scale-invariance.  Most
of this analysis is carried out under the assumption of a prescribed
statistics of the velocity field. In Chapter~III we shall analyze
passive scalar and vector fields transported by turbulent flow and
what can be inferred about their statistics from the motion of fluid
particles. In Chapter~IV, we briefly discuss the Lagrangian dynamics
in the Burgers and the Navier-Stokes equations. The statistics of the
advecting velocity is not prescribed anymore, but it results from
nonlinear dynamics. Conclusions focus on the impact of the results
presented in this review on major directions of future
research. Readers from other fields of physics interested mainly in
the breakdown of scale invariance and statistical conservation laws
may restrict themselves to Sects.
\ref{sec:nonsmooth},\,\ref{sec:multi},\,\ref{sec:passinert},\,\ref{sec:con}.

The picture presented in this review is, to a large extent, an outcome
of joint work and numerous discussions with our colleagues,
E.~Balkovsky, D.~Bernard, A.~Celani, M.~Chertkov, G.~Eyink, A.~Fouxon,
U.~Frisch, I.~Kolokolov, A.~Kupiainen, V.~Lebedev, A.~Mazzino and
A.~Noullez.  We thank K.~Khanin, P.~Muratore-Ginanneschi,
A.~Shafarenko, B.~Shraiman and the referee for valuable comments about 
the manuscript. We are indebted to R.~H.~Kraichnan whose works 
and personality have been a permanent source of inspiration.

\section{Particles in fluid turbulence} \label{sec:part} 

As explained in the Introduction, understanding the properties of
transported fields involves the analysis of the behavior of
fluid particles. We have therefore decided to first present results on
the time-dependent statistics of the Lagrangian trajectories ${\bm
R}_n(t)$ and to devote the subsequent Chapter~III to the description
of transported fields.  In the present Chapter we sequentially
increase the number of particles involved in the problem. We start
from a single trajectory whose effective motion is a simple diffusion 
at times longer than the velocity correlation time in the Lagrangian frame
(Sect.~\ref{sec:single}). We then move to two particles. The separation 
law of two close trajectories depends on the scaling properties of the 
velocity field ${\bm v}({\bm r},t)$.  If the velocity is smooth, that is 
$\,|{\bbox v}({\bbox R}_n)-{\bbox v}({\bbox R}_m)|\propto |{\bbox R}_n-{
\bbox R}_m|$, then the initial separation grows exponentially in time
(Sect.~\ref{sec:smooth}). The smooth case can be analyzed in much detail 
using the large deviation arguments presented in Sect.~\ref{sec:general}. 
The reader mainly interested in applications to transported fields might 
wish to take the final results (\ref{PDF3}) and (\ref{posit}) for granted, 
skipping their derivation and the analysis of the few solvable cases where 
the large deviations may be calculated exactly. If the velocity is nonsmooth, 
that is $\,|{\bbox v}({\bbox R}_n) -{\bbox v}({\bbox R}_m)| \propto 
|{\bbox R}_n-{\bbox R}_m|^\alpha\,$ with $\alpha<1$, then the separation 
distance between two trajectories grows as a power of time  
(Sect.~\ref{sec:nonsmooth}), as first observed by Richardson (1926). 
We discuss important implications of such a behavior on the nature of the 
Lagrangian dynamics. The difference between the incompressible flows, where 
the trajectories generally separate, and compressible ones, where they may 
cluster, is discussed in Sect.~\ref{sec:compres}. Finally, in the 
consideration of three or more trajectories, the new issue
of geometry appears. Statistical conservation laws come to light in
two-particle problem and then feature prominently in the consideration
of multiparticle configurations. Geometry and statistical conservation
laws are the main subject of 
Sect.~\ref{sec:multi}. Although we try to keep the discussion as general 
as possible, much of the insight into the trajectory dynamics is obtained 
by studying simple random ensembles of synthetic velocities where exact 
calculations are possible. The latter serve to illustrate the general 
features of the particle dynamics.

\subsection{Single-particle diffusion} 
\label{sec:single}

The Lagrangian trajectory ${\bm R}(t)$ of a fluid particle advected by
a prescribed incompressible velocity field ${\bm v}({\bm r},t)$ in $d$
space dimensions and undergoing molecular diffusion with diffusivity
$\kappa$ is governed by the stochastic equation (Taylor, 1921),
customarily written for differentials:
\begin{equation}
\label{sde}
d{\bm R}={\bm v}({\bm R},t)\,dt+\sqrt{2\kappa}\,d{\bm\beta(t)}.
\end{equation}
Here, $\bm\beta(t)$ is the $d$-dimensional standard Brownian motion
with zero average and covariance function
$\langle\beta^i(t)\hspace{0.05cm} \beta^j(t')\rangle=\delta^{ij}\,{\rm
min}(t,t')$. The solution of (\ref{sde}) is fixed by prescribing the
particle position at a fixed time, e.g. the initial position ${\bm
R}(0)$.

The simplest instance of (\ref{sde}) is the Brownian motion, where the
advection is absent. The probability density ${\cal P}(\Delta{\bm
R};t)$ of the displacement $\Delta{\bm R}(t)={\bm R}(t)-{\bm R}(0)$
satisfies the heat equation $\,(\partial_t-\kappa{\bm\nabla}^2){\cal
P}=0\,$ whose solution is the Gaussian distribution ${\cal
P}(\Delta{\bm R};t)=(4\pi\kappa t)^{-d/2} \,\exp[{-(\Delta {\bm
R})^2/(4\kappa t)}]$.  The other limiting case is pure advection
without noise.  The properties of the displacement depend then on the
specific trajectory under consideration. We shall always work in the
frame of reference with no mean flow. We assume statistical
homogeneity of the Eulerian velocities which implies that the
Lagrangian velocity ${\bm V}(t) ={\bm v}({\bm R}(t), t)$ is
statistically independent of the initial position. If, additionally,
the Eulerian velocity is statistically stationary, then so is the
Lagrangian one\footnote{This follows by averaging the expectations
involving ${\bm V}(t+\tau)$ over the initial position ${\bm R} (0)$
(on which they do not depend) and by the change of variables ${\bm
R}(0)\mapsto{\bm R}(\tau)$ under the velocity ensemble average.  The
argument requires the incompressibility of the velocity, see
Sect.\,\ref{sec:compres}.}. The single-time expectations of the
Lagrangian velocity coincide in particular with those of the Eulerian
one, e.g. $\langle{\bm V}(t)\rangle = \langle{\bm v}\rangle=0$. The
relation between the multi-time statistics of the Eulerian and the
Lagrangian velocities is however quite involved in the general case.

For $\kappa=0$, \,the mean square displacement satisfies the
differential equation:
\begin{equation}
\label{GITaylor}
{d\over dt}\,\left\langle(\Delta{\bm R}(t))^2\right\rangle\,=\,
2\int_{_0}^{^t}\bigl\langle{\bm V}(t)\cdot{\bm V}(s)\bigr\rangle\,ds\,
=\,2\int_{_0}^{^t}\bigl\langle{\bm V}(0)\cdot{\bm V}(s)\bigr\rangle\,ds\,,
\end{equation}
where the second equality uses the stationarity of ${\bm V}(t)$.
The behavior of the displacement is crucially dependent on the range
of temporal correlations of the Lagrangian velocity.  Let us define 
the Lagrangian correlation time as
\begin{equation}
\label{Lagrtime}
\tau\,=\,{\,\int_0^\infty\langle{\bm V}(0) \cdot{\bm
V}(s)\rangle\,ds\over\langle{\bm V}^2\rangle}\,.
\end{equation}
The value of $\tau$ provides a measure of the Lagrangian velocity
memory. Divergence of $\,\tau\,$ is symptomatic of persistent
correlations.  As we shall discuss in the sequel, no general relation
between the Eulerian and the Lagrangian correlation times can be
established but for the case of short-correlated velocities.  For
times $t\ll\tau$, the 2-point function in (\ref{GITaylor}) is
approximately equal to $\langle{\bm v}^2\rangle$ and the particle
transport is ballistic: $\,\langle(\Delta{\bm R})^2\rangle\simeq
\langle{\bm v}^2\rangle\, t^2$. \,When the Lagrangian correlation time
is finite, a generic situation in a turbulent flow, an effective
diffusive regime arises for $t\gg\tau$ with $\langle\left(\Delta{\bm
R}\right)^2\rangle= 2\langle{\bm v}^2 \rangle\,\tau t$ (Taylor, 1921).
The particle displacements over time segments spaced by distances much
larger than $\tau$ are indeed almost independent. At long times, the
displacement $\Delta{\bm R}$ behaves then as a sum of many independent
variables and falls into the class of stationary processes governed by
the Central Limit Theorem. In other words, the displacement for
$t\gg\tau$ becomes a Brownian motion in $d$ dimensions with
\begin{equation}
\Big\langle\Delta R^i(t)\,\Delta R^j(t)\Big\rangle \,\simeq\,
2\hspace{0.05cm} D_e^{ij}\,t\,,
\label{dif}
\end{equation}
where 
\begin{equation}
D_e^{ij}\,=\,{_1\over^2}\,
\int\limits_0^\infty\langle V^i(0)\, V^j(s)+
V^j(0)\, V^i(s)\rangle\,ds\,.
\label{eddif}
\end{equation}
The same arguments carry over to the case of a non-vanishing molecular
diffusivity. The symmetric second order tensor $D_e^{ij}$ describes
the effective diffusivity (also called eddy diffusivity). The trace of
$D_e^{ij}$ is equal to the long-time value $\langle{\bm
v}^2\rangle\tau$ of the integral in (\ref{GITaylor}), while its
tensorial properties reflect the rotational symmetries of the velocity
field. If it is isotropic, the tensor reduces to a diagonal form
characterized by a single scalar value.  The main problem of turbulent
diffusion is to obtain the effective diffusivity tensor, given the
velocity field ${\bm v}$ and the value of the diffusivity
$\kappa$. Exhaustive reviews of the problem are available in the
literature (Bensoussan {\it et al.}, 1978; Pope, 1994; Fannjiang and
Papanicolaou, 1996; Majda and Kramer, 1999).

The other general issue in turbulent diffusion is about the conditions
on the velocity field ensuring the Lagrangian correlation time $\tau$
be finite and an effective diffusion regime take place for large
enough times. A sufficient condition (Kraichnan, 1970; Avellaneda and
Majda, 1989; Avellaneda and Vergassola, 1995) is that the vector
potential variance $\langle {\bm A}^2\rangle$ is finite, where the
$3d$ incompressible velocity ${\bm v}={\bm\nabla}\times {\bm A}$.
Similar conditions are valid for any space dimension.  The condition
$\kappa\neq 0$ is essential to the validity of the previous result, as
shown by the counter-example of Rayleigh-B\'enard convective cells,
see e.g. (Normand {\it et al.}, 1977).  In the absence of molecular
noise, the particle circulates forever in the same convective cell,
with no diffusion taking place at any time. This provides an example
of subdiffusion\,: the integral in (\ref{GITaylor}) goes to zero as
$t\to \infty$ and the growth of the mean square displacement is slower
than linear. Note that any finite molecular diffusivity, however
small, creates thin diffusive layers at the boundaries of the cells;
particles can then jump from one cell to another and diffuse.
Subdiffusion is particularly relevant for static $2d$ flows, where
tools borrowed from percolation/statistical topography find most
fruitful applications (Isichenko, 1992). Trapping effects required for
subdiffusion are, generally speaking, favored by the compressibility
of the velocity field, e.g., in random potentials (Bouchaud and
Georges, 1990).  Subdiffusive effects are expected to be overwhelmed
by chaotic mixing in flows leading to Lagrangian chaos, i.e., to
particle trajectories that are chaotic in the absence of molecular
diffusion (Ottino, 1989; Bohr {\it et al.}, 1998).  This is the
generic situation for $3d$ and $2d$ time-dependent incompressible
flows.

Physical situations having an infinite Lagrangian correlation time
$\tau$ correspond to superdiffusive transport\,: divergences of the
integral in (\ref{GITaylor}) as $t\to\infty$ signal that the particle
transport is faster than diffusive. A classical example of such
behavior is the class of parallel flows presented by Matheron and de
Marsily (1980). If the large-scale components of the velocity field
are sufficiently strong to make the particle move in the same
direction for arbitrarily long periods the resulting mean square
displacement grows more rapidly than $t$. Other simple examples of
superdiffusive motion are L\'evy-type models (Geisel {\it et al.},
1985; Shlesinger {\it et al.}, 1987).  A detailed review of
superdiffusive processes in Hamiltonian systems and symplectic maps
can be found in Shlesinger {\it et al.} (1993).

Having listed different subdiffusive and superdiffusive cases, from
now on we shall be interested in random turbulent flows with finite
Lagrangian correlation times, which are experimentally known 
to occur for sufficiently high Reynolds numbers (Pope, 1994).
For the long-time description of the diffusion in such flows, it 
is useful to consider the extreme case of random homogeneous and 
stationary Eulerian velocities with a short correlation time.  
The formal way to get these processes is to change the time scale 
by taking the scaling limit $\,\lim\limits_{\mu\to\infty}\,
\mu^{_1\over^2}{\bm v}({\bm r},\mu t)$, i.e.\,\,considering the process 
as viewed in a sped-up film. We assume that the connected correlation 
functions\footnote{The connected correlation functions, also called 
cumulants, are recursively defined by the relation $\,\langle v_1\,
\dots \,v_n\rangle=\sum\limits_{\{\pi_{\alpha}\}}\prod\limits_\alpha\,
\langle\langle v_{\pi_\alpha(1)},\dots,v_{\pi_\alpha(n_\alpha)}\rangle
\rangle\,$ with the sum over the partitions of $\,\{1,\dots,n\}$.}  
decay fast enough when time differences increase.  The elementary 
consequences of those assumptions are the existence of the long-time 
asymptotic limit and the fact that it is governed by the Central Limit 
Theorem. When $\mu\to\infty$, we recover a velocity field which is 
Gaussian and white in time, characterized by the 2-point function
\begin{equation}
\label{2PF}
\langle v^i({\bm r},t)\hspace{0.05cm} v^j({\bm r}',t')\rangle 
\,=\,2\hspace{0.05cm}\delta(t-t')\,D^{ij}({\bm r}-{\bm r}')\,.
\end{equation}
The advection by such velocity fields was first considered by
Kraichnan (1968) and it is common to call the Gaussian ensemble of
velocities with 2-point function (\ref{2PF}) the Kraichnan ensemble.
For the Kraichnan ensemble, the Lagrangian velocity ${\bm V}(t)$ has
the same white noise temporal statistics as the Eulerian one ${\bm
v}({\bm r},t)$ for fixed ${\bm r}$ and the displacement along a
Lagrangian trajectory $\Delta{\bm R}(t)$ is a Brownian motion for all
times. The eddy diffusivity tensor is $D_e^{ij}=D^{ij}({\bm 0})$,
which is a special case of relation (\ref{eddif}).  In the presence of
molecular diffusion, the overall diffusivity is the sum of the eddy
contribution and the molecular value $\kappa\hspace{0.05cm}\delta^{ij}$. 

In realistic turbulent flows, the Lagrangian correlation time $\tau$
is comparable to the characteristic time scale of large eddies.
Progress in numerical simulations (Yeung, 1997) and experimental 
technique (Voth {\it et al.}, 1998; La Porta {\it et al.}, 2001;
Mordant {\it et al.} 2001) has provided information on the single 
particle statistics in the regime intermediate between ballistic 
and diffusive. Such behavior is captured by the the subtracted 
Lagrangian autocorrelation function $\,\langle{\bm V}(0)({\bm V}(0)
-{\bm V}(t))\rangle\,$ or its second time derivative that is 
the autocorrelation function of the Lagrangian acceleration. 
This information has provided stringent tests on simple stochastic
models (that eliminate velocity fields), often used in the past to describe 
the one-particle and two-particle statistics in turbulent flows (Pope, 1994). 
The Kraichnan ensemble that models stochastic velocity fields, 
certainly missrepresents the single particle statistics by suppressing 
the regime of times smaller than $\tau$. It constitutes, however, as we 
shall see in the sequel, an important theoretical laboratory for studying 
the multiparticle statistics in fluid turbulence.

\subsection{Two-particle dispersion in a spatially smooth velocity}
\label{sec:smooth}

The separation $\,{\bm R}_{12}={\bm R}_1-{\bm R}_2\,$ between two
fluid particles with trajectories $\,{\bm R}_n(t)={\bm R}(t;{\bm
r}_{n})\,$ passing at $t=0$ through the points $\,{\bm r}_{n}$
satisfies (in the absence of Brownian motion) the equation
\begin{equation}\dot{\bm R}_{12}\ =\ \bm v(\bm R_1,t)-\bm v(\bm R_2,t)\,.
\label{disp1}
\end{equation}
We consider first an incompressible flow where the particles generally
separate.  In this Section, we start from the smallest distances where
the velocity field can be considered spatially smooth due to viscous
effects. In next Section, we treat the dispersion problem for larger
distances (in the inertial interval of turbulence) where the velocity
field has a nontrivial scaling. Finally, we describe a compressible
flow and show how the separation among the particles is replaced by
their clustering as the degree of compressibility grows.

\subsubsection{General considerations}
\label{sec:general}

In smooth velocities, for separations ${\bm R}_{12}$ much smaller
than the viscous scale of turbulence, i.e.\,\,in the so-called
Batchelor regime (Batchelor, 1959), we may approximate 
$\,\bm v(\bm R_1,t)-\bm v(\bm
R_2,t)\approx\sigma(t) \, {\bm R}_{12}(t)\,$ with the Lagrangian
strain matrix $\,\sigma_{ij}(t)=\nabla_jv^i({\bm R}_2(t),t)$. \,In
this regime, the separation obeys the ordinary differential
equation \qq \dot{\bm R}_{12}(t)=\sigma(t)\,{\bm R}_{12}(t)\,,
\label{disp2}
\qqq   
leading to the linear propagation
\qq 
{\bm R}_{12}(t)\ =\ W(t)\,{\bm R}_{12}(0),
\label{batch0}
\qqq where the evolution matrix is defined as $W^{ij}(t)=\partial R^i({
\bm r};t)/\partial r^j$ with ${\bm r}={\bm r}_2$.  We shall also use 
the notation $W(t;{\bm r})$ when we wish to keep track of the initial 
point or 
$\,W(t;{\bm r},s)\,$ if the initial time $s$ is different from zero.

The equation (\ref{disp2}), with the strain treated as given, may be 
explicitly solved for arbitrary $\sigma(t)$ only in the $1d$ case
by expressing $W(t)$ as the exponential of the time-integrated
strain:
\begin{equation}
\ln[R(t)/R(0)]=\ln W(t)=\int_0^t\sigma(s)\,ds\equiv X\,.
\label{disp3}
\end{equation}
We have omitted subscripts replacing ${\bm R}_{12}$ by ${\bm
R}$.  When $t$ is much larger than the correlation time $\tau$ of the
strain, the variable $X$ behaves as a sum of many independent
equally distributed random numbers $X=\sum_1^Ny_i$ with
$N\propto t/\tau$. Its mean value $\langle X\rangle=N\langle y\rangle$
grows linearly in time. Its fluctuations $X-\langle X\rangle$
on the scale ${\cal O}(t^{1/2})$ are governed by the Central Limit
Theorem that states that $(X-\langle X\rangle)/N^{1/2}$ becomes
for large $N$ a Gaussian random variable with variance 
$\langle y^2\rangle-\langle y\rangle^2\equiv\Delta$. Finally, its 
fluctuations on the larger scale ${\cal O}(t)$ are governed by the Large
Deviation Theorem that states that the PDF of $X$ has asymptotically
the form \begin{equation}{\cal P}(X)\ \propto\ {\rm e}^{-NH(X/N-\langle 
y\rangle)}\,.\label{disp5}\end{equation}
This is an easy consequence of the exponential dependence on $N$
of the generating function $\langle {\rm e}^{\,zX}\rangle$ of the moments
of $X$. Indeed, $\langle {\rm e}^{\,zX}\rangle={\rm e}^{NS(z)}$, where 
we have denoted $\langle{\rm e}^{\,zy}\rangle\equiv {\rm e}^{S(z)}$ 
(assuming that the expectation exists for all complex $z$). The PDF 
${\cal P}(X)$ is then given by the inverse Laplace transform 
$\,{1\over2\pi i}\int{\rm e}^{-\,z\,X+NS(z)}\,dz\,$ with the
integral over any axis parallel to the imaginary one. For
$X\propto N$, the integral is dominated by the saddle point
$z_0$ such that $S'(z_0)=X/N$ and the large deviation relation (\ref{disp5})
follows with $H=-S(z_0)+z_0S'(z_0)$. The function $H$ of the variable
$\,X/N-\langle y\rangle$ is called entropy function as it appears also
in the thermodynamic limit in statistical physics (Ellis, 1985). 
A few important properties of $\,H\,$ (also called rate or Cram\'er function) 
may be established independently of the distribution ${\cal P}(y)$.
It is a convex function which takes its minimum at zero, i.e. for
$X$ taking its mean value $\langle X\rangle =NS'(0)$. The
minimal value of $H$ vanishes since $S(0)=0$.  The entropy is
quadratic around its minimum with $H''(0)=\Delta^{-1}$, where
$\Delta=S''(0)$ is the variance of $y$. The possible non-Gaussianity
of the $y$'s leads to a non-quadratic behavior of $H$ for (large) deviations
of $X/N$ from the mean of the order of $\Delta /S'''(0)$. 

Coming back to the logarithm $\ln W(t)$ of the interparticle distance 
ratio in (\ref{disp3}), its growth (or decay) rate $\,\lambda =\langle X
\rangle/t$ is called the Lyapunov exponent. The moments $\langle[R(t)]^n
\rangle$ behave exponentially 
as $\exp[\gamma(n)t]$ with $\gamma(n)$ a convex function of $n$
vanishing at the origin. Even if $\lambda=\gamma'(0)<0$, high-order 
moments of $R$ may grow exponentially in time, see, for instance, 
the behavior of the interparticle distance discussed in 
Section\,\,\ref{sec:compres}. In this case, there must be one more
zero $n_1$ of $\gamma(n)$ and a statistical integral of motion, 
$\langle R^{n_1}\rangle$, that does not depend on time at large times.

In the multidimensional case, the solution (\ref{batch0}) for $\,{\bm
R}(t)\,$ is determined by products of random matrices rather than just
random numbers. The evolution matrix $\,W(t)\,$ may be written as
\begin{equation}
{{ W}}(t)\ =\ {\cal T}\,\exp \bigg[ \int\limits_{0}^{t}
\sigma(s)\,ds\bigg]\ =\ \sum\limits_{n=0}^\infty
\int\limits_0^t\sigma(s_n)\,ds_n\ldots\int\limits_0^{s_3}
\sigma(s_2)\,ds_2\int\limits_0^{s_2}\sigma(s_1)\,ds_1\,.
\label{disp6}\end{equation}
This time-ordered exponential form is, of course, not very useful for
direct calculations except for the particular case of a
short-correlated strain, see below.  The main statistical properties
of the separation vector ${\bm R}$ needed for most physical
applications might still be established for quite arbitrary strains
with finite temporal correlations.  The basic idea goes back to
Lyapunov (1907) and Furstenberg and Kesten (1960) and it found further
development in the Multiplicative Ergodic Theorem of Oseledec (1968).
The modulus $R$ of the separation vector may be expressed via the
positive symmetric matrix $W^T W$. The main result states that in
almost every realization of the strain, the matrix $\,{1\over t}\ln
W^T W$ stabilizes as $t\to\infty$.
In particular, its
eigenvectors tend to $d$ fixed orthonormal eigenvectors ${\bm
f}_i$. To understand that intuitively, consider some
fluid volume, say a sphere, which evolves into an elongated ellipsoid
at later times. As time increases, the ellipsoid is more and more
elongated and it is less and less likely that the hierarchy of the
ellipsoid axes will change. The limiting eigenvalues
\begin{equation}
\lambda_i=\lim_{t\to\infty}t^{-1}\ln| W{\bm f}_i|\,
\label{disp7}
\end{equation}
define the so-called Lyapunov exponents. The major property of the
Lyapunov exponents is that they are
realization-independent if the strain is ergodic. 
The usual convention is to arrange the exponents in
non-increasing order.

The relation (\ref{disp7}) tells that two fluid particles separated
initially by ${\bm R}(0)$ pointing into the direction ${\bm f}_i$ will
separate (or converge) asymptotically as $\exp(\lambda_it)$. The
incompressibility constraints $\det(W)=1$ and $\sum\lambda_i=0$ imply
that a positive Lyapunov exponent will exist whenever at least one of
the exponents is nonzero. Consider indeed
\begin{equation}
E(n)=\lim_{t\to\infty}t^{-1}\ln\langle
[R(t)/R(0)]^n\rangle\ ,
\label{energy}
\end{equation}
whose slope at the origin gives the largest Lyapunov exponent
$\lambda_1$.  The function $E(n)$ obviously vanishes at the
origin. Furthermore, $E(-d)=0$, i.e. incompressibility and isotropy
make that $\langle R^{-d}\rangle$ is time-independent as $t\to \infty$
(Furstenberg, 1963; Zeldovich {\it et al.}, 1984). Negative moments of
orders $n<-1$ are indeed dominated by the contribution of directions
${\bm R}(0)$ almost aligned to the eigendirections ${\bm f}_2, \ldots
{\bm f}_d$. At $n<1-d$ the main contribution comes from a small subset
of directions in a solid angle $\propto \exp(d\lambda_dt)$ around
${\bm f}_d$. It follows immediately that $\langle
R^n\rangle\propto\exp[\lambda_d(d+n)t]$ and that $\langle
R^{-d}\rangle$ is a statistical integral of motion. Since $E(n)$ is a
convex function, it cannot have other zeroes except $-d$ and $0$ if it
does not vanish identically between those values. It follows that 
the slope at the origin,
and thus $\lambda_1$, is positive.  The simplest way to appreciate
intuitively the existence of a positive Lyapunov exponent is to
consider, following Zel'dovich {\it et al.}  (1984), the saddle-point
$2d$ flow $v_x=\lambda x,v_y=-\lambda y$. A vector initially forming
an angle $\phi$ with the $x$-axis will be stretched after time $T$ if
$\cos\phi\geq[1+\exp(2\lambda T)]^{-1/2}$, i.e.  the fraction of
stretched directions is larger than $1/2$.

A major consequence of the existence of a positive Lyapunov exponent
for any random incompressible flow is an exponential growth of the
interparticle distance $R(t)$. In a smooth flow, it is also possible
to analyze the statistics of the set of vectors ${\bm R}(t)$ and to
establish a multidimensional analog of (\ref{disp5}) for the general
case of a nondegenerate Lyapunov exponent spectrum.  The final results
will be the Large Deviation expressions (\ref{PDF3}) and (\ref{posit})
below. The idea is to reduce the $d$-dimensional problem to a set of
$d$ scalar problems excluding the angular degrees of freedom. We
describe this procedure following Balkovsky and Fouxon
(1999). Consider the matrix $I(t)=W(t)W^T(t)$, representing the tensor
of inertia of a fluid element like the above mentioned ellipsoid.  The
matrix is obtained by averaging $R^i(t)R^j(t)/ \ell^2 d$ over the
initial vectors of length $\ell$.  In contrast to $W^T\,W$ that
stabilizes at large times, the matrix $I$ rotates in every
realization. To account for that rotation, we represent the matrix as
$O^T\Lambda O$ with the orthogonal $ O$ composed of the eigenvectors
of $I$ and the diagonal $\Lambda$ having the eigenvalues ${\rm
e}^{2\rho_1},\ldots{\rm e}^{2\rho_d}$ arranged in non-increasing
order. The evolution equation $\partial_t I=\sigma I+I\sigma^T$ takes
then the form
\begin{eqnarray}&&
\partial_t \rho_i=\tilde\sigma_{ii}\,,\quad
\tilde\sigma= O\sigma  O^T\,,\label{prt}\\&&
\partial_t O=\Omega  O\,,\quad
\Omega_{ij}=\frac{{\rm e}^{2\rho_i}\tilde\sigma_{ji}+{\rm e}^{2\rho_j}
\tilde\sigma_{ij}}{{\rm e}^{2\rho_i}-{\rm e}^{2\rho_j}}\,\,,
\label{efd}\end{eqnarray}
with no summation over repeated indices. We assume isotropy so that at
large times the $SO(d)$ rotation matrix $O$ is distributed uniformly 
over the sphere. Our task is to
describe the statistics of the stretching and the contraction,
governed by the eigenvalues $\rho_i$. We see from (\ref{prt},\ref{efd}) 
that the evolution of the
eigenvalues is generally entangled to that of the angular degrees of
freedom.  As time increases, however, the eigenvalues will become
widely separated ($\rho_1\gg\ldots \gg\rho_d$) for a majority of the
realizations and $\Omega_{ij}\to \tilde\sigma_{ji}$ for $i<j$ (the
upper triangular part of the matrix follows from antisymmetry).  
The dynamics of the angular degrees of freedom
becomes then independent of the eigenvalues and the set of equations
(\ref{prt}) reduces to a scalar form. The solution $\rho_i=\int_0^t
\tilde\sigma_{ii}(s)\,ds$ \,allows the application of the large deviation
theory, giving the asymptotic PDF: \qq {\cal
P}(\rho_1,\dots,\rho_d;t)\ &\propto&\ \exp\left[ -t\,
H(\rho_1/t-\lambda_1,\ldots,\rho_{d-1}/t- \lambda_{d-1})\right]\cr&&
\times\ \theta(\rho_1-\rho_2)\ldots\theta(\rho_{d-1}-\rho_d)
\,\,\delta(\rho_1+\ldots+\rho_d)\ .
\label{PDF3}
\qqq The Lyapunov exponents $\lambda_i$ are related to the strain
statistics as $\,\lambda_i=\left\langle\tilde\sigma_{ii}
\right\rangle$ where the average is temporal. 
\,The expression (\ref{PDF3}) is not valid near the
boundaries $\rho_i=\rho_{i+1}$ in a region of order unity, negligible
with respect to $\lambda_i t$ at times $t\gg(\lambda_i
-\lambda_{i+1})^{-1}$.

The entropy function $H$ depends on the details of the strain
statistics and has the same general properties as above: it is
non-negative, convex and it vanishes at zero. Near the minimum,
$\,H({\bm x})\approx{1\over 2}(C^{-1})_{ij} x_ix_j\,$ with the
coefficients of the quadratic form given by the integrals of the
connected correlation functions of $\tilde{\sigma}$ defined in
(\ref{prt}):
\begin{eqnarray}
C_{ij}=\int\,\left\langle\left\langle
\tilde\sigma_{ii}(t),\tilde\sigma_{jj}(t')
\right\rangle\right\rangle\,\,dt'\,, \quad i,j=1,\dots,d-1.
\label{Cij}\end{eqnarray}
In the $\delta$-correlated case, the entropy is everywhere quadratic.
For a generic initial vector ${\bm r}$, the long-time asymptotics of
$\,\ln(R/r)\,$ coincides with ${\cal P}(\rho_1)=\int{\cal
P}(\rho_1,\ldots,\rho_d)\,d\rho_2\ldots d\rho_d$ which also takes the
large-deviation form at large times, as follows from (\ref{PDF3}). The
quadratic expansion of the entropy near its minimum corresponds to the
lognormal distribution for the distance between two particles
\begin{equation} 
{\cal P}(r;R;t)\ \propto\ 
\exp\left\{-\left[\ln(R/r)-\bar\lambda t\right]^2/(2\Delta t)\right\}\,,
\label{kk3} 
\end{equation}
with $r=R(0)$, $\bar\lambda=\lambda_1$ and $\Delta=C_{11}$.

It is interesting to note that under the same assumption of
nondegenerate Lyapunov spectrum one can analyze the eigenvectors ${\bm
e}_i$ of the evolution matrix $ W$ (Goldhirsch {\it et al.}, 1987).
Note the distinction between the eigenvectors ${\bm e}_i$ of $W$ and
${\bm f}_i$ of $\,W^T W$.  Let us order the eigenvectors ${\bm e}_i$
according to their eigenvalues.  Those are real due to the assumed
nondegeneracy and they behave asymptotically as
$\exp(\lambda_1t),\ldots, \exp(\lambda_dt)$.  The ${\bm e}_d$
eigenvector converges exponentially to a fixed vector and any subspace
spanned by $\{{\bm e}_{d-k},\ldots, {\bm e}_d\}$ for $0\leq k\leq d$
tends asymptotically to a fixed subspace for every realization. Remark
that the subspace is fixed in time but changes with the realization.

Molecular diffusion is incorporated into the above picture by
replacing the differential equation (\ref{disp2}) by its noisy
version
\qq d{\bm
R}(t)=\sigma(t){\bm R}(t)\,dt+2\sqrt{\kappa}\,d{\rm\beta}(t)\,.
\label{dispp}
\qqq 
The separation vector is subject to the
independent noises of two particles, hence the factor $2$ 
with respect to (\ref{sde}). 
The solution to the inhomogeneous linear stochastic equation (\ref{dispp}) is
easy to express via the matrix $W(t)$ in (\ref{disp6}).  The tensor of
inertia of a fluid element $\,I^{ij}(t)=\,{1\over \ell^2
d}\,R^i(t)R^j(t)\,$ is now averaged both over the initial vectors of
length $\ell$ and the noise, thus obtaining (Balkovsky and Fouxon,
1999): \qq I(t)\ =\ W(t)W(t)^T\,+\,{_{4\,\kappa}\over^{\ell^2
d}}\int\limits_0^t W(t)\,[W(s)^TW(s)]^{-1}W(t)^T\,ds\,.
\label{I}
\qqq The matrix $\,I(t)\,$ evolves according to $\,\partial_tI =\sigma
I+I\sigma+{4\,\kappa\over \ell^2 d}$ and the elimination of the
angular degrees of freedom proceeds as previously.  An additional
diffusive term $2\kappa\,\exp(-2\rho_i)$ appears in (\ref{prt})
and the asymptotic solution becomes
\begin{eqnarray}&&
\rho_i(t)\,=\,\int_0^t\hspace{-0.06cm}\tilde\sigma_{ii}(s)\,ds
+\frac{1}{2}\ln\left\{
1+{_{4\,\kappa}\over^{\ell^2 d}}\int_0^t\,\exp\Big[-2\int_0^{s}
\hspace{-0.06cm}\tilde\sigma_{ii}(s')
\,ds'\Big]\,ds\right\}.
\label{sol2}
\end{eqnarray}
The last term in (\ref{sol2}) is essential for the directions
corresponding to negative $\lambda_i$. The molecular noise will indeed
start to affect the motion of the marked fluid volume when the
respective dimension gets sufficiently small. If $\ell$ is the initial
size, the required condition ${\rho_i}\lesssim
-\rho^*_i=-\ln(\ell^2|\lambda_i|/\kappa)$ is typically met for times
$\,t\simeq \rho_i^* / |\lambda_i|$. For longer times, the respective
$\rho_i$ is prevented by diffusion to decrease much below $-\rho_i^*$,
while the negative $\lambda_i$ prevents it from increasing. As a
result, the corresponding $\rho_i$ becomes a stationary random process
with a mean of the order $-\rho_i^*$. The relaxation times to the
stationary distribution are determined by $\tilde\sigma$, which is
diffusion independent, and they are thus much smaller than $t$. On the
other hand, the components $\rho_j$ corresponding to non-negative
Lyapunov exponents are the integrals over the whole evolution time
$t$. Their values at time $t$ are not sensitive to the latest period
of evolution lasting over the relaxation time for the contracting
$\rho_i$. Fixing the values of $\,\rho_j$ at times $\,t\gg \rho_i^* /
|\lambda_i|\,$ will not affect the distribution of the contracting
$\,\rho_i$ and the whole PDF is factorized (Shraiman and Siggia,\,
1994; Chertkov {\it et al.}, 1997; Balkovsky and Fouxon, 1999). For
Lagrangian dynamics in $3d$ developed Navier-Stokes turbulence there
are, for instance, two positive and one negative Lyapunov exponents
(Girimaji and Pope, 1990). For times $\,t\gg \rho_3^* / |\lambda_3|$
we have then
\begin{eqnarray}&&
{\cal P}\ \propto\ \exp\left[-t\, 
H\left(\rho_1/t-\lambda_1,\rho_2/t-\lambda_2\right)
\right]\,\,{\cal P}_{\rm st}(\rho_3)\,,
\label{posit}\end{eqnarray}
with the same function $H$ as in (\ref{PDF3}) since $\rho_3$ is
independent of $\rho_1$ and $\rho_2$.  Note that the account of the
molecular noise violates the condition $\sum\rho_i=0$ as fluid
elements at scales smaller than $\sqrt{\kappa/|\lambda_3|}$ cannot be
distinguished.  To avoid misunderstanding, note that (\ref{posit})
does not mean that the fluid is getting compressible: the simple
statement is that if one tries to follow any marked volume, the
molecular diffusion makes this volume statistically growing.

Note that we have implicitly assumed $\ell$ to be smaller than the
viscous length $\eta=\sqrt{\nu/|\lambda_3|}$ but larger than the
diffusion scale $\sqrt{\kappa/|\lambda_3|}$.  Even though $\nu$ and
$\kappa$ are both due to molecular motion, their ratio widely varies
depending on the type of material. The theory of this section is
applicable for materials having large Schmidt (or Prandtl) numbers
$\nu/\kappa$.

The universal forms (\ref{PDF3}) and (\ref{posit}) for the
two-particle dispersion are basically everything we need for physical
applications. We will show in the next Chapter that the highest
Lyapunov exponent determines the small-scale statistics of a passively
advected scalar in a smooth incompressible flow. For other problems,
the whole spectrum of exponents and even the form of the entropy
function are relevant.

\subsubsection{Solvable cases}
\label{sec:solv}

The Lyapunov spectrum and the entropy function can be derived exactly
from the given statistics of $\,\sigma\,$ for few limiting cases
only. The case of a short-correlated strain allows for a complete
solution. For a finite-correlated strain, one can express analytically
$\,\bar\lambda\,$ and $\,\Delta\,$ for a $2d$ long-correlated strain
and at large space dimensionality.

\noindent i) {\bf Short-correlated strain}. \ Consider the case where
the strain $\,\sigma(t)\,$ is a stationary white-in-time Gaussian
process with zero mean and the 2-point function
\begin{equation}
\langle\sigma_{{i}{j}}(t)\,\sigma_{{k}{\ell}}(t')\rangle\ 
=\ 2\hspace{0.05cm}\delta(t-t')\,C_{{i}{j}{k}{\ell}}\,. \label{shortstrain1}
\end{equation}
This case may be considered as the long-time scaling limit
$\,\lim\limits_{\mu\to\infty}\,\mu^{_1\over^2}\sigma(\mu t)\,$ of a
general strain along a Lagrangian trajectory, provided its
temporal correlations decay fast enough.  It may be also
viewed as describing the strain in the Kraichnan ensemble of
velocities decorrelated in time and smooth in space.  In the latter
case, the matrix $\,C_{{i}{j}{k}{\ell}}=-\nabla_{j}\nabla_{\ell}
D^{{i}{k}}({\bm 0})\hspace{0.05cm},$ \,where $D^{{i}{j}} ({\bm r})$ is the
spatial part in the 2-point velocity correlation (\ref{2PF}).  We
assume $D^{{i}{j}}({\bm r})$ to be smooth in ${\bm r}$ (or at least
twice differentiable), a property assured by a fast decay of its
Fourier transform $\hat D^{{i}{j}} ({\bm k})$. \,Incompressibility,
isotropy and parity invariance impose the form $\,D^{ij}({\bm
r})=D_0\delta^{ij}-{1\over2}\,d^{ij} ({\bm r})$ with
\begin{equation}
d^{ij}({\bm r})\ =\ D_1[(d+1)\delta^{ij}\,r^2-2\,r^ir^j]\ +\ o(r^2).
\label{shortstrain2}
\end{equation}
The corresponding expression for the 2-point function of $\,\sigma$
reads
\begin{equation}
C_{{i}{j}{k}{\ell}}\,=\,D_1\,[(d+1)\delta_{{i}{k}}
\delta_{{j}{\ell}}-\delta_{{i}{j}}\delta_{{k}{\ell}}
-\delta_{{i}{\ell}}\delta_{{j}{k}}]\,,\label{shortstrain3}
\end{equation}
with the constant $D_1$ having the dimension of the inverse of time.

The solution of the stochastic differential equation (\ref{disp2}) is
given by (\ref{disp6}) with the matrix $W(t)$ involving stochastic
integrals over time. For a white-correlated strain, such integrals are
not defined unambiguously but require a regularization that reflects
finer details of the strain correlations wiped out in the scaling
limit. An elementary discussion of this issue may be found in the
Appendix.  For an incompressible strain, however, the ambiguity in the
integrals defining $W(t)$ disappears so that we do not need to care
about such subtleties.  The random evolution matrices $W(t)$ form a
diffusion process on the group $SL(d)$ of real matrices with unit
determinant. Its generator is a second-order differential operator
identified by Shraiman and Siggia (1995) as
$\,M=D_1[d\hspace{0.05cm}H^2-(d+1)J^2]$, where $H^2$ and $J^2$ are the
quadratic Casimir of $SL(d)$ and its $SO(d)$ subgroup. In other words,
the PDF of $W(t)$ satisfies the evolution equation
$\,(\partial_t-M){\cal P}(W;t)=0$.  The matrix $W(t)$ may be viewed as
a continuous product of independent random matrices. Such products in
continuous or discrete versions have been extensively studied
(Furstenberg and Kesten, 1960; Furstenberg, 1963; Le Page, 1982) and
occur in many physical problems, e.g.\,\,in $1d$ localization
(Lifshitz {\it et al.}, 1988; Crisanti {\it et al.}, 1993).

If we are interested in the statistics of stretching-contraction
variables only, then $W(t)$ may be projected onto the diagonal matrix
$\Lambda$ with positive non-increasing entries ${\rm e}^{2\rho_1},
\dots,{\rm e}^{2\rho_d}$ by the decomposition $\,W=O\,\Lambda^{_1\over^2}
O'\hspace{0.05cm},$ where the matrices $O$ and
$O'$ 
belong to the group $SO(d)$.  As observed in (Bernard {\it et
al.}, 1998; Balkovsky and Fouxon, 1999), the generator of the
resulting diffusion of $\rho_i$ is the $d$-dimensional integrable
Calogero-Sutherland Hamiltonian. The $\rho_i$ obey the stochastic
Langevin equation
\begin{equation}
\partial_t\rho_{i}\,=\,D_1 d\sum\limits_{{j}
\not={i}}\coth{(\rho_{i}-\rho_{j})}\,+\,\eta_{i}\,,
\label{LCS}
\end{equation}
where $\bm{\eta}$ is a white noise with 2-point function
\,$\langle\eta_{i}(t)\,\eta_{j}(t')\rangle=2D_1(d\delta_{{i}{j}}-1)
\,\delta(t-t')$. At long times the separation between
the $\rho_{i}$'s becomes large and we may approximate
$\coth(\rho_{ij})$ by $\pm1$. It is then easy to solve (\ref{LCS}) and
find the explicit expression of the PDF (\ref{PDF3}):
\begin{equation}
\label{qenf}
H({\bm x})\,=\,{_1\over^{4D_1 d}}\sum_{i=1}^dx_{i}^2\,,\qquad
\lambda_i\,=\,D_1d(d-2i+1)\,.
\end{equation}
Note the quadratic form of the entropy, implying that the distribution
of $\,R(t)\,$ takes the lognormal form (\ref{kk3}) with
$\bar\lambda=\lambda_1$ and $\Delta=2D_1(d-1)$. The calculation of the
long-time distribution of the leading stretching rate $\rho_1$ goes
back to Kraichnan (1974). The whole set of $d$ Lyapunov exponents was
first computed in (Le Jan, 1985), see also (Baxendale, 1986). Gamba and
Kolokolov (1996) obtained the long-time asymptotics of $\rho_i$
by a path integral calculation. The spectral decomposition of the
Calogero-Sutherland Hamiltonian, see (Olshanetsky {\it et
al.}, 1983), permits to write explicitly the PDF of $\rho_i$ for
all times.

\noindent ii) {\bf $2d$ slow strain}. \ In $2d$, one can reduce the vector
equation (\ref{disp2}) to a second-order scalar form.  Let us indeed
consider the case of a slow strain satisfying $\dot\sigma\ll\sigma^2$
and differentiate the equation $\dot{\bm R}=\sigma{\bm R}$ with
respect to time. The term with $\dot\sigma$ is negligible with respect
to $\sigma^2$ and a little miracle happens here: because of
incompressibility, the matrix $\sigma$ is traceless and $\sigma^2$ is
proportional to the unit matrix in $2d$. We thus come to a scalar
equation for the wave function $\Psi=R_x+iR_y$:
\begin{equation}
\partial_t^2\Psi=(\sigma_{11}^2+\sigma_{12}\sigma_{21})\Psi\,.
\label{loc1}
\end{equation}
This is the stationary Schr\"odinger equation 
for a particle in the random
potential $U=S^2-\Omega^2$, where $S^2=\sigma_{11}^2+(\sigma_{12}
+\sigma_{21})^2/4$ and $\Omega^2=(\sigma_{12}-\sigma_{21})^2/4$ is the
vorticity. Time plays here the role of the coordinate. Our problem is
thus equivalent to localization in the quasi-classical limit (Lifshitz
{\it et al.}, 1988) and finding the behavior of (\ref{loc1}) with
given initial conditions is similar to the computation of a $1d$ sample
resistivity, see e.g.\,\,(Abrikosov and Ryzhkin, 1978; Kolokolov,
1993). Based on these results we can assert that the modulus
$|\Psi|=R$ in random potentials grows exponentially in time, with the
same exponent that controls the decay of the localized wave function.

The problem can be solved using semi-classical methods. The flow is
partitioned in elliptic ($\Omega>S$) and hyperbolic ($S>\Omega$)
regions (Weiss, 1991), corresponding to classical allowed ($U<0$) and
forbidden ($U>0$) regions.  The wave function $\Psi$ is given by two
oscillating exponentials or one decreasing and one increasing,
respectively.  Furthermore, the typical length of the regions is the
correlation time $\tau$, much larger than the inverse of the
mean strain and vorticity $S^{-1}_{rms}$ and $\Omega^{-1}_{rms}$. It
follows that the increasing exponentials in the forbidden regions are
large and dominate the growth of $R(t)$. With exponential accuracy we
have:
\begin{equation}
\lambda(t)=\ln\left({R(t)\over R(0)}\right)={1\over t}\,{\rm
Re}\int_0^t\sqrt{U(s)}\,ds\,,
\label{slowt}\end{equation}
where the real part restricts the integration to the hyperbolic
regions. The parameters $\bar\lambda$ and $\Delta$ in the lognormal
expression (\ref{kk3}) are immediately read from (\ref{slowt}):
\begin{equation}
\bar\lambda= \left\langle{\rm
Re}\sqrt{U}
\right\rangle\,,\qquad\qquad
\Delta={\displaystyle{\int}}\Big\langle\Big\langle{\rm
Re}\sqrt{U(0)},\, {\rm Re}\sqrt{U(t')}\Big\rangle\Big\rangle\,dt'\,.
\label{slow} 
\end{equation}
Note that the vorticity gave no contribution in the
$\delta$-correlated case.  For a finite correlation time, it
suppresses the stretching by rotating fluid elements with respect
to the axes of expansion. The real part in (\ref{slow}) is indeed
filtering out the elliptic regions. Note that the Lyapunov exponent is
given by a single-time average, while in the $\delta$-correlated case
it was expressed by the time-integral of a correlation function. It
follows that $\bar\lambda$ does not depend on the correlation time
$\tau$ and it can be estimated as $S_{rms}$ for $\Omega_{rms}\lesssim
S_{rms}$. The corresponding estimate of the variance is $\,\Delta\sim
\langle S^2\rangle\tau\,$. As the vorticity increases, the rotation
takes over, the stretching is suppressed and $\bar{\lambda}$
reduces. One may show that the correlation time $\tau_s$ of the
stretching rate is the minimum between $1/\Omega_{rms}$ and $\tau$
(Chertkov {\it et al.}, 1995a). For $\Omega_{rms}\tau\gg1$ we are back
to a $\delta$-correlated case and $\bar\lambda\sim \langle S^2\rangle/
\Omega_{rms}$. All those estimates can be made systematic for a Gaussian
strain (Chertkov {\it et al.}, 1995a).

\noindent iii) {\bf Large space dimensionality}. \ The key remark for
this case is that scalar products like $R^i(t_1)R^i(t_2)$ are sums of
a large number of random terms. The fluctuations of such sums are
vanishing in the large-$d$ limit and they obey closed equations that
can be effectively studied for arbitrary strain statistics. This
approach, developed in (Falkovich {\it et al.}, 1998), is inspired
by the large-$N$ methods in quantum field theory ('t Hooft, 1974) and
statistical mechanics (Stanley, 1968). Here, we shall relate the
behavior of the interparticle distance to the strain statistics and
find explicitly $\bar\lambda$ and $\Delta$ in (\ref{kk3}). The strain
is taken Gaussian with zero mean and correlation function
$\,\langle\sigma_{ij}(t)\sigma_{k\ell}(0)\rangle =(2D/\tau
d)\,\delta_{ik}\delta_{j\ell}\,g(\zeta)$ where higher-order terms in
$1/d$ are neglected.  The integral of $g$ is normalized to unity and
$\zeta\equiv t/\tau$.  At large $d$, the correlation function $\,F=
\langle R^i(t_1)R^i(t_2)\rangle\,$ satisfies the equation
\begin{equation}
\frac{\partial^2}{\partial \zeta_1\partial \zeta_2}F(\zeta_1,\zeta_2)
=\tau^2\langle\sigma_{ij}(t_1)\sigma_{ik}(t_2)
R^j(t_1)R^k(t_2)\rangle=\beta g(\zeta_1-\zeta_2)\,F(\zeta_1,
\zeta_2) \,,
\label{k11} \end{equation}
with the initial condition $\partial_{\zeta} F(\zeta,0)=0$. The limit of large
$d$ is crucial for the factorization of the average leading to the
second equality in (\ref{k11}). The dimensionless parameter
$\beta=2D\tau$ measures whether the strain is long or
short-correlated.  Since (\ref{k11}) is linear and the coefficient on
the right-hand side depends only on the time difference, the solution
may be written as a sum of harmonics
$F_\lambda(\zeta_1,\zeta_2)=\exp[\lambda\tau{(\zeta_1+\zeta_2)}]\,\Psi(\zeta_1
-\zeta_2)$. Inserting it into (\ref{k11}), we get the Schr\"odinger
equation for the even function $\Psi(t)$:
\begin{equation}
\partial_\zeta^2\Psi(\zeta)+\bigl[-(\lambda\tau)^2+\beta\,
g(\zeta)\bigr]\,\Psi(\zeta)=0 \,.
\label{k13} \end{equation}
At large times, the dominant contribution comes from the largest exponent
$\bar\lambda$ corresponding to the ground state in the potential
$-\beta g(\zeta)$.  Note that the ``energy'' is proportional to
$-\lambda^2$ and $\bar\lambda$ is the Lyapunov exponent since $\langle
R^2\rangle=F(\zeta,\zeta)$. From quantum mechanics textbooks it is
known that the ground state energy in deep and shallow potentials is
proportional to their depth and its square, respectively.  We conclude
that $\bar\lambda\propto D$ for a fast strain (small $\beta$) and
$\bar\lambda\propto\sqrt{D/\tau}$ in the slow case (large $\beta$). At
large time differences, the potential term in (\ref{k13}) is
negligible and $\bar\lambda$ determines both the growth of $\langle
R^2\rangle$ and the decay of different-time correlation function at
$\zeta_1+\zeta_2$ fixed. That also shows that the correlation function 
becomes independent of the larger 
of the times $t_1$ and $t_2$ when their
difference exceeds $\tau$.

For the fast strain case, one can put $g(\zeta) =\delta(\zeta)$ and
the solution of (\ref{k13}) is amazingly simple: $\,F(\zeta_1,\zeta_2)
= R^2(0)\,\exp[{\beta\min(\zeta_1,\zeta_2)}]$. The Lyapunov
exponent $\bar\lambda=D$, in agreement with the result
$\,\lambda_1=D_1d^2+{\cal O}(d)\,$ obtained for the
Kraichnan ensemble. For the slow case, the stretching rate is
independent of $\tau$ at a given value of $D/\tau$ (determining the
simultaneous correlation function of the strain). The analysis of the
Schr\"odinger equation (\ref{k13}) with a deep potential also gives
the correlation time $\tau_s$ of the stretching rate, which does not
generally coincide with the strain correlation time $\tau$ (Falkovich
{\em et al.}, 1998).

\subsection{Two-particle dispersion in a nonsmooth incompressible flow}
\label{sec:nonsmooth}

In this Section we study the separation between two trajectories in
the inertial range of scales $\eta\ll r\ll L$.  The scales $\eta$ and
$L$ stand in a $3d$ turbulent flow for the viscous and the injection 
scales 
(the latter is also called integral scale in that case). 
For a $2d$ inverse energy cascade flow, they would stand for the
scales of injection and friction damping respectively. We shall
see that the behavior of the trajectories is quite different from that
in smooth flows analyzed previously.

\subsubsection{Richardson law}
\label{sec:Rich}

As discussed in the Introduction, velocity differences in the inertial
interval exhibit an approximate scaling expressed by the power law
behavior of the structure functions $\langle (\Delta_r
v)^n\rangle\propto r^{\sigma_n}$. Low-order exponents are close to the
Kolmogorov prediction $\,\sigma_n=\alpha n\,$ with $\alpha=1/3$. A
linear dependence of $\sigma_n$ on $n$ would signal the scaling
$\Delta_r v \propto r^\alpha$ with a sharp value of $\,\alpha$. A
nonlinear dependence of $\sigma_n$ indicates the presence of a whole
spectrum of exponents, depending on the space-time position in the
flow (the so-called phenomenon of multiscaling).  The $2d$ inverse and
the $3d$ direct energy cascades in Navier-Stokes equation provide
concrete examples of the two possible situations.  Rewriting
(\ref{disp1}) for the fluid particle separation as $\,\dot{\bm
R}=\Delta{\bm v}({\bm R},t)$, \,we infer that $d R^2/dt = 2\,{\bm
R}\cdot\Delta{\bm v}\,\propto\,R^{1+\alpha}$.  If the value of
$\alpha$ is fixed and smaller than unity, this is solved (ignoring the
space-time dependence in the proportionality constant) by \qq
R^{1-\alpha}(t)-\,R^{1-\alpha}(0)\,\ \propto\,\ t\,,
\label{Rich0}
\qqq implying that the dependence on the initial separation is quickly
wiped out and that $R$ grows as $t^{1/(1-\alpha)}$.  For the random
process $\,{\bm R}(t)$, \,the relation (\ref{Rich0}) is, of course, of
the mean field type and should pertain to the long-time behavior of
the averages \qq \langle R^\zeta(t)\rangle\ \propto\
t^{\zeta/(1-\alpha)}\,.
\label{ene}
\qqq That implies their superdiffusive growth, faster than the diffusive one 
$\propto t^{\zeta/2}$.  The scaling law (\ref{ene})
might be amplified to the rescaling property 
\qq {\cal P}(R;t)\ =\
\lambda\,{\cal P}(\lambda R; \lambda^{1-\alpha}t)
\label{Rich1}
\qqq  of the interparticle distance PDF. 
Possible deviations from a linear behavior in the order 
$\zeta$ of the exponents in (\ref{ene}) should be interpreted as a signal of
multiscaling of the Lagrangian velocity $\Delta{\bm
v}({\bm R}(t),t)\equiv\Delta{\bm V}(t)$.

The power-law growth (\ref{ene}) for $\zeta=2$ and $\alpha={1/3}$,
\,i.e.\,\,$\langle R(t)^2\rangle\propto t^3$, is a direct consequence
of the celebrated Richardson dispersion law (Richardson, 1926), the
first quantitative phenomenological observation in developed
turbulence. It states that \qq {d\over dt}\langle R^2\rangle\ \propto\
\langle R^2\rangle^{2/3}\,.
\label{Rich}
\qqq The law (\ref{Rich}) seems to be confirmed also by later
experimental data, see Chapter 24 of Monin and Yaglom (1979) and
(Jullien {\it et al.}, 1999), and by the numerical simulations
(Zovari {\it et al.}, 1994; Elliott Jr. and Majda, 1996;
Fung and Vassilicos, 1998; Boffetta {\it et al.}, 1998). The more
general property of self-similarity (\ref{Rich1}) (with $\alpha=1/3$)
has been observed in the inverse cascade of two-dimensional turbulence
(Jullien {\it et al.}, 1999; Boffetta and Sokolov, 2000; Boffetta and Celani,
2000). It is likely that (\ref{Rich}) is exact in that situation, while it 
may be only approximately correct in $3d$, although the experimental data 
do not allow yet to test it with sufficient confidence.

It is important to remark that, even assuming the validity of the
Richardson law (\ref{Rich}), it is impossible to establish general 
properties of
the PDF ${\cal P}(R;t)$ such as those in Sect.\,\ref{sec:general} for
the single particle PDF. The physical reason is easy to understand
if one writes \qq {d\langle R^2\rangle\over
dt}\,=\,2\,\tau_t\hspace{0.15cm}\langle \left(\Delta {\bm
V}\right)^2\rangle\,,\label{ttl} \qqq
similarly to (\ref{GITaylor}) and (\ref{Lagrtime}). 
Here $\,\tau_t=\int_0^t\langle
\Delta {\bm V}(t)\cdot \Delta{\bm V}(s) \rangle\,ds\,/ \langle
\left(\Delta {\bm V}\right)^2\rangle\,$ is the correlation time of the
Lagrangian velocity differences.  If $\,d\langle R^2\rangle/dt\,$ is
proportional to $\,\langle R^2 \rangle^{2/3}$ and $\,\langle
\left(\Delta {\bm V}\right)^2 \rangle\,$ behaves like $\,\langle
R^2\rangle^{1/3}$ then $\,\tau_t$ grows as $\langle
R^2\rangle^{1/3}\propto t$, \,i.e. the random process $\Delta {\bm
V}(t)$ is correlated across its whole span.  The absence of
decorrelation explains why the Central Limit Theorem and the Large
Deviation Theory cannot be applied.  In general, there is no {\it a
priori} reason to expect ${\cal P}(R;t)$ to be Gaussian with respect
to a power of $\,R\,$ either, although, as we shall see, this is what
essentially happens in the Kraichnan ensemble.

\subsubsection{Breakdown of the Lagrangian flow}
\label{sec:break}

It is instructive to contrast the exponential growth (\ref{energy}) of
the distance between the trajectories within the viscous range with
the power law growth (\ref{ene}) in the inertial range. In the viscous
regime the closer two trajectories are initially, the longer time is
needed to reach a given separation.  As a result, infinitesimally
close trajectories never separate and trajectories in a fixed
realization of the velocity field are continuously labeled by the
initial conditions. Small deviations of the initial point are
magnified exponentially, though. This sensitive dependence is usually
considered as the defining feature of the chaotic
behavior. Conversely, in the inertial interval the trajectories
separate in a finite time independently of their initial distance
$R(0)$, \,provided the latter was also in the inertial interval. The
speed of this separation may depend on the detailed structure of the
turbulent velocities, including their fine geometry (Fung and
Vassilicos, 1998), but the very fact of the explosive separation is
related to the scaling behavior $\Delta_r v\propto r^\alpha$ with
$\alpha<1$. For high Reynolds numbers the viscous scale $\,\eta\,$ is
negligibly small, a fraction of a millimeter in the turbulent
atmosphere. Setting it to zero (or equivalently the Reynolds number to
infinity) is an appropriate abstraction if we want to concentrate on
the behavior of the trajectories in the inertial range. In such a
limit, the power law separation between the trajectories extends down
to arbitrarily small distances: infinitesimally close trajectories
still separate in a finite time. This makes a marked difference in
comparison to the smooth chaotic regime, clearly showing that
developed turbulence and chaos are fundamentally different phenomena.
As stressed in (Bernard {\it et al.}, 1998), the explosive separation
of the trajectories results in a breakdown of the deterministic
Lagrangian flow in the limit $Re\to\infty$, see also (Frisch {\it et
al.}, 1998; Gaw\c{e}dzki, 1998 and 1999). The effect is dramatic since
the trajectories cannot be labeled anymore by the initial
conditions. Note that the sheer existence of the Lagrangian
trajectories ${\bm R}(t;{\bm r})$ depending continuously on the
initial position ${\bm r}$ would imply that $\,\,\lim\limits_{{\bm
r}_{1}\to{\bm r}_{2}}\,\langle\, \vert{\bm R}(t;{\bm r}_{1})-{\bm
R}(t;{\bm r}_{2}) \vert^\zeta\,\rangle\,=\,0\,\,$. That would
contradict the persistence of a power law separation of the Richardson
type for infinitesimally close trajectories. Remark also that the
breakdown of the deterministic Lagrangian flow does not violate the
theorem about the uniqueness of solutions of the ordinary differential
equation $\,\dot{{\bm R}} ={\bm v}({\bm R},t)$. Indeed, the theorem
requires the velocity to be Lipschitz in $\bm r$, i.e.\,\,that
$\Delta_r v \leq{\cal O}(r)$. As first noticed by Onsager (1949), the
velocities for $Re=\infty$ are actually only H\"{o}lder continuous:
$\,\Delta_r v\simeq{\cal O} (r^\alpha)\,$ with the exponent $\alpha<1$
(in Kolmogorov's phenomenology $\alpha={1/ 3}$).  The simple
equation $\,\dot{x}=|x|^\alpha\,$ provides a classical example with
two solutions $\,x=[(1-\alpha)t]^{1/(1-\alpha)}\,$ and $\,x=0,$ \,both
starting from zero, for the non-Lipschitz case $\,\alpha<1$.  It is
then natural to expect the existence of multiple Lagrangian
trajectories starting or ending at the same point. Such a possibility
was first noticed and exploited in a somewhat different context in the
study of weak solutions of the Euler equations (Brenier, 1989;
Shnirelman, 1999).  Does then the Lagrangian description of the fluid
break down completely at $Re=\infty$?

Even though the deterministic Lagrangian description is inapplicable,
the statistical description of the trajectories is still possible.  As
we have seen above, probabilistic questions like those about the
averaged powers of the distance between initially close trajectories
still have well defined answers. We expect that for a typical velocity
realization, one may maintain at $Re=\infty$ a probabilistic
description of the Lagrangian trajectories. In particular, objects such 
as the PDF $\,p({\bm r},s;{\bm R},t\,\vert\,{\bm v})\,$ of the time $t$ 
particle position ${\bm R}$, given its time $s$ position ${\bm r}$, 
should continue to make sense. For a regular velocity with
deterministic trajectories, \qq p({\bm r},s;{\bm R},t\,\vert\,{\bm
v})\ =\ {\bm\delta}({\bm R}- {\bm R}(t;{\bm r},s))\,,
\label{dtp}
\qqq where ${\bm R}(t;{\bm r},s)$ denotes the unique Lagrangian
trajectory passing at time $s$ through ${\bm r}$.  In the presence of
a small molecular diffusion, equation (\ref{sde}) for the Lagrangian
trajectories has always a Markov process solution in each fixed
velocity realization, irrespective of whether the latter be Lipschitz
or H\"{o}lder continuous (Stroock and Varadhan, 1979). The resulting
Markov process is characterized by the transition probabilities
$p({\bm r},s;{\bm R},t\,\vert\,{\bm v})$ satisfying the
advection-diffusion equation\footnote{For $\kappa>0$ and smooth
velocities, the equation results from the It\^{o} formula generalizing
(\ref{Itof}) applied to (\ref{dtp}) and averaged over the noise.}  \qq
(\partial_t-{\bm\nabla}_{_{\hspace{-0.07cm}{\bbox R}}}\cdot{\bm
v}({\bm R},t) - \kappa{\bm\nabla}^2_{_{\hspace{-0.07cm}{\bm R}}})
\,\,p({\bm r},s;{\bm R},t\,\vert\,{\bm v})\ =\ 0\,,
\label{ade}
\qqq for $t>s$.  The mathematical difference between smooth and rough
velocities is that in the latter case the transition probabilities are
weak rather than strong solutions. What happens if we turn off the
molecular diffusion?  If the velocity is Lipschitz in ${\bm r}$, then
the Markov process describing the noisy trajectories concentrates on
the deterministic Lagrangian trajectories and the transition
probabilities converge to \,(\ref{dtp}). It has been conjectured in
(Gaw\c{e}dzki, 1999) that, for a generic $Re=\infty$ turbulent flow,
the Markov process describing the noisy trajectories still tends to a
limit when $\,\kappa\to0$, but the limit stays diffused, see
Fig.~\ref{fig:panache}. In other words, the transition probability
converges to a weak solution of the advection equation \qq
(\partial_t-{\bm\nabla}_{_{\hspace{-0.07cm}{\bm R}}}\cdot{\bm v} ({\bm
R},t))\,\,p({\bm r},s;{\bm R},t\,\vert\,{\bm v})\ =\ 0\,,
\label{ade0}
\qqq which does not concentrate on a single trajectory, as it was the
case in (\ref{dtp}). We shall then say that the limiting Markov
process defines a stochastic Lagrangian flow. This way the roughness
of the velocity would result in the stochasticity of the particle
trajectories persisting even in the limit $\kappa\to0$. To avoid
misunderstanding, let us stress again that, according to this claim,
the Lagrangian trajectories behave stochastically already in a fixed
realization of the velocity field and for negligible molecular
diffusivities, i.e. the effect is not due to the molecular noise or to
random fluctuations of the velocities. This spontaneous stochasticity
of fluid particles seems to constitute an important aspect of
developed turbulence. It is an unescapable consequence of the
Richardson dispersion law and of the Kolmogorov-like scaling of
velocity differences in the limit $Re\to\infty$ and it provides for a
natural mechanism assuring the persistence of dissipation in the
inviscid limit: $\,\lim\limits_{\nu\to0}\nu\langle|{\bm\nabla}{\bbox
v}|^2\rangle\not=0$.

\subsubsection{The example of the Kraichnan ensemble}
\label{sec:Kraich}

The general conjecture about the existence of stochastic Lagrangian
flows for generic turbulent velocities, e.g.\,\,for weak solutions
of the incompressible Euler equations locally dissipating energy, as
discussed by Duchon and Robert (2000), has not been mathematically
proven.  The conjecture is known, however, to be true for the
Kraichnan ensemble (\ref{2PF}), as we are going to discuss in this
Section.

We should model the spatial part $D^{ij}$ of the 2-point function
(\ref{2PF}) so that it has proper scalings in the viscous and inertial
intervals. This can be conveniently achieved by taking its Fourier
transform \qq \hat D^{ij}({\bm k})\ \propto\ \left(\delta^{ij}-{k^i
k^j\over k^2}\right) \ {{\rm e}^{-(\eta{\bm k})^2}\over({\bm
k}^2+L^{-2})^{(d+\xi)/2}},
\label{sfe}
\qqq with $0\leq \xi\leq 2$. In physical space, \qq D^{ij}({\bm r}) \
=\ D_0\,\delta^{ij}\,-\ {_1\over^2}\,d^{ij}({\bm r})\,,
\label{sfe1}
\qqq where $\,d^{ij}({\bm r})\,$ scales as $\,r^\xi\,$ in the inertial
interval $\,\eta\ll r\ll L$, as $\,r^2\,$ in the viscous range
$\,r\ll\eta\,$ and tends to $\,2D_0\,\delta^{ij}\,$ at very large
scales $\,r\gg L$.  As we discussed in Sect.\,\ref{sec:single},
$\,D_0$ gives the single-particle effective diffusivity.  Notice that
$\,D_0={\cal O}(L^{\xi})\,$ indicating that turbulent diffusion is
controlled by the velocity fluctuations at large scales of order $L$.
On the other hand, $\,d_{ij}({\bm r})\,$ describes the statistics of
the velocity differences and it picks up contributions of all
scales. In the limits $\eta\to0$ and $L\to\infty$, it takes the
scaling form: \qq \lim\limits_{\eta\to0\atop
L\to\infty}\,\,d^{ij}({\bm r}) \ =\
D_1\,r^{\xi}[(d-1+\xi)\,\delta^{ij}\,-\,\xi\,{r^ir^j\over r^2}]\,,
\label{tobe1}
\qqq where the normalization constant $D_1$ has the dimensionality
of $\,length^{2-\xi}\,time^{-1}$.

For $0<\xi<2$ and $\eta>0$, the typical velocities are smooth in space
with the scaling behavior $r^{\xi}$ visible only for scales much
larger than the viscous cutoff $\eta$. When 
the cutoff is set to zero,
however, the velocity becomes nonsmooth.  The Kraichnan ensemble is
then supported on velocities that are H\"{o}lder-continuous with the
exponent $\xi/2-0$. That mimics the major property of turbulent
velocities at the infinite Reynolds number.  The limiting case $\xi=2$
describes the Batchelor regime of the Kraichnan model: the velocity
gradients are constant and the velocity differences are linear in
space. This is the regime that the analysis of
Sect.\,\,\ref{sec:solv}(i) pertains to.  In the other limiting case
$\xi=0$, the typical velocities are very rough in space
(distributional). For any $\xi$, the Kraichnan velocities have even
rougher behavior in time. We may expect that the temporal roughness
does not modify largely the qualitative picture of the trajectory
behavior as it is the regularity of velocities in space, and not in
time, that is crucial for the uniqueness of the trajectories (see,
however, below).

For time-decorrelated velocities, both terms on the right hand side of
the Lagrangian equation (\ref{sde}) should be treated according to the
rules of stochastic differential calculus. The choice of the
regularization is irrelevant here even for compressible velocities,
see Appendix.  The existence and the properties of solutions of such
stochastic differential equations were extensively studied in the
mathematical literature for velocities smooth in space, see
e.g.\,\,(Kunita, 1990).  Those results apply to our case as long as
$\eta>0$ both for positive or vanishing diffusivity. The
advection-diffusion equation (\ref{ade}) for the transition
probabilities also becomes a stochastic equation for white-in-time
velocities. The choice of the convention, however, is important here
even for incompressible velocities: the equation should be interpreted
with the Stratonovich convention, see Appendix. The equivalent It\^{o}
form contains an extra second-order term that amounts to the
replacement of the molecular diffusivity by the effective diffusivity
$\,(D_0+\kappa)\,$ in (\ref{ade}). The It\^{o} form of the equation
explicitly exhibits the contribution of the eddy diffusivity, hidden
in the convention for the Stratonovich form.  As pointed out by Le Jan
and Raimond (1998 and 1999), the regularizing effect of $\,D_0$ permits
to solve the equation by iteration also for the nonsmooth case giving
rise to transition probabilities $\,p({\bm r},s;{\bm R},t\,\vert
\,{\bm v})\,$ defined for almost all velocities of the Kraichnan
ensemble. Moreover, the vanishing diffusivity limit of the transition
probabilities exist, defining a stochastic Lagrangian flow.

The velocity averages over the Kraichnan ensemble of the transition 
probabilities $\,p({\bm r},s;{\bm R},t\,\vert\,{\bm v})\,$ are 
exactly calculable. We shall use a formal functional integral 
approach (Chertkov, 1997; Bernard {\it et al.}, 1998). In the phase space 
path integral representation of the solution of (\ref{ade}),
\qq
p({\bm r},s;{\bm R},t\,\vert\,{\bm v})\ = 
\int\limits_{{\bm r}(s)={\bm r}\atop
{\bm r}(t)={\bm R}}{\rm e}^{\,\mp\int\limits_s^t[\,i\,{\bm p}(\tau)
\cdot(\dot{{\bm r}}(\tau)-{\bm v}({\bm r}(\tau),\tau))\,+\,
\kappa\,{\bm p}^2(\tau)]\,d\tau}\,{\cal D}{\bm p}\,{\cal D}{\bm r}\,,
\label{pir}\qqq for $s{_<\atop^>}t$, the Gaussian average over 
the velocities is easy to perform. It replaces the exponent in 
(\ref{pir}) by $\,\mp\int\limits_s^t [\,i\,{\bm
p}(\tau) \cdot\dot{{\bm r}}(\tau)+(D_0+\kappa) \,{\bm
p}^2(\tau)]\,d\tau\,$ and results in the path integral represention 
of the heat kernel of the Laplacian for which we shall use the operator 
notation $\,{\rm e}^{\,|t-s|(D_0+\kappa) {\bm\nabla}^2}
\hspace{-0.05cm}({\bm r};{\bm R})$. \,In other words, the average of
(\ref{pir}) is the solution of the heat equation (with diffusivity
$D_0+\kappa$) equal to ${\bm\delta}({\bm R} -{\bm r})$ at time
$s$. \,The above calculation confirms then the result discussed at the
end of Sect.\,\ref{sec:single} about the all-time diffusive behavior
of a single fluid particle in the Kraichnan ensemble.

In order to study the two-particle dispersion, one should examine the
joint PDF of the equal-time values of two fluid particles averaged
over the velocities \qq \Big\langle\,p({\bm r}_{1},s;{\bm
R}_1,t\,\vert\,{\bm v})\ p({\bm r}_{2},s;{\bm R}_2,t\,\vert\,{\bm
v})\,\Big\rangle\ \equiv \ {\cal P}_{_2}({\bm r}_{1},{\bm
r}_{2};\,{\bm R}_1,{\bm R}_2;t-s)\,.
\label{gex}
\qqq
The latter is given for the Kraichnan ensemble by the heat kernel 
$\,{\rm e}^{\,\vert t-s\vert\,{\cal M}_2}
({\bm r}_{1},{\bm r}_{2};\,{\bm R}_1,{\bm R}_2)\,$
of the elliptic second-order differential operator
\qq
{\cal M}_2\ =\ \sum\limits_{n,n'=1}^2D^{ij}({\bm r}_n-{\bm r}_{n'})
\,\nabla_{{r_{_n}^i}}\nabla_{{r_{n'}^j}}\,+\,\kappa\sum\limits_{n=1}^2
\,{\bm\nabla}^2_{{{\bm r}_n}}\,\,.
\label{m2}
\qqq 
In other words, the PDF $\,{\cal P}_{_2}$ satisfies the equation
$(\partial_t-{\cal M}_2){\cal P}_2= \delta(t-s)\delta({\bm R}_1-{\bm
r}_{1})\delta({\bm R}_2-{\bm r}_{2})$, a result which goes back to the
original work of Kraichnan (1968).  Indeed, the Gaussian expectation
(\ref{gex}) is again easily computable in view of the fact that the
velocity enters through the exponential function in (\ref{pir}). The
result is the path integral expression \qq \int\limits_{{\bm
r}_n(s)={\bm r}_{n}\atop {\bm r}_n(t)={\bm R}_n}\hspace{-0.5cm}{\rm
e}^{\,\mp\int\limits_s^t \Big(\sum\limits_{n=1}^2[\,i\,{\bm
p}_n(\tau)\cdot\dot{{\bm r}}_n(\tau)\, +\,\kappa\,{\bm
p}_n^2(\tau)]\,+\hspace{-0.1cm}\sum\limits_{n, n'=1}^2D^{ij}({\bm
r}_{n}(\tau)-{\bm r}_{n'}(\tau))\,\,p_{ni}(\tau)
\,p_{n'j}(\tau)\Big)\,d\tau}\,\prod\limits_n\,{\cal D}{\bm p}_n\,
{\cal D}{\bm r}_n \label{hk} \qqq for the heat kernel of $\,{\cal
M}_2$.

Let us concentrate on the relative separation $\,{\bm R}={\bm
R}_1 -{\bm R}_2\,$ of two fluid particles at time $t$, given their
separation ${\bm r}$ at time zero. The relevant PDF $\,\widetilde{\cal
P} ({\bm r};{\bm R};t)\,$ is obtained by averaging over the
simultaneous translations of the final (or initial) positions of the
particles. Explicitly, it is given by the heat kernel of the operator
$\,\widetilde{\cal M}=(d^{ij}({\bm r})
+2\kappa\,\delta^{ij})\,\nabla_{r^i}\nabla_{r^j}\,$ equal to the
restriction of $\,{\cal M}_2\,$ to the translationally invariant
sector. \,Note that the eddy diffusivity $\,D_0$, dominated by the
integral scale, drops out from $\,\widetilde{\cal M}$. \,The above result
shows that the relative motion of two fluid particles is an effective
diffusion with a distance-dependent diffusivity tensor scaling like
$\,r^\xi\,$ in the inertial range. This is a precise realization of
the scenario for turbulent diffusion put up by Richardson (1926).

Similarly, the PDF of the distance $\,R\,$ between two particles is
given by the heat kernel $\,{\rm e}^{\,|t|\,M}(r;R),$ \,where \,$M\,$
is the restriction of $\,{\cal M}_2\,$ to the homogeneous and
isotropic sector. Explicitly, \qq M\ =\ {1\over
r^{d-1}}\,\partial_r\Big[(d-1)D_1\,r^{d-1+\xi}
+2\kappa\,r^{d-1}\Big]\partial_r
\label{m3}
\qqq in the scaling regime and its heat kernel may be readily
analyzed. In the Batchelor regime $\xi=2$ and for $\kappa\to0$, the
heat kernel of $\,M\,$ reproduces the lognormal distribution
(\ref{kk3}) with $\Delta=2D_1(d-1)$ and $\bar\lambda=D_1d(d-1)$, see
Sect.\,\,\ref{sec:solv}(i).

The simple criterion allowing to decide whether the Markov process
stays diffused as $\,\kappa\to0\,$ is to control the limit $\,r\to0\,$
of the PDF \,${\cal P} (r;R;t)$ (Bernard {\it et al.}, 1998). For
smooth velocities, it follows from (\ref{kk3}) that \qq
\lim\limits_{r\to0\atop\kappa\to0}\ {\cal P}(r;R;t) \ =\
\delta(R).\label{sexp}\qqq In simple words, when the initial points
converge, so do the endpoints of the process. Conversely, for
$0\leq\xi<2$ we have \qq \lim\limits_{r\to0\atop\kappa\to0}\ {\cal
P}(r;R;t) \ \,\propto\ \,{{R^{d-1}}\over{|t|^{d/(2-\xi)}}}\,\,
{\exp\left[-{\rm const}.\,{R^{2-\xi}\over|t|}\right]}\ ,
\label{oprn}
\qqq in the scaling limit $\eta=0$, $L=\infty$.  
That confirms the diffused character of the limiting
process describing the Lagrangian trajectories in fixed non-Lipschitz
velocities: the endpoints of the process stay at finite distance even
if the initial points converge. If we set the viscous cutoff to zero
keeping $L$ finite, the behavior (\ref{oprn}) crosses over for $\,R\gg
L\,$ to a simple diffusion with diffusivity $2D_0\,$: at such large
distances the particle velocities are essentially independent and the
single particle behavior is recovered.

The stretched-exponential PDF (\ref{oprn}) has the scaling form
(\ref{Rich1}) for $\,\alpha=\xi-1\,$ and implies the power law growth
(\ref{ene}) of the averaged powers of the distance between
trajectories. The PDF is Gaussian in the rough case $\xi=0$.  Note
that the Richardson law $\,\langle R^2(t)\rangle\,\propto\,t^3\,$ is
reproduced for $\,\xi=4/3\,$ and not for $\,\xi=2/3\,$ (where the
velocity has the spatial H\"{o}lder exponent $1/3$). The reason is
that the velocity temporal decorrelation cannot be ignored and
the mean-field relation (\ref{Rich0}) should be replaced by 
$\,R^{1-\xi/2}(t)-R^{1-\xi/2}(0)
\propto\beta(t)\,$ with the Brownian motion $\beta(t)$. Since
$\beta(t)$ behaves as $t^{1/2}$, the replacement changes the power and
indeed reproduces the large-time PDF (\ref{oprn}) up to a geometric
power-law prefactor. In general, the time dependence of the velocities
plays a role in determining whether the breakdown of deterministic
Lagrangian flow occurs or not. Indeed, the relation (\ref{ttl})
implies that the scale-dependence of the correlation time $\,\tau_t\,$
of the Lagrangian velocity differences may change the time behavior of
$\langle R^2\rangle$. In particular, $\langle R^2\rangle$ ceases to
grow in time if $\,\tau_t\propto\langle R^2\rangle^\beta$ and
$\,\langle\left(\Delta {\bm V}\right)^2\rangle\propto\langle R^2
\rangle^\alpha\,$ with $\beta\geq1-\alpha$. It has been recently shown
in (Fannjiang {\it et al.}, 2000) that the Lagrangian trajectories are
deterministic in a Gaussian ensemble of velocities with H\"{o}lder
continuity in space and such fast time decorrelation on short
scales. The Kolmogorov values of the exponents $\alpha=\beta=1/3$
satisfy, however, $\,\beta<1-\alpha$.

Note the special case of the average $\langle R^{2-\xi-d}\rangle$ in 
the Kraichnan velocities. Since $\,M\,r^{2-\xi-d}\,$ is a contact term
$\propto\delta(r)$ for $\kappa=0$, one has $\,\partial_t\langle
R^{2-\xi-d}\rangle\propto{\cal P}(r;0;t)$. The latter is zero in the smooth
case so that $\langle R^{-d}\rangle$ is a true integral of motion. In
the nonsmooth case, $\langle R^{2-\xi-d}\rangle\propto t^{1-d/(2-\xi)}$ 
and is not conserved due to a nonzero probability density to find 
two particles at the same place even when they started apart.

As stated, the result (\ref{oprn}) holds when the molecular
diffusivity is turned off in the velocity ensemble with no viscous
cutoff, i.e.\,\,for vanishing Schmidt number $\,Sc={\nu/\kappa}$,
\,where $\nu$ is the viscosity defined in the Kraichnan model as
$D_1\eta^\xi$. The same result holds also when $\nu$ and $\kappa$ are
turned off at the same time with $Sc={\cal O}(1)$, provided the
initial distance $r$ is taken to zero only afterwards (E and Vanden
Eijnden, 2000b). This confirms that the explosive separation of close
trajectories persists for finite Reynolds numbers as long as their
initial distance is not too small, as anticipated by Bernard {\it et
al.}\,\,(1998).

\subsection{Two-particle dispersion in a compressible flow}
\label{sec:compres}

Discussing the particle dispersion in incompressible fluids and
exposing the different mechanisms of particle separation, we paid
little attention to the detailed geometry of the flows, severely
restricted by the incompressibility. The presence of compressibility
allows for more flexible flow geometries with regions of ongoing
compression effectively trapping particles for long times and
counteracting their tendency to separate. To expose this effect and
gauge its relative importance for smooth and nonsmooth flows, we start
from the simplest case of a time-independent $1d$ flow
$\,\dot{x}=v(x)\,$. In $1d$, any velocity is potential:
$\,v(x)=-\partial_x\phi(x)\,$, and the flow is the steepest descent in
the landscape defined by the potential $\phi$.  The particles are
trapped in the intervals where the velocity has a constant sign and
they converge to the fixed points with lower value of $\phi$ at the
ends of those intervals.  In the regions where $\partial_xv$ is
negative, nearby trajectories are compressed together.  If the flow is
smooth the trajectories take an infinite time to arrive at the fixed
points (the particles might also escape to infinity in a finite time).
Let us consider now a nonsmooth version of the velocity, e.g.\,\,a
Brownian path with H\"{o}lder exponent $1/2$.  At variance with
the smooth case, the solutions will take a finite time to reach the
fixed points at the ends of the trapping intervals and will stick to
them at subsequent times, as in the example of the equation
$\,\dot{x}=|x-x_0| ^{1/2}$. The nonsmoothness of the velocity clearly
amplifies the trapping effects leading to the convergence of the
trajectories. A time-dependence of the velocity changes somewhat the
picture.  The trapping regions, as defined for the static case, start
wandering and they do not enslave the solutions which may cross their
boundaries. Still, the regions of ongoing compression effectively trap
the fluid particles for long time intervals. Whether the tendency of
the particles to separate or the trapping effects win is a matter of
detailed characteristics of the flow.

In higher dimensions, the behavior of potential flows is very similar
to the $1d$ case, with trapping totally dominating in the
time-independent case, its effects being magnified by the
nonsmoothness of the velocity and blurred by the time-dependence.  The
traps might of course have a more complicated geometry.  Moreover, we
might have both solenoidal and potential components in the
velocity. The dominant tendency for the incompressible component is to
separate the trajectories, as we discussed in the previous Sections.
On the other hand, the potential component enhances trapping in the
compressed regions.  The net result of the interplay between the two
components depends on their relative strength, spatial smoothness and
temporal rate of change.

Let us consider first a smooth compressible flow with a homogeneous
and stationary ergodic statistics.  Similarly to the incompressible
case discussed in Sect.\,\ref{sec:general}, the stretching-contraction
variables $\rho_i$, $i=1,\dots,d$, behave asymptotically as
$t\lambda_i$ with the PDF of large deviations $x_i=\rho_i/t-\lambda_i$
determined by an entropy function $H(x_1,\dots,x_d)$. The asymptotic
growth rate of the fluid volume is given by the sum of the Lyapunov
exponents $s=\,\sum\limits_{i=1}^d\lambda_i\,$. Note that density
fluctuations do not grow in a statistically steady compressible flow
because the pressure provides feedback from the density to the
velocity field. That means that $s$ vanishes even though the $\rho_i$
variables fluctuate. However, to model the growth of density
fluctuations in the intermediate regime, one can consider an idealized
model with a steady velocity statistics having nonzero $s$.  This
quantity has the interpretation of the opposite of the entropy
production rate, see Section~\ref{sec:density} below, and it is
necessarily $\leq0$ (Ruelle, 1997). Let us give here the argument due
to Balkovsky {\it et al.}\,\,(1999a) which goes as follows.  In any
statistically homogeneous flow, incompressible or compressible, the
distribution of particle displacements is independent of their initial
position and so is the distribution of the evolution matrix
$\,W_{ij}(t;{\bm r})=\partial\,R^i(t;{\bm r})/\partial {r^j}$.
\,Since the total volume $V$ (assumed finite in this argument) is
conserved, the average $\,\langle\det W \rangle\,$ is equal to unity
for all times and initial positions although the determinant
fluctuates in the compressible case.  The average of $\,\det W={\rm
e}^{\,\sum \rho_i}$ is dominated at long times by the saddle-point
${\bm x}^*$ giving the maximum of $\,\sum(\lambda_i+x_i)-H({\bm x})\,$,
which has to vanish to conform with the total volume
conservation. Since $\sum x_i -H({\bm x})$ is concave and vanishes at
$x=0$, its maximum value has to be non-negative. We conclude that the sum
of the Lyapunov exponents is non-positive. The physics behind this result
is transparent: there are more Lagrangian particles in the contracting
regions, leading to negative average gradients in the Lagrangian
frame. Indeed, the volume growth rate tends at large times to the
Lagrangian average of the trace of the strain $\,\sigma={\bm
\nabla}{\bm v}$:
\begin{eqnarray}&&
{1\over t}\Big\langle \ln{\det{(W({\bm r};t))}}
\Big\rangle\longrightarrow\int{\rm tr}\,\sigma({\bbox R}(t;{\bm r}),t)
\,\,\frac{_{d{\bm r}}}{^{V}} \ =\ \int{\rm tr}\,\sigma({\bm
R},t)\,\,\frac{_{d{\bm R}}}{^{V\, \det{(W({\bm r};t))}}}.
\label{compr}\end{eqnarray}
The Lagrangian average generally coincides with the Eulerian one $\int
d\bbox r \,\,{\rm tr}\,\sigma(t,{\bbox r}) /V$, only in the
incompressible case (where it is zero). For compressible flow, the
integrals in (\ref{compr}) vanish at the initial time (when we set the
initial conditions for the Lagrangian trajectories so that the measure
was uniform back then). The regions of ongoing compression with
negative ${\rm tr}\,\sigma$ acquire higher weight in the average in
(\ref{compr}) than the expanding ones.  Negative values of ${\rm tr}
\,\sigma\,$ suppress stretching and enhance trapping and that is the
simple reason for the volume growth rate to be generally negative.
Note that, were the trajectory ${\bm R}({\bm r};t)$ defined by its
final (rather than initial) position, the sign of the average strain
trace would be positive. Let us stress again the essential difference
between the Eulerian and the Lagrangian averages in the compressible
case: an Eulerian average is uniform over space, while in a Lagrangian
average every trajectory comes with its own weight determined by the
local rate of volume change. For the corresponding effects on the
single-particle transport, the interested reader is referred to
(Vergassola and Avellaneda, 1997).

In the particular case of a short-correlated strain one can take $t$
in (\ref{compr}) larger than the correlation time $\tau$ of the
strain, yet small enough to allow for the expansion $\,\det{(W({\bm
r};t))}^{-1} \approx 1-\int_0^{t}{\rm tr}\,\sigma(\bbox r,t')\,dt'\,$
so that (\ref{compr}) becomes equal to $\,-\int\limits_0^t\langle{\rm
tr}\,\sigma({\bbox r},t)\, {\rm tr}\,\sigma({\bbox
r},t')\rangle\,dt'$. More formally, let us introduce the compressible
generalization of the Kraichnan ensemble for smooth velocities. Their
(non-constant part of the) pair correlation function is defined as
\begin{eqnarray}
d^{ij}(\bbox r)\,=\,D_1\left[(d+1-2\wp)\,\delta^{ij}\,r^2\,+\,2
(\wp d-1)\,r^ir^j\right]\ +\ o(r^2)\,,
\label{fastcomp}
\end{eqnarray}
compare to (\ref{shortstrain2}). The degree of compressibility
$\,\wp\equiv\langle(\nabla_i v^i)^2\rangle/\langle(\nabla_i
v^j)^2\rangle\,$ 
is between 0 and 1 for the isotropic case at hand, with the the
two extrema corresponding to the incompressible and the potential
cases. The corresponding strain matrix $\sigma=\nabla {\bm v}$ has the
Eulerian mean equal to zero and 2-point function \qq
\langle\,\sigma_{ij}(t)\,\sigma_{k\ell}(t')\,\rangle=
2\,\delta(t-t')\,D_1 \left[(d+1-2\wp)\,\delta_{ik}\delta_{j\ell} +(\wp
d-1)\bigl(\delta_{ij}\delta_{k\ell}+\delta_{i\ell}\delta_{jk}\bigr)
\right]\,.  \qqq The volume growth rate
$\,-\int\limits_0^t\langle\,\sigma_{ii}(t)\,
\sigma_{jj}(t')\rangle\,dt'$ is thus strictly negative, in agreement
with the general discussion, and equal to $-\wp\,D_1\,d(d-1)(d+2)$
if we set $\,\int_{_0}^{^\infty}\delta(t)\,dt=1/2$.
\,The same result is obtained more systematically by considering the
It\^{o} stochastic equation $\,dW=\sigma\,dt\,W\,$ for the evolution
matrix and applying the It\^{o} formula to $\,\ln\det(W)$, see
Appendix. One may identify the generator 
of the process $W(t)$ and proceed as for the
incompressible case calculating the PDF ${\cal
P}(\rho_1,\ldots,\rho_d;t)$.  It takes again the large deviation form
(\ref{PDF3}), with the entropy function and the Lyapunov exponents
given by \qq &&H({\bbox
x})\,=\,{_1\over^{4D_1(d+\wp(d-2))}}\,\Big[\sum\limits_{i=1}^d
x_i^2\,+\,{_{1-\wp d}\over^{\wp(d-1)(d+2)}}\Big(\sum\limits_{i=1}^d
x_i\Big)^2\Big]\,,\label{Hfct}\\
&&\lambda_i\,=\,D_1\left[d(d-2i+1)-2\wp\,(d+(d-2)i) \right]\,,
\label{complambda}
\qqq to be compared to (\ref{qenf}). Note how the form (\ref{Hfct}) of
the entropy imposes the condition $\,\sum x_i=0\,$ in the
incompressible limit. The interparticle distance $\,R(t)\,$ has the
lognormal distribution (\ref{kk3}) with $\,\bar\lambda=\lambda_1
=D_1(d-1)(d-4\wp)\,$ and $\,\Delta=2D_1(d-1)(1+2\wp)$. Explicitly,
$t^{-1}\ln\langle R^n\rangle \propto n[n+d+2\wp(n-2)]$ (Chertkov {\it
et al.}, 1998). The quantity $R^{(4\wp-d)/(1+2\wp)}$ is thus
statistically conserved.  The highest Lyapunov exponent $\bar\lambda$
becomes negative when the degree of compressibility is larger than
$d/4\,$ (Le Jan, 1985; Chertkov {\it et al.}, 1998). Low-order moments
of $R$, including its logarithm, would then decrease while high-order
moments would grow with time.

It is instructive to decompose the strain into its
``incompressible and compressible parts'' $\sigma_{ij}-{1\over
d}\,\delta_{ij}\,{\rm tr}\,\sigma$ and ${1\over d}\,\delta_{ij}\,{\rm
tr}\,\sigma\,$. From the equality
$\lambda_i=\langle\tilde\sigma_{ii}\rangle$, see (\ref{PDF3}), it
follows that the Lyapunov exponents of the incompressible part (having
$\bar\lambda>0$) get uniformly shifted down by the Lagrangian average
of $\,{\rm tr}\,\sigma /d\,$. \,In $1d$, where the compressibility is 
maximal\footnote{One-dimensional results are recovered from our formulae 
by taking $\wp=1$ and $D_1\propto1/(d-1)$.}, $\bar\lambda<0$. The lowering 
of the Lyapunov exponents when $\wp$ grows clearly signals the increase 
of trapping. The regime with $\,\wp>d/4\,$, with all the Lyapunov 
exponents becoming negative, is the one where trapping effects dominate. 
The dramatic consequences for the scalar fields advected by such flow 
will be discussed in Sect.\,\ref{sec:scal}.

As it was clear from the $1d$ example, we should expect even stronger
effects of compressibility in nonsmooth velocity fields, with an
increased tendency for the fluid particles to aggregate in finite
time. This has, indeed, been shown to occur when the velocity has a
short correlation time, i.e. for the nonsmooth version of the
compressible Kraichnan ensemble (Gaw\c{e}dzki and Vergassola,
2000). The expression (\ref{sfe}) for the Fourier transform of the
2-point correlation function is easily modified. The functional
dependence on $k^2$ remains the same, the solenoidal projector is
simply multiplied by $1-\wp$ and one adds to it the compressible
longitudinal component $\wp(d-1) k^i k^j/k^2$. This gives for the
non-constant part of the 2-point function \qq \label{Rehovot}
d^{ij}({\bm r})\ =\ D_1[(d-1+\xi-\wp\xi)\,\delta^{ij}\,r^{\xi}\,+\,\xi
(\wp d-1)\,r^ir^j\,r^{\xi-2}]\,. \qqq For $\xi=2$, (\ref{Rehovot})
reproduces (\ref{fastcomp}) without the $o(r^2)$ corrections.  Most of
the results discussed in Sect.\ref{sec:nonsmooth} for the
incompressible version of the model still go through, including the
construction of the Markov process describing the noisy trajectories
and the heat-kernel form of the joint PDF of two particle
positions. The restriction $\,M$ of $\,{\cal M}_2$ to the homogeneous
and isotropic sector, whose heat kernel gives the PDF $\,{\cal
P}(r;R;t)\,$ of the distance between two particles, takes now the form
\qq \label{ciccio}M\ =\ \bigg[{D_1(d-1)(1+\wp\xi)\over r^{d-1-\gamma}}
\,\partial_r\,r^{d-1+\xi-\gamma}\,+\, {2\kappa\over
r^{d-1}}\,\partial_r\,r^{d-1}\bigg]\partial_r\,, \qqq where
$\,\gamma={\wp\xi(d+\xi)/(1+\wp\xi)}$. \,As found in (Gaw\c{e}dzki and
Vergassola, 2000), see also (Le Jan and Raimond, 1999; E and Vanden
Eijnden, 2000b), depending on the smoothness exponent $\xi$ and the
degree of compressibility $\wp$, two different regimes arise in
the limit $\kappa\to0$.

For weak compressibility $\,\wp<\wp_c\equiv{d\over\xi^2}$, the situation 
is very much the same as for the incompressible case and
\qq
\lim\limits_{r\to0\atop\kappa\to0}\ {\cal P}(r;R;t)
\ \,\propto\ \,{{R^{d-\gamma-1}}\over{|t|^{(d-\gamma)/(2-\xi)}}}\,\,
{\exp\Big[-{\rm const}.\,{R^{2-\xi}\over|t|}\Big]}\,.
\label{oprm}
\qqq We still have an explosive separation of trajectories but, in
comparison to the incompressible situation, the power prefactor
$R^{-\gamma}$ with $\gamma\propto\wp$ suppresses large separations and
enhances small ones.  When $\wp$ crosses
$\,\tilde\wp_c={d+\xi-2\over2\xi},$ \, the particle-touching event
$R=0$ becomes recurrent for the Markov process describing the distance
between the two particles (Le Jan and Raimond, 1999). In other words,
for $\tilde\wp_c\leq\wp\leq\wp_c$ a pair of Lagrangian trajectories
returns infinitely often to a near touch, a clear sign of increased
trapping.

When $\,\wp\,$ crosses $\,\wp_c$, the singularity at $R=0$ of the
right hand side of (\ref{oprm}) becomes non-integrable and a different
limit is realized. Indeed, in this regime, \qq
\lim\limits_{\kappa\to0}\ {\cal P}(r;R;t)\ =\ {\cal P}^{reg}(r;R;t)\
+\ p(r;t)\,\delta(R),
\label{oprr}
\qqq with the regular part $\,{\cal P}^{reg}\,$ tending to zero and
$\,p\,$ approaching unity when $\,r\to0$. This reproduces for $\eta=0$
the result (\ref{sexp}), always holding when the viscous cutoff
$\eta>0$ smoothes the velocity realizations. In other words, even
though the velocity is nonsmooth, the Lagrangian trajectories in a
fixed velocity field are determined by their initial positions.
Moreover, the contact term in (\ref{oprr}) signals that trajectories
starting at a finite distance $r$ collapse to zero distance and stay
together with a positive probability growing with time (to unity if
the integral scale $L=\infty$).  The strongly compressible regime
$\wp>\wp_c$ is clearly dominated by trapping effects leading to the
aggregation of fluid particles, see Fig.~\ref{fig:compress}. The same 
results hold if we turn off the diffusivity $\kappa$ and the viscosity 
$\nu$ at the same time, with the notable exception of the intermediate 
regime $\tilde\wp_c\leq\wp<\wp_c$. In this interval, if the Schmidt 
number $\,Sc\,$ diverges fast enough during the limiting process, the 
resulting PDF of the distance takes the form (\ref{oprr}) rather than 
that of (\ref{oprn}) arising for bounded $Sc$ (E and Vanden Eijnden,
2000b). Sufficiently high Schmidt numbers thus lead to the particle
aggregation in this case. Note that in the limit of smooth velocities
$\xi\to2$, the intermediate interval shrinks to the point
$\,\wp=d/4\,$ where the highest Lyapunov exponent crosses zero.

As was mentioned, the aggregation of fluid particles can take place
only as a transient process. The back reaction of the density on the
flow eventually stops the growth of the density fluctuations. The
transient trapping should, however, play a role in the creation of the
shocklet structures observed in high Mach number compressible flows
(Zeman, 1990). There is another important physical situation that may
be modeled by a smooth compressible random flow with a nonzero sum of
the Lyapunov exponents.  Let us consider a small inertial particle of
density $\rho$ and radius $a$ in a fluid of density $\rho_0$. Its
movement may be approximated by that of a Lagrangian particle in an
effective velocity field provided that $a^2/\nu$ is much smaller than
the velocity time scale in the Lagrangian frame. The inertial
difference between the effective velocity ${\bbox v}$ of the particle
and the fluid velocity ${\bm u}({\bm r},t)$ is proportional to the
local acceleration: $\,{\bbox v}={\bbox u}+(\beta-1)\,\tau_s\,d{\bm
u}/dt$, where $\beta=3\rho/(\rho+2\rho_0)$ and $\tau_s=a^2/3\nu\beta$
is the Stokes time. Considering such particles distributed in the
volume, one may define the velocity field $\,{\bbox v}({\bbox
r},t)\,$, whose divergence $\propto {\bm\nabla}[({\bm
u}\cdot{\bm\nabla}) {\bm u}]$ does not vanish even if the fluid flow
is incompressible.  As discussed above, this leads to a negative
volume growth rate and the clustering of the particles (Balkovsky {\it
et al.}, 2001).

\subsection{Multiparticle dynamics, statistical conservation laws 
and breakdown of scale invariance}
\label{sec:multi} 

This subsection is a highlight of the review. We describe here the
time-dependent statistics of multiparticle configurations with the
emphasis on conservation laws of turbulent transport.  As we have seen
in the previous subsections, the two-particle statistics is
characterized by a simple behavior of the single separation vector. In
nonsmooth velocities, the length of the vector grows by a power law,
while the initial separation is forgotten and there are no statistical
integrals of motion. In contrast, the many-particle evolution exhibits
non-trivial statistical conservation laws that involve geometry and
are proportional to positive powers of the distances.  The distance
growth is balanced by the decrease of the shape fluctuations in those
integrals. The existence of multiparticle conservation laws indicates
the presence of a long-time memory and is a reflection of the coupling
among the particles due to the simple fact that they all are in the
same velocity field. The conserved quantities may be easily built for
the limiting cases. For very irregular velocities, the fluid particles
undergo independent Brownian motions and the interparticle distances
grow as $\langle R_{nm}^2(t)\rangle=R_{nm}^2(0)+Dt$. \,Here,
examples of statistical integrals of motion are $\langle
R_{nm}^2-R_{pr}^2\rangle$ and $\langle
2(d+2)R_{nm}^2R_{pr}^2-{d}(R_{nm}^4+R_{pr}^4)\rangle$, \,and an
infinity of similarly built harmonic polynomials where all the powers
of $t$ cancel out. Another example is the infinite-dimensional case,
where the interparticle distances do not fluctuate. The two-particle
law (\ref{Rich0}), $R_{nm}^{1-\alpha}(t)-R_{nm}
^{1-\alpha}(0)\propto\,t$, implies then that the expectation of any
function of $R_{nm}^{1-\alpha}-R_{pr}^{1-\alpha}$ does not change
with time.  A final example is provided by smooth velocities, where
the particle separations at long times become aligned with the
eigendirections of the largest Lyapunov exponent of the evolution
matrix $W(t)$ defined in (\ref{disp6}). All the interparticle
distances $R_{nm}$ will then grow exponentially and their ratios
$R_{nm}/R_{kl}$ do not change.  Away from the degenerate limiting
cases, the conserved quantities continue to exist, yet they cannot be
constructed so easily and they depend on the number of particles and
their configuration geometry. The very existence of conserved
quantities is natural; what is generally nontrivial is their precise
form and their scaling. The intricate statistical conservation laws of
multiparticle dynamics were first discovered for the Kraichnan
velocities. That came as a surprise since the Kraichnan velocity
ensemble is Gaussian and time-decorrelated, with no structure built
in, except for the spatial scaling in the inertial range.  The
discovery has led to a new qualitative and quantitative understanding
of intermittency, as we shall discuss in detail in
Sect.\,\ref{sec:passKrai}. Even more importantly, it has pointed to
aspects of the multiparticle evolution that seem both present and
relevant in generic turbulent flows.  Note that those aspects are
missed by simple stochastic processes commonly used in numerical
Lagrangian models. There is, for example, a long tradition to take for
each trajectory the time integral of a $d$-dimensional Brownian motion
(whose variance is $\propto t^3$ as in the Richardson law) or an
Ornstein-Uhlenbeck process. Such models, however, cannot capture
correctly the subtle features of the $\,N$-particle dynamics such as
the statistical conservation laws.

\subsubsection{Absolute and relative evolution of particles}
\label{sec:absrel}

As for many-body problems in other branches of physics, e.g.\,\,in
kinetic theory or in quantum mechanics, multiparticle dynamics brings
about new aspects due to cooperative effects.  In turbulence, such
effects are mediated by the velocity fluctuations with long
space-correlations. Consider the joint PDF's of the equal-time
positions $ \underline{\bm R}=({\bm R}_{_1},\dots,{\bm R}_{_N})$ of
\,$N\,$ fluid trajectories \qq \Big\langle\prod\limits_{n=1}^N p({\bm
r}_n,s; \,{\bm R}_n,t\,| {\bm v})\Big\rangle\ \equiv\ {\cal
P}_{_{N}} (\underline{\bm r};\,\underline{\bm R};\,t-s),
\label{pn}
\qqq with the average over the velocity ensemble, see 
Fig.~\ref{fig:particelle}. More generally, one
may study the different-time versions of (\ref{pn}). Such PDF's, 
called multiparticle Green functions, account for the overall
statistics of the many-particle systems.

For statistically homogeneous velocities, it is convenient to separate
the absolute motion of the particles from the relative one, as in
other many-body problems with spatial homogeneity. For a single
particle, there is nothing but the absolute motion which is diffusive
at times longer that the Lagrangian correlation time
(Sect.\,\,\ref{sec:single}).  For $N$ particles, we may define the
absolute motion as that of the mean position $ \overline{\bm R}=
\sum\limits_n{\bm R}_n/N$. When the particles separate beyond the
velocity correlation length, they are essentially independent.  The
absolute motion is then diffusive with the diffusivity $N $ times
smaller than that of a single particle. The relative motion of $N$
particles may be described by the versions of the joint PDF's
(\ref{pn}) averaged over rigid translations: \qq \widetilde{\cal
P}_{_{N}}(\underline{\bm r};\,\underline{\bm R};\,t) \ =\ \int{\cal
P}_{_{N}}(\underline{\bm r};\,\underline{\bm R}
+\underline{\bbox\rho};\,t)\,\,d{\bm\rho}\,,
\label{pdfrel}
\qqq where $\underline{\bm\rho}=({\bm\rho},\dots,{\bm\rho})$. The PDF
in (\ref{pdfrel}) describes the distribution of the particle
separations $ {\bm R}_{nm}={\bm R}_{n}-{\bm R}_{m}$ or of the
relative positions $\widetilde{\underline{\bm R}}=({\bm R}_{_1}-
\overline{\bm R},\dots,.{\bm R}_{_N}-\overline{\bm R})$.

As for two particles, we expect that when $\kappa\to0$ the
multiparticle Green functions ${\cal P}_{_{N}}$ tend to (possibly
distributional) limits that we shall denote by the same symbol. The
limiting PDF's are again expected to show a different short-distance
behavior for smooth and nonsmooth velocities. For smooth velocities,
the existence of deterministic trajectories leads for $\kappa=0$ to
the collapse property \qq \lim\limits_{{\bm r}_{_N}\to{\bm
r}_{_{N-1}}}\,{\cal P}_{_N} (\underline{\bm r};\,\underline{\bm
R};\,t)\ =\ {\cal P}_{_{N-1}} (\underline{\bm r}';\,\underline{\bm
R}';\,t)\ \, {\bbox\delta}({\bm R}_{_{N-1}}-{\bm R}_{_N}),
\label{colla}
\qqq where $\underline{\bm R}'=({\bm R}_{_1},\dots,{\bm
R}_{_{N-1}})$ and similarly for the relative PDF's. If all the
distances among the particles are much smaller than the viscous
cutoff, the velocity differences are approximated by linear
expressions and \qq \widetilde{\cal P}_{_N}(\underline{\bm
r};\,\underline{\bm R};\,t)\ =\
\int\Big\langle\,\prod\limits_{n=1}^N{\bm\delta}({\bm R}_n+{\bm\rho}
-W(t)\,{\bm r}_n)\,\Big\rangle\,\,d{\bm \rho}\,.
\label{raex}
\qqq The evolution matrix $W(t)$ was defined in (\ref{disp6}) and
the above PDF's clearly depend only on its statistics which 
has been discussed in Sect.\,\ref{sec:smooth}.

\subsubsection{Multiparticle motion in Kraichnan velocities}
\label{sec:multKr}
The great advantage of the Kraichnan model is that the statistical
Lagrangian integrals of motion can be found as zero modes of explicit
evolution operators.  Indeed, the crucial simplification lies in the
Markov character of the Lagrangian trajectories due to the velocity
time decorrelation. In other words, the processes $\underline{\bm
R}(t)$ and $\widetilde{\underline{\bm R}}(t)$ are Markovian and
the multiparticle Green functions ${\cal P}_{_{N}}$ and
$\widetilde{\cal P}_{_N}$ give, for fixed $N$, their transition
probabilities.  The process $\underline{\bm R}(t)$ is
characterized by its second-order differential generator ${\cal
M}_{_N}$, whose explicit form may be deduced by a straightforward
generalization of the path integral representation (\ref{hk}) to $N$
particles.  The PDF ${\cal P}_{_{N}}(\underline{\bm
r};\,\underline{\bm R};\,t) \ =\ {\rm e}^{\,\vert t-s\vert\,{\cal
M}_{_N}} (\underline{\bm r};\,\underline{\bm R})\,$ with \qq {\cal
M}_{_N}\ =\ \sum\limits_{n,m=1}^ND^{ij}({\bm r}_{nm})
\,\nabla_{r_{n}^i}\nabla_{r_{m}^j}\,+\,\kappa\sum\limits_{n=1}^N
\,{\bm\nabla}^2_{{\bm r}_n}\,.
\label{mN}
\qqq For the relative process $\widetilde{\underline{\bm R}}(t)$,
the operator ${\cal M}_{_N}$ should be replaced by its
translation-invariant version \qq \widetilde{\cal M}_{_N}\ =\
-\sum\limits_{n<m}\left(d^{ij}({\bm r}_{nm})
+2\kappa\,\delta^{ij}\right)\,\nabla_{r_{n}^i} \nabla_{r_m^j},
\label{mNrel}
\qqq with $\,d^{ij}$ related to $D^{ij}$ by (\ref{sfe1}).  Note the
multibody structure of the operators in (\ref{mN}) and
(\ref{mNrel}). The limiting PDF's obtained for $\kappa\to 0$ define
the heat kernels of the $\kappa=0$ version of the operators that are
singular elliptic and require some care in handling (Hakulinen, 2000).

As we have seen previously, the Kraichnan ensemble may be used to
model both smooth and H\"{o}lder continuous velocities.  In the first
case, one keeps the viscous cutoff $\eta$ in the two-point
correlation (\ref{sfe}) with the result that $d^{ij}({\bm r})={\cal
O}(r^2)$ for $r\ll\eta$ as in (\ref{shortstrain2}), or one sets
$\xi=2$ in (\ref{tobe1}).  The latter is equivalent to the
approximation (\ref{raex}) with $W(t)$ becoming a diffusion
process on the group $SL(d)$ of unimodular matrices, with an
explicitly known generator, as discussed in
Sect.\,\,\ref{sec:solv}(i). The right hand side of (\ref{raex}) may
then be studied by using the representation theory (Shraiman and
Siggia, 1995 and 1996; Bernard {\it et al.}, 1998), see also
Sect.\,\,\ref{sec:pert} below. It exhibits the collapse property
(\ref{colla}).

 From the form (\ref{mNrel}) of the generator of the process
$\widetilde{\underline{\bm R}}(t)$ in the Kraichnan model, we
infer that $N$ fluid particles undergo an effective diffusion with
the diffusivity depending on the interparticle distances. In the
inertial interval and for a small molecular diffusivity $\kappa$, the
effective diffusivity scales as the power $\xi$ of the interparticle
distances.  Comparing to the standard diffusion with constant
diffusivity, it is intuitively clear that the particles spend longer
time together when they are close and separate faster when they become
distant.  Both tendencies may coexist and dominate the motion of
different clusters of particles. It remains to find a more analytic
and quantitative way to capture those behaviors. The effective
short-distance attraction that slows down the separation of close
particles is a robust collective phenomenon expected to be present
also in time-correlated and non-Gaussian velocity fields. We believe
that it is responsible for the intermittency of scalar fields
transported by high Reynolds number flows, as it will be discussed in
the second part of the review.

As for a single particle, the absolute motion of $N$ particles is
dominated by velocity fluctuations on scales of order $L$. \,In
contrast, the relative motion within the inertial range is
approximately independent of the velocity cutoffs and it is convenient
to take directly the scaling limit $\eta=0$ and $L=\infty$. \,We
shall also set the molecular diffusivity to zero.  In these limits,
$\widetilde{\cal M}_{_N}$ has the dimension $length^{\xi-2}$,
implying that $time$ scales as $length^{2-\xi}$ and \qq
\widetilde{\cal P}_{_N}(\lambda\underline{\bm r};\,\underline{\bm
R};\,t) \ =\ \lambda^{-(N-1)d}\,\,\widetilde{\cal
P}_{_N}(\underline{\bm r}; \,\lambda^{-1}\underline{\bm
R};\,\lambda^{\xi-2}t)\,.
\label{Nscal}
\qqq The relative motion of $N$ fluid particles may be tested by
tracing the time evolution of the Lagrangian averages \qq
\Big\langle\,f\left(\underline{\bm R}(t)\right)\,\Big\rangle\ = \ \int
f(\underline{\bm R})\ \,\widetilde{\cal P}_{_{N}} (\underline{\bm
r};\,\underline{\bm R};\,t)\ d\underline{\bm R}'
\label{tav}
\qqq of translation-invariant functions $f$ of the simultaneous
particle positions.  Think about the evolution of $N$ fluid
particles as that of a discrete cloud of marked points in physical
space.  There are two elements in the evolution of the cloud:
the growth of its size and the change of its shape.  We shall define
the overall size of the cloud as $R=({1\over N}
\sum\limits_{n<m}R_{nm}^{\,2})^{1/2}$ and its ``shape'' as
$\underline{\widehat{\bm R}}=\widetilde{\underline{\bm R}}/R$.
\,For example, three particles form a triangle in space (with labeled
vertices) and the notion of shape that we are using includes the
rotational degrees of freedom of the triangle.  The growth of the size
of the cloud might be studied by looking at the Lagrangian average of
the positive powers $R^{^\zeta}$. More generally, \,let $f$ be
a scaling function of dimension $\zeta$, \,i.e.\,\,such that
$f(\lambda\underline{\bm R}) =\lambda^\zeta\,f(\underline{\bm R})$.
The change of variables $\underline{\bm R}\mapsto
t^{1\over2-\xi}\underline{\bm R}$, the relation (\ref{Nscal}) and
the scaling property of $f$ allow to trade the Lagrangian PDF
$\widetilde{\cal P}_{_N}$ in (\ref{tav}) for that at unit time
$t^{\zeta\over2-\xi}\,\widetilde{\cal P}_{_{N}}
(t^{-{1\over2-\xi}}\underline{\bm r};\,\underline{\bm R},\,1)$.  \,As
for two points, the limit of $\widetilde{\cal P}_{_N}$ when the
initial points approach each other is nonsingular for nonsmooth
velocities and we infer that \qq \Big\langle\,f\left(\underline{\bm
R}(t)\right)\,\Big\rangle\ =\ t^{\zeta\over2-\xi}\,\int
f(\underline{\bm R})\ \, \widetilde{\cal P}_{_{N}}(\underline{\bm
0};\,\underline{\bm R};\,1) \ \,d\underline{\bm R}'\,\ +\
\,o(t^{\zeta\over2-\xi})\,.
\label{scalcol}
\qqq In particular, we obtain the $N$-particle generalization 
of the Richardson-type behavior (\ref{ene}):
$\langle\,R(t)^\zeta\,\rangle\propto\,\hspace{0.05cm}
t^{\zeta\over2-\xi}$. \,Hence, in the Kraichnan model the size of the
cloud of Lagrangian points grows superdiffusively
$\propto\,t^{1\over2-\xi}$. \,What about its shape?

\subsubsection{Zero modes and slow modes}
\label{sec:0slow}

In order to test the evolution of the shape of the cloud one might
compare the Lagrangian averages of different scaling functions. The
relation (\ref{scalcol}) suggests that at large times they all scale
dimensionally as $t^{\zeta\over{2-\xi}}$. \,Actually, all do but
those for which the integral in (\ref{scalcol}) vanishes. The latter
scaling functions, whose evolution violates the dimensional
prediction, may thus be better suited for testing the evolution of the
shape of the cloud. Do such functions exist?  Suppose that $f$ is
a scaling function of non-negative dimension $\zeta$ annihilated by
$\widetilde{\cal M}_{_N}$, \,i.e.\,\,such that $\widetilde{\cal
M}_{_N}f=0$. Its Lagrangian average, rather than obeying the
dimensional law, does not change in time: $
\Big\langle\,f\left(\underline{\bm R}(t)\right)\Big\rangle\ =\
f({\underline{\bm r}})$. Indeed, $\partial_t \widetilde{\cal
P}_{_{N}} (\underline{\bm r};\,\underline{\bm R};\,t)=\widetilde{\cal
M}_{_N} \widetilde{\cal P}_{_{N}} (\underline{\bm r};\,\underline{\bm
R};\,t)$. Therefore, the time-derivative of the right hand side of
(\ref{tav}) vanishes since it brings down the (Hermitian) operator
$\widetilde{\cal M}_{_N}$ acting on $f$. \,Thus, the zero modes of
$\widetilde{\cal M}_{_N}$ are conserved in mean by the Lagrangian
evolution. The importance of such conserved modes for the transport
properties by $\delta$-correlated velocities has been recognized
independently by Shraiman and Siggia (1995), Chertkov {\it et al.}
(1995b), and Gaw\c{e}dzki and Kupiainen (1995 and 1996).

The above mechanism may be easily generalized (Bernard {\it et al.},
1998). Suppose that $f_k$ is a zero mode of the $(k+1)^{\rm st}$
power of $\widetilde{\cal M}_{_N}$ (but not of a lower one), with
scaling dimension $\zeta+ (2-\xi)k$.  Then, its Lagrangian average is
a polynomial of degree $k$ in time since its $(k+1)^{\rm st}$ time
derivative vanishes. Its temporal growth is slower than the
dimensional prediction $t^{{\zeta\over2-\xi}+k}$ if $\zeta>0$ so that
the integral coefficient in (\ref{scalcol}) must vanish. We shall call
such scaling functions slow modes. The slow modes may be organized
into ``towers'' with the zero modes at the bottom\footnote{Note that
$(\widetilde{\cal M}_{_N})^{k}f_k$ is a zero mode of scaling
dimension $\zeta$.}.  One descends down the tower by applying the
operator $\widetilde{\cal M}_{_N}$ which lowers the scaling dimension
by $(2-\xi)$. The zero and the slow modes are natural candidates for
probes of the shape evolution of the Lagrangian cloud.  There is an
important general feature of those modes due to the multibody
structure of the operators: the zero modes of $\widetilde{\cal
M}_{_{N-1}}$ are also zero modes of $\widetilde{\cal M}_{_{N}}$ and
the same for the slow modes. Only those modes that depend
non-trivially on all the positions of the $N$ points may of course
give new information on the $N$-particle evolution which cannot be
read from the evolution of a smaller number of particles. We shall
call such zero and slow modes irreducible.

To get convinced that zero and slow modes do exist, let us first
consider the limiting case $\xi\to0$ of very rough velocity
fields. In this limit $d^{ij}({\bm r})$ reduces to
$D_1(d-1)\delta^{ij}$, \,see (\ref{tobe1}), and the operator
$\widetilde{\cal M}_{_N}$ becomes proportional to
$\underline{\bm\nabla}^2$, \,the $(Nd)$-dimensional Laplacian
restricted to the translation-invariant sector. The relative motion of
the particles becomes a pure diffusion. If $R$ denotes the
size-of-the-cloud variable then \qq \underline{\bm\nabla}^2\ =\
R^{-d_{_N}+1}\,\partial_{_R}\,
R^{\,d_{_N}-1}\,\partial_{_R}\,+\,R^{-2}\,\underline{\widehat{\bm\nabla}}^2\,,
\label{lapdec}
\qqq where $d_{_N}\equiv(N-1)d$ and
$\underline{\widehat{\bm\nabla}}^2$ is the angular Laplacian on
the $(d_{_N}-1)$-dimensional unit sphere of shapes
$\underline{\widehat{\bm R}}$. The spectrum of the latter may be
analyzed using the properties of the rotation group.  Its eigenvalues
are $\,-j(j+d_{_N}-2),$ \,where $j=0,1,\dots \,$ is the angular
momentum. The functions $f_{j,0}=R^j\phi_{j}(\underline{\widehat {\bm
R}})$, where $\phi_{j}$ is an angular momentum $j$ eigenfunction, are
zero modes of the Laplacian with scaling dimension $j$. The
contributions coming from the radial and the angular parts in
(\ref{lapdec}) indeed cancel out.  The polynomials
$f_{j,k}=R^{2k}\,f_{j,0}$ form the corresponding (infinite) tower of
slow modes. All the scaling zero and slow modes of the Laplacian are
of that form. The mechanism behind the special behavior of their
Lagrangian averages is that mentioned at the beginning of the section.
Beside the constant, the simplest zero mode has the form of the
difference ${\bm R}_{12}^{\,2}-{\bm R}_{13}^{\,2}$. Both terms are
slow modes in the tower of the constant zero mode and their Lagrangian
averages grow linearly in time with the same leading
coefficient. Their difference is thus constant. A similar mechanism
stands behind the next example, the difference $2(d+2){\bm
R}_{12}^{\,2}{\bm R}_{34}^{\,2}-{d}({\bm R}_{12}^{\,4} +{\bm
R}_{34}^{\,4})$, whose Lagrangian average is conserved due to the
cancellation of linear and quadratic terms in time, and so on.

It was argued by Bernard {\it et al.} (1998) that the slow modes exist
also for general $\xi$ and show up in the asymptotic behavior of the
multiparticle PDF's when the initial points get close: \qq
\widetilde{\cal P}_{_N}(\lambda\underline{\bm r};\,\underline{\bm
R};\,t) \ =\
\sum\limits_{{a}}\sum\limits_{k=0}^\infty\,\,\lambda^{\zeta_{a}
+(2-\xi)k}\ f_{{a},k}(\underline{\bm r})\ \,g_{{a},k}(\underline{ \bm
R},t),
\label{asex}
\qqq for small $\lambda$. \,The first sum is over the zero modes
$f_{{a},0}\equiv f_{{a}}$ with scaling dimensions $\zeta_{a}$, while
higher $k$ give the corresponding towers of slow modes. The functions
in (\ref{asex}) may be normalized so that $f_{{a},k-1}=\widetilde{\cal
M}_{_N}f_{{a},k}$ and $g_{a,k+1}=\widetilde{\cal M}_{_N}g_{a,k}
=\partial_tg_{a,k}$. \,The leading term in the expansion comes from the
constant zero mode $f_{0,0}=1$. The corresponding $g_{0,0}$
coincides with the PDF of $N$ initially overlapping particles. The
asymptotic expansion (\ref{asex}) is easy to establish for vanishing
$\xi$ and for $N=2$ with arbitrary $\,\xi\,$. In the general case, it
has been obtained under some plausible, but yet unproven,
regularity assumptions. Note that, due to (\ref{Nscal}), the expansion
(\ref{asex}) describes also the asymptotics of the multiparticle PDF's
when the final points get far apart and the times become large. The
use of (\ref{asex}) allows to extract the complete long-time
asymptotics of the Lagrangian averages: \qq
\Big\langle\,f\left(\underline{\bm R}(t)\right)\Big\rangle\ = \
\sum\limits_{{a}}\sum\limits_{k=0}^\infty\,\,
t^{{\zeta-\zeta_{a}\over2-\xi}-k}\ f_{{a},k}(\underline{\bm r})\, \int
f(\underline{\bm R})\ g_{{a},k}(\underline{\bm R},1) \ d\underline{\bm
R}'\,\,,
\label{detail}
\qqq which is a detailed refinement of (\ref{scalcol}), corresponding
just to the first term. Note that the pure polynomial-in-time behavior
of the Lagrangian averages of the slow modes implies partial
orthogonality relations between the slow modes and the $g$ modes.

\subsubsection{Shape evolution}
\label{sec:shape}

The qualitative mechanism behind the preservation of the Lagrangian
average of the zero modes is the compensation between its increase due
to the size growth and its depletion due to the shape evolution. The
size and the shape dynamics are mixed in the expansion (\ref{asex})
that, together with (\ref{Nscal}), describes the long-time
long-distance relative evolution of the Lagrangian cloud. To get more
insight into the cooperative behavior of the particles and the
geometry of their configurations, it is useful to separate the shape
evolution following Gat and Zeitak (1998), see also (Arad and
Procaccia, 2000).  The general idea is to trade time in the relative
$\,N$-particle evolution $\,\widetilde{\underline{\bm R}}(t)\,$ for
the size variable $\,R(t)$. \,This may be done as follows. Let us
start with $N$ particles in a configuration of size $r$ and shape
$\underline{\widehat{\bm r}}$.  Denote by $\,\underline{\widehat{\bm
R}}(R)\,$ the shape of the particle configuration the first time its
size reaches $R\ge r$. Varying $R$, one obtains a description of the
evolution of the shape of the particle cloud with its size (which may
be discontinuous if the size does not grow at all moments of
time). For scale invariant velocity fields, the PDF of the shapes
$\,\underline{\widehat{\bm R}}(R)$, i.e.\,\,of the first passages of
$\,\widetilde{\underline{\bm R}}(t)\,$ through the sphere of size $R$,
depends only on the ratio $R/r$. We shall denote it by
$\,P_{_N}(\widehat{\underline{\bm r}}; \underline{\widehat{\bm
R}};\,r/R)\,$.  The shape evolution $\,\underline{\widehat{\bm
R}}(R)\,$ is still a Markov process: in order to compute the
probability of the first passage through the sphere of size $\,R$,
\,one may condition with respect to the first passage through a sphere
of an intermediate size. As a result, the PDF's $\,P_{_N}\,$ obey a
semigroup Chapman-Kolmogorov relation.  As observed by Gat and Zeitak
(1998), the eigenmode expansion of the (generally non-Hermitian)
Markov semigroup $\,P_{_N}(\lambda)\,$ involves the zero modes
$\,f_a\,$: \qq P_{_N}(\underline{\widehat{\bm
r}};\underline{\widehat{\bm R}}; \lambda)\ =\
\sum\limits_{a}\,\,\lambda^{\zeta_{a}}\,\,f_{{a}}
(\underline{\widehat{\bm r}})\,\,h_{{a}}(\underline{\widehat{ \bm
R}})\,.
\label{efex}
\qqq The formal reason is as follows. The statistics of the first passage 
through a given surface may be obtained by imposing the Dirichlet boundary 
condition on the surface in the differential generator of the process. 
For the case at hand, if $\,G_{_N}(\underline{\bm r},
\underline{\bm R})\,$ denotes the kernel of the inverse of 
$\,-\widetilde{{\cal M}}_{_N}\,$ and $\,G_{_N}^{\rm Dir}\,$
refers to its version with the Dirichlet conditions at $\,R=1$, \,then
\qq P_{_N}(\underline{\widehat{\bm
r}};\underline{\widehat{\bm R}};\lambda) \ =\
{_1\over^2}\,\sum\limits_{n,m=1}^N
\Big[\nabla_{_{\hspace{-0.06cm}R_n^i}}\hspace{-0.08cm}R\Big]\,
\Big[\,d^{ij}({\bm R}_{nm})\,\nabla_{_{\hspace{-0.06cm}R^j_m}}
\hspace{-0.1cm}G^{\rm Dir}_{_N}
(\lambda\underline{\widehat{\bm r}},\underline{\bm R})\Big]\,,
\label{elpth}
\qqq with the derivatives taken on the sphere of unit radius. The
potential theory relation (\ref{elpth}) expresses the simple fact that
the probability of a first passage through a given surface is the
normal component of the probability current (the expression in second
parenthesis on the right hand side). On the other hand, by integrating
the asymptotic expansion (\ref{asex}), one obtains the expansion \qq
G_{_N}(\lambda\underline{\bm r},\underline{\bm R})\ = \
\int\limits_0^\infty\widetilde{\cal P}_{_N}(\lambda\underline{\bm
r};\underline{\bm R};t)\,dt=\sum\limits_{a}\,\,\lambda^{\zeta_{a}}
\,\,f_{a}(\underline{\bm r})\, \,\tilde g_{{a}}(\underline{\bm R})\,,
\label{inrs}
\qqq where $\,\tilde g_{{a}}=\int\limits_0^\infty
g_{{a},0}\,dt$. \,Note that the slow modes do not appear since
$g_{{a},k+1}$ is the time derivative of $g_{{a},k}$, that vanishes at
the boundary of the integration interval for nonzero $\,\underline{\bm
R}$. The Dirichlet boundary condition at $R=1$ should not affect the
dependence on the asymptotically close positions
$\lambda\underline{\bm r}$. An expansion analogous to (\ref{inrs})
should then hold for $\,G_{_N}^{\rm Dir}\,$ with the same zero modes
and modified functions $\,\tilde g$. The potential theory formula
(\ref{elpth}) together with the Dirichlet version of (\ref{inrs}) give
(\ref{efex}).

Let us now consider the ``shape only'' version of the Lagrangian
average (\ref{tav}). Substituting the expansion (\ref{efex}), we
obtain \qq \Big\langle f\left(\underline{\widehat{\bm
R}}(R)\right)\hspace{-0.07cm} \Big\rangle\ =\
\sum\limits_a\,\,(R/r)^{-\zeta_a}\,\, f_a(\underline{\widehat{\bm
r}})\,\int\hspace{-0.04cm}f(\underline { \widehat{\bm
R}})\,\,h_a(\underline{\widehat{\bm R}})\,\,d \underline{\widehat{\bm
R}}\,.
\label{shav}
\qqq The interpretation of (\ref{shav}) is simple: when the size
increases, the average of a generic function of the shape relaxes as a
combination of negative powers of $\,R\,$ to a constant, with the zero
modes $\,f_{a}(\underline{\widehat{\bm r}})\,$ giving the modes of
relaxation (Gat and Zeitak, 1998). On the other hand, due to the
orthogonality of the left and right eigenfunctions of the semigroup
$\,P_N(\lambda)$, \,the shape averages of the zero modes decay:
$\,\Big\langle f_{a}\left(\underline{ \widehat{\bm
R}}(R)\right)\hspace{-0.07cm}\Big\rangle=
(R/r)^{-\zeta_{a}}\,f_{{a}}(\underline{\widehat{\bm r}})$.  \,Since
the size grows as $t^{{1\over{2-\xi}}}$, this quantitatively
illustrates the decrease of the shape average responsible for the
conservation in time of the Lagrangian average of the zero modes.  A
vivid and explicit example of the compensation is provided by the case
of three particles. The contour lines of the relevant zero mode as a
function of the shape of the triangle are shown in
Fig.~\ref{fig:threepartcls}. The function tends to decrease for
configurations where all the interparticle distances are comparable.
It is then clear that the decrease in the shape average is simply due
to particle evolving toward symmetrical configurations with aspect
ratios of order unity (Pumir, 1998; Celani and Vergassola, 2001).

The relative motion of the particles in the limit $\,\xi\to0\,$
becomes a pure diffusion. It is then easy to see that the zero modes
of the Laplacian play indeed the role of relaxation modes of the
$\xi=0$ shape evolution. Pure diffusion is the classical case of
potential theory: $\,G_{_N}(\underline {\bm r},\underline{\bm
R})\propto|\underline{\bm r}-\underline{\bm R} |^{-d_{_N}+\,2}$ is the
potential induced at $\,\underline{\bm r}\,$ by a unit charge placed
at $\,\underline{\bm R}\,$ (the absolute value is taken in the sense
of the size variable). The Dirichlet version $\,G^{\rm Dir}_{_N}\,$
corresponds to the potential of a unit charge inside a grounded
conducting sphere and it is obtained by the image charge method:
$\,G^{\rm Dir}_{_N}(\underline{\bm r}, \underline{\bm
R})=G_{_N}(\underline{\bm r},\underline{\bm R})
-R^{-d_{_N}+\,2}\,G_{_N}(\underline{\bm r},\underline{\bm R}/R^2)$.
\,The potential theory formula (\ref{elpth}) gives then for the
shape-to-shape transition probability
$\,\,P_{_N}(\underline{\widehat{\bm r}};\underline{\widehat{\bm R}};
\lambda)\,\propto\,(1-\lambda^2) \,|\underline{\widehat{\bm
R}}-\lambda\hspace{0.05cm}\underline{ \widehat{\bm r}}|^{-d_{_N}}\,$,
with the proportionality constant equal to the inverse volume of the
sphere. On the other hand, calculating $\,G^{\rm Dir}_{_N}\,$ in each
angular momentum sector by (\ref{lapdec}), one easily describes the
generator of the Markov semigroup $\,P_{_N}(\lambda)$.  \,It is the
pseudo-differential Hermitian operator with the same eigenfunctions
$\,\phi_j\,$ as the angular Laplacian
$\,\underline{\widehat{\bm\nabla}}^2\,$ but with eigenvalues $-j$.
The expansion (\ref{efex}) with $\,h_{a}=f_{a}$ follows (recall that
the functions $\,R^j\phi_{j}\,$ form the zero modes of the
Laplacian). The Markov process $\,\underline{\widehat{\bm R}} (R)\,$
lives on distributional realizations (and not on continuous ones).

It is instructive to compare the shape dynamics of the Lagrangian
cloud to the imaginary-time evolution of a quantum mechanical
many-particle system governed by the Hamiltonian
$\,H_{_N}=\sum\limits_n{p_n^2\over2m}
+\hspace{-0.08cm}\sum\limits_{n<m}\hspace{-0.09cm}V({\bm r}_{nm})$.
\,The (Hermitian) imaginary-time evolution operators $\,{\rm
e}^{-t\,H_{_N}}\,$ decompose in the translation-invariant sector as
$\,\sum\limits_{{a}}{\rm e}^{-t\,E_{_{N,{a}}}}
|\psi_{_{N,{a}}}\rangle\langle\psi_{_{N,{a}}}|\,$. The ground state
energy is $\,E_{_{N,0}}$ and the sum is replaced by an integral for
the continuous part of the spectrum. An attractive potential between
the particles may lead to the creation of bound states at the bottom
of the spectrum of $\,H_{_N}$. Breaking the system into subsystems of
$\,N_i\,$ particles by removing the potential coupling between them
would then raise the ground state energy: $\,E_{_{N,0}}<\sum\limits_i
E_{_{N_i,0}}$.  A very similar phenomenon occurs in the stochastic
shape dynamics. Consider indeed an even number of particles and denote
by $\,\zeta_{_{N,0}}\,$ the lowest scaling dimension of the
irreducible zero mode invariant under translations, rotations and
reflections. For two particles there is no invariant irreducible zero
mode and its role is played by the first slow mode $\propto r^{2-\xi}$
and $\,\zeta_{_{2,0}}=2-\xi$. Suppose now that we break the system
into subsystems with an even number $\,N_i\,$ of particles by removing
in $\,\widetilde{\cal M}_{_N}\,$ the appropriate terms $\,d({\bm
r}_{nm})\nabla_{{\hspace{-0.02cm}\bm r}_n}
\hspace{-0.12cm}\nabla_{{\hspace{-0.02cm}\bm r}_m}\,$ coupling the
subsystems, see (\ref{mNrel}).  For $\,N_i\geq4$, the irreducible zero
mode for the broken system factorizes into the product of such modes
for the subsystems. If $N_i=2$ for some $i$, the factorization still
holds modulo terms dependent on less variables. In any case, the
scaling dimensions simply add up. The crucial observation, confirmed
by perturbative and numerical analyses discussed below, is that the
minimal dimension of the irreducible zero modes is raised:
$\,\zeta_{_{N,0}}<\sum\limits_i \zeta_{_{N_i,0}}$. \,In particular,
$\,\zeta_{_{N,0}}$ is smaller than ${N\over2}(2-\xi)$.  One even
expects that $\,\zeta_{_{N,0}}\,$ is a concave function of (even) $N$
and that for $N\gg d$ its dependence on $N$ saturates, see
Sects.\,\,\ref{sec:inst},\,\ref{sec:num},\,\ref{sec:kra}.  By analogy
with the multibody quantum mechanics, we may say that the irreducible
zero modes are bound states of the shape evolution of the Lagrangian
cloud.  The effect is at the root of the intermittency of a passive
scalar advected by nonsmooth Kraichnan velocities, as we shall see in
Sect.\,\ref{sec:passKrai}. It is a cooperative phenomenon exhibiting a
short-distance attraction of close Lagrangian trajectories superposed
on the overall repulsion of the trajectories. There are indications
that similar bound states of the shape evolution persist in more
realistic flows and that they are still responsible for the scalar
intermittency, see Sect.\,\ref{sec:num} and (Celani and Vergassola,
2001).

\subsubsection{Perturbative schemes}
\label{sec:pert}

The incompressible Kraichnan model has two parameters $d\in[2,\infty)$ 
and $\xi\in[0,2]$. It is then natural to ask if the problem is simplified 
at their limiting values and if perturbative methods might be used to 
get the zero modes near the limits. No significant simplification has 
been recognized for $d=2$ at arbitrary $\xi$. The other limits do allow 
for a perturbative treatment since the particle interaction is weak 
and the anomalous scaling disappears there. The perturbation theory is
essentially regular for $\xi\to0$ and $d\to\infty$. Conversely, the
perturbation theory for $\xi\to2$ is singular for two reasons. First,
the advection by a smooth velocity field preserves the configurations
with the particles aligned on a straight line. A small roughness of
the velocity has little effect on the particle motion almost
everywhere but for quasi-collinear geometries. A separate treatment of
those regions and a matching with the regular perturbation expansion
for a general geometry is thus needed. Second, for almost smooth
velocities, close particles separate very slowly and their collective
behavior is masked by this effect which leads to an accumulation of
zero modes with very close scaling dimensions.
We shall start by the more regular cases of small $\,\xi\,$ and large
$\,d$. The scaling of the irreducible four-point zero mode with the
lowest dimension was first calculated to the linear order in $\,\xi\,$
by Gaw\c{e}dzki and Kupiainen (1995) by a version of degenerate
Rayleigh-Schr\"{o}dinger perturbation theory. In parallel, a similar
calculation in the linear order in $1/d$ was performed by Chertkov
{\it et al.}\,\,(1995b). Bernard {\it et al.}\,\,(1996) streamlined
the small $\,\xi\,$ analysis and generalized it to any even order,
following a similar generalization by Chertkov and Falkovich (1996)
for the $1/d$ expansion. We sketch here the main lines of those
calculations.

As we discussed in Section~\ref{sec:0slow}, the operator
$\,\widetilde{\cal M}_{_N}\,$ is reduced to the Laplacian
(\ref{lapdec}) for $\xi=0$.  The zero modes of the 
Laplacian depend on the size of the particle
configuration as $R^j$ and on its shape as the eigenfunctions of
$\,\underline{\widehat{\bm\nabla}}^2\,$ with angular momentum
$j$. The zero modes invariant under $\,d$-dimensional translations, 
rotations and reflections can be reexpressed as polynomials 
in $\,{\bm R}_{nm}^{\,2}$. For even $\,N$, \,the irreducible
zero modes with the lowest scaling dimension have the form \qq
f_{_{N,0}}(\underline{\bm R})\ =\ {\bm R}_{_{12}}^{\,2} \,{\bm
R}_{_{34}}^{\,2} \,\dots\,{\bm R}_{_{(N-1)\,N}}^{\,2}\ +\ \
[\,\dots\,]
\label{00m}
\qqq where $\,\,[\,\dots\,]\,\,$ denotes a combination of terms that
depend on the positions of $\,(N-1)\,$ or less particles.  For four
particles, the zero mode is $\,2(d+2){\bm R}_{12}^{\,2} {\bm
R}_{34}^{\,2}-{d}({\bm R}_{12}^{\,4}+{\bm R}_{34}^{\,4})$, \,our
recurrent example. The terms $\,[\,\dots\,]\,$ are not uniquely
determined since any degree $\,N\,$ zero mode for a smaller number of
points might be added. Furthermore, permutations of the points in
$f_{_{N,0}}$ give other zero modes so that we may symmetrize the above
expressions and look only at the permutation-invariant modes.  The
scaling dimension $\zeta_{_{N,0}}$ of $f_{_{N,0}}$ is clearly equal to
$N$. This linear growth signals the absence of attractive effects
between the particles diffusing with a constant diffusivity (no
particle binding in the shape evolution). As we shall see in
Sect.\,\ref{sec:passKrai}, this leads to the disappearance of the
intermittency in the advected scalar field, that becomes Gaussian in
the limit $\xi\to0$.

To the linear order in $\,\xi$, the operator $\,\,\widetilde{\cal
M}_{_{N}}$ will differ from the Laplacian by a second order
differential operator $-\xi\,V$, involving logarithmic terms $\propto
\ln(r_{nm})$. The zero mode and its scaling dimension are
expanded as $f_0+\xi\,f_1$ and $N+\xi\hspace{0.05cm} \zeta_1$,
respectively. The lowest order term $\,f_0\,$ is given by the
symmetrization of (\ref{00m}).  As usual in such problems, the
degeneracy hidden in $\,[\,\dots\,]\,$ may be lifted by the
perturbation that fixes $f_0$ for each zero mode, see below. At the 
first order in $\,\xi$, \,the equations that define the zero modes 
and their scaling dimension reduce to the relations 
\qq
\underline{\bm\nabla}^2\,f_1\ =\ V\,f_0\,, \qquad\quad
(R\,\partial_{_R}-\,N)\,f_1\ =\ \zeta_1\,f_0\,.
\label{2eqs}
\qqq Given an arbitrary zero mode $\,f_0$, \,one shows that the first
equation admits a solution of the form $\,f_1=h+\sum
\limits_{n<m}h_{nm}\,\ln{(r_{nm})}\,$ with $\,O(d)$-invariant, degree
$N$ polynomials $\,h_{nm}\,$ and $\,h$, the latter being determined up
to zero modes of $\underline{\bm\nabla}^2$. Note that the function
$\,(R\,\partial_{_R}-\,N)\,f_1=\sum\limits_{n<m}h_{nm}\,$ is also
annihilated by the Laplacian: \qq
\label{primordine}
\underline{\bm\nabla}^2\,(R\,\partial_{_R}-\,N)\,f_1\ =\
(R\,\partial_{_R}-\,N+2)\,\underline{\bm\nabla}^2\,f_1 \ =\
(R\,\partial_{_R}-\,N+2)\,V\hspace{0.05cm}f_0\,\,\cr =\
(\,[R\,\partial_{_R},\,V]+2V)\,f_0\,+\,V\,(R\,\partial_{_R} -N)\,f_0\
=\ 0\,. \qqq The last equality follows from the scaling of $\,f_0$ and
the fact that the commutators of $\,R\,\partial_{_R}\,$ with
$\widetilde{\cal M}_{_N}$ and $V$ are $(\xi-2)\,\widetilde{\cal
M}_{_N}$ and $-\,\underline{\bm\nabla}^2\,-\,2\,V$, respectively.  One
obtains this way a linear map $\,\Gamma\,$ on the space of the degree
$N$ zero modes of the Laplacian: $\Gamma\,f_0= (R\,\partial_{_R}-\,N)
\,f_1$. \,The second equation in (\ref{2eqs}) states that $\,f_0\,$
must be chosen as an eigenstate of the map $\,\Gamma$. \,Furthermore,
the function should not belong to the subspace of unit codimension of
the zero modes that do not depend on all the points. It is easy to see
that such subspace is preserved by the map $\,\Gamma$. \,As the
result, the eigenvalue $\,\zeta_1\,$ is equal to the ratio between the
coefficients of $\,{\bm R}_{_{12}}^{\,2}{\bm R}_{_{34}}^{\,2}\dots{\bm
R}_{_{(N-1) \,N}}^{\,2}\,$ in $\,\Gamma f_0\,$ and in $\,f_0$. \,The
latter is easy to extract, see (Bernard {\it et al.}, 1996) for the
details, and yields the result $\,\,\zeta_1=-{N(N+d)\over2(d+2)}\,\,$
or, equivalently, \qq \zeta_{_{N,0}}\ =\ {N\over2}(2-\xi)\ -\
{N(N-2)\over2(d+2)}\,\xi\, \ +\ {\cal O}(\xi^2),
\label{pn12}
\qqq giving the leading correction to the scaling dimension of the
lowest irreducible zero mode. Note that to that order $\zeta_{_{N,0}}$
is a concave function of $\,N$. Higher-order terms in $\xi$ have been
analyzed in (Adzhemyan {\it et al.}, 1998) (the second order) and in
(Adzhemyan {\it et al.}, 2001) (the third order). The latter papers
used a renormalization group resummation of the small $\xi$
perturbative series for the correlation functions of the scalar
gradients in conjunction with an operator product expansion, see
Sec.\,\,\ref{sec:passKrai}.  The expression (\ref{pn12}) may be easily
generalized to the compressible Kraichnan ensemble of compressibility
degree $\,\wp$. The correction $\,\zeta_1\,$ for the tracer exponent
picks up an additional factor $\,(1+2\wp)\,$ (Gaw\c{e}dzki and
Vergassola, 2000). Higher order corrections may be found in (Antonov
and Honkonen, 2000).  The behavior of the density correlation
functions was analyzed in (Adzhemyan and Antonov, 1998; Gaw\c{e}dzki
and Vergassola, 2000; Antonov and Honkonen, 2000).

For large dimensionality $\,d$, \,it is convenient to use the
variables $\,x_{_{nm}}=R_{nm}^{2-\xi}\,$ as the independent
coordinates\footnote{Their values are restricted only by the triangle
inequalities between the interparticle distances.} to make the
$d$-dependence in $\,\widetilde{\cal M}_{_N}\,$ explicit.  Up to
higher orders in $1/d$, the operator $\,\widetilde{\cal
M}_{_N}\propto{\cal L}-{1\over d}\,U$, where $\,{\cal
L}=d^{-1}\sum\limits_{n<m}[(d-1)\partial_{_{x_{_{nm}}}}+
(2-\xi)x_{_{nm}}\partial_{_{x_{_{nm}}}}^{\,2}]\,$ and $\,U\,$ is a
second order $d$-independent differential operator mixing derivatives
over different $\,x_{_{nm}}$. We shall treat $\,\cal L\,$ as the
unperturbed operator and $\,-{1\over d}\,U\,$ as a perturbation. The
inclusion into $\,\cal L\,$ of the diagonal terms $\propto{1\over d}$
makes the unperturbed operator of the same (second) order in
derivatives as the perturbation and renders the perturbative expansion
less singular. The irreducible zero modes of $\,{\cal L}\,$ with the
lowest dimension are given by an expression similar to (\ref{00m}):
\qq f_{_{N,0}}(\underline{\bm R})\ =\ x_{_{12}}
\,x_{_{34}}\,\dots\,x_{_{(N-1)\,N}}\ +\ \ [\,\dots\,]\,,
\label{00d}
\qqq and the permutations thereof. Their scaling dimension is
$\,{N\over2}(2-\xi)$. For $\,N=4$, one may for example take \,$f_{4,0}
=x_{_{12}}x_{_{34}}-{d-1\over2(2-\xi)}\,(x_{_{12}}^2+\,x_{_{34}}^2)$.
\,As in the $\xi$-expansion, in order to take into account the
perturbation $\,U$, \,one has to solve the equations \qq {\cal
L}\,f_1\ =\ Uf_0\,,\qquad\quad \Big(\sum\limits_{n<m}x_{_{nm}}
\partial_{_{x_{_{nm}}}}\hspace{-0.15cm}-\,{_N\over^2}\Big)\,f_1\ =\
{_{\zeta_{_1}}\over^{2-\xi}}\,f_0\,.  \qqq One checks again that
$\,\Gamma f_0\equiv(\sum\limits_{n<m}x_{_{nm}}
\partial_{_{x_{_{nm}}}}\hspace{-0.15cm}-\,{N\over2})\,f_1\,$ is
annihilated by $\,{\cal L}$.  \,In order to calculate $\,\zeta_1$,
\,it remains to find the coefficient of $\,x_{_{12}}\dots
x_{_{(N-1)\,N}}\,$ in $\,\Gamma f_0$.  In its dependence on the
$\,x$'s, the function $\,Uf_0$ scales with power
$\,({N\over2}-1)$. \,One finds $\,f_1\,$ by applying the inverse of
the operator $\,{\cal L}\,$ to it.  $\,\Gamma f_0\,$ is then obtained
by gathering the coefficients of the logarithmic terms in $\,f_1$,
\,see (Chertkov {\it et al.}, 1995b) for the details. When
$\,d\to\infty$, the operator $\,{\cal L}\,$ reduces to the first order
one $\,{\cal L}'\equiv\sum\limits_{n<m}
\partial_{_{x_{_{nm}}}}$. \,This signals that the particle evolution
becomes deterministic at $d=\infty$, with all $\,x_{_{nm}}\,$ growing
linearly in time. If one is interested only in the ${1\over d}$
correction to the scaling exponent and not in the zero mode, then it
is possible to use directly the more natural (but more singular)
decomposition $\,\widetilde{\cal M}_{_N}\propto{\cal L}'-\,{1\over
d}\,U'$.  The leading zero modes of $\,{\cal L}'\,$ also have the form
(\ref{00d}). Noting that ${\cal L}'$ is a translation operator, the
zero mode $\,\Gamma' f_0\,$ may be obtained as the coefficient of the
logarithmically divergent term in $\,
\displaystyle{\int_0^\infty}U'f_0(x_{_{nm}}\hspace{-0.08cm}-t)\,\,dt\,$,
\,\,see (Chertkov and Falkovich, 1996). In both approaches, the final
result is \qq \zeta_{_{N,0}}\ =\ {N\over2}(2-\xi)-{N(N-2)\over 2d}\xi\
+\ {\cal O}({1\over d^2})\,,
\label{pnd2}
\qqq
which is consistent with the small $\xi$ expression (\ref{pn12}).

The non-isotropic zero modes, as well as those for odd $N$, may be
studied similarly. The zero modes of fixed scaling dimension form a
representation of the rotation group $\,SO(d)\,$ which may be
decomposed into irreducible components.  In particular, one may
consider the components corresponding to the symmetric tensor products
of the defining representation of $\,SO(d)\,$, labeled by the angular
momentum $\,j\,$ (the multiplicity of the tensor product). For two
particles, no other representations of $\,SO(d)\,$ appear. The 2-point
operator $\,\,\widetilde{\cal M}_2\,$ becomes in each angular momentum
sector an explicit second-order differential operator in the radial
variable. It is then straightforward to extract the scaling dimensions
of its zero modes: \qq \label{aniso} \zeta_{_{2,0}}^{\,j}\ =\
-{d-2+\xi\over 2}+{1\over 2}\sqrt
{ (d-2+\xi)^2+{4\,(d-1+\xi)\,j\,(j+d-2)\over d-1}}\,. \qqq
Note that $\zeta^{\,1}_{_{2,0}}=1$ in any $d$ corresponding to linear zero 
modes. For the 3-point operator, the lowest scaling dimensions are
$\,\zeta_{_{3,0}}^{0}=4-{2(d-2)\over d-1}\xi +{\cal O}(\xi^2)\,$
(Gat {\it et al.}, 1997a,b) and $\,\zeta_{_{3,0}}^{1}= 3-{d+4\over
d+2}\,\xi+{\cal O}(\xi^2)\,$ (Pumir, 1996 and 1997) or
$\,\zeta_{_{3,0}}^{1}=3-\xi-{2\xi\over d}+{\cal O}({1\over d^2})\,$
(Gutman and Balkovsky, 1996). \,For even $\,N\,$ and $\,j$, Arad, L'vov {\it et
al.} (2000) have obtained the generalization of (\ref{pn12}) in the form 
\qq \zeta_{_{N,0}}^{\,j}\ =\
{N\over2}(2-\xi)\ -\ \bigg({N(N-2)\over2(d+2)}\,-\,
{j(j+d-2)(d+1)\over2(d+2)(d-1)}\bigg)\,\xi\, \ +\ {\cal O}(\xi^2)\,.
\label{pn1l}
\qqq 

The effective expansion parameter in the small $\xi$ or large $\,d\,$
approach turns out to be $\,{N\xi\over(2-\xi)\,d}$ so that neither of
them is applicable to the region of the almost smooth velocity fields.
This region requires a different perturbative technique exploiting the
numerous symmetries exhibited by the multiparticle evolution in the
limiting case $\xi=2$. Those symmetries were first noticed and
employed to derive an exact solution for the zero modes by Shraiman
and Siggia (1996). The expression of the multiparticle operators at
$\xi=2$ reads \qq \widetilde{\cal M}_{_N}\ =\
D_1\left(d\,H^2-(d+1)J^2\right)\,,
\label{dfg}
\qqq with $\,H^2=\sum\limits_{ij}H_{ij}H_{ji}$ and
$\,J^2=-\sum\limits_{i<j} J_{ij}^2$ denoting the Casimir operators of
the group $\,SL(d)\,$ and of its $\,SO(d)\,$ subgroup acting on the
index $\,i=1,\dots,d\,$ of the particle positions $\,r^i_n$. The
corresponding generators are given by
$\,H_{ij}=\sum\limits_n\Big(-r_n^i \nabla_{r_n^j}+\,{1\over
d}\delta_{ij}(r_n^k\nabla_{r_n^k})\Big)\,$ and
$\,J_{ij}=H_{ij}-H_{ji}$. \,The relation (\ref{dfg}), that may be
easily checked directly, is consistent with the expression
(\ref{raex}) for the heat kernel of $\,\widetilde{\cal M}_{_N}$.  As
mentioned in Sect.\,\,\ref{sec:smooth}(i), the right hand side of
(\ref{dfg}) is indeed the generator of the diffusion process
$\,W(t)\,$ on the group $\,SL(d)$.  \,In their analysis, Shraiman and
Siggia (1995) employed an alternative expression for the multiparticle
operators, exhibiting yet another symmetry of the smooth case: \qq
\widetilde{\cal M}_{_N}\ =\
D_1\left(d\,G^2-(d+1)J^2+{_{d-N+1}\over^{N-1}}
\,\Lambda(\Lambda+d_{_N})\right)\,, \qqq where
$\,G^2=\sum\limits_{n,m=1}^{N-1}G_{nm}G_{mn}\,$ is the quadratic
Casimir of $\,SL(N-1)\,$ acting on the index $\,n=1,\dots,N-1\,$ of
the difference variables $\,r^i_{_{nN}}\,$ and
$\,\Lambda=\sum\limits_{i,\,n}r_n^i\nabla_{r_n^i}\,$ is the
generator of the overall dilations. For three points, one may then
decompose the scaling translationally invariant functions into the
eigenfunctions of $\,G^2$, $\,L^2$ and other generators commuting with
the latter and with $\,\Lambda$. \,The zero modes of
$\,\widetilde{\cal M}_{_3}\,$ at $\,\xi=2\,$ have the lowest scaling
dimension equal to unity and vanishing in the angular momentum sectors
$\,j=1\,$ and $\,j=0\,$, see (Pumir {\it et al.}, 1997) and
(Balkovsky {\it et al.}, 1997a).  They have infinite multiplicity since
their space carries an infinite-dimensional representation of
$\,SL(2)$. \,Similar zero modes exist for any higher scaling
dimension. In $2d$, for example, the scaling dimension of a
three-particle zero mode may be raised multiplying it by a power of
$\,\det(r^i_{_{n3}})$. \,The continuous spectrum of the dimensions and
the infinite degeneracy of the zero modes in the smooth case is one
source of the difficulties.  Another related difficulty is that for
$\,\xi=2\,$ the principal symbol of $\,\widetilde{\cal M}_3\,$ looses
positive-definiteness not only when two of the points coincide, but
also when all the three points become collinear. That leads to the
domination for quasi-collinear geometries of the perturbative terms in
$\,\widetilde{ \cal M}_{_3}\,$ over the unperturbed ones.  The problem
requires a boundary layer approach developed first by Pumir {\it et
al.}\,\,(1997) for the $\,j=1\,$ sector with the conclusion that the
minimal scaling dimension of the zero mode behaves as
$\,1+o(2-\xi)$. \,Similar techniques led Balkovsky {\it et
al.}\,\,(1997a) to argue for a $\,{\cal O}(\sqrt{2-\xi})\,$ behavior of
the minimal scaling dimension of the isotropic zero modes.  The
three-particle zero mode equation was solved numerically for the whole
range of values of $\xi$ by Pumir (1997) for $\,j=1\,$ and $\,d=2,3\,$
and by Gat {\it et al.}\,\,(1997a,b) in the isotropic case for
$\,d=2,3,4\,$.  Their results are compatible with the perturbative
analysis around $\,\xi=0\,$ and $\,\xi=2\,$, with a smooth
interpolation for the intermediate values (no crossing between
different branches of the zero modes). Analytical non-perturbative
calculations of the zero modes were performed for the passive scalar
shell models, where the degrees of freedom are discrete. We refer the
interested reader to the original works (Benzi {\it et al.}, 1997;
Andersen and Muratore-Ginanneschi, 1999) and to (Bohr {\it et al.},
1998) for an introduction to shell models.

\section{Passive Fields}
\label{sec:fields}

The results on the statistics of Lagrangian trajectories presented in
Chapter~II will be used here to analyze the properties of passively
advected scalar and vector fields. The qualification ``passive'' means
that we disregard the back-reaction of the advected fields on the
advecting velocity. We shall treat both a scalar per unit
mass (a tracer field), satisfying the equation
\begin{equation}
\partial_t\theta+{\bbox v}\cdot{\bm\nabla}\theta=\kappa{\bm\nabla}^2\theta\ ,
\label{tracer}
\end{equation}
and the density per unit volume, whose evolution is governed by
\begin{equation}
\partial_t n+{\bm\nabla}\cdot(n\,{\bbox v})=\kappa{\bm\nabla}^2n\ .
\label{density}
\end{equation}
For incompressible flows, the two equations are obviously 
coinciding. Examples of passive vector fields are provided by the 
gradient of a tracer ${\bbox \omega}=\nabla\theta$, obeying 
\begin{equation}
\partial_t{\bbox \omega}+{\bm\nabla}({\bbox v}\cdot{\bbox \omega})
=\kappa\nabla^2 {\bbox \omega},
\label{gradient}
\end{equation}
and the divergenceless magnetic field evolving in incompressible flow
according to
\begin{equation}
\partial_t{\bbox B}+{\bbox v}\cdot{\bm\nabla}{\bbox B}-
{\bbox B}\cdot{\bm\nabla}{\bbox v}=\kappa{\bm\nabla}^2
{\bbox B}\ .
\label{magn}
\end{equation}
Two broad situations will be distinguished\,: forced and unforced.
The evolution equations for the latter are
(\ref{tracer})-(\ref{magn}). They will be analyzed in
Sect.\,\,\ref{sec:unforced}. For the former, a pumping mechanism such
as a forcing term is present and steady states might be
established. The rest of the Chapter treats steady cascades of passive
fields under the action of a permanent pumping. As we shall see below,
the advection equations may be easily solved in terms of the
Lagrangian flow, hence the relation between the behavior of the
advected fields and of the fluid particles.  In particular, the
multipoint statistics of the advected fields will appear to be closely
linked to the collective behavior of the separating Lagrangian
particles. An introduction to the passive advection 
problem may be found in (Shraiman and Siggia, 2000).

\subsection{Unforced evolution of passive scalar and vector fields} 
\label{sec:unforced}

The physical situation of interest is that the initial passive field
or its distribution is prescribed and the problem is to determine the
field distribution at a later time $t$. The simplest question to
address is which fields have their amplitudes decaying in time and
which growing, assuming the velocity field to be statistically
steady. A tracer field always decays because of 
dissipative effects,
with the law of decay depending on the velocity properties. The
fluctuations of a passive density may grow in a compressible flow,
with this growth saturated by diffusion after some time.  The
fluctuations of both ${\bbox \omega}$ and ${\bbox B}$ may grow
exponentially as long as diffusion is unimportant. After 
diffusion comes into play, their destinies are different\,: $\,{\bbox
\omega}\,$ decays, while the magnetic field continues to grow.  This
growth is known as dynamo process and it continues until saturated by the
back-reaction of the magnetic field on the velocity.  Another
important issue here is the presence or absence of a dynamic
self-similarity\,: for example, is it possible to present the
time-dependent PDF ${\cal P}(\theta;t)$ as a function of a single
argument?  In other words, does the form of the PDF remain invariant
in time apart from a rescaling of the field? We shall show that for
large times the scalar PDF tends to a self-similar limit when the
advecting velocity is nonsmooth, while self-similarity is broken in
smooth velocities.

\subsubsection{Backward and forward in time Lagrangian description}
\label{sec:back}

If the advecting velocities are smooth and if the diffusive terms are
negligible,\footnote{Recall that the Schmidt number $\nu/\kappa$, also
called Prandtl number when considering temperature or magnetic fields,
is assumed to be large.} the advection equations may be easily solved
in terms of the Lagrangian flow. To calculate the value of a passively
advected field at a given time one has to trace the field evolution
backwards along the Lagrangian trajectories. This is to be contrasted
with the description of the particles in the previous Chapter, which
was developed in terms of the forward evolution.  The tracer
$\,\theta\,$ stays constant along the Lagrangian trajectories: \qq
\theta({\bm r},t)\ =\ \theta({\bm R}(0;{\bm r},t),0),
\label{tracer1}
\qqq where $\,{\bm R}(\,\cdot\,;{\bm r},t)\,$ denotes the Lagrangian
trajectory passing at time $\,t\,$ through the point $\,{\bm r}$. The
density $\,n\,$ changes along the trajectory as the inverse of the
volume contraction factor. Let us consider the matrix $\tilde W(t;{\bm
r})\ =\ W(t;{\bm R}(0;{\bm r},t))\,$, where $\,W(t;{\bm r})\,$ is
given by (\ref{batch0}) and describes the forward evolution of small
separations of the Lagrangian trajectories starting at time zero near
$\bm r$. \,The volume contraction factor is $\,\det(\tilde W(t;{\bm
r}))\,$ and \qq n({\bm r},t)\ =\ [\det(\tilde W(t;{\bm
r}))]^{-1}\,\,n({\bm R} (0;{\bm r},t),0)\,.
\label{density1}
\qqq Note that the matrix $\tilde W(t;{\bm r})\,$ is the inverse of
the backward-in-time evolution matrix $\,W'(t;{\bm r})\,$ with the
matrix elements $\,{\partial R^i(0;{\bm r},t)/\partial r^j}$.  \,This
is indeed implied by the identity $\,{\bm R}(t;{\bm R}(0; {\bm
r},t),0)={\bm r}\,$ and the chain rule for differentiation.  The
solution of the evolution equation for the gradient of the tracer is
obtained by differentiating (\ref{tracer1}): \qq {\bm\omega}({\bm
r},t)\ =\ (\tilde W(t;{\bm r})^{-1})^T\, {\bm\omega}({\bm R}(0;{\bm
r},t),0)\,.
\label{gradient1}
\qqq Finally, the magnetic field satisfies \qq {\bm B}({\bm r},t)\ =\
\tilde W(t;{\bm r})\,\,{\bm B}({\bm R}(0;{\bm r},t),0)\,.
\label{magn1}
\qqq The relations (\ref{tracer1}) to (\ref{magn1}) give, in the absence
of forcing and diffusion, the solutions of the initial value problem
for the advection equations in a given realization of a smooth velocity.

For non-zero $\,\kappa$, \,the solutions of the scalar equations are
given 
essentially by the same expressions. However, $\,{\bm R} (\,\cdot\,;{\bm
r},t)\,$ denotes now a noisy Lagrangian trajectory satisfying the
stochastic equation (\ref{sde}) and passing through ${\bm r}$ at time
$t$ and the right hand sides of the equations
(\ref{tracer1}) to (\ref{magn1}) should be averaged over the noise using
the It\^{o} formula of stochastic calculus discussed in Appendix. 
These solutions may be rewritten using the transition PDF's $\,p({\bm r},s;
{\bm R},t\,|\,{\bm v})\,$ introduced in
Sect.\,\,\ref{sec:nonsmooth}, see (\ref{dtp}) and describing the
probability density to find the noisy particle at time $\,t\,$ at
position $\,{\bm R}\,$, given its time $\,s\,$ position $\,{\bm r}$.
One has \qq \theta({\bm r},t)=\int p({\bm r},t;{\bm R},0\,| \,{\bm
v})\,\,\theta({\bm R},0)\,\,d{\bm R}\,,\quad\ n({\bm r},t)=\int p({\bm
R},0;{\bm r},t\,|\,{\bm v})\,\, n({\bm R},0)\,\,d{\bm R}\,.
\label{trsol}
\qqq The two PDF's appearing in these formulae, one backward and the
other forward in time, coincide for incompressible velocities but they
are generally unequal for the compressible cases.  For nonsmooth
velocities, those PDF's continue to make sense and we shall use
(\ref{trsol}) to define the solutions of the scalar advection
equations in that case. As for the vector fields, their properties
depend both on the noisy Lagrangian trajectory endpoints and on the
matrices $\,\tilde W(t;{\bm r})$, that are well defined only in smooth
velocities. The formal procedure for nonsmooth velocities is to first
impose a viscous cutoff, smoothing the velocity behavior at small
scales, and then removing it.  When this is done, some properties of
the field remain well defined and may be analyzed (see, for instance,
Section~\ref{sec:magnana}).

In random velocity fields, the advected quantities become random fields
whose statistics may be probed by considering the equal-time correlation 
functions. In particular, those of a tracer evolve according to
\qq
C_N(\un{\bm r},t)\,\equiv\,\Big\langle\theta({\bm r}_{_1},t)\,\dots\,
\theta({\bm r}_{_N},t)\Big\rangle\ =\ \int
{\cal P}_{_{N}}(\un{\bm r};\,\un{\bm R};-t)\,\,
\theta({\bm R}_{_1},0)\,\dots\,
\theta({\bm R}_{_N},0)\,\,d\un{\bm R}\,.
\label{thetaN}
\qqq Here, as in Sect.\,\,\ref{sec:nonsmooth}, the Green functions
$\,\,{\cal P}_{_{N}}\,$ are the joint PDF's of the equal-time
positions of $\,N\,$ fluid particles, see (\ref{pn}). For the
correlators of the density $n$, the backward propagator in
(\ref{thetaN}) should be replaced by its forward version, in agreement
with (\ref{trsol}).  If the initial data are random and independent of
the velocities, they may be easily averaged over. For a tracer with a
Gaussian, mean zero initial distribution:
\begin{equation}
\label{gaussiano}
C_{2n}\left(\underline{\bm r},t\right)\ =\ \int {\cal
P}_{2n}\left(\underline{\bm r};\,\underline{\bm R};-t\right)
\Big[\,C_2\left({\bm R}_{12},0\right)\,\ldots\,C_2\left({\bm
R}_{(2n-1)\,2n},0\right)+\,\,\ldots\,\,\Big]\,\,d\underline{\bm R}\,,
\end{equation}
where, according to the Wick rule, the dots stand for the
other pairings of the $2n$ points.

Let us now briefly discuss the compressible case, where the statistics
of the matrices $\,\tilde{W}\,$ and $\,W\,$ generally do not
coincide.  As we have already discussed in Section~\ref{sec:compres},
every trajectory then comes with its own weight determined by the
local rate of volume change and exhibited by the Lagrangian average of
a function $f(\tilde W)$: \qq \int f(\tilde W(t;{\bm r}))\,\,{_{d\bbox
r}\over^V} \,=\,\int f(W(t;{\bm R}))\,\,\det(W(t;{\bm R}))\,\,
{_{d\bbox R}\over^V}\, .\label{con} \qqq The relation in (\ref{con})
simply follows from the definition of $\tilde W$. \,The volume
change factor $\,\det(W)=\exp\sum\rho_i\,$, with the same notation as in
Sect.\,\,\ref{sec:smooth}. Recall that only the average of the
determinant is generally equal to unity for compressible flow.  The
averages of the $SO(d)$-invariant functions of $\,\tilde W\,$ are
described for large times by the large-deviation function $\,\tilde
H=H-\sum\rho_i/t$, with the last term coming from the volume factor.
The corresponding Lyapunov exponents $\tilde{\lambda_i}$ are
determined by the extremum of $\,\tilde H\,$ (Balkovsky {\it et al.},
1999a).  The exponents generally depend on the form of the entropy
function $\,H\,$ and cannot be expressed via the Lyapunov exponents
$\lambda_i$ only.  Since the matrix $\,W'\,$ of the backward evolution
is the inverse of $\,\tilde W$, the backward Lyapunov exponents are
given by $-\tilde\lambda_{d-i+1}$ and not by the na\"{\i}ve guess
$-\lambda_{d-i+1}$.  In particular, the interparticle distance
diverges backward in time with the exponent $\,-\tilde\lambda_d$.  The
same way as we have shown in Sect.\ref{sec:compres} that
$\sum\lambda_i\leq0$ in a compressible flow, one shows that $\sum\tilde
\lambda_i\geq0$ (implying $\tilde \lambda_1\geq0$). For the forward
Lagrangian evolution we thus have an average compression of volumes,
whereas passive fields rather feel an average expansion.  Indeed, as
we go away from the moment where we imposed a uniform Lagrangian
measure, the rate of change of the volume is becoming negative in a
fluctuating compressible flow.

The forward and backward in time Lyapunov exponents coincide if the
statistics of the velocities is time-reversible, i.e.\,\,if $\,{\bm
v}({\bm r},t)\,$ and $\,-{\bm v}({\bm r},-t)\,$ are identically
distributed. More generally, the entire distributions of the forward
and the backward in time stretching rates coincide in that case: \qq
H(\rho_1/t-\lambda_1,\ldots, \rho_d/t-\lambda_d)\ =\
H(-\rho_d/t-\lambda_1,\ldots,-\rho_1/t-
\lambda_d)\,+\,\sum\limits_i\rho_i/t\,.
\label{GK}
\qqq This is an example of the time-reversibility symmetry of the
large deviation entropy function that was described by Gallavotti
and Cohen (1995). The symmetry holds also for a $\delta$-correlated
strain, although the above finite-volume argument (\ref{con}) does not 
apply directly to this case. Recall that in that instance the entropy 
function
(\ref{Hfct}) describes the large deviations of the stretching rates of
the matrix $\,W(t)\,$ given by the It\^{o} version of (\ref{disp6}).
For the inverse evolution, the strain $\,\sigma(s)\,$ should be
replaced by $\,\sigma'(s)=-\sigma(t-s)\,$ and the matrix $\,W'(t)\,$
has the same distribution as $W(t)$.  The matrix $\,\tilde
W(t)=W'(t)^{-1}\,$ is then given by (\ref{disp6}) with the
anti-It\^{o} regularization and the relation between the conventions
(see \ref{difsto12}) implies $\,\tilde W(t)=W(t)\, {\rm
e}^{-2\sum\lambda_it/d}\,$.  Realistic turbulent flows are
irreversible because of the dissipation so that the symmetry
(\ref{GK}), that was confirmed in an experimental situation (Ciliberto
and Laroche, 1998), may be at most approximate.

\subsubsection{Quasi-Lagrangian description of the advection} 
\label{sec:qLag}

Important insights into the advection mechanisms are obtained by
eliminating global sweeping effects and describing the advected fields
in a frame whose origin moves with the fluid. This picture of the
hydrodynamic evolution, known under
the name of quasi-Lagrangian description, 
is intermediate between the
static Eulerian and the dynamic Lagrangian ones (Monin, 1959; 
Belinicher and L'vov, 1987). Specifically,
quasi-Lagrangian fields are defined as\qq \tilde\psi({\bm r},t)\ =\
\psi({\bm r}+{\bm R}(t;{\bm r_0},0),t),
\label{qLag} 
\qqq where $\,\psi\,$ stands for any Eulerian field, scalar or vector,
introduced previously and $\,{\bm R}(t;{\bm r}_0,0)\,$ is the
Lagrangian trajectory passing through $\,{\bm r}_0\,$ at time
zero. The quasi-Lagrangian fields satisfy the same evolution equations
as the Eulerian ones except for the replacement of the advective term
by $\,[\tilde{\bm v} ({\bm r},t)-\tilde{\bm v}({\bm
0},t)]\cdot{\bm\nabla}$.  If the incompressible velocity and the
initial values of the advected field are statistically homogeneous, the
equal-time statistics of the quasi-Lagrangian and the Eulerian fields
coincide. The equal-time statistics is indeed independent of the
initial point $\,{\bm r}_0$. The equality of the equal-time Eulerian
and quasi-Lagrangian distributions follows then by averaging first
over $\,{\bm r}_0$ and then over the velocity and by changing the
variables $\,{\bm r}_0\mapsto{\bm R}(t;{\bm r}_0,0)\,$.  The equality
does not hold if the initial values of the advected fields are
non-homogeneous.

It will be specially convenient to use the quasi-Lagrangian picture
for distances $\,r\,$ much smaller than the viscous scale, i.e.\,\,in
the Batchelor regime (Falkovich and Lebedev, 1994).  The variations in
the velocity gradients may then be ignored so that $\,\tilde{\bm
v}({\bm r},0)-\tilde{\bm v}({\bm 0},t)\approx \sigma(t){\bm r}$.  \,In
this case, the velocity field enters into the advection equations only
through the time dependent strain matrix. For the tracer, one obtains
then the evolution equation \qq \partial_t\tilde\theta+(\sigma\,{\bm
r})\cdot{\bm\nabla}\tilde\theta= \kappa{\bm\nabla}^2\tilde\theta .
\label{qLtracer}
\qqq This may be solved as before using now the noisy Lagrangian
trajectories for a velocity linear in the spatial variables: \qq
\tilde\theta({\bm r},t)\ =\ \overline{ \theta\Big(W(t)^{-1}{\bm
r}-\sqrt{2\kappa}\int_0^t W(s)^{-1}\,d{\bm\beta}(s),\,0\Big)},
\label{ttheta}
\qqq where $\,W(t)\,$ is the evolution matrix of (\ref{disp6}). The
overline denotes the average over the noise which is easily performed
for incompressible velocity fields using the Fourier representation:
\qq &&\tilde\theta({\bm r},t)\ =\ \int\hat\theta(W(t)^T{\bm k},\,0)
\,\,\exp\Big[\,i\,{\bm k}\cdot{\bm r}\,-\,{\bm k} \cdot Q(t)\,{\bm
k}\,\Big]\,\,{_{d{\bm k}}\over^{(2\pi)^d}}\label{qls}\\ &&{\rm
with}\qquad Q(t)\,=\,\kappa \int_0^t W(t)[W(s)^TW(s)]^{-1}W(t)^T\,.
\label{Fourier}
\qqq

\subsubsection{Decay of tracer fluctuations} 
\label{sec:dec}

For practical applications, e.g. in the diffusion of pollution, the
most relevant quantity is the average $\langle \theta({\bm
r},t)\rangle$. It follows from (\ref{thetaN}) that the average
concentration is related to the single particle propagation discussed
in Sect.\,\,\ref{sec:single}. For times longer than the Lagrangian
correlation time, the particle diffuses and $\langle\theta\rangle$
obeys the effective heat equation
\begin{equation}
\partial_t\Big\langle \theta({\bm r},t)\Big\rangle=D^{ij}_e\,\nabla_i
\nabla_j\Big\langle \theta({\bm r},t)\Big\rangle\,,\label{eulspot}
\end{equation}
with the effective diffusivity $D^{ij}_e$ given by (\ref{eddif}).  The
decay of higher-order moments and multipoint correlation functions involves
multiparticle propagation and it is sensitive to the degree of
smoothness of the velocity field. 

The simplest decay problem is that of a uniform scalar spot of size
$\ell$ released in the fluid. Another relevant situation is that where
a homogeneous statistics with correlations decaying on the scale
$\ell$ is initially prescribed.  The corresponding decay problems are
discussed hereafter for the two cases of smooth and nonsmooth
incompressible flow.

\noindent i) {\bf Smooth velocity}. \ Let us consider an initial
scalar configuration given in the form of a single spot of size
$\ell$.  Its average spatial distribution at later times is given by
the solution of (\ref{eulspot}) with the appropriate initial
condition. On the other hand, the decay of the scalar in the spot as
it is carried with the flow corresponds to the evolution of
$\,\tilde\theta\,$, defined by (\ref{qLag}).  We assume the
Schmidt/Prandtl number $\nu/\kappa$ to be large so that the viscous
scale $\eta$ is much larger than the diffusive scale $r_d$. We shall
be considering the $3d$ situation, where the diffusive scale $r_d=
\sqrt{\kappa/|\lambda_3|}$ and $\lambda_3$ is the most negative
Lyapunov exponent defined in Section~\ref{sec:general}.  As shown
there, for times $t\ll t_d=|\lambda_3|^{-1} \ln(\ell/r_d)$, the
diffusion is unimportant and the values of the scalar field inside the
spot do not change.  At later times, the width in the direction of the
negative Lyapunov exponent ${\lambda}_{3}$ is frozen at $r_d$, while
the spot keeps growing exponentially in the other two
directions\footnote{As in Section~II.B, we consider the case of two
non-negative Lyapunov exponents; the arguments are easily modified for
two non-positive exponents.}. The freezing of the contracting
direction at $r_d$ thus results in an exponential growth of the volume
$\propto \exp(\rho_1+\rho_2)$. Hence, the scalar moment of order
$\alpha$
measured at locations inside the spot will decrease as the average of
$\,\exp[-\alpha(\rho_1+\rho_2)]$. The resulting decay laws
$\exp(-\gamma_\alpha t)$ may be calculated using the PDF (\ref{posit})
of the stretching variables $\rho_i$. More formally, the scalar
moments inside the spot are captured by the quasi-Lagrangian
single-point statistics.  Following Balkovsky and Fouxon (1999), let us
take in (\ref{qls}) a Gaussian initial configuration
$\,\hat\theta({\bm k},0)=\exp[-{3\over4}(\ell{\bm k})^2]$. As a result,
\qq \tilde\theta({\bm 0},t)\ =\ \int\exp\Big[-{_3\over^4}\ell^2\, {\bm
k}\cdot I(t){\bm k}\,\Big] \,\,{_{d{\bm k}}\over^{(2\pi)^3}}\ \propto\
\det\, I(t)^{-1/2}\,=\,{\rm e}^{-\sum\rho_i}\,, \qqq where $I(t)$ is
the mean tensor of inertia introduced in Sect.\,\,\ref{sec:general},
see (\ref{I}). Using the PDF (\ref{posit}), one obtains then
\begin{equation}
\Big\langle{\tilde\theta}^{\,^\alpha}(t)\Big\rangle\ \propto\ \int
\exp\left[-\alpha(\rho_1+\rho_2)-tH(\rho_1/t-\lambda_1,\rho_2/t-\lambda_2)
\right ]\,\,d\rho_1d\rho_2\ .\label{spot1}
\end{equation}
At large times, the integral is determined by the saddle point.  At
small $\alpha$, it lies within the parabolic domain of $\,H$ and the
decay rate $\gamma_\alpha$ increases quadratically with the order
$\alpha$. At large enough orders, the integral is dominated by the
rare realizations where the volume of the spot does not grow in time
and the growth rates become independent of the order (Shraiman and
Siggia, 1994; Son, 1999; Balkovsky and Fouxon, 1999).  That conclusion
is confirmed experimentally (Groisman and Steinberg, 2000).

An alternative way to describe the decay of
$\langle{\tilde\theta}^N(t)\rangle$ is to take $N$ fluid particles
that come to the given point ${\bm r}$ at time $t$ and to track them
back to the initial time.  The realizations contributing to the
moments are those for which all the particles were initially inside
the original spot of size $\ell$, see (\ref{ttheta}). Looking backward
in time, we see that the molecular noise splits the particles by small
separations of order ${\cal O}(r_d)$ during a time interval of the
order $t_d=r_d^2/\kappa \approx|\lambda_3^{-1}\vert$ near $t$.  After
that, the advection takes over. The realizations contributing to the
moments are those with the interparticle separations almost orthogonal
to the (backward) expanding direction $\rho_3$ of
$\,\tilde{W}^{-1}$. More exactly, they should form an angle
$\,\lesssim\,(\ell/r_d)\,{\rm e}^{\rho_3}\,$ with the plane orthogonal
to the expanding direction. Such separations occupy a solid angle
fraction of the same order. Since we now track particles moving due to
the advection (the molecular noise is accounted for by the finite
splitting) then $\sum\rho_i=0$ and (\ref{spot1}) follows.

The same simple arguments lead to the result for the case of a random
initial condition with zero mean. Let us first consider the case when
it is Gaussian with correlation length $\ell$. It follows from
(\ref{gaussiano}) that the realizations contributing to
$\,\langle{\tilde\theta}^{^{\,2n}}
\rangle=\langle\theta^{^{\,2n}}\rangle\,$ are those where $n$
independent pairs of particles are separated by distances smaller than
$\ell$ at time $t=0$. The moments are therefore given by
\begin{equation}
\Big\langle \theta^{2n}(t)\Big\rangle\ \propto\ \int 
\exp\left[-n(\rho_1+\rho_2)-tH(\rho_1/t-\lambda_1,\rho_2/t-\lambda_2)
\right]\,\,d\rho_1d\rho_2\ .
\label{spot2}
\end{equation}
Note that the result is in fact independent of the scalar field
initial statistics. Indeed, for a non-Gaussian field we should average
over the Lagrangian trajectories the initial correlation function
$C_{2n}(\underline{\bm R}(0),0)$ that involves a non-connected and a
connected part. The latter is assumed to be integrable with respect to
the $2n-1$ separation vectors (and thus to depend on them). Each
dependence brings an $\exp\left[-\rho_1-\rho_2\right]$ factor and the
connected part will thus give a subleading contribution with respect
to the non-connected one. The above results were first obtained by
Balkovsky and Fouxon (1999) using different arguments (see Sections
III.4 and III.5). Remark the square root of the volume factor
appearing in (\ref{spot2}) as distinct from (\ref{spot1}).  In the
language of spots, this is explained by the mutual cancellations of
the tracer from different spots and the ensuing law of large numbers.
Indeed, different blobs of size $\ell$ with initially uncorrelated
values of the scalar will overlap at time $t$ and the r.m.s. value of
$\,\theta$ will be proportional to the square root of the number of
spots $\propto \exp(\rho_1+\rho_2)$.  The same qualitative conclusions
drawn previously about the decay rates $\gamma_\alpha$ may be obtained
from (\ref{spot2}). In particular, Balkovsky and Fouxon (1999)
performed the explicit calculation for the short-correlated case
(\ref{shortstrain1}). The result is
\begin{eqnarray}&&
\gamma_{\alpha}\,=\,-\lim_{t\to\infty}{1\over t}\ln\Big\langle
|\theta|^{\alpha}\Big\rangle
\ \propto\ \alpha\left(1-\frac{\alpha}{8}\right)\,,
\label{Son}\end{eqnarray}
for $\,\alpha<4\,$ and $\,\gamma_{\alpha}={\rm const.}\,$ for
$\,\alpha>4$.  \,An important remark is that the PDF of the decaying
scalar is not self-similar in a smooth velocity field. The PDF is
indeed becoming more and more intermittent with time, as signaled by
the growth of the kurtosis
$\,\Big\langle|\theta|^\alpha\Big\rangle/\Big\langle\theta^2\Big
\rangle^{\alpha/2}$ for $\alpha>2$. The previous arguments may be
easily generalized to the case of compressible flow.

\noindent ii) {\bf Nonsmooth velocity}. \ For the decay in
incompressible nonsmooth flow, we shall specifically consider the case
of a time reversible Kraichnan velocity field.  The comments on the
general case are reserved to the end of the section.  The simplest
objects to investigate are the single-point moments $\,\Big\langle
\theta^{2n}(t)\Big\rangle\,$ and we are interested in their long-time
behavior $t\gg \ell^{2-\xi}/D_1$. Here, $\ell$ is the correlation
length of the random initial field and $D_1$ enters the velocity
2-point function as in (\ref{tobe1}).  Using (\ref{thetaN}) and
the scaling property (\ref{Nscal}) of the Green function we obtain
\begin{equation}
\label{blues}
\Big\langle \theta^{2n}(t)\Big\rangle\ =\ \int {\cal P}_{2n}
\left(\underline{\bm 0};\,\underline{\bm R};-1\right)\,C_{2n}
\left(t^{1\over2-\xi}\underline{\bm R}
,0\right)\,d\underline{\bm R} .
\end{equation}
There are two universality classes for this problem, corresponding to either
non-zero or vanishing value of the so-called Corrsin integral $J_0=\int
C_2({\bm r},t)\,d{\bm r}$.  Note that the integral is generally
preserved in time by the passive scalar dynamics.

We concentrate here on the case $J_0\neq 0$ and refer the interested
reader to the original paper by Chaves {\it et al.} (2001) for more
details. For $J_0\neq 0$, the function $\,t^{d\over2-\xi}
\,C_2(t^{1\over2-\xi}{\bm r},0)\,$ tends to $\,J_0\,\delta({\bm r})\,$
in the long-time limit and (\ref{blues}) is reduced to
\begin{equation}
\label{blues2}
\langle \theta^{2n}(t)\rangle\ \approx\ (2n-1)!!\,{J_0^n\over t^{nd
\over2-\xi}}\int{\cal P}_{2n}\left(\underline{\bm 0};\,{\bm R}_1,{\bm
R}_1,\ldots {\bm R}_n,{\bm R}_n;-1\right)\,d\underline{\bm R},
\end{equation}
for a Gaussian initial condition. A few remarks are in order. First,
the previous formula shows that the behavior in time is
self-similar. In other words, the single point PDF ${\cal
P}(t,\theta)$ takes the form
$\,t^{d\over2(2-\xi)}Q(t^{d\over2(2-\xi)}\theta)$. That means that the
PDF of $\theta/\sqrt{\bar\epsilon}$ is asymptotically time-independent as 
was hypothetized by Sinai and Yakhot (1989), with 
$\bar\epsilon(t)=\kappa\langle(\nabla\theta)^2\rangle$ being time-dependent
(decreasing) dissipation rate. This should 
be contrasted with the lack of self-similarity found previously 
for the smooth case.
Second, the result is asymptotically independent of the initial
statistics (of course, within the universality class
$J_0\neq 0$). As in the previous subsection, this follows from the
fact that the connected non-Gaussian part of $C_{2n}$ depends on more
than $n$ separation vectors. Its contribution is therefore decaying 
faster than $t^{-{nd\over2-\xi}}$. Third, it follows 
from (\ref{blues2}) that the
long-time PDF, although universal, is generally non-Gaussian. Its
Gaussianity would indeed imply the factorization of the probability
for the $2n$ particles to collapse in pairs at unit time. Due to the
correlations existing among the particle trajectories, this is
generally not the case, except for $\xi=0$ where the particles are
independent. The degree of non-Gaussianity is thus expected to
increase with $\xi$, as confirmed by the numerical simulations
presented in (Chaves {\it et al.}, 2001).

Other statistical quantities of interest are the structure functions
$S_{2n}({\bm r},t)=\langle[ \theta({\bm r},t)-\theta({\bm
0},t)]^{2n}\rangle\,$ related to the correlation functions by
\begin{equation}
S_{2n}({\bm r},t)=\int_0^1\ldots \int_0^1\partial_{\mu_{_1}}\ldots
\partial_{\mu_{_{2n}}} C_{2n}(\mu_{_1}{\bm r},\ldots,\mu_{_{2n}}{\bm r},t)
\,\,\prod d\mu_{_i}\ \equiv\ \Delta({\bm r})\,C_{2n}(\cdot)\,.
\label{soleil}
\end{equation}
To analyze their long-time behavior, we proceed similarly as in
(\ref{blues}) and use the asymptotic expansion (\ref{asex}) to obtain
\qq S_{2n}({\bm r},t)\,&=&\,\int\Delta(t^{-{1\over2-\xi}}{\bm r})\,
{\cal P}_{2n}(\,\cdot\,;\,{\bm R};-1)\,\,\,C_{2n}(t^{1\over2-\xi}
\underline{\bm R},0)\,\,d\underline{\bm R}\cr
\,&\approx&\,{\Delta({\bm r})f_{2n,0}(\cdot)\over
t^{\zeta_{2n,0}\over2-\xi}} \int g_{2n,0}(\underline{\bm
R},-1)\,\,C_{2n}(t^{1\over2-\xi}\underline{\bm
R},0)\,\,d\underline{\bm R}\ \ \propto\,\
\left(r\over\ell(t)\right)^{\zeta_{2n}}
\langle\theta^{2n}\rangle(t)\,. \qqq Here, $f_{2n,0}$ is the
irreducible zero mode in (\ref{asex}) with the lowest dimension and
the scalar integral scale $\ell(t)\propto t^{1\over2-\xi}$. As we
shall see in Sect.\,\,\ref{sec:passKrai}, the quantities
$\,\zeta_{2n}=\zeta_{2n,0}\,$ give also the scaling exponents of the
structure function in the stationary state established in the forced
case.

Let us conclude this subsection by briefly discussing the scalar decay
for velocity fields having finite correlation times. The key
ingredient for the self-similarity of the scalar PDF is the rescaling
(\ref{Nscal}) of the propagator. Such property is generally expected
to hold (at least for large enough times) for self-similar velocity
fields regardless of their correlation times. This has been confirmed
by the numerical simulations in (Chaves {\it et al.}, 2001).  For an
intermittent velocity field the presence of various scaling exponents
makes it unlikely that the propagator possesses a rescaling property
like (\ref{Nscal}). The self-similarity in time of the scalar
distribution might then be broken.

\subsubsection{Growth of density fluctuations 
in compressible flow}
\label{sec:density}

The evolution of a passive density field $n({\bm r},t)$ is governed by
the equation (\ref{density}). In smooth velocities and in the absence
of diffusion, its solution is read from (\ref{density1}), where we
shall take the initial field on the right hand side to be uniform.
This gives $\,n({\bm r},t)=[\det(\tilde W(t;{\bm r}))]^{-1}$.
\,Performing the velocity average and recalling the long-time
asymptotics of the $\tilde W$ statistics, we obtain \qq \langle
n^\alpha(t)\rangle \ \propto\ \int \exp\left[\,(1-\alpha)\sum_i\rho_i
-tH(\rho_1/t-\lambda_1,\ldots,\rho_d/t-\lambda_d) \right]\,\,\prod
d\rho_i\,.  \qqq The moments at long times may be calculated by the
saddle-point method and they are generally behaving as $\propto
\exp(\gamma_\alpha t)$. The growth rate function $\gamma_{\alpha}$ is
convex, due to H\"older inequality, and vanishes both at the origin
and for $\alpha=1$ (by the total mass conservation).  This leads to
the conclusion that $\gamma_\alpha$ is negative for $0<\alpha<1$ and
is otherwise positive: low-order moments decay, whereas high-order and
negative moments grow.  For a Kraichnan velocity field, the large
deviations function $\,H\,$ is given by (\ref{Hfct}) and the density
field becomes lognormal with $\,\gamma_{\alpha} \propto \alpha
(\alpha-1)\,$ (Klyatskin and Gurarie, 1999). Note that the asymptotic
rate $\langle\ln{n}\rangle/t$ is given by the derivative at the origin
of $\,\gamma_\alpha$ and it is equal to $-\sum\tilde\lambda_i\leq0$.
The density is thus decaying in almost any realization if the sum of
the Lyapunov exponents is nonzero.  Since the mean density is
conserved, it has to grow in some (smaller and smaller) regions, which
implies the growth of high moments.  The amplification of negative
moments is due to the growth of low density regions. The positive
quantity $-\sum\lambda_i$ has the interpretation of the mean (Gibbs)
entropy production rate per unit volume. Indeed, if we define the
Gibbs entropy $S(n)$ as $\,-\int (\ln{n})n\,d{\bm r}$ then, by
(\ref{con}), $\,S(n)=\int\ln\,\det(W(t;{\bm r})) \,d{\bm r}$. \,Since
$\,\ln\,\det(W)=\sum\rho_i$, \,the entropy transfered to the
environment per unit time and unit volume is $\,-\sum\rho_i/t\,$ and
it is asymptotically equal to $-\sum\lambda_i>0$, see (Ruelle, 1997).

The behavior of the density moments discussed above is the effect 
of a linear but random hyperbolic stretching and contracting 
evolution (\ref{batch0}) of the trajectory separations. 
In a finite volume, the linear evolution is
eventually superposed with non-linear bending and folding effects.  In
order to capture the combined impact of the linear and the non-linear
dynamics at long times, one may observe at fixed time $t$ the density
produced from an initially uniform distribution imposed at much
earlier times $\,t_0$. When $t_0\to-\infty$ and if $\lambda_1>0$, the
density approaches weakly, i.e.\,\,in integrals against test
functions, a realization-dependent fractal density $n_*({\bm r},t)$ in
almost all the realizations of the velocity. The resulting density
field is the so-called SRB (Sinai-Ruelle-Bowen) measure, see
e.g.\,\,(Kiffer, 1988). The fractal dimension of the SRB measures may
be read from the values of the Lyapunov exponents (Frederikson {\it et
al.}, 1983). For the Kraichnan ensemble of smooth velocities, the
SRB measures were first discussed by (Le Jan, 1985).  In $2d$, they have
a fractal dimension equal to $\,1+{1-2\wp\over1+2\wp}\,$ if
$\,0<\wp<{1\over2}\,$. In $3d$, the dimension is $\,2+{1-3\wp\over
1+2\wp}\,$ if $\,0<\wp\leq{1\over3}\,$ and $\,1+{3-4\wp\over5\wp}\,$
if $\,{1\over3}\leq\wp<{3\over4}$, \,where $\,\wp\,$ is the
compressibility degree.

The above considerations show that, as long as one can neglect
diffusion, the passive density fluctuations grow in a random
compressible flow. One particular case of the above phenomena is the
clustering of inertial particles in an incompressible turbulent flow,
see (Balkovsky {\em et al.}, 2001) where the theory for a general
flow and the account of the diffusion effects that eventually stops
the density growth were presented.

\subsubsection{Gradients of the passive scalar in a smooth velocity}
\label{sec:grad}

For the passive scalar gradients ${\bbox\omega}={\bm\nabla}\theta$ in
an unforced incompressible situation, the equation to be solved is
(\ref{gradient}). The initial distribution is assumed statistically 
homogeneous with a finite correlation length.  As discussed previously,
one may treat diffusion either by adding a Brownian motion to the
backward Lagrangian trajectories or by using the Fourier transform
method (\ref{qls}).  For pedagogical reasons, we choose here the
latter and solve (\ref{gradient}) by simply taking the gradient of the
scalar expression (\ref{qls}).  The long-time limit is independent of
the initial scalar statistics (Balkovsky and Fouxon, 1999) and it is
convenient to take it Gaussian with the 2-point function $\propto
\exp[-{1\over 2d}(r/\ell)^2]$. \,The averaging over the initial
statistics for the generating function ${\cal Z}({\bbox y})=
\left\langle\exp\left[i{\bm y}\cdot {\bm \omega}\right]\right\rangle$
reduces then to Gaussian integrals involving the matrix $\,I(t)\,$ determined 
by (\ref{I}). The inverse Fourier transform is given by another Gaussian
integral over ${\bm y}$ and one finally obtains for the PDF of
$\,{\bbox \omega}\,$:
\begin{eqnarray}&&
{\cal P}({\bm \omega})\ \propto\ \left\langle (\det
I)^{d/4+1/2}\,\exp\left[-{\rm const.}\,\sqrt{\det I}\,({\bbox \omega},
I{\bbox \omega})\right] \right\rangle\,.
\label{Zw}\end{eqnarray}
As may be seen from (\ref{I}), 
during the initial period $t<t_d=|\lambda_d^{-1}|\ln(\ell/r_d)$, the
diffusion is unimportant, the contribution of the matrix
$\,Q\,$ to $\,I\,$ is negligible, the determinant of the latter is
unity and $\omega^2$ grows as the trace of $I^{-1}$.  In other words,
the statistics of $\,\ln \omega\,$ and of $\,-\rho_d\,$ coincide in
the absence of diffusion. The statistics of the gradients can
therefore be immediately taken over from Section~\ref{sec:smooth}.
The growth rate $(2t)^{-1}\langle\ln{\bm \omega}^2\rangle$ approaches
$|\lambda_d|$ while the gradient PDF depends on the entropy function.
For the Kraichnan model (\ref{shortstrain1}), the PDF is lognormal
with the average $\,D_1d(d-1)t\,$ and the variance $\,2D_1(d-1)t\,$
read directly from (\ref{qenf}). This result was obtained by Kraichnan
(1974) using the fact that, without diffusion, $\omega$ satisfies the
same equation as the distance between two particles, whose PDF is
given by (\ref{kk3}).

As time increases, the wavenumbers (evolving as $\,\dot{\bm
k}=\sigma^T{\bm k}\,$) reach the diffusive scale $\,r_d^{-1}$ and the
diffusive effects start to modify the PDF, propagating to lower and
lower values of $\,\omega$. High moments first and then lower ones
will start to decrease. The law of decay at $t\gg t_d$ can be deduced
from (\ref{Zw}).  Considering this expression in the eigenbasis of the
matrix $I$, we observe that the dominant component of ${\bm \omega}$
coincides with the largest eigendirection of the $I^{-1}$ matrix,
i.e.\,\,the one along the $\,\rho_d$ \,axis. Recalling from the
Section~\ref{sec:smooth} that the distribution of $\rho_d$ is
stationary, we infer that $\Big\langle
\vert\omega\vert^{\alpha}(t)\Big\rangle \ \propto\ \left\langle(\det
I)^{-\alpha/4}\right\rangle\,$.  The comparison with (\ref{spot2})
shows that the decay laws for the scalar and its gradients coincide
(Son, 1999; Balkovsky and Fouxon, 1999). This is qualitatively
understood by estimating $\,\omega\sim \theta/\ell_{\rm min}$, \,where
$\,\ell_{\rm min}$ is the smallest size of the spot.  Noting that
$\theta$ and $\,\ell_{\rm min}$ are independent and that $\,\ell_{\rm
min}\approx {\rm e}^{\,\rho_d\,}\ell\,$ at large times has a
stationary statistics concentrated around $r_d$, it is quite clear that
the decrease of $\omega$
is due to the decrease of $\,\theta$.

\subsubsection{Magnetic dynamo}
\label{sec:dyn}

In this Section, we consider the generation of inhomogeneous magnetic
fluctuations below the viscous scale of incompressible turbulence. The
question is relevant for astrophysical applications as the magnetic
fields of stars and galaxies are thought to have their origin in the
turbulent dynamo action (Moffatt, 1978; Parker, 1979; Zeldovich {\it
et al.}, 1983; Childress and Gilbert, 1995). In this problem, the
magnetic field can be treated as passive.  Furthermore, the
viscosity-to-diffusivity ratio is often large enough for a sizable
interval of scales between the viscous and the diffusive cut-offs to
be present. That is the region of scales with the fastest growth rates
of the magnetic fluctuations. Their dynamics, modeled by the passive
advection of magnetic field by a large-scale (smooth) velocity field,
will be described here.

The dynamo process is caused by the stretching of fluid elements
already extensively discussed above and the major new point to be
noted is the role of the diffusion.  In a perfect conductor, when the
diffusion is absent, the magnetic field satisfies the same equation as
the infinitesimal separation between two fluid particles (\ref{disp2}): 
$d{\bm B}/dt=\sigma\,{\bm B}$.  Any chaotic flow would then
produce dynamo, with the growth rate
\begin{equation}
\label{crescita}
\bar\gamma=\lim_{t\to\infty}(2t)^{-1}\langle\ln B^2\rangle,
\end{equation}
equal to the highest Lyapunov exponent $\lambda_1$. Recall that the
gradients of a scalar grow with the growth rate $-\lambda_3$ during
the diffusionless stage. In fact, any real fluid has a nonzero
diffusivity and, even though it can be very small, its effects may be
dramatic. The long-standing problem solved by Chertkov, Falkovich 
{\em et al.} (1999) was whether the presence of a small, yet finite, 
diffusivity could stop the dynamo growth process at large times 
(as it is the case for the gradients of a scalar).

Our starting point is (\ref{magn1}), expressing the magnetic field in
terms of the stretching matrix $\tilde{W}$ and the backward Lagrangian
trajectory. In incompressible flow, matrices $\tilde{W}$ and $W$ are 
identically distributed and we do not distinguish them here. 
For example, the second-order correlation function is given by
\begin{equation}
\label{secsec}
{\cal C}_{2}^{ij}({\bm r}_{12},t)\,\equiv\,\Big\langle B^i({\bm r}_1,t)
\,B^j({\bm r}_2,t)\Big\rangle\ =\ \Big\langle W^{il}\,W^{jm}\,
{\cal C}_{2}^{lm}({\bm R}_{12}(0;{\bm r}_{12},t),0)\Big\rangle\,,
\end{equation}
with the average taken over the velocity and the molecular noise.  For
the sake of simplicity, we assume that the initial statistics of
$\,{\bm B}$ is homogeneous, Gaussian, of zero mean and of correlation
length $\ell$. We concentrate on the behavior at scales
$r_{12}\ll\ell$.  For times less than $t_d=|\lambda_3|^{-1}\ln
(\ell/r_d)$, the Lagrangian separations $R_{12}\ll \ell$ and the
magnetic field is stretched by the $W$ matrix as in a perfect
conductor, see (\ref{secsec}).  For longer times, the separation
$R_{12}$ can reach $\ell$, irrespective of its original value. This is
the long-time asymptotic regime of interest, where the destinies of
the scalar gradients and the magnetic field are different.

It follows from (\ref{secsec}) that the correlations are due to those
realizations where $R_{12}(0)\lesssim\ell$.  As in the
Section~\ref{sec:dec}, the initial separation ${\bm r}_{12}$ should
then be quasi-orthogonal to the expanding direction $\rho_3$ of
$\,W^{-1}$ and the fraction of solid angle occupied by those
realizations is $\,\propto (\ell/r_{12})\,{\rm e}^{\rho_3}$.  Together with
the $\,{\rm e}^{2\rho_1}\,$ factor coming from the perfect
conductor amplification, we thus obtain for the trace of the
correlation function\,:
\begin{equation}
\label{due222}
{\rm tr}\,{\cal C}_{2}({\bm r},t)\ \propto\ \int {\cal
P}(\rho_1,\rho_2,t)\,\,{\rm e}^{\rho_1-\rho_2}\,\,d\rho_1\,d\rho_2,
\end{equation}
with ${\cal P}(\rho_1,\rho_2;t)$ as in (\ref{PDF3}).  The integration
is constrained by $-\ln(\ell/r_{12})\lesssim\rho_2 $, required for the
separation along $\,{\rho_2}$ to remain smaller than $\ell$. Note that
the gradients of a scalar field are stretched by the same $W^{-1}$
matrix that governs the growth of the Lagrangian separations. It is
therefore impossible to increase the
stretching factor of the gradient and keep the particle separation within
the correlation length $\ell$ at the same time. 
That is why diffusion eventually kills all the gradients while
the component ${B}^i$  that points into the direction of stretching 
survives and grows with $\nabla B^i$ perpendicular to it.
This simple picture also explains the absence
of dynamo in $2d$ incompressible flow, where the stretching in one
direction necessarily means the contraction in the other one.

Let us now consider the single-point moments $\langle
B^{2n}\rangle(t)$. The $2n$ particles, all at the same point at time
$t$, are split by the molecular diffusion by small separations of
length ${\cal O}(r_d)$ in a time of the order $t_d=r_d^2/\kappa$ near
$t$. For the subsequent advection not to stretch the separations
beyond $\ell$, the ``diffusive'' separations at time $t-t_d$ should be
quasi-orthogonal to the expanding direction $\rho_3$. More exactly,
they should form an angle $\,\lesssim (\ell/r_d)\,{\rm e}^{\rho_3}$
with the plane orthogonal to the expanding direction. Together with
the pure conductor stretching factor, we are thus left with a
contribution $\,\propto \exp[n(2\rho_1+\rho_3)]$.  Two possible
classes of Lagrangian trajectories should now be distinguished,
depending on whether the angle formed by the ``diffusive'' separations
with the $\rho_2$ direction is arbitrary or constrained to be small
(see also Molchanov {\em et al.}, 1985). For the former, the
contribution is simply given by the average of the expression
$\,\exp[n(\rho_1-\rho_2)]\,$ derived previously, with the constraint
$-\ln \ell/r_d\lesssim\rho_2 $ ensuring the control of the particle
separation along $\rho_2$.  For the latter, the contribution is
proportional to the average of $\,\exp[n(\rho_1+2\rho_2)]$.  Indeed,
the condition of quasi-orthogonality to the $\rho_2$ direction
contributes a $\,n\rho_2$ term in the exponent and the remaining
$\,2n\rho_2$ term is coming from the solenoidality condition ${\bm
\nabla}\cdot{\bm B}=0$. The magnetic field correlation is in fact
proportional to the solenoidal projector and the component stretched
by the $W$ matrix, see (\ref{secsec}), is $\,{\cal C}_{2}^{11}\propto
(1-({\bm\nabla}^2)^{-1} \nabla^2_1){\cal C}(r)$. The realizations
having the particle separations precisely aligned with the $\rho_1$
direction will therefore not contribute. For separations almost
aligned to $\rho_1$, one may show that the square of the angle with
respect to $\rho_1$ appears in ${\cal C}_{2}^{11}$, thus giving the
additional small factor $(\ell/r_d\,{\rm e}^{\rho_2})^2$.

Which one of the two previous classes of Lagrangian trajectories
dominates the moments depends on the specific form of the entropy
function. For the growth rate (\ref{crescita}), the situation is
simpler as the average is dominated by the region around
$\rho_i=\lambda_i$. The average of the logarithm is indeed obtained by
taking the limit $n\to 0$ in $\langle B^{2n}-1\rangle/2n$ and the
saddle point at large times sits at the minimum of the entropy
function.  The two previous classes of Lagrangian trajectories
dominate for positive and negative $\lambda_2$, respectively. Using
the identity $\sum\lambda_i=0$, we finally obtain
\begin{equation}
\label{FINE}
\bar\gamma={\rm min} \{\left(\lambda_1-\lambda_2\right)/2,\,\,
\left(\lambda_2-\lambda_3\right)/2\}.
\end{equation}
The validity of this formula is restricted, first, by the condition
that $\,\ell\exp(\lambda_1t)\,$ is still less than the viscous scale.
The stretching in the third direction also imposes a restriction: the
finiteness of the maximal possible
size $\ell_0$ of the initial fluctuations gives the
constraint $\ell_0\exp(\lambda_3t)>r_d$. 
At larger time, $\langle\log |\bbox B|^2\rangle$ decays.

The most important conclusion coming from (\ref{FINE})
is that the growth rate is always non-negative for a chaotic
incompressible flow. Note
that the growth rate vanishes if two of the Lyapunov exponents
coincide, corresponding to the absence of dynamo for axially symmetric
cases. For time-reversible and two-dimensional flows, the intermediate 
Lyapunov exponent vanishes and $\bar\gamma=\lambda_1/2$. Note that $3d$ 
magnetic field does grow in a $2d$ flow; when, however,  both the flow 
and the field are two-dimensional, one finds $\bar\gamma=-\lambda_1/2$. 
For isotropic Navier-Stokes turbulence, numerical data suggest 
$\lambda_2\approx\lambda_1/4$ (Girimaji and Pope, 1990) and the 
long-time growth rate is then $\bar\gamma\approx 3\lambda_1/8$.

The moments of positive order all grow in a random incompressible flow
with a nonzero Lyapunov exponent. Indeed, the curve $E_n=\ln \langle
B^{2n}\rangle /2t$ is a convex function of $n$ (due to H\"older
inequality) and it vanishes at the origin, where its derivative
coincides by definition with the non-negative growth rate.  Even when
$\bar\gamma=0$, the growth rates for $n>0$ are positive if the entropy
function has a finite width. For $n=1$ this was stated in (Gruzinov
{\em et al.}, 1996). As discussed previously, the behavior of the
growth rate curve $E_n$ is nonuniversal and it depends on the specific
form of the entropy function. For the Kraichnan case, we can use the
result (\ref{qenf}) for the entropy function and the calculation is
elementary.  The dominant contribution is coming from the average of
$\,\exp[n(\rho_1-\rho_2)]\,$ and the $\rho_2$ integration is dominated
by the lower bound $-\ln \ell/r_d$.  The answer $\,E_2=3\lambda_1\,$
was first obtained by Kazantsev (1968). The general result is
$\,E_n=\lambda_1 n(n+4)/4$, \,to be compared with the perfect
conductor result $\,\lambda_1 n(2n+3)/2$. \,The difference between
them formally means that the two limits of large times and small
diffusivity do not commute (what is called ``dissipative
anomaly'', see next Section).  Multipoint correlation functions were
calculated by Chertkov, Falkovich {\it et al.}\,\,(1999).  They reflect 
the prevailing strip structure of the magnetic field. An initially
spherical blob evolves indeed into a strip structure, with the
diffusive effects neutralizing one of the two directions that are
contracted in a perfect conductor.  The strips induce strong angular
dependences and anomalous scalings similar to those described in
Sect.\,\ref{sec:dirlarge} below.

\subsubsection{Coil-stretch transition for polymer molecules in a random flow}

At equilibrium, a polymer molecule coils up into a spongy ball whose
typical radius is kept at $R_0$ by thermal noise.  Being placed in a
flow, such molecule is deformed into an elongated ellipsoid which can
be characterized by its end-to-end extension ${\bf R}$.  
As long as the elongation is much smaller than the
total length of the molecule, the entropy is quadratic in $R$ so that
the molecule is brought  back to its equilibrium shape by a damping
linear in ${\bf R}$.  The equation for the elongation is as
follows (Hinch, 1977)
\begin{eqnarray} &&
\partial_t {\bm R}+{\bm v}\cdot{\bm\nabla}{\bm R} ={\bm
R}\cdot{\bm\nabla}{\bm v}-\frac{1}{\tau} {\bm R} +{\bm \eta}\,.
\label{Tlhs} \end{eqnarray}
The left hand side describes advection of the molecule as a whole, the
first term on the right hand side is responsible for stretching,
$\tau$ is the relaxation time and ${\bm \eta}$ is the thermal noise
with
$\langle\eta^i(t)\eta^j(0)\rangle=\delta^{ij}\delta(t)R_0^2/\tau$.
Since the size of the molecules is always much smaller than the
viscous length then $\nabla{\bm v}=\sigma$ and one can solve
(\ref{Tlhs}) using the evolution matrix $W$ introduced in
Sect.\,~\ref{sec:general}.  At long enough times (when the initial
condition is forgotten) the statistics of the elongation is given by
${\bm R}=\int^{\infty}_0 ds\,W(s){\bm
\eta}(s)\exp\left[-t/\tau\right]$. We are interested in the tail of
the PDF ${\cal P}(R)$ at $R\gg R_0$. The events contributing to it are 
related to the realizations with a long-time history of stretching
where the variable $\rho_1$
(corresponding to the largest Lyapunov exponent $\lambda_1$) is large.
The tail of the PDF is estimated analyzing the behavior of
$R/R_0=\int^{\infty}_0 \exp(\rho_1(s)-s/\tau)\,ds$.   The realizations
dominating the tail are those where $\rho_1(s)-s/\tau$ takes a sharp
maximum at some time $s_*$ before relaxing to its typical
negative values. The probability of those events is read from the
large deviation expression (\ref{PDF3}): $\ln {\cal P}\sim -s_*
H({s_*}^{-1}{\rho_1(s_*)}-\lambda_1)$, where $H$ is the entropy
function. With logarithmic accuracy one can then replace
$\rho_1(s_*)=\ln(R/R_0)+s_*/\tau $ and what is left is just to find
the maximum with respect to $s_*$.  The extremum value $X_*\equiv
(s_*)^{-1}\ln(R/R_0)$ is fixed by the saddle-point condition that
$H-X_*H'$ should vanish at $X_*+{\tau}^{-1}-\lambda_1$. The final
answer for the PDF is as follows
\begin{eqnarray} &&
{\cal P}( R)\propto R_0^\alpha\,\, R^{-1-\alpha}\quad{\rm with}\quad
\alpha=H'\left(X_*+{\tau}^{-1}-\lambda_1\right) \,.\label{powert}
\label{alpha} \end{eqnarray}  
The convexity of the entropy function ensures that
$\alpha$ is positive if $\lambda_1<1/\tau$.  

In accordance with (\ref{powert}), the exponent $\alpha$ decreases
when $\lambda_1$ increases and it tends to zero as
$\lambda_1\to1/\tau$.  In this region, the entropy function is
quadratic and the exponent is expressed via the average value of
$\rho_1$ and its dispersion only:
$\alpha={2}(1-\lambda_1\tau)/\tau{\Delta}$.  The integral of the PDF
diverges at large $R$ as $\alpha$ tends to zero. The transition at
$\lambda_1\to 1/\tau$ is called the coil-stretch transition as the
majority of the polymer molecules got stretched. This stretching can
be stopped by non-linear elastic effects or by the back reaction of
the polymers onto the flow. The understanding of the coil-stretch
transition goes back to the works by Lumley (1972, 1973).  The
power-law tail (\ref{powert}) has been derived by Balkovsky {\em et
al.}  (2000). The influence of non-linear effects on the statistics of
the elongation was examined by Chertkov (2000).

\subsection{Cascades of a passive scalar}
\label{sec:batch}

This Section describes forced turbulence of a passive tracer
statistically stationary in time and homogeneous in space.  We
consider the advection-diffusion equation
\begin{equation}\partial_t\theta+(\bbox v\cdot{\bm\nabla})\theta=
\kappa {\bm\nabla}^2 \theta+\varphi
\label{theta}\end{equation}
with the pumping $\,\varphi\,$ assumed stationary, homogeneous,
isotropic, Gaussian, of zero mean and with covariance \qq
\,\Big\langle\varphi({\bm r},t)\,\varphi({\bm 0},0)\Big\rangle\ =\
\delta(t)\,\Phi(r/\ell)\,.
\label{forzacorr}
\qqq The function $\,\Phi\,$ is taken constant for $r/\ell\leq 1$ and
decaying rapidly for large ratios.  The following considerations are
valid for a pumping finite-correlated in time provided its correlation
time in the Lagrangian frame is much smaller than the stretching time
from a given scale to the pumping correlation scale $\ell$. Note that
in most physical situations the sources do not move with the fluid so
that the Lagrangian correlation time of the pumping is the minimum
between its Eulerian correlation time and $\ell/V$, where $V$ is the
typical fluid velocity. Most of the general features of the advection
are however independent of the details of the pumping mechanism and
its Gaussianity and $\delta$-correlation are not a very serious
restriction, as it will be shown in Section~\ref{sec:passKrai}.

Equation (\ref{theta}) implies for incompressible velocities the
balance relation for the ``scalar energy'' density $\,e=\theta^2/2$: \qq
\partial_te+{\bm\nabla}\cdot{\bm j}=-\epsilon+\phi\,,
\label{enbal}
\qqq where $\epsilon=\kappa({\bm\nabla}\theta)^2$ is the rate of
dissipation, $\phi=\varphi\theta$ is that of energy injection, and
${\bm j}={_1\over^2}\theta^2{\bm v}-\kappa\theta{\bm\nabla} \theta\,$
is the flux density. In a steady state, the injection must be balanced
by the diffusive dissipation, while the stretching and the contraction
by the velocity provide for a steady cascade of the scalar from the
pumping scale $\ell$ to the diffusion scale $r_d$ (where diffusion
is comparable to advection).

The advection-diffusion dynamics induces the Hopf equations
of evolution for the equal-time correlation functions. For a
white-in-time pumping, one obtains \qq
&&\partial_t\,\Big\langle\theta_{_1}\,\dots\, \theta_{_N}\Big\rangle\
+\ \sum\limits_{n=1}^N \Big\langle\theta_{_1}\,\dots\,{\bm v}_{_n}
\cdot{\bm\nabla}_{_n}\theta_{_n}\,\dots\, \theta_{_N}\Big\rangle\,\cr
&&=\,\kappa\sum\limits_{n=1}^N\Big\langle \theta_{1}\,\dots\,
{\bm\nabla}^2_{_n}\theta_{_n}\,\dots\,
\theta_{_N}\Big\rangle\,+\,\sum\limits_{n,m}\Big\langle
\theta_{_1}\mathop{.\dots.}  \limits_{\widehat n\,\ \widehat
m}\theta_{_N}\Big\rangle\,\,\Phi_{_{nm}}
\label{Hopf}
\qqq in the shorthand notation $\,\theta_{_n}\equiv\theta({\bm
r}_{_n},t),$ $\,\Phi_{_{nm}}\equiv\Phi({\bm r}_{nm}/\ell)$
\,etc. These equations are clearly not closed since the left hand side
involves the mixed correlators of the advected fields and the
velocities. An exception is provided by the case of the Kraichnan
ensemble of velocities where the mixed correlators may be expressed in
terms of those containing only the advected fields, see
Sect.\,\,\ref{sec:passKrai} below.  The stationary version of the
2-point Hopf equation may be written in the form \qq
\Big\langle({\bbox v}_1\cdot \nabla_1+{\bbox v}_2\cdot\nabla_2)\,
\theta_1\theta_2\Big\rangle\ +\ 2\,\kappa\,\Big\langle{\bm\nabla}_{_1}
\theta_1\cdot{\bm\nabla}_{_2}\theta_2\Big\rangle\ =\ \Phi_{_{12}}\,.
\label{2pHopf}
\qqq The relative strength of the two terms on the left hand side
depends on the distance. For velocities scaling as $\,\Delta_r v\propto
r^\alpha\,$, the ratio of advection and diffusion terms 
$\,Pe(r)=\Delta_r v\,r /\kappa$ may be
estimated as $r^{\alpha+1}/\kappa$.  \,In particular, 
$\,Pe\equiv Pe(\ell)$ is called the P\'eclet
number, and the diffusion scale $\,r_d\,$ is
defined by the relation $Pe(r_d)=1\,$.

In the ``diffusive interval'' $\,r_{12}\ll r_d\,$, the diffusion term
dominates in the left hand side of (\ref{2pHopf}). Taking the limit of
vanishing separations, we infer that the mean dissipation rate is
equal to the mean injection rate
$\bar\epsilon\equiv\langle\kappa({\bm\nabla}\theta)^2\rangle
={1\over2}\Phi(0)$. This illustrates the aforementioned phenomenon of
the ``dissipative anomaly'': the limit $\kappa\to0$ of the mean
dissipation rate is non-zero despite the explicit $\kappa$ factor in
its definition. The ``convective interval'' $\,r_d\ll r_{12}\ll\ell$
widens up at increasing P\'eclet number. There, one may drop the
diffusive term in (\ref{2pHopf}) and thus obtain
\begin{equation}
\Big\langle({\bbox v}_1\cdot{\bm\nabla}_1+{\bbox
v}_2\cdot{\bm\nabla}_2)
\,\theta_1\theta_2\Big\rangle\,\approx\,\Phi(0)\ .
\label{Yaglom}\end{equation}
The expression (\ref{Yag}) may be derived from the general flux
relation (\ref{Yaglom}) by the additional assumption of isotropy.  The
relation (\ref{Yaglom}) states that the mean flux of $\,\theta^2$ stays
constant within the convective interval and expresses analytically the
downscale scalar cascade. For the velocity scaling $\,\Delta_r
v\propto r^\alpha\,$, dimensional arguments suggest that
$\,\Delta_r\theta\propto r^{(1-\alpha)/2}\,$ (Obukhov, 1949; Corrsin,
1951). This relation gives a proper qualitative understanding that the
degrees of roughness of the scalar and the velocity are complementary,
yet it suggests a wrong scaling for the scalar structure functions of
order higher than the second one, see Sect.\,\,\ref{sec:passKrai}.

Let us now derive the exact Lagrangian expressions for the scalar
correlation functions.
The scalar field along the Lagrangian trajectories ${\bm R}(t)$
changes as
\begin{equation}
\label{eva}
{d\over dt}\,\theta({\bm R}(t),t)=\varphi({\bm R}(t),t).
\end{equation}
The $N$-th order scalar correlation
$\,\,\Big\langle\,\prod\limits_{p=1}^N \theta({\bm
r_n},t)\,\Big\rangle\equiv C_{_N}(\un{\bm r},t)\,\,$ is then given by
\begin{equation}
\label{forzato}
C_{_N}(\,\underline{\bm r},t)=\Big\langle 
\int_0^t\varphi({\bm R}_1(s_{_1}),s_{_1})\,\,ds_{_1}\,\,\ldots\,\int_0^t
\varphi({\bm R}_{_N}(s_{_N}),s_{_N})\,\,ds_{_N}\Big\rangle\,, 
\end{equation}
with the Lagrangian trajectories satisfying the final conditions ${\bm
R}_n(t)={\bm r}_n$. For the sake of simplicity, we have written down
the expression for the case where the scalar field was vanishing at
the initial time. If some of the distances among the particles get
below the diffusive scale, the molecular noises in the Lagrangian
trajectories become relevant and the averaging of (\ref{forzato}) over
their statistics is needed.

The average over the Gaussian pumping in (\ref{forzato}) gives for 
the pair correlation function:
\begin{equation}
C_2({\bm r}_{12},t)\,=\,\left\langle\int_0^t \Phi\left({\bm
R}_{12}(s)/\ell\right)\,ds\right\rangle . \label{Gauss}
\end{equation} 
The function $\Phi$ essentially restricts the integration to the time
interval where $\,R_{12}\,$ is smaller than the injection length
$\ell$. \,If the Lagrangian trajectories separate, the pair
correlation reaches at long times the stationary form given by the
same formula with $\,t=\infty$.  Simply speaking, the stationary pair
correlation function of a tracer is proportional to the average time
that two particles spent in the past within the correlation scale of
the pumping (Falkovich and Lebedev, 1994). Similarly, the pair
structure function $\,S_2(r)=\langle(\theta_1-\theta_2)^2\rangle=
2\left[ C_2({\bm 0})-C_2({\bm r})\right]\,$ is proportional to the
time it takes for two coinciding particles to separate to a distance
$\,r$. \,This is proportional to $\,r^{1-\alpha}$ for a
scale-invariant velocity statistics with $\,\Delta_r v\propto r^\alpha$,
see (\ref{Rich0}), so that $S_2(r)$ is in agreement with the
Obukhov-Corrsin dimensional prediction.

Higher-order equal-time correlation functions are expressed similarly 
by using the Wick rule to average over the Gaussian forcing:
\qq
C_{2n}(\un{\bm r},t)\ =\ 
\Big\langle\int_{0}^t\Phi({\bm R}_{12}(s_1))\,ds_1\,\,\ldots\,
\int_0^t\Phi({\bm R}_{(2n-1)\,2n}(s_n))\, ds_n\Big\rangle\,+\ \ldots\ ,
\label{n-point}
\qqq where the remaining average is over the velocity and the
molecular noise ensemble and the dots stand for the other possible
pairings of the $2n$ points.  The correlation functions may be obtained from
the generating functional \qq \Big\langle\exp\Big[i\int\theta({\bm
r},t)\,\chi({\bm r})\,\,d{\bm r}\Big] \Big\rangle\ =\
\exp\Big[-{_1\over^2}\int_0^t\hspace{-0.07cm}ds\int\int \Phi({\bm
R}_{12}(s))\,\,\chi({\bm r}_1)\,\chi({\bm r}_2)\,\,d{\bm r}_1 d{\bm
r}_2\Big]\,.
\label{genfc}
\qqq 
They probe the statistics of times spent by fluid
particles at distances $R_{ij}$ smaller than $\ell$. In
nonsmooth flows, the correlation functions at small scales, 
$r_{ij}\ll \ell$, are dominated by the single-point contributions, 
corresponding to
initially coinciding particles. This is the signature of the explosive
separation of the trajectories. To pick up a strong dependence on the
positions $\un{\bm r}$, one has to study the structure functions which are
determined by the time differences between different initial
configurations. Conversely, the correlation functions at scales larger
than $\ell$ are strongly dependent on the positions, as it will be
shown in Sect.\,\ref{sec:dirlarge}.

\subsubsection{Passive scalar in a spatially smooth velocity} 
\label{sec:scal}

In the rest of Section \ref{sec:batch}, all the scales are
supposed much smaller than the viscous scale of turbulence so that we
may assume the velocity field to be spatially smooth and use the
Lagrangian description developed in Sects.\,\,\ref{sec:smooth} and
\ref{sec:compres}.  In the Batchelor regime, the backward evolution of
the Lagrangian separation vector is given by $\,{\bm R}_{12}(0)=\tilde
W(t)^{-1} \,{\bm r}_{12}\,$ (if we ignore diffusion) and it is
dominated by the stretching rate $\,\rho_d$ at long times.  The
equation (\ref{Gauss}) takes then the asymptotic form \qq C_2({\bbox
r},t)\ \approx\ \int\limits_{0}^tds\int \Phi\left({\rm
e}^{-\tilde\rho_d(s)}\,r/\ell\right) \,\widetilde{\cal
P}(\rho_1,\ldots,\rho_d;s)\,\,d\rho_1\ldots d\rho_d \,.\label{pair}
\qqq The behavior of the interparticle distance crucially depends on
the sign of $\,\tilde\lambda_d$. For $\tilde\lambda_d<0$, the
backward-in-time evolution separates the particles and leads in the
limit $t\to \infty$ to a well-defined steady state with the
correlation function
\begin{equation}\langle\theta(t,0)\,\theta(t,\bbox r)\rangle\approx
\vert\tilde\lambda_d\vert^{-1}\Phi(0)\ln(\ell/r)\,,\label{log}\end{equation}
for $r\lesssim\ell$. This corresponds to the direct cascade.
Conversely, if $\tilde\lambda_d>0$ the particles contract and the pair
correlation function grows proportionally to $t$. Note that the
growing part is independent of $r$.  This means that, in a flow
contracting backwards in time, tracer fluctuations grow at larger
and larger scales, which is a signature of the so-called inverse 
cascade of a passive tracer.

If the velocity ensemble is time-reversible, as it is the case for the
$\delta$-correlated model (\ref{fastcomp}), then
$\,\tilde\lambda_i=-\lambda_{d-i+1}\,$ and $\lambda_1$ and
$\tilde\lambda_d$ have opposite sign.  They will thus both change sign
at the same value of the degree of compressibility $\,\wp=d/4$, see
(\ref{complambda}).  This is peculiar for a short-correlated case and
does not hold for an arbitrary velocity statistics. There, the change
from stretching to contraction in the forward Lagrangian dynamics does
not necessarily correspond to the change in the direction of the
cascade for the passive tracer, related to the backward in time
Lagrangian dynamics.

\subsubsection{Direct cascade, small scales}
\label{sec:dirsmall} 

We consider here the case $\tilde\lambda_d<0$ (that includes smooth
incompressible flows) when the particles do separate backward in time
and a steady state exists. We first treat the convective interval of
distances between the diffusion scale $r_d$ and the pumping scale
$\ell$. Deep inside the convective interval where $r\ll \ell$, the
statistics of the passive scalar tends to become Gaussian. Indeed, the
reducible part in the $2n$-point correlation function
$\,\Big\langle\Phi({\rm e}^{-\tilde
\rho_d(s_1)}\,r_{12}/\ell)\,\ldots\,\Phi({\rm e}^{-\tilde
\rho_d(t_n)}\,r_{(2n-1)\,2n}/\ell)\Big\rangle$, \,see
(\ref{n-point}) and (\ref{pair}), dominates the irreducible one for
$\,n\ll n_{cr}\approx(\vert\tilde\lambda_d\vert\tau_s)^{-1}
\ln(\ell/r)$. The reason is that the logarithmic factors are smaller
for the irreducible than for the reducible contribution (Chertkov {\it
et al.}, 1995a). The critical order $n_{cr}$ is given by the ratio
between the time for the particles to separate from a typical distance
$r$ to $\ell$ and the correlation time $\,\tau_s\,$ of the stretching
rate fluctuations.  Since $\ell\gg r$, the statistics of the passive
tracer is Gaussian up to orders $n_{cr}\gg 1$. For the single-point
statistics, the scale appearing in the expression of the critical
order $n_{cr}$ should be taken as $r=r_d$.  The structure functions are
dominated by the forced solution rather than the zero modes in the
convective interval: $S_{2n}=\langle[\theta({\bm 0})-\theta({\bm
r})]^{2n}\rangle\propto \ln^n(r/r_d)$ for $n\ll \ln (r/r_d)$. A
complete expression (the forced solution plus the zero modes) for the
four-point correlation function in the Kraichnan model can be found 
in Balkovsky, Chertkov {\it et al.} (1995).

Let us show now that the tails of the tracer PDF decay exponentially
(Shraiman and Siggia, 1994; Chertkov {\it et al.}, 1995a; Bernard {\it
et al.}, 1998; Balkovsky and Fouxon, 1999). The physical reasons
behind this are transparent and most likely they apply also for a
nonsmooth velocity. First, large values of the scalar can be achieved
only if during a long interval of time the pumping works uninterrupted
by stretching events that eventually bring diffusion into play. We are
interested in the tails of the distribution, i.e. in intervals much
longer than the typical stretching time from $r_d$ to $\ell$. Those
rare events can be then considered as the result of a Poisson random
process and the probability that no stretching occurs during an
interval of length $\,t\,$ is $\,\propto\exp(-ct)$.  Second, the
values achieved by the scalar in those long intervals are Gaussian
with variance $\Phi(0) t$. Note that this is also valid for a
non-Gaussian and finite-correlated pumping, provided $t$ is larger
than its correlation time. By integrating over the length of the
no-stretching time intervals with the pumping-produced distribution of
the scalar we finally obtain: $\,{\cal P}(\theta)\propto \int
dt\,\exp(-ct-\theta^2/2\Phi(0) t) \propto\exp(-\theta\sqrt{2c/
\Phi(0)})$. This is valid for $t<L^2/\kappa$ that is for 
$\theta<\sqrt{c\Phi(0)}\,L^2/\kappa$. Interval of exponential behavior thus increases
with the Peclet number. For a smooth case, the calculations have been carried out 
in detail and the result agrees with the previous arguments.
Experimental data in (Jullien {\em et al.}, 2000) confirm both the
logarithmic form of the correlation functions and the exponential
tails of the scalar PDF. In some experimental set-ups the
aforementioned conditions for the exponential tails are not satisfied
and a different behavior is observed, see for example (Jayesh and
Warhaft, 1991).  The physical reason is simple to grasp. The injection
correlation time in those experiments is given by $L/V$, where $L$ is
the velocity integral scale and $V$ is the typical velocity. The
no-stretching times involved in the tail of the scalar distribution
are of the order of $W/V$, where $W$ is the width of the channel where
the experiment is performed. Our previous arguments clearly require
$W\gg L$.  As the width of the channel is increased, the tails indeed
tend to become exponential.

For a $\delta$-correlated strain, the calculation of the generating
functional of the scalar correlators may be reduced to a quantum
mechanical problem. In the Batchelor regime and in the limit of
vanishing $\,\kappa\,$, the exponent in the generating functional
(\ref{genfc}) may indeed be rewritten as \qq -\int_0^t
V_{_\chi}(\tilde
W(s))\,\,ds\,=\,-{_1\over^2}\int_0^t\,ds\,\int\int\Phi(\tilde
W^{-1}(s){\bm r}_{12}) \,\,\chi({\bm r}_1)\,\chi({\bm r}_2)\,\,d{\bm
r}_1d{\bm r}_2\,.
\label{potent}
\qqq Recall that the matrices $\,\tilde W\,$ form a stochastic process
describing a diffusion on $\,GL(d)$ (or on $SL(d)$ in the
incompressible case) with a generator $\,\tilde M$. The above formula
may thus be interpreted as the Feynman-Kac expression for the integral
$\,\displaystyle{\int}{\rm e}^{\,t\,(\tilde M-V_{_\chi})} (1,\tilde
W)\,d\tilde W\,$ of the heat kernel of $\tilde M$ perturbed by the
positive potential $\,V_{_\chi}$. As long as the trajectories separate
backward in time, i.e. for $\,\tilde\lambda_d<0$, the generating
functional has a stationary limit, given by (\ref{potent}) with the
time integral extending to infinity. The Feynman-Kac formula may be
used to find the exponential rate of decay of the PDF $\,{\cal
P}(\theta)\propto{\rm e}^{-b\vert\theta\vert}\,$.  As shown by Bernard
{\it et al.}  (1998), this involves the quantum-mechanical Hamiltonian
$\,-\tilde M-a^2V_{_\chi}\,$, where the positive operator $-\tilde
M\,$ has its spectrum starting at a strictly positive value and the
negative potential tends to produce a bound state as the parameter $a$
is increased. In the incompressible case, the decay rate $b$ is
characterized by the property that the ground state of the Hamiltonian
has zero energy.  For isotropic situations, the potential is
only a function of the stretching rates of $\tilde W$ and the
quantum-mechanical problem reduces to the perturbation of the
Calogero-Sutherland Hamiltonian by a potential, see
Sect.\,\ref{sec:solv}.

\subsubsection{Direct cascade, large scales}
\label{sec:dirlarge}

We consider here the scales $r\gg\ell$ in the steady state established
under the condition $\tilde\lambda_d<0$ (Balkovsky {\it et
al.}, 1999b). From a general physical viewpoint, it is of interest to
understand the properties of turbulence at scales larger than the
pumping scale. A natural expectation is to have there an equilibrium
equipartition with the effective temperature determined by the
small-scale turbulence (Forster {\it et al.}, 1977; Balkovsky, Falkovich 
{\it et al.}, 1995). The peculiarity of our problem is that we consider
scalar fluctuations at scales larger than that of pumping yet smaller
than the correlation length of the velocity field. This provides for
an efficient mixing of the scalar even at those large scales. Although
one can find the simultaneous correlation functions of different
orders, it is yet unclear if such a statistics can be described by any
thermodynamical variational principle.

The correlation functions of the scalar are proportional to the time
spent by the Lagrangian particles within the pumping scale. It follows
that the statistics at $r\gg \ell$ is related to the probabilities of
initially distant particles to come close.  For spatially smooth
random flow, such statistics turns out to be strongly intermittent and
non-Gaussian. Another unexpected feature in this regime is a total
breakdown of scale invariance: not only the scaling exponents are
anomalous and do not grow linearly with the order of the correlation
function, but even fixed correlation functions are generally not scale
invariant. The scaling exponents depend indeed on the angles between
the vectors connecting the points. Note that the large-scale
statistics of a scalar is scale-invariant in a nonsmooth velocity,
see (Balkovsky {\it et al.}, 1999b) and Sect.\,\ref{sec:passKrai}\,v.

What is the probability for the vector ${\bbox R}_{12}(t)$, that was
once within the pumping correlation length $\ell$, to come exactly to
the prescribed value ${\bbox r}$ at time $t$?  The advection makes a
sphere of ``pumping'' volume $\ell^d$ evolve into an elongated ellipsoid
of the same volume.  Ergodicity may be assumed provided that the
stretching time $\bar\lambda^{-1}\ln(r/\ell)$ is larger than the
strain correlation time. It follows that the probability for two
points separated by $r$ to belong to a ``piece'' of scalar originated
from the same pumping sphere behaves as the volume fraction
$(\ell/r)^d$. That gives the law of decrease of the two-point
function: $C_2\propto r^{-d}$.

The advection by spatially smooth velocities preserves straight
lines. To determine the correlation functions of an arbitrary order
when all the points lie on a line, it is enough to notice that the
history of the stretching is the same for all the particles.  Looking
backward in time we may say that when the largest distance among the
points was smaller than $\ell$, then all the other distances were as
well. It follows that the correlation functions for a collinear
geometry depend on the largest distance $r$ among the points so that
$\,C_{2n}\propto r^{-d}$. This is true also when different pairs of
points lie on parallel lines. Note that the exponent is
$n$-independent which corresponds to a strong intermittency and an
extreme anomalous scaling.  The fact that $\,C_{2n}\gg C_2^n$ is due
to the strong correlation of the points along their common line.

The opposite takes place for non-collinear geometries, namely the
stretching of different non-parallel vectors is generally
anti-correlated in the incompressible case due to the volume
conservation.  The ${d}$-volume $\,\epsilon_{i_1i_2\ldots
i_d}\,R_{12}^{i_1}\ldots R_{1d}^{i_d}$ is indeed preserved for $(d+1)$
Lagrangian trajectories ${\bbox R}_n(t)$ and, for $d=2$ and any three
trajectories, the area $\,\epsilon_{ij}\,{\bm R}_{12}^i{\bm R}_{13}^j$
of the triangle defined by the three particles remains constant.  The
anti-correlation due to the area conservation may then be easily
understood and the scaling for non-collinear geometries at $d=2$ may be
determined.  Since the area of any triangle is conserved, three points
that form a triangle with area $\,A\gg \ell^2$ will never come within
the pumping correlation length. In the presence of a triple correlator
$\,\Phi_3$ for a non Gaussian $\delta$-correlated pumping, the triple
correlation function of a scalar
\begin{eqnarray} &&
\Big\langle\theta({\bm r}_1)\,\theta({\bm r}_2)\,\theta({\bm r}_3)
\Big\rangle=\bigg\langle\int_0^{\infty}\Phi_3({\bm R}_{12}(s),
{\bm R}_{13}(s))\,\,ds\bigg\rangle
\label{tri1} \end{eqnarray}
is determined by the asymptotic behavior of $\,\Phi_3$ at $\,r_{ij}
\gg\ell$. \,For example, if $\Phi_3$ has a Gaussian tail, then
$\,C_3\propto\exp(-A/\ell^2)$. On the other hand, the correlation functions
decrease as $r^{-2}$ for a collinear geometry. We conclude that $C_3$
as a function of the angle between the vectors $\,{\bbox
r}_{12}$ and $\,{\bbox r}_{13}$ has a sharp maximum at zero and
rapidly decreases within an interval of width of the order
$\ell^2/r^2\ll 1 $.

Similar considerations apply for the fourth-order correlation
function. Note that, unlike for the 3-point function, there
are now reducible contributions.  Consider, for instance, that coming
from $\,\Big\langle\int\limits_0^\infty \int\limits_0^\infty\Phi({\bm
R}_{12}(s_1))\,\Phi({\bm R}_{34}(s_2)) \,ds_1ds_2\Big\rangle$. Since
the area of the polygon defined by the four particles is conserved
throughout the evolution, the answer is again crucially dependent on
the relation between the area and $\,\ell^2$.  The events contributing
to the correlation function $\,C_4$ are those where, during the evolution,
$\,R_{12}$ became of the order $\ell$ and then, at some other moment
of time, $R_{34}$ reached $\ell$.  The probability for the first event
to happen is $\,\ell^2/r_{12}^2$. When this happens, the area
preservation makes $\,R_{34}\sim r_{12}r_{34}/\ell$. The probability
for this separation to subsequently reduce to $\ell$ is $\propto
\ell^4/r_{12}^2r_{34}^2$. The total probability can be thus estimated
as $\,\ell^6/r^6\,$, where $r$ is the typical value of the separations
$r_{ij}$. Remark that the naive Gaussian estimation $\,\ell^4/r^4\,$
is much smaller than the collinear answer and yet much larger than the
non-collinear one.

The previous arguments can be readily extended to an arbitrary number
of non-collinear pairs.  In accordance with (\ref{n-point}), the
realizations contributing to the correlation are those where the
separations $\,R_{ij}$ reduce down to $\,\ell\,$ during the evolution
process.  Suppose that this happens first for $\,R_{12}$. Such process
was already explained in the consideration of the pair correlation
function and occurs with probability $\,(\ell/r_{12})^2$. All the
remaining separations will then be larger than their initial values by
a factor $\,r_{12}/\ell$, due to the conservation law of the
triangular areas. Next, we should reduce, say, $\,R_{34}$ from
$\,r_{34}r_{12}/\ell\,$ down to the integral scale $\,\ell$. \,Such a
process occurs with probability $\,(\ell^2/(r_{12}r_{34}))^2$. When
this happens, all the other separations are larger than their initial
values by a factor $\,r_{34}/\ell\,$.  Repeating the process, we come
to the final answer $\,C_{2n}\propto(\ell/r)^{4n-2}$, where $r$ is
again the typical value of the separations $r_{ij}$.

The above analysis is easy to generalize for arbitrary geometries. The
points are divided into sets consisting of the pairs of points with
parallel separations ${\bbox r}_{ij}$ (more precisely, forming angles
smaller than $\ell^2/r^2$). The points within a given set behave as a
single separation during the Lagrangian evolution. The order $n$ in
the previous formulae should then be replaced by the (minimal) number
of sets.  The estimates obtained above are supported by the rigorous
calculations in (Balkovsky {\it et al.}, 1999b).

In $2d$, the area conservation allowed to get the scaling without
calculations. This is related to the fact that there is a single
Lyapunov exponent. When $d>2$, we have only the conservation of
${d}$-dimensional volumes and hence more freedom in the dynamics.  For
example, the area of a triangle can change during the evolution and
the three-point correlation function for a non-collinear geometry is
not necessarily suppressed.  Nevertheless, the anti-correlation
between different Lagrangian trajectories is still present and the
3-point exponent is expected to be larger than the naive estimate
$2{d}$. The answer for the Kraichnan model $d+(d-1)\sqrt{d/(d-2)}$ is
determined by the whole hierarchy of the Lyapunov exponents (Balkovsky
{\it et al.}, 1999b).  In the limit of large dimensions, the
anti-correlation tends to disappear and the answer approaches $2d$.
The four-point correlation function is also determined by the joint
evolution of two distances, which results for large $d$ in the same
value of the exponent.

To conclude this section, we briefly comment on the case
$\tilde\lambda_d>0$ when the particles approach rather than separate
backward in time. Here, an inversion of what has been described for
the direct cascade takes place: the scalar correlation functions are
logarithmic and the PDF has a wide Gaussian core at $r>\ell$, while
the statistics is strongly non-Gaussian at small scales (Chertkov {\it
et al.}, 1998). Since the scalar fluctuations injected at $\ell$
propagate upscale, small-scale diffusion is negligible and some
large-scale damping (say, by friction) is needed to provide for a
steady state, see also Sect.\,\,\ref{sec:cascades} below.

\subsubsection{Statistics of the dissipation}
\label{sec:dis}

We now describe the PDF's of the scalar gradients ${\bm
\omega}=\nabla\theta$ and of the dissipation $\epsilon=\kappa\,\omega^2$
in the steady state of a direct cascade arising under the action of a
large-scale pumping.  We consider a smooth velocity field,
i.e.\,\,both the Schmidt/Prandtl and the P\'eclet numbers are assumed
large. As we remarked in Sect.\,\ref{sec:grad}, the scalar gradients
can be estimated as $\,\theta\,{\rm e}^{-\rho_d}/\ell\,$, where $\theta$
is the scalar value and $\,{\rm e}^{\rho_d}\ell$ is the smallest
(diffusive) scale. The tails of the gradient PDF are controlled by the
large values of $\theta$ and $-\rho_d$. The statistics of the former
depends both on the pumping and the velocity and that of the latter
only on the velocity.  The key remark for solving the problem was made
by (Chertkov {\it et al.}, 1997 and 1998): since $\,\theta\,$ and
$-\rho_d\,$ fluctuate on very separated time scales
($\,\vert\lambda_d \vert^{-1}\ln(\ell/r_d)\,$ and $\,\lambda_d^{-1}$,
respectively), their fluctuations may be analyzed separately.  The PDF
of the scalar has been shown in Sect.\,\ref{sec:dirsmall} to decay
exponentially.  On the other hand, large negative values of $\rho_d$
are determined by the tail of its stationary distribution, see
(\ref{posit}).  For a Gaussian short-correlated strain this tail is
$\,\propto\exp[-{\rm const}.\,{\rm e}^{-2\rho_d}]\,$ (Chertkov {\it et
al.}, 1998) and the moments of the gradients are
$\,\langle\omega^n
\rangle\propto\langle\theta^n\rangle\langle\exp(-n\rho_d)\rangle\propto
n^{3n/2}$. The ensuing behavior $\,\langle\epsilon^n \rangle\propto
n^{3n}$ corresponds to a stretched-exponential tail for the PDF of the
dissipation
\begin{equation}
\ln{\cal P}(\epsilon)\ \propto\ -\epsilon^{1/3}\ .
\label{dissip}
\end{equation}
The detailed calculation for the Kraichnan model as well as a
comparison with numerical and experimental data can be found in
(Chertkov, Kolokolov {\it et al.}, 1998; Chertkov, Falkovich {\it et al.},
1998; Gamba \& Kolokolov, 1998). The
general case of a smooth flow with arbitrary statistics was considered
in (Balkovsky and Fouxon, 1999).

It is instructive to compare the stretched-exponential PDF of the
gradients in a steady state with the lognormal PDF described in
Sect.\,\ref{sec:grad} for the initial diffusionless growth.
Intermittency builds up during the initial stage, i.e. the higher the
moment, the faster it grows. On the other hand, the higher the moment,
the shorter is the breakdown time of the diffusionless approximation.
This time behaves for example as $\,(n+2)^{-1}$ in the Kraichnan
model.  Since higher moments stop growing earlier than lower ones, the
tails of the PDF become steeper and the intermittency is weaker in
the steady state.

\subsection{Passive fields in the inertial interval of turbulence}
\label{sec:passinert}

For smooth velocities, the single-point statistics of the advected
quantities could be inferred from the knowledge of the stretching
rates characterizing the Lagrangian flow in the infinitesimal
neighborhood of a fixed trajectory. This was also true for the
multipoint statistics as long as all the scales involved were smaller
than the viscous scale of the velocity, i.e.\,\,in the Batchelor
regime.  In this Section we shall analyze advection phenomena, mostly
of scalars, in the inertial interval of scales where the velocities
become effectively nonsmooth.  As discussed in
Sect.\,\,\ref{sec:nonsmooth}, the explosive separation of the
trajectories in nonsmooth velocities blows up interparticle
separations from infinitesimal to finite values in a finite time.
This phenomenon plays an essential role in maintaining the dissipation
of conserved quantities nonzero even when the diffusivity $\,\kappa\to
0\,$. The statistics of the advected fields is consequently more
difficult to analyze, as we discuss below.

\subsubsection{Passive scalar in the Kraichnan model}
\label{sec:passKrai}

The Kraichnan ensemble of Gaussian white-in-time velocities permits
an exact analysis of the nonsmooth case and a deeper insight into
subtle features of the advection, like intermittency and anomalous
scaling. Those aspects are directly related to the collective
behavior of the particle trajectories studied in the first part of the
review.  Important lessons learned from the model will be discussed in
next Sections in a more general context.

\noindent{\bf i). Hopf equations.} \ The simplifying feature
associated to the Kraichnan velocities is a reduction of the
corresponding Hopf equations to a closed recursive system involving
only correlators of the advected fields.  This is due to the temporal
decorrelation of the velocity and the ensuing Markov property of the
Lagrangian trajectories.  Let us consider for example the evolution
equation (\ref{theta}) for a scalar field. For the Kraichnan model, it
becomes a stochastic differential equation. As mentioned in
Sect.\,\,\ref{sec:solv}, one may view white-in-time velocities as the
scaling limit of ensembles with short time correlations.  The very
fact that $\,{\bm v}(t)dt\,$ tends to become of the order
$\,(dt)^{1/2}$ calls for a regularization.  For velocity ensembles
invariant under time reversal, the relevant convention is that of
Stratonovich (see the Appendix). Interpreting (\ref{theta}) within
this convention and applying the rules of stochastic differential
calculus, one obtains the equation for the scalar correlation functions: 
\qq
\partial_t\,C_{_N}(\un{\bm r})\ =\ {\cal M}_{_N}\,C_{_N} (\un{\bm r})\
+\ \sum\limits_{n<m}C_{_{N-2}}({\bm r}_{_1},
\mathop{.\dots.}\limits_{\widehat n\,\ \widehat m}, {\bm
r}_{_N})\,\,\Phi({\bm r}_{nm}/\ell)\,.
\label{KrHopf}
\qqq Here, the differential operator $\,{\cal M}_{_N}\,$ is the
same\footnote{The $n=m$ terms would drop out of the expression for
$\,{\cal M}_{_N}$ in the It\^{o} convention for (\ref{theta}).} as in
(\ref{mN}) and it may be formally obtained from the second term on the
left hand side of (\ref{Hopf}) by a Gaussian integration by
parts. Note the absence of any closure problem for the triangular
system of equations (\ref{KrHopf}): once the lower-point functions
have been found, the $N$-point correlation function satisfies a
closed equation.  For spatially homogeneous situations, the operators
$\,{\cal M}_{_N}\,$ may be replaced by their restrictions
$\,\widetilde{\cal M}_{_N}\,$ to the translation-invariant sector.  It
follows from their definition (\ref{mNrel}) and the velocity
correlation function (\ref{tobe1}) that the equations (\ref{KrHopf})
are then invariant with respect to the rescalings \qq{\bm
r}\to\lambda\,{\bm r}\,,\quad \ell\to\lambda\hspace{0.025cm}\ell\,,
\quad
t\to\lambda^{2-\xi}\,t\,,\quad\kappa\to\lambda^\xi\,\kappa\,,\quad
\theta\to\lambda^{-{2-\xi\over2}}\,\theta\,.
\label{chs}
\qqq This
straightforward observation implies scaling relations between the
stationary correlators: $C_{_N}(\lambda{\bm
r};\lambda^\xi\kappa,\lambda \ell)\ =\
\lambda^{N(2-\xi)\over2}\,C_{_N}({\bm r};\kappa,\ell)$.

\noindent{\bf ii). \ Pair correlator.} \ For the isotropic pair 
correlation function, the equation (\ref{KrHopf})  takes the form:
\qq \partial_t
C_2(r)-r^{1-d}\,\partial_r\Big[(d-1)D_1
\,r^{d-1+\xi}+2\kappa\,r^{d-1}\Big]\,\partial_r\,C_2(r)=\Phi(r/\ell)\,,
\label{pairc}
\qqq see (\ref{m3}). The ratio of the advective and the diffusive
terms is of order unity at the diffusion scale $\,r_d\equiv[{2\kappa/
(d-1)D_1}]^{1/\xi}$. For the Kraichnan model, the P\'eclet number
$Pe\equiv{(d-1)D_1}\ell^\xi/2\kappa\gg 1$ as we assume the scale of
pumping much larger than that of diffusion.  The stationary form of
(\ref{pairc}) becomes an ordinary differential equation (Kraichnan,
1968) that may be easily integrated with the two boundary conditions
of zero at infinity and finiteness at the origin: \qq C_2(r)\ =\
{_1\over^{(d-1)D_1}} \int\limits_r^\infty{x^{1-d}\,dx\over x^\xi+
r_d^\xi}\,\int\limits_0^x\Phi(y/\ell)\,\,y^{d-1}\,dy\,.
\label{2pf}
\qqq

Even without knowledge of the explicit form (\ref{2pf}), it is easy to
draw from (\ref{pairc}) general conclusions, as for time-correlated
velocities. Taking the limit $\,r\to0\,$ for $\,\kappa>0$,\,we infer
the mean scalar energy balance
\qq \partial_t\bar e+\bar\epsilon=\Phi(0)/2\,,
\label{enbalKr}
\qqq where $\bar e=\langle\theta^2\rangle/2$ and $\bar\epsilon=
\langle\kappa(\nabla\theta)^2\rangle$. In the stationary state, 
the dissipation balances the injection. On the other hand, 
for $r\gg r_d$ (or for any $r>0\,$ in the $\kappa\to0$ limit) 
we may drop the diffusive term in the
Hopf equation (\ref{pairc}). For $r\ll \ell$, we thus obtain the
Kraichnan model formulation of the Yaglom relation (\ref{Yaglom})
expressing the constancy of the downscale flux: \qq
-\,(d-1)D_1\,{1\over r^{d-1}}\,\partial_r\,r^{d-1+\xi}\,\partial_r
\,C_2(r,t)\ \approx\ \Phi(0)\,.
\label{KrYag}
\qqq To obtain the balance relation (\ref{enbalKr}) for vanishing 
$\,\kappa\,$ as the $\,r\to0\,$ limit of (\ref{pairc}), one has to 
define the limiting  dissipation field by the operator product expansion 
\qq
\lim\limits_{\kappa\to0}\,\,\kappa({\bm\nabla}\theta)^2({\bm r})\ =\
{_1\over^2}\,\lim\limits_{{\bm r}'\to{\bm r}} \,\,d^{ij}({\bm r}-{\bm
r'})\,\nabla_i\theta({\bm r})\,\nabla_j \theta({\bm r}')\,.
\label{limdiss}
\qqq
The relation (\ref{limdiss}), encoding the dissipative
anomaly, holds in general correlation functions away from other 
insertions (Bernard {\it et al.}, 1996).

Let us discuss now the solution (\ref{2pf}) in more detail.  There are
three intervals of distinct behavior. First, at large scales $\,r\gg
\ell$, the pair correlation function is given by 
\qq C_2(r)\ \approx\
{_1\over^{(d+\xi-2)(d-1)D_1}}\,\bar\Phi\,\ell^d\,
r^{2-\xi-d}\,,\label{RJ} \qqq where $\,\bar\Phi=\int_0^\infty
y^{d-1}\Phi(y)\,dy$. \,This may be thought of as the Rayleigh-Jeans
equipartition $\,\langle\theta({\bm k})\,\theta({\bm k}')\rangle=
\delta({\bm k}+{\bm k}')\,\bar\Phi\,\ell^d/\Omega(k)\,$ with
$\,\Omega(k)\propto k^{2-\xi}\,$ and the temperature proportional to
$\bar\Phi\ell^d$. Note that the right hand side of (\ref{RJ}) is a 
zero mode of $\widetilde{\cal M}_2$ away from the origin:
$r^{1-d}\partial_rr^{d-1+\xi}\partial_rr^{2-\xi-d}\propto\delta(r)$.
\,Second, in the convective interval $\,r_d\ll
r\ll\ell$, the pair correlator is equal to a constant (the genuine
zero mode of $\,\widetilde{\cal M}_2$) plus an inhomogeneous part: 
\qq C_2(r)\ \approx\ {\cal A}_2\,\ell^{2-\xi}\ -\
{_1\over^{(2-\xi)d(d-1)D_1}}\,\Phi(0)\,r^{2-\xi}\,, \qqq 
where ${\cal A}_2=\Phi(0)/(2-\xi)(d+\xi-2)(d-1)D_1$. 
The leading constant term drops out of the structure
function: \qq S_2(r)=2\left[C_2(0)-2C_2(r)\right] \ \approx\
{_2\over^{(2-\xi)d(d-1)D_1}}\,\Phi(0)\,r^{2-\xi}\,.
\label{2sf}
\qqq Note that the last expression is independent of both $\,\kappa\,$ 
and $\,\ell\,$ and it depends on the pumping through the mean
injection rate only, i.e.  $S_2(r)$ is universal. Its scaling exponent
$\,\zeta_2=2-\xi\,$ is fixed by the dimensional rescaling properties
(\ref{chs}) or, equivalently, by the scaling of the separation time of
the Lagrangian trajectories, see (\ref{oprn}).  As remarked before,
the degrees of roughness of the scalar and the velocity turn out to be
complementary: a smooth velocity corresponds to a rough scalar and
{\it vice versa}.  \,Finally, in the diffusive interval $\,r\ll r_d$,
the pair correlation function is dominated by a constant and the 
structure function $\,S_2(r)\approx{1\over{2\kappa
d}}\,\Phi(0)\,r^2$. Note that $S_2(r)$ is not analytic at the
origin, though. Its expansion in $r$ contains noninteger powers of
order higher than the second, due to the non-smoothness of the
velocity down to the smallest scales. The analyticity is recovered if
we keep a finite viscous cutoff for the velocity.

In the limit $\kappa\to0$, the diffusive interval disappears and the
pair correlator is given by (\ref{2pf}) with the diffusion scale $r_d$
set to zero. The mean square of the scalar $\,C_2(0)\,$ remains 
finite for finite $\,\ell\,$ but diverges in the $\ell\to\infty$ limit 
that exists only for the structure function.  Recall from 
Sect.\,\ref{sec:scal} that $C_2(r)$ has the interpretation of the mean 
time that two Lagrangian trajectories take to separate from distance 
$r$ to $\ell$. The finite value of the correlation function at the 
origin is therefore another manifestation of the explosive separation 
of the Lagrangian trajectories.

The solution (\ref{2pf}) for the pair correlator and most of the above
discussion remain valid also for $\xi=2$, i.e. for smooth Kraichnan
velocities. A notable difference should be stressed, though. In smooth
velocities and for $\kappa\to0$, the mean time of separation of two
Lagrangian trajectories diverges logarithmically as their initial
distance tends to vanish. The pair correlator has a logarithmic
divergence at the origin, implying that $\langle\theta^2\rangle$ is
infinite in the stationary state with $\kappa=0$. Indeed, it is the
$\partial_tC_2(0)$ term that balances the right hand side of
(\ref{pairc}) at finite times and $r=0$. As the diffusivity vanishes,
the variance $\langle\theta^2\rangle$ keeps growing linearly in time
with the rate $\Phi(0)$ and the mean dissipation tends to zero: no
dissipative anomaly is present at finite times. The anomaly occurs
only in the stationary state that takes longer and longer to achieve
for smaller $r$. In mathematical terms:
$\lim\limits_{t\to\infty}\,\lim\limits_{\kappa\to0}\ \bar\epsilon\,\
\not=\,\ \lim\limits_{\kappa\to0}\,\lim\limits_{t\to\infty}\
\bar\epsilon\,$, with the left hand side vanishing and the right hand
side equal to the mean injection rate.  The physics behind this
difference is clear. The dissipative anomaly for nonsmooth velocities
is due to the non-uniqueness of the Lagrangian trajectories, see
Sect.\,\ref{sec:nonsmooth}. The incompressible version of
(\ref{trsol}) implies that, in the absence of forcing and diffusion,
\qq \int \theta^2({\bm r}',0)\,d{\bm r}'\,-\,\int \theta^2({\bm
r}',t)\, d{\bm r}'=\int d{\bm r}\int p({\bm r},t;{\bm R},0\,|\,{\bm
v})\Big[\theta({\bm R},0)-\theta({\bm r},t)\Big]^2\,d{\bm R}\,\geq
0\,.  \qqq The equality holds if and only if, for almost all ${\bm
r}$, the scalar is constant on the support of the measure $\,p({\bm
r},t;{\bm R},0\,|\,{\bm v}) \,d{\bm R}\,$ giving the distribution of
the initial positions of the Lagrangian trajectories ending at ${\bm
r}$ at time $t$. In plain language, $\int \theta^2\,d{\bm r}$ is
conserved if and only if the Lagrangian trajectories are uniquely
determined by the final condition. This is the case for smooth
velocities and no dissipation takes place for $\kappa=0$ as long as
$\int \theta^2\,d{\bm r}$ is finite. When the latter becomes infinite
(as in the stationary state), the above inequalities become void and
the dissipation may persist in the limit of vanishing $\kappa$ even
for a smooth flow.

For $\xi=0$, \,the equation (\ref{2pf}) still gives the stationary
pair correlation function if $\,d\geq3$. The distinction between the 
behavior in the convective and the diffusive regimes disappears.  
The overall behavior becomes diffusive with the stationary equal-time
correlation functions coinciding with those of the forced diffusion
$\,\partial_t\theta=({1\over 2}(d-1)D_1+\kappa){\bm\nabla}^2
\theta+\varphi\,$ (Gaw\c{e}dzki and Kupiainen, 1996). In $\,d=2$, the
pair correlation function has a constant contribution growing 
logarithmically in time but the structure function does stabilize, as 
in forced diffusion.

\noindent{\bf iii). \ Higher correlators and zero modes.}  Let us consider 
the evolution of higher-order scalar correlation functions $C_{_N}$,
assumed to decay rapidly in the space variables at the initial
time. At long times, the correlation functions will then approach a 
stationary form given by the recursive relation \qq C_{_N}(\un{\bm r})\ 
=\ \int\,G_{_N}(\un{\bm r},\un{\bm R}) \,\sum\limits_{n<m}C_{_{N-2}}(
{\bm R}_{_1},\mathop{.\dots.\,} \limits_{\widehat n\,\ \widehat m},{\bm
R}_{_N})\,\,\Phi({\bm R}_{_{nm}}/\ell) \,\,d\un{\bm R}
\label{grfct}
\qqq for even $\,N$ and vanishing for odd $\,N\,$ (by the
$\,\theta\mapsto -\theta\,$ symmetry). Here,
$\,G_{_N}=\int\limits_0^\infty{\rm e}^{\,t \widetilde{\cal
M}_{_N}}\,dt\,$ are the operators inverse to $\,-\widetilde{\cal
M}_{_N}$. \,The above formulae give specific solutions of the
stationary Hopf equations (\ref{KrHopf}) that, alone, determine
solutions only up to zero modes of operators $\,{\cal M}_{_N}$. 
\,We are interested in the scaling properties of the stationary correlation 
function $\,C_{_N}\,$ in the convective interval. If the correlation 
functions were becoming independent of $\,\kappa\,$ and $\,\ell\,$ in 
this interval (mathematically, if the limits $\,\kappa\to0\,$ and 
$\,\ell\to\infty\,$ of the functions existed), the scaling behavior 
would follow from the dimensional relation (\ref{chs}): 
\qq C_{_N}(\lambda\,\un{\bm r})\ =\ \lambda^{(2-\xi)N\over 2}\,\, 
C_{_N}(\un{\bm r})\,.\label{KOC}
\qqq This would be the Kraichnan-model version of the normal
Kolmogorov-Obukhov-Corrsin scaling (Obukhov, 1949; Corrsin, 1951). The
$\,\kappa\to0\,$ limit of the stationary correlation functions does exist 
and the $\,\kappa$-dependence drops out of the expressions in the
convective interval, as for the pair correlator. The limit is given by
the formulae (\ref{grfct}) with the $\,\kappa=0\,$ versions of
$\,G_{_N}\,$ (Hakulinen, 2000). Note in passing that the advection
preserves any power of the scalar so that dissipative anomalies are
present also for orders higher than the second. The existence of the
zero diffusivity limit means that possible violations of the
normal scaling in the convective interval may only come from a
singularity of the limit $\ell\to\infty$. In fact, this was already the
case for $C_2$, dominated by the constant term that diverged as $\ell$
increases. The constant dropped out, however, from the pair
structure function (\ref{2sf}) that did not depend on $\,\ell\,$
and, consequently, scaled dimensionally. Concerning
higher-order scalar structure functions, Kraichnan (1994) was the
first to argue in favor of their anomalous scaling.  His paper steered
a renewed interest in the problem which led to the discovery by
Chertkov {\it et al.}\,\,(1995b), Gaw\c{e}dzki and Kupiainen (1995) and
Shraiman and Siggia (1995) of a simple mechanism to avoid normal
scaling: the domination of the correlation functions by scaling zero
modes of the operators $\,\widetilde{ \cal M}_{_N}\,$. For small
$\,\xi\,$, Gaw\c{e}dzki and Kupiainen (1995) and Bernard {\it et
al.}\,\,(1996) showed that in the convective interval \qq
C_{_N}(\un{\bm r})\ =\ {\cal A}_{_N}\,\ell^{\Delta_{_N}}\,f_{_{N,0}}
(\un{\bm r})\ +\ C_{_N}'(\un{\bm r})\ +\ o(\ell)\ +\ [\,\dots\,]\,.
\label{inconint}
\qqq Above, $\,f_{_{N,0}}\,$ is the irreducible isotropic zero mode of
scaling dimension $\,\zeta_{_{N,0}} ={N\over2}(2-\xi)-\Delta_{_N}$,
\,see Sect\,\,\ref{sec:0slow}, the term $\,C_{_N}'\,$ is scaling with
the normal dimension $\,{N\over2} (2-\xi)$, and $\,[\,\dots\,]\,$
stands for reducible contributions depending only on a subset of
points. The anomalous corrections $\,\Delta_{_N}={N(N-2)\over
2(d+2)}\,\xi\, +\,{\cal O}(\xi^2)$ are positive for small $\,\xi\,$,
see (\ref{pn12}).  A similar result $\,\Delta_{_N}={N(N-2)\over 2d}\,
+\,{\cal O}({1\over{d^2}})$ was established by Chertkov {\it et
al.}\,\,(1995b) and Chertkov and Falkovich (1996) for large space
dimensionalities. For $\,\Delta_{_N}>0$, the first term on the right
hand side of (\ref{inconint}) is dominating the second one for large
$\,\ell\,$ or, equivalently, at short distances for
$\,\kappa=0$. \,The analytic origin of the zero-mode dominance of the
stationary correlation functions (\ref{grfct}) \,lies in the asymptotic
short-distance expansion (\ref{inrs}) of the kernels of $\,G_{_N}$
(Bernard {\it et al.}, 1998).  The dominant zero mode $\,f_{_{N,0}}\,$
is the irreducible term in (\ref{inrs}) with the lowest scaling
dimension.  The reducible terms $\,[\,\dots\,]\,$ drop out of the
correlators of scalar differences, e.g. in the $\,N$-point structure
functions. The latter are dominated by the contribution from
$\,f_{_{N,0}}\,$: \qq S_{_N}(r)=\langle[\theta({\bm r}) -\theta({\bm
0})]^N\rangle \propto\ \ell^{\Delta_{_N}}\,r^{\zeta_{_N}}\,,
\label{ansc}
\qqq with $\,\zeta_{_N}=\zeta_{_{N,0}}$. The physical meaning of zero-mode
dominance is transparent. Any structure function is a difference between 
the terms with different number of particles coming at the points 1 and 2 
at time $t$, like, for instance, $S_3=3\langle \theta_1^2\theta_2-\theta_1
\theta_2^2\rangle$. Under the (backward-in-time) Lagrangian evolution, 
this difference decreases as $\,(r/\ell)^{\zeta_{_{N,0}}}$ because of shape 
relaxation with the slowest term due to the irreducible zero mode. The 
structure function is thus given by the total temporal factor $\ell^{N
(2-\xi)/2}$ multiplied by $(r/\ell)^{\zeta_{_{N,0}}}$.

The scaling exponents in (\ref{ansc}) are universal in the sense that they 
do not depend on the shape of the pumping correlation functions $\Phi(r)$. 
The coefficients $\,{\cal A}_{_N}$ in (\ref{inconint}), as well as 
the proportionality constants in (\ref{ansc}), are, however, non-universal.  
Numerical analysis, see Sect.\,\,\ref{sec:num}, indicates that the previous 
framework applies for all $\,0<\xi<2\,$ at any space dimensionality,
with the anomalous corrections $\,\Delta_{_N}$ continuing to be strictly 
positive for $\,N>2$.  That implies the small-scale intermittency of the
scalar field: the ratios $\,S_{2n}/S_2^n\,$ grow as $r$ decreases. At
orders $\,N\gg (2-\xi)d/\xi$, \,the scaling exponents $\,\zeta_{N}\,$
tend to saturate to a constant, see Sects.\,\,\ref{sec:inst} and
\ref{sec:num} below.

In many practical situations the scalar is forced in an anisotropic
way. Shraiman and Siggia (1994, 1995) have proposed a simple way to
account for the anisotropy. They subtracted from the scalar field an
anisotropic background by defining $\,\theta'({\bm r})=\theta({\bm
r})-\,{\bm g}\cdot{\bm r}\,$, with $\,{\bm g}\,$ a fixed vector. It
follows from the unforced equation (\ref{tracer}) that \qq
\partial_t\theta'+{\bbox v}\cdot{\bm\nabla}\,\theta'-
\kappa\,{\bm\nabla}^2\theta'= -{\bm g}\cdot{\bm v},
\label{tracerg}
\qqq with the term on the right hand side giving the effective
pumping.  In Kraichnan velocities, the translation invariance of 
the equal-time correlators of $\,\theta'$ is preserved by the evolution
with the Hopf equations taking the form \qq
\partial_t\,C_{_N}(\un{\bm r})\,&=&\,\widetilde{\cal M}_{_N}\,
C_{_N}(\un{\bm r})\ +\ 2\sum\limits_{n<m}C_{_{N-2}} ({\bm
r}_{_1},\mathop{.\dots.}\limits_{\widehat n\ \, \widehat m},{\bm
r}_{_N})\,\,g_i\,g_j\, D^{ij}({\bm r}_{nm})\cr &&-\
\sum\limits_{n,m}\,g_i\,\,d^{ij} ({\bm
r}_{nm})\,\nabla_{r_m^j}\,C_{_{N-1}}({\bm r}_{_1},
\mathop{\dots}\limits_{\widehat n},{\bm r}_{_N})
\label{KrHopfg}
\qqq in the homogeneous sector. The stationary correlation functions
of $\,\theta'$ which arise at long times if the initial correlation
functions decay in the space variables, may be analyzed as before. In
the absence of the $\,\theta'\mapsto-\theta'\,$ symmetry, the odd
correlators are not anymore constrained to be zero. Still, the
stationary 1-point function vanishes so that the scalar mean is
preimposed: $\,\langle\theta({\bm r})\rangle={\bm g}\cdot{\bm
r}$. \,For the 2-point function, the solution remains the same as in
the isotropic case, with the forcing correlation function simply
replaced by $\,2\,g_i g_j\,D^{ij}({\bm r})\,$ and approximately equal
to the constant $\,2\,D_0\,{\bm g}^2\,$ in the convective
interval. The 3-point function is \qq C_3(\un{\bm r})\ =\
-\int\,G_3(\un{\bm r},\un{\bm R})\,\,
\sum\limits_{n,m}g_i\,\,d^{ij}(\bm R_{nm})\,\nabla_{R_m^j}C_{2} ({\bm
R}_{1},\mathop{\dots}\limits_{\widehat n},{\bm R}_{3})\,\, d\un{\bm
R}\,.  \qqq The dimensional scaling would imply that
$\,C_3(\lambda\un{\bm r})= \lambda^{3-\xi}\,C_3(\un{\bm r})\,$ in the
convective interval since $\,{\bm\nabla}C_2({\bm r})\,$ scales there
as $\,r^{1-\xi}$. \,Instead, for $\,\xi\,$ close to $\,2$, \,the
3-point function is dominated by the angular momentum $\,j=1\,$ zero
mode of $\,{\cal M}_3\,$ with scaling dimension $\,2+o(2-\xi)$, as
shown by Pumir {\it et al.}\,\,(1997). A similar picture arises from
the perturbative analysis around $\,\xi=0\,$ (Pumir, 1996 and 1997),
around $\,d=\infty\,$ (Gutman and Balkovsky, 1996), and from the
numerical study of the whole interval of $\,\xi$ values for $\,d=2\,$
and $\,d=3\,$ (Pumir, 1997), see Sect.\,\,\ref{sec:pert}. As will be
discussed in Section~\ref{sec:kra}, the zero mode mechanism is likely
to be responsible for the experimentally observed persistence of the
anisotropies, see e.g. (Warhaft, 2000).

It is instructive to analyze the limiting cases $\,\xi=0,2\,$ and
$\,d=\infty\,$ from the viewpoint of the statistics of the
scalar. Since the field at any point is the superposition of
contributions brought from $\,d\,$ directions, it follows from the
Central Limit Theorem that the scalar statistics becomes Gaussian as
the space dimensionality $\,d\,$ increases.  In the case $\,\xi=0$,
\,an irregular velocity field acts like Brownian motion. The
corresponding turbulent transport process is normal diffusion and the
Gaussianity of the scalar statistics follows from that of the
input. What is general in the both previous limits is that the degree
of Gaussianity, as measured, say, by the flatness $\,S_4/S_2^{\,2}$,
is scale-independent. Conversely, we have seen 
in Sect.\,\,\ref{sec:scal} that $\ln(\ell/r)$ is the parameter 
of Gaussianity in the Batchelor limit with the statistics becoming 
Gaussian at small scales whatever the input statistics. 
The key here is in the temporal rather than the
spatial behavior. Since the stretching in a smooth velocity field is
exponential, the cascade time is growing logarithmically as the scale
decreases. That leads to the essential difference: at small yet
nonzero $\,\xi/d\,$ the degree of non-Gaussianity increases
downscales, while at small $\,(2-\xi)\,$ it first decreases downscales
until $\,\ln(\ell/r) \approx 1/(2-\xi)\,$ and then it starts to
increase. Note that the interval of decrease grows as the Batchelor
regime is approached. Already that simple reasoning suggests that the
perturbation theory is singular in the limit $\,\xi\to2$, \,which is
formally manifested in the quasi-singularities of the many-point
correlation functions for collinear geometries (Balkovsky, Chertkov {\it et
al.}, 1995).

The anomalous exponents determine also the moments of the dissipation
field $\,\epsilon=\kappa({\bm\nabla}\theta)^2$. \,A straightforward
analysis of (\ref{grfct}) indicates that
$\,\langle\epsilon^n\rangle=c_n
\,\bar\epsilon^n\,(\ell/r_d)^{\Delta_{2n}}\,$ (Chertkov {\it et
al.}, 1995b; Chertkov and Falkovich, 1996), where $\,\bar\epsilon$ is
the mean dissipation rate.  The dimensionless constants $\,c_n$ are
determined by the fluctuations of the dissipation scale and, most
likely, they are of the form $\,n^{qn}\,$ with yet unknown $\,q$.
\,In the perturbative domain $\,n\ll (2-\xi)d/\xi$, the anomalies
$\,{\Delta_{2n}}\,$ are a quadratic function of the order and the
corresponding part of the dissipation PDF is close to lognormal
(Chertkov and Falkovich, 1996). The form of the distant tails
of the PDF are still unknown.

\noindent{\bf iv). Operator product expansion}. \ 
While the irreducible zero mode dominates the respective structure
function, all the zero modes may be naturally incorporated into
an operator product expansion (OPE) of the scalar correlation functions.
There has been many attempts to use this powerful tool of quantum 
field theory (Wilson, 1969) in the context of turbulence, see
(Eyink, 1993; Adzhemyan {\it et al.}, 1996; Polyakov, 1993 and 1995).
We briefly describe here a general direction for accomplishing that
for the problem of scalar advection (Chertkov and Falkovich, 1996; 
Adzhemyan {\it et al.}, 1998; Zamolodchikov {\it et al.}, 2000). 
Let $\{{\cal O}_a\}$ be a set of local observables (which contains 
all spatial derivatives of any field already included). The 
existence of OPE presumes that
\begin{equation}
{\cal O}_a ({\bm r})\,{\cal O}_b({\bm r}') = \sum_{k} C_{ab}^c ({\bm r}
-{\bm r}') \, {\cal O}_c ({\bm r}'),\label{OPE}
\end{equation}
which is understood as the following relations among the correlation 
functions 
\begin{equation}
\langle{\cal O}_a ({\bm r})\, {\cal O}_b ({\bm r}')\,\ldots\rangle\ 
=\ \sum_k C_{ab}^c
({\bm r}-{\bm r}') \ \langle{\cal O}_c( {\bm r}')\,\ldots\rangle\,.
\end{equation}
The sum represents the correlation function in the left-hand side 
if $|{\bm r}-{\bm r}'|$ is small enough.  Renormalization symmetry 
$\,\varphi\to\Lambda\varphi$, $\theta\to\Lambda\theta\,$ allows one 
to classify the operators (fluctuating fields) by degrees: 
${\cal O}^{(n)}$ has degree $n$ if $\,{\cal O}^{(n)}\to 
\Lambda^n\,{\cal O}^{(n)}\,$ under the transformation. The OPE conserves 
the degree and is supposed to be scale invariant in the convective
interval. This means that one may choose a basis of the observables in
such a way that ${\cal O}_a$ has ``dimension'' $d_a$, and the OPE is
invariant under the transformation $\,{\cal O}_a ({\bm r}) \to\lambda^{d_a}
\,{\cal O}_a (\lambda\,{\bm r})$ so that its coefficient functions scale:
$\,C_{ab}^c (\lambda({\bm r}-{\bm r}'))=\lambda^{d_c - d_a - d_b
}C_{ab}^c({\bm r}-{\bm r}')\,$.
Besides, functions $\,C_{ab}^c$ are supposed to be pumping-independent 
with the whole dependence on pumping carried by the expectation values 
$\,\langle {\cal O}_c\rangle\propto\ell^{-d_c}$. 
\,The scale invariance can be (``spontaneously'') broken at the level 
of correlation functions if some of the fields with nonzero dimension 
develop nonzero expectation values. The dimension of $\,\theta^N$ is
$\,N(\xi-2)/2$.  \,The operators can be organized into strings,
each with the primary operator $\Theta_a$ with the lowest dimension
$d_a$ and its descendants with the dimensionalities $d_a+n(2-\xi)$. A
natural conjecture is that there is one-to-one correspondence
between the primary operators of degree $N$ and the zero modes of
$\,\widetilde{M}_{_N}$. \,The dimensions of such primaries are minus the 
anomalous dimensions $\Delta$'s of the zero modes $f_{a}$ 
and are therefore negative. 
By fusing $N-1$ times one gets $\,\theta_{_1}\ldots\theta_{_N}=\sum
f_{a}({\bm r}_{_1},\ldots,{\bm r}_{_N})\,\Theta_a({\bm r}_{_N})+\ldots\,$, 
\,where the dots include the derivatives and descendants of 
$\,\Theta_a$. 

For $n=2$, one has only one primary field $\theta^2$, and its
descendents $|\nabla^m \theta|^2$.  For $n=4$, there is an infinity of
primaries. Only $\theta^4$, $\epsilon^2$ and $\epsilon\theta\nabla^2
\theta-d^2\epsilon/(d^2+2)$ have nonzero expectation values. The
operators with zero expectation values correspond to the operators
with more derivatives than twice the degree (that is with the order of
the angular harmonic in the respective zero mode being larger than the
number of particles, in terms of Sect.\,\ref{sec:multi}).  Building an
OPE explicitly and identifying its algebraic nature remains a task for
the future.

\noindent{\bf v). Large scales}. \ The scalar correlation functions at
scales larger than that of the pumping decay by power laws. The pair
correlation function is given by (\ref{RJ}). Recall that applying ${\cal
M}_2$ on it, we obtain a contact term $\propto \delta({\bm
r})$. Concerning higher-order correlation functions, straight lines are 
not preserved in a nonsmooth flow and no strong angular dependencies of
the type encountered in the smooth case are thus expected. To
determine the scaling behavior of the correlation functions, it is 
therefore enough to focus on a specific geometry.  Consider for instance 
the equation $\,\widetilde{\cal M}_4C_4(\un{\bm r})=
\sum\chi(r_{ij})\,C_2(r_{kl})\,$ for the fourth order function.  
A convenient geometry to analyze is that with one distance among the
points, say $r_{12}$, much smaller than the other $r_{1j}$, whose
typical value is $R$. At the dominant order in $r_{12}/R$, the
solution of the equation is $C_4\propto C_2(r_{12}) C_2(R)\sim
\left(r_{12}\,R\right)^{2-\xi-d}$. Similar arguments apply to
arbitrary orders. We conclude that the scalar statistics at $r\gg
\ell$ is scale-invariant, i.e. $C_{2n}(\lambda\un{\bm
r})=\lambda^{n(2-\xi-d)}\,C_{2n}(\un{\bm r})$ as $\lambda\to\infty$.
Note that the statistics is generally non-Gaussian when the
distances between the points are comparable. As $\xi$ increases from
zero to two, the deviations from the Gaussianity starts from zero and
reach their maximum for the smooth case described in
Sect.\,\ref{sec:dirlarge}.

\noindent{\bf vi). Non-Gaussian and finite correlated pumping}. \
The fact that the scalar correlation functions in the 
convective interval are dominated by zero modes indicates that the 
hypotheses of
Gaussianity and $\delta$-correlation of the pumping are not
crucial. The purpose of this Section is to give some more details on
how they might be relaxed. The new point to be taken into account is
that the pumping has now a finite correlation time $\tau_p$ and
irreducible contributions are present. The situation with the second
order is quite simple. The injection rate of $\theta^2$ is $\langle
\phi\,\theta\rangle $ and its value defines the mean dissipation rate
$\bar\epsilon$ at the stationary state. The only difference is that
its value cannot be estimated {\it a priori} as $\Phi(0)$.  Let us
then consider the behavior at higher orders, whose typical example is
the fourth.  Its general flux relation, derived similarly as
(\ref{2pHopf}), reads \qq \Big\langle({\bbox v}_1\cdot \nabla_1+{\bbox
v}_2\cdot\nabla_2)\, \theta_1^2\theta_2^2\Big\rangle\ +\
\,\kappa\,\Big\langle
\theta_1\theta_2\left[{\bm\nabla}_{_1}^2+{\bm\nabla}_{_2}^2\right]
\theta_1\theta_2\Big\rangle\ =\ \langle
\varphi_1\theta_1\theta_2^2+\varphi_2\theta_2\theta_1^2\rangle\ .
\label{fl02} \qqq
Taking the limit of coinciding points, we get the production rate of
$\theta^4$. It involves the usual reducible contribution
$3\bar\epsilon C_2(0)$ and an irreducible one.  The ratio of the two
is estimated as $C_2(0) /\tau_p$. The non-Gaussianity of the pumping
is irrelevant as long as $\tau_p$ is smaller than the time for the
particles to separate from the diffusive to the integral scale. The
smooth and nonsmooth cases need to be distinguished. For the former,
the separation time is logarithmically large and the previous
condition is always satisfied. Indeed, the reducible part of the
injection rate in (\ref{fl02}) necessarily contains
$2\langle\theta_1\theta_2\rangle\left[\langle
\varphi_1\theta_2\rangle+\langle \varphi_2\theta_1\rangle\right]
\simeq 2{\bar\epsilon}^2\tilde\lambda_d^{-1}\ln(\ell/r_{12})$.  Since
the correlation function grows as $r_{12}$ decreases, one can always
neglect the constant irreducible contribution for small enough
separations.  Similarly, the input rate of all even moments up to
$N\simeq\ln (\ell/r_d)$ is determined by $\bar\epsilon$. The fact that
the fluxes of higher integrals are not constant in the convective
interval was called the effect of ``distributed pumping'' in 
(Falkovich, 1994; Falkovich and Lebedev, 1994).

In the nonsmooth case, the cascade time is finite and the irreducible
contributions might be relevant. They affect the statistics of the
scalar yet, of course, not the scaling of the zero modes.  The fourth
order correlation function $C_4$ acquires for example extra terms
proportional to $\sum r_{ij}^{2-\xi}$. They contribute to the fourth
order cumulant but not to the structure function $S_4$. The existence
of those extra terms in the correlation function affects the matching
conditions at the pumping scale, though.  We conclude that the
numerical coefficients ${\cal A}_{_N}$ in the structure functions
$S_{_N}={\cal A}_{_N}r^{\zeta_{_N}} \ell^{\Delta_{_N}}$ generally
depend on all the irreducible pumping contributions of order $m\leq
N$.

\subsubsection{Instanton formalism for the Kraichnan model}
\label{sec:inst}

Since the perturbative approaches in Section~\ref{sec:pert} are all
limited to finite orders, it is natural to look for alternative
methods to capture the scaling exponents in the non-perturbative
domain ${N\xi}\gg(2-\xi) d$. As in many other instances in field
theory or statistical physics, such a non-perturbative formalism is
expected to result from a saddle-point technique applied to the path
integral controlling the statistics of the field. Physically, that
would correspond to finding some optimal fluctuation responsible for a
given structure function.  Not any structure function can be found by
this approach but only those with $N\gg1$, related to the PDF tails
which are indeed controlled by rare events. This is a general idea of
the instanton formalism (see, e.g., Coleman, 1977)
adapted for turbulence by Falkovich {\it et
al.}\,\,(1996). The case in question is so complicated though that an
effective analysis (carried out by Balkovsky and Lebedev, 1998) is
possible only with yet another large parameter, $(2-\xi)d\gg1$, which
guarantees that the Lagrangian trajectories are almost deterministic.
The relation between ${N\xi}$ and $(2-\xi) d$ is now arbitrary so one
is able to describe both the perturbative and non-perturbative
domains.  Unfortunately, a straightforward application of this
approach to the path integral over the velocity field does not work
because of a usual problem in saddle-point calculations: the existence 
of a soft mode makes the integrand non-decaying in some direction in the
functional space. One ought to integrate over the soft mode before the
saddle-point approximation is made.  Balkovsky and Lebedev identified
the soft mode as that responsible for the slow variations of the
direction of the main stretching. Since the structure functions are
determined only by the modulus of the distance then an effective
integration over the soft mode simply corresponds to passing from the
velocity to the absolute value of the Lagrangian separation as the
integration variable in the path integral.  This can be conveniently
done by introducing the scalar variable
\begin{equation} 
\eta_{12}\equiv(2-\xi)^{-1}\partial_t R_{12}^{2-\xi}= 
R_{12}^{-\xi}R_{12}^i
(v_{1}^i-v_{2}^i) \,.  
\label{la31} \end{equation} 
For the Kraichnan velocity field, $\eta_{12}$ has the nonzero mean
$\langle\eta_{12}\rangle=-D$ and the variance
\begin{equation}
\langle\langle\eta_{12}(t_1)\eta_{34}(t_2)\rangle\rangle 
=\frac{2D}{{d}\,}q_{12,34}\,\delta(t_1-t_2)\,.  
\label{la33} \end{equation} 
The explicit dependence of the $\,q_{12,34}$ function 
on the particle distances
will not be needed here and can be found in the original paper
(Balkovsky and Lebedev, 1998).  Any average over the statistics of the
Lagrangian distances can be written in terms of a Martin-Siggia-Rose
path integral $\,\int {\cal D}R\,{\cal D}m\,\exp(\imath {\cal I}_R)$, 
\,with the action \qq {\cal I}_R=\int\limits_{-\infty}^0 dt\, \int d
{\bm r}_1\,d{\bm r}_2\, m_{12}\left[ \left({\partial_t
R_{12}^{2-\xi}\over 2-\xi}+D\right) +\frac{iD}{{d}} d{\bm
r}_3\,d{\bm r}_4\,q_{12,34}\,m_{34}\right] \,.
\label{la36} \qqq
The auxiliary field conjugated to $R_{12}$ is denoted by $m_{12}\equiv
m(t,{\bbox r}_1,{\bbox r}_2)$. Note that the second (nonlinear) term
in (\ref{la36}) vanishes both as $(2-\xi)d\to\infty$ and $\xi\to0$.
The moments of any linear functional of the scalar $\vartheta=\int
d{\bm r}\beta({\bm r})\theta({\bm r})$ are then expressed as \qq
\langle|\vartheta|^N\rangle=\int\frac{{d}y\,{d}\vartheta}{2\pi} \int
{\cal D}R\,{\cal D}m\,\, {\rm e}^{\,i{\cal I}_R-{\cal F}_\lambda
-iy\vartheta+N\ln|\vartheta|} \,,
\label{bb57} \qqq
where ${\cal F}_\lambda= \frac{y^2}{2}\int {d}t\,{d}{\bbox
r}_1\,{d}{\bbox r}_2\, \chi(R_{12})\beta({\bbox r}_1)\beta({\bbox
r}_2)$.  To obtain the structure functions, one should in principle
take for $\beta$ differences of $\delta$ functions. This would however
bring diffusive effects into the game. To analyze the scaling
behavior, it is in fact enough to consider any observable where the
reducible components in the correlation functions are filtered out. A
convenient choice is $\beta({\bbox r}_1)= \delta_\Lambda\left({\bbox
r}_1-\frac{{\bbox r}}{2}\right) -\delta_\Lambda\left({\bbox
r}_1+\frac{{\bbox r}}{2}\right)$, where the smeared function
$\delta_\Lambda({\bbox r})$ has a width $\Lambda^{-1}$ and satisfies
the normalization condition $\int {d}{\bbox r}\,\delta_\Lambda({\bbox
r})=1$. The diffusive effects may be disregarded provided the width is
taken much larger than $r_d$.

The saddle-point equations for the integral (\ref{bb57}) are 
\begin{eqnarray}
\partial_t R_{12}^{2-\xi}&=&-(2-\xi)D\left[1+{2i\over d}
\int {d}{\bbox r}_3\,
{d}{\bbox r}_4\,q_{12,34}m_{34}\right]\,,
\label{la37} \\ 
-iR_{12}^{1-\xi}\partial_tm_{12}&=&{2D \over d}\int {d}{\bbox
r}_3\,{d}{\bbox r}_4\,{\partial q_{12,34}\over \partial
R_{12}}\left[m_{12}m_{34}+2m_{13}m_{24}\right]+\frac{
y^2}{2}\chi'(R_{12})\beta({\bbox r}_1)\beta({\bbox r}_2)
\,,
\label{la38}
\end{eqnarray}
with the two extremal conditions on the parameters $\vartheta$ and
$y$: \qq \vartheta=iy\int {d}t\,{d}{\bbox r}_1\,{d}{\bbox
r}_2\,\chi(R_{12}) \beta({\bbox r}_1)\beta({\bbox r}_2)\,, \quad
iy=N/\vartheta \,.
\label{qq2} \qqq
The two boundary conditions are $R_{12}(t=0)=|{\bm r}_1-{\bm r}_2|$
and $m_{12}\to0$ as $t\to-\infty$. The variables $R_{12}$ and $m_{12}$
are {\it a priori} two fields, i.e. they depend on both $t$ and ${\bm
r}$. In fact, the problem can be shown to reduce effectively to two
degrees of freedom: $R_-$, describing the separation of two points,
and $R_+$, describing the spreading of a cloud of size $\Lambda$
around a single point. It follows from the analysis in (Balkovsky and
Lebedev, 1998) that there are two different regimes, depending on the
order of the moments considered. At $N<(2-\xi)d/(2\xi)$ the values of
$R_+$ and $R_-$ are very close during most of the evolution and
different fluid particles behave similarly.  For higher moments, $R_+$
and $R_-$ differ substantially throughout the evolution. The fact that
different groups of fluid particles move in a very different way might
be interpreted as the signature of the strong fronts in the scalar
field that are discussed in Section\,\ref{sec:kra}.  The final result
for the scaling exponents is:
\begin{eqnarray}
&&\zeta_{N}=N(2-\xi)/2-2\xi N^2/2d \qquad{\rm at}\quad
N<(2-\xi)d/(2\xi)\,,\label{exp1}\\&&
\zeta_{N}=(2-\xi)^2d/(8\xi)\qquad{\rm at}\quad N>(2-\xi)d/(2\xi)\,.
\label{exp2}
\end{eqnarray}
These expressions are valid when the fluctuations around instanton
give negligible contribution which requires $N\gg1$ and $(2-\xi)d\gg1$, while the
relation between $d$ and $N$ is arbitrary.  The exponents depend
quadratically on the order and then saturate to a constant. The
saturation for the Kraichnan passive scalar model had been previously
inferred from a qualitative argument by Yakhot (1997) and from an
upper bound on $\zeta_{_N}$ by Chertkov (1997).  The relevance of the
phenomenon of saturation for generic scalar turbulence is discussed in
Section\,\ref{sec:kra}.

\subsubsection{Anomalous scaling for magnetic fields}
\label{sec:magnana}

Magnetic fields transported by a Kraichnan velocity field display
anomalous scaling already at the level of the second-order correlation
functions.  For a scalar, $\langle\theta^2\rangle$ is conserved and
its flow across the scales fixes the dimensional scaling of the
covariance found in Section~\ref{sec:passKrai}\,{\bf ii)}. For a
magnetic field, this is not the case. The presence of an anomalous
scaling for the covariance ${\cal C}_{2}^{ij}({\bm r},t)=\langle
B^i({\bm r},t) \,B^j({\bm 0},t)\rangle$ becomes quite intuitive from
the Lagrangian standpoint.  We have seen in Section~\,\ref{sec:multi}
that the zero modes are closely related to the geometry of the
particle configurations. For a scalar field, the single distance
involved in the two-particle separation explains the absence of
anomalies at the second order.  The magnetic field equation
(\ref{magn}) for $\kappa=0$ is the same as for a tangent vector, i.e.
the separation between two infinitesimally close particles. Although
${\cal C}_{2}^{ij}$ involves again two Lagrangian particles, each of
them is now carrying its own tangent vector.  In other words, we are
somehow dealing with a four-particle problem, where the tangent
vectors bring the geometric degrees of freedom needed for the
appearance of nontrivial zero modes.  
The compensation mechanism leading to the respective
two-particle integral of motion is now due to the interplay between
the interparticle distance and the angular correlations of the vectors
carried by the particles. The attractive feature of the problem is
that the anomaly can be calculated nonperturbatively.

Specifically, consider the equation (\ref{magn}) for the solenoidal
magnetic field ${\bm B}({\bm r},t)$ and assume the Kraichnan
correlation function (\ref{tobe1}) for the Gaussian incompressible
velocity. We first analyze the isotropic sector (Vergassola, 1996).
The first issue to be addressed is the possibility of a stationary
state. For this to happen, there should be no dynamo effect,
i.e.\,\,an initial condition should relax to zero in the absence of
injection. Let us show that this is the case for $\xi<1$ in $3d$. As for
a scalar, the $\delta$-correlation of the velocity leads to a closed
equation for the pair correlation function $\partial_t{\cal
C}_{2}^{ij}={\cal M}^{kl}_{ij}\,{\cal C}_{2}^{kl}$.  The isotropy and
the solenoidality of the magnetic field permit to write ${\cal
C}_{2}^{ij}({\bm r},t)$ in terms of its trace $H({\bm r},t)$ only.
This leads to an imaginary time Schr\"odinger equation for the 
``wave function'' $\,\psi(r,t)={\kappa+D_1\,r^\xi\over r}\,\int_0^r
H(\rho,t)\,\rho^2\,d\rho\,$ (Kazantsev, 1968). The energy eigenstates 
$\,\psi(r)\,{\rm e}^{-Et}$, \,into which $\,\psi(r,t)\,$ may be 
decomposed, satisfy the stationary equation 
\begin{equation}
\label{schr}
{d^2\psi(r)\over dr^2}+m(r)\left[E-U(r)\right]\psi(r)=0
\end{equation}
of a quantum particle of variable mass $m(r)$ in the  potential $U(r)$.
The presence of a dynamo effect is
equivalent to the existence of negative energy levels. Since the mass
is everywhere positive, it is enough to look for bound states in the
effective potential $V=mU$. The detailed expressions of the mass and
the potential can be found in (Vergassola, 1996). Here, it is enough
to remark that $V(r)$ is repulsive at small scales and has a quadratic
decay in the inertial range with a prefactor $2-3/2\xi-3/4\xi^2$. It
is known from quantum mechanics textbooks that the threshold for bound
states in an attractive potential $-c/r^2$ is $c=1/4$. The
absence of a dynamo effect for $\xi< 1$ follows immediately from the
expression of the potential prefactor.

The stability just described implies that, in the presence of a forcing
term in (\ref{magn}), the magnetic field covariance will relax to a
time-independent expression at long times. We may then study the
spatial scaling properties at the stationary state. The energy $E$ in
(\ref{schr}) should be set to zero and we should add the corresponding
forcing term in the first equation. Its precise form is not important
here as any anomalous scaling is known to come from the zero modes of
the operator $M={d^2/dr^2}-V(r)$. The behavior $-c/r^2$ of the
potential in the inertial range implies that the operator is
scale-invariant and its two zero modes behave as power laws with
exponents $1/2\left[1\pm \sqrt{1-4c}\right]$.  The zero mode with the
smaller exponent is not acceptable due to a singular behavior in the
dissipation range of the corresponding correlation function.  The
remaining zero mode dominates the inertial-range behavior $\,H(r)\propto
r^{\gamma_2}\,$ with
\begin{equation}
\label{anomalia}
\gamma_2=-(3+\xi)/2+3/2\sqrt{1-\xi(\xi+2)/3}.
\end{equation}
Note that dimensional arguments based on a constant flux of $A^2$
(with the vector potential defined by ${\bm B}={\bm \nabla}\times{\bm
A}$) would give $\gamma_2=-\xi$. This is indeed the result in the
limits of small $\xi$ and large space dimensionality and for the $2d$
case. For the latter, the vector potential is reduced indeed to a
scalar whose equation coincides with the advection-diffusion equation
(\ref{tracer}). Note that $\gamma_2\leq -\xi$, that is the zero mode
provides for correlation functions that are $(\ell/r)^{-\gamma_2-\xi}$
times larger than what dimensional arguments would suggest.

The zero mode dominating the stationary 2-point function of ${\bm B}$
is preserved by the unforced evolution. In other words, if 
$\,{\cal C}_{2}^{ij}({\bm r},0)$ is taken
as the zero mode of ${\cal M}$ then the correlation function does not
change with time.  A counterpart to that is the existence of a
statistical Lagrangian invariant that contains both the distance
between the fluid particles and the values of the fields:
$I(t)=\langle B^k({\bf R}_1)B^l({\bf R}_2)Z^{+}_{kl}({\bf R}_{12})
\rangle$ (Celani and Mazzino, 2000). Here $Z^+_{kl}$ is a zero mode 
of the operator adjoint to ${\cal M}$. The scaling dimension of such
zero mode is $\gamma_2+2>0$.  The appearance of the adjoint operator
has a simple physical reason. To calculate the correlation functions
of the magnetic field, the tangent vectors attached to the particles
evolve forward in time while, as for the scalar, their trajectories
should be traced backward in time. Adjoint objects naturally appear
when we look for invariants where all the quantities run in the same
direction of time.  The conservation of $\,I(t)\,$ is due to the 
power-law growth of $Z^+$ with time being offset by the decorrelation 
between the directions of the ${\bm B}$ vectors along the separating 
trajectories.

Up to now, we have been considering the covariance in the isotropic
sector. The scaling exponents in the nonisotropic sectors can also be
calculated nonperturbatively (Lanotte and Mazzino, 1999; Arad, Biferale 
{\it et al.}, 2000). The problem is analogous to that for the scalar in
Section~\ref{sec:pert}, solved by the expression (\ref{aniso}). Here,
the calculation is more involved since in each sector $j,m$ of SO(3)
there are nine independent tensors into which ${\cal C}_2^{ij}$
may be decomposed. Their explicit expression may be found in (Arad 
{\it et al.}, 1999). As in the isotropic case, it is shown that no 
dynamo takes place for $\xi<1$ and the scaling properties at the 
stationary state can then be calculated. For odd $j$, one has for 
example:\begin{equation} \label{anisotropia}
\gamma_2^j=-(3+\xi)/2+1/2\sqrt{1-10\xi+\xi^2+2j(j+1)(\xi+2)}\,.
\end{equation}
The expression for the other sectors can be found in (Lanotte and
Mazzino, 1999; Arad, Biferale {\it et al.}, 2000). An important point 
(we shall come back to it in Section~\ref{sec:kra}) is that the 
exponents increase with $j$: the more anisotropic the contribution, 
the faster it decays going toward the small scales.

Higher-order correlation functions of the magnetic field also obey
closed equations.  Their analysis proceeds along the same lines as for
the scalar, with the additional difficulty of the tensorial
structure. The extra-terms in the equations for the correlation
functions of the magnetic field come from the presence of the
stretching term ${\bm B}\cdot{\bm\nabla}{\bm v}$.  Since the gradient
of the velocity appears, they are all proportional to $\xi$. On the
other hand, the stabilizing effective-diffusivity terms do not vanish
as $\xi\to 0$. We conclude that the forced correlation functions at
any arbitrary yet finite order will relax to time-independent
expressions for sufficiently small $\xi$. It makes then sense to
consider their scaling properties at the stationary state, as it was
done in (Adzhemyan and Antonov, 1998).  For the correlation functions
$\langle\left({\bm B}({\bm r},t) \cdot{\bm B}({\bm
0},t)\right)^N\rangle\propto r^{\gamma_{_N}}$, the perturbative
expression in $\xi$ reads $\gamma_{_N}=-N\xi-{2N(N-1)\xi\over d+2}+
{\cal O}(\xi^2)$, demonstrating the intermittency of the magnetic 
field distribution.

Let us conclude by discussing the behavior of the magnetic helicity
$\langle {\bm A}\cdot{\bm B}\rangle$, considered in (Borue and Yakhot,
1996). This quantity is conserved in the absence of the molecular
diffusivity, as can be easily verified using (\ref{magn}) and the
equation for ${\bm A}$:
\begin{equation}
\label{vectorpot}
\partial_t{\bm A}={\bm v}{\bm \times}{\bm B}-{\bm\nabla}\phi
+\kappa{\bm\nabla}^2{\bm A}.
\end{equation}
The function $\phi$ may be fixed by the choice of a specific gauge,
e.g. ${\bm\nabla}\cdot{\bm A}=0$.  The spatial behavior of the
helicity correlation functions $\langle{\bm A}({\bm r},t) \cdot{\bm
B}({\bm 0},t)\rangle$ is derived using (\ref{magn}) and
(\ref{vectorpot}) and averaging over the Gaussian velocity with the
variance (\ref{tobe1}). The resulting equation coincides with that for
the scalar covariance (\ref{pairc}), implying the dimensional scaling
$r^{2-\xi}$ and a constant helicity flux. We conclude that it is
possible to have coexistence of normal and anomalous scaling for
different components of the correlation tensor of a given order.  Note
also that the helicity correlation functions relax to a stationary
form even for $\xi>1$, i.e. when the magnetic correlation functions do
not. The increase of the magnetic field magnitude is indeed
accompanied by a modification of its orientation and the
quasi-orthogonality between ${\bm A}$ and ${\bm B}$ ensures the
stationarity of the helicity correlation functions. For a helical
velocity, considered in (Rogachevskii and Kleeorin, 1999), the
magnetic and the helicity correlators are coupled via the so-called
$\alpha$-effect, see e.g. (Moffatt, 1978), and the system is unstable
in the limit $\kappa\to 0$ considered here.

\subsection{Lagrangian numerics}
\label{sec:Lagnum}

The basic idea of the Lagrangian numerical strategy is to calculate
the scalar correlation functions using the particle trajectories. 
The expressions (\ref{thetaN}) and (\ref{forzato}) naturally
provide for such a Lagrangian Monte-Carlo formulation: the $N$
Lagrangian trajectories are generated by integrating the stochastic
equations (\ref{sde}), the right hand side of (\ref{thetaN}) and
(\ref{forzato}) is calculated for a large ensemble of realizations and
averaged over it.  If we are interested in correlation functions of finite
order, the Lagrangian procedure involves the integration of a few
differential equations. This is clearly more convenient than having to
deal with the partial differential equation for the scalar field. The
drawback is that quantities involving a large number of particles,
such as the tails of the PDF's, are not accessible.  Once the
correlation functions have been measured, their appropriate
combinations will give the structure functions.  For the second-order
$S_2(r,t)$ two different configurations of particles are needed. One
corresponding to $\langle\theta^2\rangle$, where the particles are at
the same point at time $t$, and another one corresponding to
$\langle\theta({\bbox r})\theta(0)\rangle$, where they are spaced by
${\bm r}$. For the $2n$-th order structure function,
$n+1$ particle configurations are needed.

Another advantage of the Lagrangian method is that it gives direct
access to the scaling in $\ell$ of the structure functions
(\ref{ansc}), that is to the anomalous dimensions
$\,\Delta_N=N\zeta_2/2-\zeta_{N}\,$.
The quantities
(\ref{thetaN}) and (\ref{forzato}) for various $\ell$'s can  indeed be
calculated along the same Lagrangian trajectories.  That is more
efficient than measuring the scaling in $r$, i.e. changing the final
positions of the particles and generating a new ensemble of
trajectories. 

\subsubsection{Numerical method}

The Lagrangian method as presented up to now might be applied to any
velocity field. The situation with the Kraichnan model is simpler
in two respects. First, the velocity statistics is time-reversible and
the Lagrangian trajectories can be generated forward in time. Second,
the velocity fields at different times are independent. Only $(N-1)d$
random variables are needed at each time step, corresponding to the
velocity increments at the location of the $N$ particles.  The major
advantage is that there is no need to generate the whole velocity
field. Finite-size effects, such as the space periodicity for
pseudo-spectral methods, are thus avoided.

The Lagrangian trajectories for the Kraichnan model are conveniently
generated as follows. The relevant variables are the interparticle
separations, e.g. ${\bm R}_{nN}={\bm R}_n-{\bm R}_N$ for $n=1,\ldots
N-1$. Their equations of motion are easily derived from (\ref{sde})
and conveniently discretized by the standard Euler-It\^o scheme of
order $1/2$ (Kloeden and Platen, 1992)
\begin{equation}
\label{EulIto}
{\bm R}_{nN}(t+\Delta t)-{\bm R}_{nN}(t)=\sqrt{\Delta t}\left(
{\bm V}_n+\sqrt{2\kappa}{\bm W}_n\right)\,,
\end{equation}
where $\Delta t$ is the time step. The quantities ${\bm V}_n$ and
${\bm W}_n$ are $d$-dimensional Gaussian, independent random vectors
generated at each time step. Both have zero mean and their covariance
matrices follow directly from the definition (\ref{tobe1}) of the
Kraichnan velocity correlation:
\begin{eqnarray}
\label{cheneso}
\langle V_n^iV_m^j\rangle=d^{ij}({\bm R}_{nN}) +d^{ij}({\bm
R}_{mN})-d^{ij}({\bm R}_{nN}-{\bm R}_{mN}),\qquad \langle
W_n^iW_m^j\rangle=(1+\delta_{nm})\delta^{ij}\,.
\end{eqnarray}
The most convenient numerical procedure to generate the two sets of
vectors is the classical Cholesky decomposition method (Ralston and
Rabinowitz, 1978). The covariance matrices are triangularized in the
form $M\,M^T$ and the lower triangular matrix $M$ is then multiplied by a
set of $(N-1)d$ Gaussian, independent random variables with zero mean
and unit variance. The resulting vectors have
the appropriate correlations.

Various possibilities to extract the anomalous scaling exponents are
available for the Kraichnan model. The straightforward one, used in
(Frisch {\it et al.}, 1998), is to take the forcing correlation close
to a step function (equal to unity for $r/\ell<1$ and vanishing
otherwise).  The correlation functions (\ref{forzato}) involve then
the products of the average residence times of couples of particles at
distances smaller than $\ell$.  An alternative method is based on the
shape dynamics discussed in Sect.\,\,\ref{sec:shape}.  Measuring
first-exit times and not residence times gives an obvious advantage in
computational time.  As stressed in (Gat {\it et al.}, 1998), the
numerical problem here is to measure reliably the contributions of the
irreducible zero modes, masked by the fluctuations of the reducible
ones. The latter were filtered out by Celani {\it et al.}\,\,(1999)
taking various initial conditions and combining appropriately the
corresponding first-exit times. A relevant combination for the fourth
order is for example ${\cal T}_\ell({\bm 0},{\bm 0},{\bm 0},{\bm 0})-
4{\cal T}_\ell({\bm r}_0,{\bm 0},{\bm 0},{\bm 0})+3{\cal T}_\ell({\bm
r}_0,{\bm r}_0,{\bm 0},{\bm 0})$, where ${\cal T}_\ell$ is the first
time the size of the particle configuration reaches $\ell$. 

\subsubsection{Numerical results}
\label{sec:num}

We shall now present the results for the Kraichnan model obtained by
the Lagrangian numerical methods just discussed.

The fourth-order anomaly $2\zeta_2-\zeta_4$ {\it vs} the exponent
$\xi$ of the velocity field is shown in Fig.~\ref{fig:anomaly} for
both $2d$ and $3d$ (Frisch {\it et al.}, 1998 and 1999). A few remarks
are in order. First, the comparison between the $3d$ curve and the
prediction $4\xi /5$ for small $\xi$'s provides direct support for the
perturbation theory discussed in Sect.\,\,\ref{sec:pert}. Similar
numerical support for the expansion in $1/d$ has been obtained in
(Mazzino and Muratore-Ginanneschi, 2001). Second, the curve close to
$\xi=2$ is fitted with reasonable accuracy by $a(2-\xi)
+b(2-\xi)^{3/2}$ for $a=0.06$ and $b=1.13$.  That is compatible with
an expansion in powers of $(2-\xi)^{1/2}$ (Shraiman and Siggia, 1996),
where the first term is ruled out by the Schwartz inequality
$\zeta_4\le 2\zeta_2=2(2-\xi)$.  Third, remark that the anomalies are
stronger in $2d$ than in $3d$ and their maximum shifts towards smaller
$\xi$ as the dimension decreases. The former remark is in agreement
with the general idea which emerged in previous Sections that
intermittency is associated with the particles staying close to each
other for times longer than expected.  It is indeed physically quite
sensible that those processes are favored by lowering the space
dimensionality.  The second remark can be qualitatively interpreted as
follows (Frisch {\it et al.}, 1998).  Consider scalar fluctuations at
a given scale. The smaller-scale components of the velocity act like
an effective diffusivity whilst its larger-scale components affect the
scalar as in the Batchelor regime. Neither of them leads to any
anomalous scaling of the scalar. Those non-local interactions are
dominant as $\xi\to 0$ and $\xi\to 2$, respectively.  For intermediate
values of $\xi$ the velocity components having a scale comparable to
that of the scalar fluctuations become important and intermittency is
produced. The strongest anomalies are attained when the relevant
interactions are mostly local.  To qualitatively explain how the
maximum of the anomalies moves with the space dimensionality, it is
then enough to note that the effective diffusivity increases with $d$
but not the large-scale stretching. As for the dependence on the order
of the moments, the maximum moves toward smaller $\xi$ as $N$
increases, see the $3d$ curves for the sixth-order anomaly
$3\zeta_2-\zeta_6$ (Mazzino and Muratore-Ginanneschi, 2001). It is
indeed natural that higher moments are more sensitive to
multiplicative effects due to large-scale stretching than to additive
effects of small-scale eddy diffusivity.

Let us now discuss the phenomenon of saturation, i.e. the fact that
$\zeta_N$ tend to a constant at large $N$.  The orders where
saturation is taking place are expected to increase with $\xi$ and
diverge as $\xi\to 0$. It is then convenient to consider small values
of $2-\xi$. On the other hand, approaching the Batchelor limit too
closely makes non-local effects important and the range of scales
needed for reliable measurements becomes huge.  A convenient trade-off
is that considered in (Celani {\em et al.}, 2000a) with the $3d$ cases
$2-\xi=0.125, 0.16$ and $0.25$. For the first value of $\xi$ it is
found there that the fourth and the sixth-order exponents are the same
within the error bars.  The curves for the other $\xi$ values show
that the order of saturation increases with $2-\xi$, as expected. How
do those data constrain the $\zeta_N$ curve? It follows from the
H\"older inequalities that the curve for $N>6$ must lie below the
straight line joining $\zeta_4$ and $\zeta_6$.  Furthermore, from the
results in Section~\ref{sec:dec} we know that the spatial scaling
exponents in the forced and the decaying cases are the same and
independent of the scalar initial conditions.  For the unforced
equation (\ref{tracer}), the maximum value of $\theta$ cannot increase
with time. Taking an initial condition with a finite maximum value, we
have the inequality $S_N(r,t)\leq \left(2\,{\rm max}\,\theta\right)^p
S_{N-p}(r,t)$. We conclude that the $\zeta_N$ curve cannot decrease
with the order. The presence of error bars makes it, of course,
impossible to state rigorously that the $\zeta_N$ curves tend to a
constant. It is however clear that the combination of the numerical
data in (Celani {\em et al.}, 2000a) and the theoretical arguments
discussed in Section~\ref{sec:inst} leaves little doubt about the
saturation effect in the Kraichnan model. The situation with an
arbitrary velocity field is the subject of Section~\ref{sec:kra}.

\subsection{Inverse cascade in the compressible Kraichnan model}
\label{sec:cascades}

The uniqueness of the Lagrangian trajectories discussed in
Section~\ref{sec:compres} for the strongly compressible Kraichnan
model has its counterpart in an inverse cascade of the scalar field,
that is in the appearance of correlations at larger and larger
scales. Moreover, the absence of dissipative anomaly allows to
calculate analytically the statistics of scalar increments and to show
that intermittency is suppressed in the inverse cascade regime. In
other words, the scalar increment PDF tends at long times to a
scale-invariant form.

Let us first discuss the simple physical reasons for those results.
The absence of a dissipative anomaly is an immediate consequence of
the expression (\ref{thetaN}) for the scalar correlation functions.  
If the trajectories are unique, particles that
start from the same point will remain together throughout the
evolution and all the moments $\langle\theta^N\rangle(t)$ are
preserved. Note that the conservation laws are statistical: the
moments are not dynamically conserved in every realization, but
their average over the velocity ensemble are.

In the presence of pumping, fluctuations are injected and a
flux of scalar variance toward the large scales is established. As
explained in Section~\ref{sec:dirlarge}, scalar correlation functions at very
large scales are related to the probability for initially
distant particles to come close.  In strongly compressible flow, the
trajectories are typically contracting, the particles tend to approach
and the distances will reduce to the forcing correlation length $\ell$
(and smaller) for long enough times. Strong correlations at larger and
larger scales are therefore established as time increases, signaling
the inverse cascade process.

The absence of intermittency is due to the fact that the $N$-th order
structure function is dynamically related to a two-particles
process. Correlation functions of the $N$-th order are generally
determined by the evolution of $N$-particle configurations. However,
the structure functions involve initial configurations with just two
groups of particles separated by a distance $r$.  The particles
explosively separate in the incompressible case and we are immediately
back to an $N$-particle problem.  Conversely, the particles that are
initially in the same group remain together if the trajectories are
unique. The only relevant degrees of freedom are then given by the
intergroup separation and we are reduced to a two-particle
dynamics. It is therefore not surprising that the scaling behaviors at
the various orders are simply related in the inverse cascade regime.

Specifically, let us consider the equations for the single-point
moments $\langle\theta^N\rangle(t)$. Since the moments are conserved by 
the advective term, see (\ref{thetaN}), their behavior in the limit
$\kappa\to 0$ (nonsingular now) is the same as for the equation
$\partial_t\theta=\varphi$. It follows that the single-point
statistics is Gaussian, with $\langle\theta^2\rangle$ coinciding with
the total injection $\Phi(0)t$ by the forcing.

The equation for the structure functions $S_N(r,t)$ is derived from
(\ref{KrHopf}). That was impossible in the incompressible case since
the diffusive terms could not be expressed in terms of $S_N$ (another
sign of the dissipative anomaly).  No such anomaly exists here so we
can disregard the diffusion term and simply derive the equation for
$S_N$ from (\ref{KrHopf}). This is the central technical point allowing
for the analytical solution. The equations at the various
orders are recast in a compact form via the generating function
$Z(\lambda\,;{\bm r},t)=\langle {\rm
e}^{i\lambda\,\Delta_r\theta}\rangle$ for the scalar increments
$\Delta_r\theta=\theta({\bm r},t)-\theta({\bm 0},t)$. The equation for
the generating function is
\begin{equation}
\label{santabarbara}
\partial_t Z(\lambda\,;{\bm
r},t) =M\, Z(\lambda\,;{\bm
r},t) +\lambda^2\left(\Phi(0)-\Phi(r/\ell)\right)Z(\lambda\,;{\bm
r},t),
\end{equation}
where the operator $M$ was defined in (\ref{ciccio}) and $Z=1$ at the
initial time. Note that $M$ is the restriction of ${\cal M}_2$,
signaling the two-particle nature of the dynamics at any order.  The
stationary solution for $Z$ depends on the pumping, but
two different regions with a universal behavior can be identified.

\noindent{\bf i). Large scales $r\gg \ell$}. \ In this region, 
$\Phi(r)$ in (\ref{santabarbara}) can be neglected and the generating 
function is an eigenfunction of $M$ with eigenvalue
$\lambda^2\,\Phi(0)$. Introducing a new variable $r^{(2-\xi)/2}$, the
equation (\ref{santabarbara}) is transformed into a Bessel form and
solved analytically. The solution takes a scale-invariant form
$Z(|\lambda|\,r^{(2-\xi)/2})$ whose detailed expression may be found in
(Gaw\c{e}dzki and Vergassola, 2000). It follows that $S_2(r)\propto
r^{2-\xi}$. The associated scalar flux is calculated as in
(\ref{KrYag}) and turns out to be constant in the upscale direction,
the footprint of the inverse cascade.  The scale-invariant form of
the generating function signals that no anomalous scaling is
present in the inverse cascade regime. As we shall discuss in
Section~\ref{sec:2dcasc}, the phenomenon is not accidental and other
physical systems with inverse cascades share the same property.
Note that, despite its
scale-invariance, the statistics of the scalar field is
strongly non-Gaussian. The expression for the scalar increment PDF is
obtained by the Fourier transform of the generating function.  The
tails of the PDF decay algebraically with the power $-(2b+1)$, where
$b=1+(\gamma-d)/(2-\xi)$ and $\gamma$ was defined in
(\ref{ciccio}). The slow decay of the PDF renders moments of order
$N>2b$ divergent (as $t^{N/2-b}\,r^{b(2-\xi)}$) when time increases.
A special case is that of smooth velocities $\xi=2$, considered in
(Chertkov {\it et al.}, 1998).  The PDF of scalar increments reduces
then to the same form as in the direct cascade at small scales. For
amplitudes smaller than $\ln (r/\ell)$ the PDF is Gaussian.  The
reason is the same as in Section~\ref{sec:dirsmall}: the time to reach
the integral scale $\ell$ from an initial distance $r\gg \ell$ is
typically proportional to $\ln(r/\ell)$; the fluctuations of the travel 
time are Gaussian.  For larger amplitudes, the PDF has an exponential tail
whose exponent depends on the whole hierarchy of the Lyapunov
exponents, as in the smooth incompressible case.

\noindent{\bf ii). Small scales $r\ll \ell$}. \ Contrary to large
scales, the scalar increments are now strongly intermittent. The
structure functions of integer orders are all dominated by the zero
mode of the $M$ operator scaling as $r^{b(2-\xi)}$, with $b$ defined
in the previous paragraph.  The exponent of the constant
flux solution $\propto r^{2-\xi}$ crosses that of the zero mode at the
threshold of compressibility $b=1$ for the inverse cascade.

Let us now consider the role of an infrared cutoff in the inverse
cascade dynamics. The natural motivation is the quasi-stationarity of
the statistics: due to the excitation of larger and larger scales,
some observables do not reach a stationary form. It is therefore of
interest to analyze the effects of physical processes, such as
friction, acting at very large scales. The corresponding equation
of motion is
\begin{equation} 
\partial_t \theta+{\bm
v}\cdot{\bm\nabla}\theta+\alpha\,\theta-\kappa{\bm\nabla}^2\theta
=\varphi , \label{friction}
\end{equation}
and we are interested in the limit $\alpha\to 0$.  For nonsmooth
velocities, the friction and the advection balance at a scale
$\eta_{_f}\sim \alpha^{-1/(2-\xi)}$, much larger than $\ell$ as
$\alpha\to 0$.  The smooth case $\xi=2$ is special as no such scale
separation is present and it will be considered at the end of the
Section. For nonsmooth flows, the energy is injected at the integral
scale $\ell$, transferred upwards in the inertial range and finally
extracted by friction at the scale $\eta_{_f}$.  

The friction term in (\ref{friction}) is taken into account by noting
that the field $\exp(\alpha\,t)\,\theta$ satisfies the usual passive
scalar equation with a forcing $\exp (\alpha\, t)\, f$. It follows
that the previous Lagrangian formulae can be carried over by just
introducing the appropriate weights. The expression (\ref{grfct}) for
the $N$-point correlation function becomes, for instance,
\begin{equation}
C_N({\bm r},t) = \int_0^t\hspace{-0.1cm}ds\int
\hspace{-0.07cm}{\rm e}^{-(t-s)\, N\alpha}\,
P_N^{t,s}({\bm r};\,{\bm r'}) \sum_{n<m}C_{N-2}({\bm
r}^{'}_1,\mathop{\dots\dots}\limits_{\widehat{n}\ \ \widehat{m}},{\bm
r}^{'}_N\,;s)\,\Phi({\vert {\bm r}^{'}_n-{\bm r}^{'}_m\vert}/\ell)\,d
{\bm r}^{'}.
\label{Nc1a}
\end{equation}
From (\ref{Nc1a}), one can derive the equations for
$\langle\,\theta^N\,\rangle (t)$ and the structure functions
$S_{N}(r,t)$ and analyze them (Gaw\c{e}dzki and Vergassola, 2000). The
single-point moments are finite and the scalar distribution is
Gaussian with $\langle\theta^2\rangle=\Phi(0)/2\alpha$.  The structure
functions of order $N<2b$ are not affected by friction. The orders
that were previously diverging are now finite and they all scale as
$r^{b(2-\xi)}$ in the inertial range, with a logarithmic correction
for $N=2b$. The algebraic tails that existed without friction are
replaced by an
exponential fall off for amplitudes larger than $\sqrt{{\Phi(0)/
\alpha}}$.  The saturation of the exponents comes from the fact that
the moments of order $N/2\geq b$ are all dominated by the contribution
near the cut.

Let us conclude by considering the smooth case $\xi=2$, where the velocity
increments scale linearly with the distance. The advective term ${\bm
v}\cdot{\bm\nabla}$ has zero dimension, like the friction term. As
first noted by Chertkov (1998, 1999), this naturally leads to an anomalous
scaling and intermittency. Let us consider for example the
second-order correlation function $C_2(r,t)=\langle\theta({\bm r},t)
\theta({\bm 0},t)\rangle$.  
Its governing equation is derived from (\ref{Nc1a}):
\begin{equation}
\label{duedue}
\partial_t\,C_2=MC_2+\Phi(r)C_2-2\alpha C_2,
\end{equation}
with the same $M$ operator as in (\ref{santabarbara}). At large
scales $r\gg \ell$, the forcing term is negligible and we look for a
stationary solution.  Its nontrivial decay $C_2\propto(r/\ell)^{-{\Delta}_2}$ 
is due to the zero mode arising from the balance between the $\,M\,$
operator and the friction term:
\begin{equation}
\label{frican}
\Delta_2={1\over 2}\left[
\sqrt{(\gamma-d)^2+{8\alpha \over
(d-1)(1+2\wp)}}-\left(\gamma-d\right)\right],
\end{equation}
where $\gamma$ is as in (\ref{ciccio}) and $\wp$ is the
compressibility degree of the velocity.  The notation $\Delta_2$ is
meant to stress that dimensional arguments would predict an exponent
zero. Higher-order connected correlation functions also exhibit
anomalous decay laws. Similar mechanisms for anomalous scaling and
intermittency for the $2d$ direct enstrophy cascade in the presence of
friction are discussed in Section~\ref{sec:vort}.

\subsection{Lessons for general scalar turbulence} 
\label{sec:kra}

The results for spatially nonsmooth flows have mostly been derived
within the framework of the Kraichnan model. Both the forcing and the
velocity were Gaussian and short-correlated in time. As discussed in
Section \ref{sec:passKrai}, the conditions on the forcing are not
crucial and may be easily relaxed. The scaling properties of the
scalar correlation functions are universal with respect to the
forcing, i.e. independent of its details, while the constant
prefactors are not. The situation with the velocity field is more
interesting and nontrivial. Even though a short-correlated flow might
in principle be produced by an appropriate forcing, all the cases of
physical interest have a finite correlation time. The very existence
of closed equations of motion for the particle propagators, which we
heavily relied upon, is then lost.  It is therefore natural to ask
about the lessons drawn from the Kraichnan model for scalar turbulence
in the generic situation of finite-correlated flows. The existing
numerical evidence is that the basic mechanisms for scalar
intermittency are quite robust: anomalous scaling is still associated
with statistically conserved quantities and the expansion (\ref{asex})
for the multiparticle propagator seems to carry over. The specific
statistics of the advecting flow affects only quantitative details,
such as the numerical values of the exponents.  The general
consequences for the universality of the scalar statistics and the
decay of the anisotropies are presented in what follows. We also show
that the phenomenon of saturation, discussed in
Sections~\ref{sec:inst} and \ref{sec:num} for the Kraichnan model, is
quite general.

A convenient choice of the velocity ${\bm v}$ to investigate the
previous issues is a $2d$ flow generated by an inverse energy cascade
(Kraichnan, 1967). The flow is isotropic, it has a constant upscale
energy flux and is scale-invariant with exponent $1/3$,
i.e. without intermittency corrections as shown both by 
experiments and numerical simulations (Paret and Tabeling, 1998;
Smith and Yakhot, 1993; Boffetta {\it et al.}, 2000).  The inverse
cascade flow is thus akin to the Kraichnan ensemble, but its
correlation times are finite.

Let us first discuss the preserved Lagrangian structures. The simplest
nontrivial case to analyze anomalous scaling is the third-order
correlation function $C_3(\underline{\bm r})= \langle \theta({\bm
r}_1,t)\,\theta({\bm r}_2,t)\,\theta({\bm r}_3,t)\rangle$. The
function is non-zero only if the symmetry $\theta\mapsto -\theta$ is
broken, which often happens in real systems via the presence of a mean
scalar gradient $\langle\theta\rangle={\bm g}\cdot{\bm r}$. The
function $C_3(\underline{\bm r})$ depends then on the size, the
orientation with respect to $\hat{\bm g}$ and the shape of the
triangle defined by ${\bm r}_1$, ${\bm r}_2$ and ${\bm r}_3$. For a
scalar field advected by the $2d$ inverse energy cascade flow, the
dependence on the size $R$ of the triangle is a power law with
anomalous exponent $\zeta_3=1.25$, smaller than the dimensional
prediction $5/3$ (Celani {\it et al.}, 2000a).  To look for
statistical invariants under the Lagrangian dynamics, let us take a
translation-invariant function $f(\underline{\bm r})$ of the $N$
points ${\bm r}_n$ and define its Lagrangian average as in
(\ref{tav}), i.e. as the average of the function calculated along the
Lagrangian trajectories. In the $2d$ inverse energy cascade, the
distances grow as $|t|^{3/2}$ and the Lagrangian average of a scaling
function of positive degree $\sigma$ is expected to grow as
$|t|^{3\sigma /2}$.  The numerical evidence presented in (Celani and
Vergassola, 2001) is that the anomalous scaling is again due to
statistical integrals of motion: the Lagrangian average of the
anomalous part of the correlation functions remains constant in time.
The shape of the figures identified by the tracer particles plays
again a crucial role: the growth $\propto |t|^{3\zeta_3 /2}$ of the
size factor $R^{\zeta_3}$ in $C_3(\underline{\bm r})$ is compensated
by its shape dependence.  As we indicated in Section~\ref{sec:shape},
anomalous scaling reflects slowed-down separations among subgroups of
particles and the fact that triangles with large aspect ratios live
much longer than expected.  It is also immediate to provide an example
of slow mode, see (\ref{asex}), for the two particle dynamics.  The
Lagrangian average of $\left({\bm g}\cdot{\bm r}_{12}\right)$ is
obviously preserved as its time derivative is proportional to $\langle
\left({\bm v}_1- {\bm v}_2\right)\rangle=0$. Its first slow mode is
given by $\left({\bm g}\cdot{\bm r}_{12}\right)\,{r_{12}^{2/3}}$,
whose Lagrangian average is found to grow as $|t|$ (that is much slower than
$|t|^{5/2}$) at
large times. In the presence of a finite volume and boundaries, the
statistical conservation laws hold as intermediate asymptotic
behaviors, as explicitly shown in (Arad {\it et al.}, 2001).

An important consequence is about the decay rate of the
anisotropies. As already mentioned, isotropy is usually broken by the
large-scale injection mechanisms. One of the assumptions in the
Oboukhov-Corrsin reformulation of the
Kolmogorov 1941 theory for the passive scalar 
is that the statistical isotropy of the scalar
is restored at sufficiently small scales. Experiments do not confirm
those expectations. Consider, for example, the case where a mean
scalar gradient ${\bm g}$ is present. A quantitative measure of the
degree of anisotropy is provided by odd-order structure functions or
by odd-order moments of $\hat{\bm g}\cdot \nabla\theta$. All these
quantities are identically zero for isotropic fields. The predictions
of the Oboukhov-Corrsin theory for the anisotropic situations are the
following. The hyperskewness of the scalar increments
$S_{2n+1}(r)/S_2^{n+1/2}(r)$ should decay with the separation as
$r^{2/3}$. The corresponding behavior of the scalar gradient
hyperskewness with respect to the P\'eclet number should be ${\rm
Pe}^{-1/2}$. In fact, the previous quantities are experimentally found
to remain constant or even to increase with the relevant parameter
(Gibson {\it et al.}, 1977; Mestayer, 1982; Sreenivasan, 1991;
Mydlarski and Warhaft, 1998). There is therefore no restoration of
isotropy in the original Kolmogorov sense and the issue of the role of
anisotropies in the small-scale scalar statistics is naturally raised
(Sreenivasan, 1991; Warhaft, 2000).  The analysis of the same problem
in the Kraichnan model is illuminating and permits to clarify the
issue in terms of zero medes and their scaling exponents.  For
isotropic velocity fields, the correlation functions may be decomposed
according to their angular momentum $j$, as in Section~\ref{sec:pert}
Each of those contributions is characterized by a scaling exponent
$\zeta_{_N}^{j}$.  The general expectation, confirmed in all the
situations where the explicit calculations could be performed, is that
$\zeta_{_N}^{ j\neq 0}>\zeta_{_N}^{j=0}$ and that the exponents
increase with $j$. As their degree of anisotropy increases, the
contributions are less and less relevant at small scales. Note that,
in the presence of intermittency, the inequality $\zeta_{_N}^{j\neq
0}< {N\over2}\zeta_{_2}$ is still possible and anisotropies might then
have dramatic effects for observables whose isotropic dominant
contribution is vanishing, such as $S_{2n+1}/S_2^{n+1/2}$. Rather than
tending to zero, they may well increase while going toward the small
scales, blatantly violating the restoration of isotropy in the
original Kolmogorov sense. Remark that no violation of the hierarchy
in $j$ is implied though. In other words, the degree of anisotropy of
every given moment does not increase as the scale decreases; if
however one measures odd moments in terms of the appropriate power of
the second one (as is customary in phenomenological approaches) then
the degree of anisotropy may grow downscales. The previous arguments
are quite general and compatible with all the existing experimental
and numerical data for passive scalar turbulence. For Navier-Stokes
turbulence, the use of the rotation symmetries and the existence of a
hierarchy among the anisotropic exponents were put forward and
exploited in (Arad {\it et al.} 1998, 1999). Numerical evidence for
persistence of anisotropies analogous to those of the scalar fields
was first presented in (Pumir and Shraiman, 1995).

Let us now discuss the phenomenon of saturation. A snapshot of the
scalar field advected by the $2d$ inverse cascade flow is shown in
Fig.~\ref{fig:snap}. A clear feature is that strong scalar
gradients tend to concentrate in sharp fronts separated by large
regions where the variations are weak. The scalar jumps across the
fronts are of the order of
$\theta_{rms}=\sqrt{\langle\theta^2\rangle}$, i.e.  comparable to the
largest values of the field itself. Furthermore, the minimal width of
the fronts reduces with the dissipation scale, pointing to their
quasi-discontinuous nature. If the probability of having such
$\theta_{rms}$ jumps across a separation $r$ decreases as
$r^{\zeta_{\infty}}$, then phenomenological arguments of the
multifractal type suggest a saturation to the asymptotic value
$\zeta_{\infty}$, see (Frisch, 1995). The presence of fronts in scalar
turbulence is a very well established fact, both in experiments
(Gibson {\it et al.}, 1977; Mestayer, 1982; Sreenivasan, 1991;
Mydlarski and Warhaft, 1998) and in numerical simulations (Holzer and
Siggia, 1994; Pumir, 1994). It is shown in the latter work that fronts
are formed in the hyperbolic regions of the flow, where distant
particles are brought close to each other. The other important remark
is that fronts appear also in the Kraichnan model (Fairhall {\it et
al.}, 1997; Vergassola and Mazzino, 1997; Chen and Kraichnan, 1998),
despite the $\delta$-correlation of the velocity.  What matters for
bringing distant particles close to each other are indeed the effects
cumulated in time. The integral of a $\delta$-correlated random
process behaves as a Brownian motion in time, whose sign is known to
have strong persistence properties (Feller, 1950). Even a
$\delta$-correlated flow might then efficiently compress the particles
(locally) and this naturally explains how fronts may be formed in the
Kraichnan model.  It is also clear from the previous arguments that a
finite correlation time favors the formation of fronts and that the
Kraichnan model is somehow the most unfavorable case in this respect.
The fact that fronts still form and that the saturation takes place
points to generality of the phenomenon for scalar turbulence. The
order of the moments and the value where the $\zeta_N$ curve flattens
out might depend on the statistics of the advecting velocity, but the saturation
itself should generally hold. Direct evidence for the advection by a
$2d$ inverse cascade flow is provided in (Celani {\it et al.},
2000a,b).  Saturation is equivalent to the scalar increment PDF taking
the form ${\cal P}(\Delta_r\theta)=r^{\zeta_{\infty}}q(\Delta_r\theta/
\theta_{rms})$ for amplitudes larger than $\theta_{rms}$. The tails at
various $r$ can thus be all collapsed by plotting
$r^{-\zeta_{\infty}}{\cal P}$, as shown in
Fig.~\ref{fig:pdfcollapse}. Note finally that the saturation exponent
$\zeta_{\infty}$ coincides with the fractal codimension of the fronts,
see (Celani {\it et al.}, 2001) for a more detailed discussion.

As far as the compressible Kraichnan model is concerned,
applying even qualitative predictions requires much more care than in
the incompressible case.  Indeed, the compressibility of the flow
makes the sum of the Lyapunov exponents nonzero and leads to the
permanent growth of density perturbations described in
Sect.\,\,\ref{sec:density}.  In a real fluid, such growth is stopped
by the back-reaction of the density on the velocity, providing for a
long-time memory of the divergence ${\bm\nabla}\cdot{\bm v}$ of the
velocity along the Lagrangian trajectory. This shows that some
characteristics of the Lagrangian velocity may be considered
short-correlated (like the off-diagonal components of the strain
tensor), while other are long correlated (like the trace of the
strain).

In summary, the situation with the Kraichnan model and general passive
scalar turbulence is much like the motto at the beginning of the
review. The interest was originally stirred by the closed equations of
motion for the correlation functions and the possibility of deriving
an explicit formula for the anomalous scaling exponents, that are
quite specific features. Actually, the model turned out to be much
richer and capable of capturing the basic properties of the Lagrangian
tracer dynamics in generic turbulent flow. The major lessons drawn
from the model, such as the statistical integrals of motion, the
geometry of the particle configurations, the dynamics in smooth flow,
the importance of multipoint correlations, the persistence of
anisotropies, all seem to have quite general validity.

\section{Burgers and Navier-Stokes equations}

All the previous sections were written under the assumption that the
velocity statistics (whatever it is) is given. In this Chapter, we
shall describe what one can learn about the statistics of the velocity
field itself by considering it in the Lagrangian frame. We start from the
simplest case of Burgers turbulence whose inviscid version describes
a free propagation of fluid particles, while viscosity
provides for a local interaction. We then consider an incompressible
turbulence where the pressure field provides for a nonlocal
interaction between infinitely many particles.

\subsection{Burgers turbulence} 
\label{sec:burg}

The $d$-dimensional Burgers equation (Burgers, 1974) is a
pressureless version of the Navier-Stokes equation (\ref{NS}): \qq
\partial_t{\bm v}+{\bm v}\cdot{\bm\nabla}{\bm v}-\nu{\nabla}^2 {\bm
v}={\bm f}
\label{bur}
\qqq for irrotational (potential) velocity ${\bm v}({\bm r},t)$ and
force ${\bm f}({\bm r},t)$. It is used to describe a variety of
physical situations from the evolution of dislocations in solids to
the formation of large scale structures in the universe, see
e.g. (Krug and Spohn, 1992; Shandarin and Zel'dovich, 1989). Involving
a compressible $\bm v$, it allows for a meaningful (and nontrivial)
one-dimensional case that describes small-amplitude perturbations of
velocity, density or pressure depending on a single spatial coordinate
in the frame moving with sound velocity, see e.g.\,\,(Landau and
Lifshitz, 1959). Without force, the evolution described by (\ref{bur})
conserves total momentum $\int{\bm v}\,d{\bm r}$. By the
substitutions ${\bm v}={\bm\nabla} \phi$ and ${\bm f}={\bm\nabla}g$
the Burgers equation is related to the KPZ equation \qq
\partial_t\phi+{1\over2}(\nabla\phi)^2-\nu\,\nabla^2\phi=g\,.
\label{KPZ}
\qqq governing an interface growth (Kardar {\em et al}., 1986).

Already the one-dimensional case of (\ref{bur}) illustrates the themes
that we discussed in previous Chapters: turbulent cascade, Lagrangian
statistics and anomalous scaling. Under the action of a large-scale
forcing (or in free decay of large-scale initial data) a cascade of
kinetic energy towards the small scales takes place. The nonlinear
term provides for steepening of negative gradients and the viscous
term causes energy dissipation in the fronts that appear this way. In
the limit of vanishing viscosity, the energy dissipation stays finite
due to the appearance of velocity discontinuities called shocks. The
Lagrangian statistics is peculiar in such an extremely nonsmooth flow
and can be closely analyzed even though it does not correspond to a
Markov process. Forward and backward Lagrangian statistics are
different, as it has to be in an irreversible flow. Lagrangian
trajectories stick to the shocks. That provides for a strong
interaction between the particles and results in an extreme anomalous
scaling of the velocity field. A tracer field passively advected by
such a flow undergoes an inverse cascade.

At vanishing viscosity, the Burgers equation may be considered as
describing a gas of particles moving in a force field.  Indeed, in the
Lagrangian frame defined for a regular velocity by $\dot{\bm R}={\bm
v}({\bm R},t)$, relation (\ref{bur}) becomes the equation of motion of
non-interacting unit-mass particles whose acceleration is determined
by the force: \qq \ddot{\bm R}\,=\,{\bm f}({\bm R},t)\ .
\label{freep}
\qqq In order to find the Lagrangian trajectory ${\bm R}(t;{\bm r})$ 
passing at time zero through $\bm r$ it is then enough to solve the 
second order equation (\ref{freep}) with the initial conditions 
${\bm R}(0)={\bm r}$ and $\dot{\bm R}(0)={\bm v}({\bm r},0)$. 
For sufficiently short times such trajectories do not cross and
the Lagrangian map $\,{\bm r}\mapsto{\bm R}(t;{\bm r})\,$ is invertible.
One may then reconstruct $\,\bm v\,$ at time $t$ from the relation 
$\,v({\bm R}(t),t)=\dot{\bm R}(t)$. The velocity stays potential 
if the force is potential. At longer times, however, the particles
collide creating velocity discontinuities, i.e.\,\,shocks.  The
nature and the dynamics of the shocks may be understood by treating the
inviscid equation as the limit of the viscous one.  Positive viscosity
removes the singularities. As is well known, the KPZ equation
(\ref{KPZ}) may be linearized by the Hopf-Cole substitution
$Z=\exp[-\phi/2\nu]$ that gives rise to the heat equation in an
external potential: \qq
[\partial_t-\nu{\nabla}^2+{_1\over^{2\nu}}\,g]\,Z=0\,.
\label{heat}
\qqq The solution of the initial value problem for the latter may be
written as the Feynman-Kac path-integral \qq Z({\bm
r},t)=\int\limits_{{\bm R}(t)={\bm r}}\exp\left[-{_1\over^{2\nu}}
\,S({\bm R})\right]\,Z({\bm R}(0),0)\ \,{\cal D}{\bm R}\,.  \qqq with
the classical action $\,S({\bm R})=\int\limits_0^t\Big[{1\over2}
{\dot{{\bm R}}}^2+g(\tau,{\bm R})\Big]\,d\tau\,$ of a path ${\bm
R}(\tau)$. The limit of vanishing viscosity selects the least-action
path: \qq \phi({\bm r},t)\ =\mathop{\rm min}\limits_{_{{\bm
R}\,:\,\,{\bm R}(t) ={\bm r}}}\ \Big\{S({\bm R})\,+\,\phi({\bm
R}(0),0)\Big\}\,.
\label{shsol}
\qqq Equating to zero the variation of the minimized expression, one
infers that the minimizing path $\,{\bm R}(\tau)\equiv{\bm
R}(\tau;{\bm r},t)\,$ is a solution of (\ref{freep}) such that
$\dot{\bm R}(0)={\bm v}({\bm R}(0),0)$.  Taking the gradient of
(\ref{shsol}), one also infers that ${\bm v}({\bm r},t)=\dot{\bm
R}(t)$. The above procedure extends the short-time construction of the
solutions of the inviscid Burgurs equation to all times at the cost of
admitting shocks where $\,{\bm v}\,$ is discontinuous. The velocity
still evolves along the solutions of (\ref{freep}) with $\dot{\bm
R}(0)$ given by the initial velocity field, but if there are many such
solutions reaching the same point at the same moment, the ones that do
not realize the minimum (\ref{shsol}) should be disregarded. The
shocks arise when there are several minimizing paths. At fixed time,
shocks are, generically, hypersurfaces which may intersect, have
boundaries, corners, etc. (Vergassola {\it et al.}, 1994; Frisch {\it
et al.}, 1999; Frisch and Bec, 2000). This may be best visualized in
the case without forcing where the equation (\ref{shsol}) takes the
form \qq \phi({\bm r},t)\ =\ \mathop{\rm min}\limits_{{\bm r}'}\
\left\{ {_1\over^{2t}}\,({\bm r}-{\bm r}')^2\,+\,\phi({\bm
r}',0)\,\right\}
\label{sheqf}
\qqq with a clear geometric interpretation: $\,\phi({\bm r},t)\,$ is
the height $C$ of the inverted paraboloid $\,C-{1\over2t} ({\bm
r}-{\bm r'})^2\,$ touching but not overpassing the graph of
$\,\phi({\bm r'},0)$.  The points of contact between the paraboloid
and the graph correspond to ${\bm r}'$ in (\ref{sheqf}) on which the
minimum is attained. Shock loci are composed of those $\bm r$ for
which there are several such contact points.

The simplest situation occurs in one dimension with ${\bm r}\equiv x$.
Here, shocks are located at points $x_i$ where the velocity jumps,
i.e.\,\,where the two-sided limits $\,v^\pm(x,t)\equiv\lim\limits_{
\epsilon\to0}v(x\pm\epsilon,t)\,$ are different. The limits correspond
to the velocities of two minimizing paths that do not cross except at
$x_i$. The shock height $\,s_i=v^+(x_i,t)-v^-(x_i,t)\,$ has to be
negative. Once created shocks never disappear but they may merge so
that they form a tree branching backward in time. For no forcing, the
shock positions $x_i$ correspond to the inverse parabolas
$\,C-{1\over2t}({x_i}-{x'})^2\,$ that touch the graph of
$\,\phi(x',0)\,$ in (at least) two points ${x'_i}^\pm$, $\,{x'_i}^+
>{x'_i}^-$, such that $\,v^\pm(x_i,t)={x_i-{x'_i}^\pm\over t\,}$, see
Fig.~\ref{fig:burgers}. Below, we shall limit our discussion to the 
one-dimensional case that was most extensively studied in the literature, 
see (Burgers, 1974; Woyczy\'{n}ski, 1998; E and Vanden Eijnden, 2000a). 
In particular, the asymptotic long-time large-distance behavior of 
freely decaying initial data with Gaussian finitely correlated velocities 
or velocity potentials has been intensively studied. In the first case,
the asymptotic solution for $\,x^{3/2}t^{-1}={\cal O}(1)\,$ has the
form (\ref{sheqf}) with $\phi(x',0)$ representing a Brownian
motion. In the second case, the solution settles under the diffusive
scaling with a logarithmic correction $x^2t^{-1}\ln^{1/4}x={\cal
O}(1)$ to the form \qq \phi(x,t)\ =\ \mathop{\rm min}\limits_{j}\
\left\{ {_1\over^{2t}} (x-y_j)^2\,+\,\phi_j\right\}\,,
\label{kida}
\qqq where $(y_j,\phi_j)$ is the Poisson point process with intensity
${\rm e}^\phi\,dyd\phi$. In both cases explicit calculations of the
velocity statistics have been possible, see, respectively, (Burgers,
1974; Frachebourg and Martin, 2000) and (Kida, 1979; Woyczy\'{n}ski,
1998). Other asymptotic regimes of decaying Gaussian initial data were
analyzed in (Gurbatov {\it et al.}, 1997).

The equation of motion of the one-dimensional shocks $x_i(t)$ is easy
to obtain even in the presence of forcing. To this aim, note that
along the shock there are two minimizing solutions determining the
same function $\phi$ and that \qq {d\over
dt}\,\phi(x_i,t)\,=\,\partial_t\phi(x_i,t)
+\dot{x_i}\partial_x\phi(x_i,t)\,=\,-{1\over 2}v^\pm(x_i,t)^2
\,+\,g(x_i,t)\,+\,\dot{x_i}v^\pm(x_i,t)\,.  \qqq Equating both
expressions, we infer that $\dot{x_i}
={1\over2}[v^+(x_i,t)+v^-(x_i,t)]\equiv\bar v(x_i,t)\equiv\bar v_i\,$,
\,i.e.\,\,that the shock speed is the mean of the velocities on both
sides of the shock. The crucial question for the Lagrangian
description of the Burgers velocities is what happens with the fluid
particles after they reach shocks where their equation of motion
$\dot{x}=v(x,t)$ becomes ambiguous. The question may be easily
answered by considering the inviscid case as a limit of the viscous
one where shocks become steep fronts with large negative velocity
gradients. It is easy to see that the Lagrangian particles are trapped
within such fronts and keep moving with them. We should then define
the inviscid Lagrangian trajectories as solutions of the equation
$\dot x\,=\,\bar v(x,t)$, with $\dot x$ understood as the right
derivative. Indeed, the solutions of that equation clearly move with
the shocks after reaching them. In other words, the two particles
arriving at the shock from the right and the left at a given moment
aggregate upon the collision. Momentum is conserved so that their
velocity after the collision is the mean of the incoming ones (recall
that the particles have unit mass) and is equal to the velocity of the
particles moving with the shock that have been absorbed at earlier
times. Note that in the presence of shocks the Lagrangian map becomes
many-to-one, compressing whole space-intervals into the shock
locations.

It is not difficult to write field evolution equations that take into
account the presence of shocks (Vol'pert, 1967; Bernard and
Gaw\c{e}dzki, 1998; E and Vanden Eijnden, 2000a).  We shall do it for
local functions of the velocity of the form ${\rm e}^{\lambda
v(x,t)}$.  \,For positive viscosity, these functions obey the equation
of motion \qq
(\partial_t+\lambda\,\partial_\lambda\,\lambda^{-1}\partial_x-\lambda\,f)
\, {\rm e}^{\,\lambda\,v}\ =\ -\lambda^2\,\epsilon(\lambda)\,,
\label{sheq}
\qqq where
$\,\epsilon(\lambda)=\nu[(\partial_xv)^2-\lambda^{-2}\partial_x^2]\,
{\rm e}^{\,\lambda\, v}\,$ is the contribution of the viscous term in
(\ref{bur}).  In the limit $\nu\to 0$ the dissipation becomes
concentrated within the shocks. Using the representation $\,{\rm
e}^{\,\lambda\, v(x,t)}={\rm
e}^{\,\lambda\,v^+(x,t)}\,\theta(x-x_i(t))+{\rm e}^{\,
\lambda\,v^-(x,t)}\,\theta(x_i(t)-x)\,$ around the shocks, it is easy
to check that (\ref{sheq}) still holds for $\nu=0$ with \qq
\epsilon(\lambda;x,t)\ =\ \sum\limits_{i}\,F(\bar v_i,s_i) \
\delta(x-x_i(t))\,,
\label{epl}
\qqq
concentrated at shock locations. The ``form-factor'' $\,F(\bar v,s)
=-2\,{\rm e}^{\,\lambda\,\bar v}\,\lambda^{-1}\,\partial_\lambda\,
\lambda^{-1}\sinh{\lambda\,s\over2}$.

When the forcing and/or initial data are random, the equation of
motion (\ref{sheq}) induces the Hopf evolution equations for the
correlation functions. For example, in the stationary homogeneous
state $\,\langle f\,{\rm e}^{\,\lambda\,v}\rangle=\lambda\,\langle
\epsilon(\lambda)\rangle$. \,Upon expanding into powers of $\lambda$,
this relates the single-point expectations $\,\langle v^n\rangle\,$ to
the shock statistics. The first of these relations says that the
average force should vanish and the second gives the energy balance
$\,\langle f v\rangle=\bar\epsilon_v\,$, \,where the mean energy
dissipation rate $\,\bar\epsilon_v=\langle\epsilon(0)\rangle
=-\rho\,\langle s^3\rangle/12$ and
$\,\rho=\Big\langle\sum\limits_i\delta(x-x_i(t))\Big\rangle\,$ is the
mean shock density.  Similarly, for the generating function of the
velocity increments $\langle\exp(\lambda\,\Delta v)\rangle\,$ with
$\Delta v=v_1-v_2$, one obtains \qq
\lambda^2\,\partial_\lambda\,\lambda^{-2}\partial_1\,\Big\langle {\rm
e}^{\,\lambda\,\Delta v}\Big\rangle\,-\,\lambda\,\Big\langle\Delta f\,
{\rm e}^{\,\lambda\,\Delta v}\,\Big\rangle\ =\ -\lambda^2\,\Big\langle
\epsilon(\lambda)_1\,{\rm e}^{-\lambda\,v_2}+{\rm
e}^{\,\lambda\,v_1}\, \epsilon(-\lambda)_2\Big\rangle\,.
\label{gf2}
\qqq For a Gaussian force with $\langle f(x,t)\,f(0,0)\rangle=
\delta(t)\,\chi(x/L)$, where $L$ is the injection scale,
$\langle\Delta f\, {\rm e}^{\lambda\Delta
v}\rangle=[\chi(0)-\chi(\Delta x/L)]\,\langle {\rm e}^{\lambda\Delta
v}\rangle$. For a spatially smooth force, the second term in
(\ref{gf2}) tends to zero when the separation $\Delta x$ (taken
positive) shrinks.  In the same limit, the quantity inside the
expectation value on the right hand side tends to a local operator
concentrated on shocks, as in (\ref{epl}), but with the
height-dependent form-factor $\,-4\,{\rm e}^{\lambda
s/2}\lambda^{-1}\partial_\lambda \lambda^{-1}\sinh{\lambda\,
s\over2}=-\partial_\lambda \lambda^{-2}({\rm
e}^{\,\lambda\,s}-1-\lambda\,s)$.  \,Comparing the terms, one infers
that for $N=3,4\dots$, \qq \lim\limits_{x_2\to x_1}\
\partial_1\,\Big\langle(\Delta v)^N \Big\rangle\ =\ \rho\,\langle
s^N\rangle\,.  \qqq The first of these equalities is the
one-dimensional version of the Kolmogorov flux relation
(\ref{45}). The higher ones express the fact that, for small $\Delta
x$, the higher moments of velocity increments are dominated by the
contribution from a single shock of height $s$ occurring with
probability $\,\rho\,\Delta x$. \,That implies the anomalous scaling of
the velocity structure functions \qq S_{_N}(\Delta x)\equiv
\Big\langle(\Delta v)^N\Big\rangle\ =\ \rho\,\langle
s^N\rangle\,\Delta x\ +\ o(\Delta x)\, ,
\label{p>1}
\qqq that is a signature of an extreme intermittency. In fact, due to 
the shock contributions, all moments of velocity increments of order 
$p\geq 1$ scale with exponents $\zeta_p=1$. In contrast, for 
$\,0\leq p\leq1$, $\,\zeta_p=p$ since the fractional moments are 
dominated by the regular contributions to velocity. Indeed, denoting 
by $\xi$ the regular part of the velocity gradient, one obtains 
(E {\it et al.}, 1997; E, Khanin {\it et al.} 2000; E and Vanden 
Eijnden, 2000a) \qq \langle
(|\Delta v|)^p\rangle=\langle|\xi|^p\rangle\,(\Delta x)^p 
+o((\Delta x)^p)\,.\label{p<1}
\qqq The shock contribution proportional to $\Delta x$ is subleading
in that case.

The Lagrangian interpretation of these results can be based on the
fact that the velocity is a Lagrangian invariant of the unforced
inviscid system.  In the presence of the force, \qq v(x,t)\ =\
v(x(0),0)\,+\int\limits_0^t f(x(s),s)\,ds\, , \qqq along the
Lagrangian trajectories. The velocity is an active scalar and the
Lagrangian trajectories are evidently dependent on the force that
drives the velocity. One cannot write a formula like (\ref{Gauss})
obtained by two independent averages over the force and over the
trajectories.  Nevertheless, the main contribution to the
distance-dependent part of the 2-point function $\,\langle
v(x,t)\,v(x',t)\rangle\,$ is due, for small distances, to realizations
with a shock in between the particles. It is insensitive to a
large-scale force and hence approximately proportional to the time
that the two Lagrangian trajectories ending at $x$ and $x'$ take to
separate backwards to the injection scale $L$. With a shock in between
$x$ and $x'$ at time $t$, the initial backward separation is linear so
that the second order structure function becomes proportional to
$\,\Delta x$, in accordance with (\ref{p>1}).  Other structure
functions may be analyzed similarly and give the same linear
dependence on the distance (all terms involve at most two
trajectories). This mechanism of the anomalous scaling is similar to
that of the collinear case in Sect.\,\ref{sec:dirlarge}.

Following numerical observations of Chekhlov and Yakhot (1995, 1996),
a considerable effort has been invested to understand the shape of
the PDF's ${\cal P}(\Delta v)\,$ of the velocity increments and
${\cal P}(\xi)\,$ of the regular part of $\,\partial_xv$. The stationary
expectation of the exponential of $\xi$ satisfies for a white-in-time
forcing the relation \qq
\left(\lambda\,\partial_\lambda^2-\partial_\lambda
-D\,\lambda^2\right)\,\Big\langle{\rm e}^{\,\lambda\,\xi}\Big\rangle
=\langle\rho_\lambda\rangle\,
\label{fex}
\qqq that may established the same way as (\ref{gf2}). Here
$\,D=-{1\over2}\,\chi''(0)\,L^{-2}\,$ and $\,\rho_\lambda\,$ is an
operator supported on shocks with the form-factor $\,{s\over 2}({\rm
e}^{\,\lambda\,\xi^+}+{\rm e}^{\,\lambda \,\xi^-})$. \,For the PDF of
$\,\xi\,$ this gives the identity (E and Vanden Eijnden, 1999, 2000a):
\qq \left(\,D\,\partial_\xi^2+\xi^2\partial_\xi+3\xi\right)\,
{\cal P}(\xi)\,+\,\int{_{d\lambda}\over^{2\pi i}}\,{\rm e}^{-\lambda\,\xi}
\,\langle\rho_\lambda\rangle\ =\ 0\,.
\label{ttak}\qqq 
Various closures for this equation or for (\ref{gf2}) have been
proposed (Polyakov, 1995; Bouchaud and M\'{e}zard, 1996; Boldyrev,
1997; Gotoh and Kraichnan, 1998). They all give the right tail
$\,\propto{\rm e}^{-\xi^3/(3D)}\,$ of the distribution, as determined
by the first two terms with derivatives in (\ref{ttak}), with the
power-law prefactors depending on the details of the closure.  That
right tail was first obtained by Feigel'man (1980) and also reproduced
for ${\cal P}(\Delta v)\,$ by an instanton calculation in (Gurarie and
Migdal, 1996).  The instanton appears to be a solution of a
deterministic Burgers equation with a force linear in space, see also
(Balkovsky {\it et al.}, 1997b). The right tail may be also understood
from the stochastic equation along the trajectory $\,{d^2\over
dt^2}\Delta x=\Delta f$. \,For $\Delta x$ much smaller than the
injection scale $L$, the force increment may be linearized $\,\Delta
f=\sigma(t)\,\Delta x$. \,In the first approximation, $\,\sigma(t)\,$
is a white noise and we obtain a problem familiar from the
one-dimensional localization in a $\delta$-correlated potential. In
particular, $\,z={d\over dt}(\Delta x) \Big/\Delta x\,$ satisfies
$\,\dot{z}=\sigma(t)-z^2$. \,This is a Langevin dynamics in the
unbounded from below potential $\propto z^3\,$ (Bouchaud and
M\'{e}zard, 1996). Upon conditioning against escape to $\,z=-\infty$,
one gets a stationary distribution for $z$ with the right tail
$\,\propto{\rm e}^{-z^3/(3D)}$ and the left tail $\,\propto
z^{-2}$. \,In reality, due to the shock creation, $\,\Delta f/\Delta
x\approx \partial_xf(x(t),t)\,$ is a white noise only if we fix the
value of $\,v\,$ at the end point of the Lagrangian trajectory. This
introduces subtle correlations which effect the power-law factors in 
${\cal P}(\xi)$, \,in particular its left tail $\propto|\xi|^{-\alpha}$. 
Large negative values of $\,\xi\,$ appear in the vicinity of preshocks, 
as stressed in (E {\it et al.}, 1997) where the value $\alpha=7/2$ was 
argued for, based on a geometric analysis of the preshocks,
the birth points of the shock discontinuities. E and Vanden Eijnden 
(2000a), have proved by analysis of realizability conditions for the 
solutions of (\ref{ttak}) that $\alpha>3\,$ and made a strong
case for $\alpha=7/2$, assuming that shocks are born with vanishing
heights and that preshocks do not accumulate.  The exponent $\alpha=7/2$
was subsequently found in the decaying case with random large scale 
initial conditions both in 1d (Bec and Frisch, 2000) and in higher
dimensions (Frisch {\it et al.}, 2001), and in the forced case when 
the forcing consists of deterministic large scale kicks repeated 
periodically (Bec {\it et al.}, 2000). In those cases, the statistics 
of the shocks and there creation process are easier to control than for 
the $\delta$-correlated forcing. The numerical analysis of the latter
case clearly confirms, however, the prediction $\alpha=7/2$.
As to the PDF ${\cal P}(\Delta v)\,$ of the velocity increment
in the decaying case and, possibly, in the forced case, it exhibits 
a crossover from the behavior characteristic for the 
velocity gradients to the one reproducing the behaviors (\ref{p>1}) 
and (\ref{p<1}). Note that the single-point velocity PDF also has cubic 
tails $\ln {\cal P}(v)\propto -|v|^3$ (Avellaneda {\it et al.}, 1995; 
Balkovsky {\it et al.}, 1997b). The same is true for white-driven 
Navier-Stokes equation and may be generalized for a non-Gaussian force 
(Falkovich and Lebedev, 1997).

The Lagrangian picture of the Burgers velocities allows for a simple
analysis of advection of scalar quantities carried by the flow. In 
the inviscid and diffusionless limit, the advected tracer satisfies 
the evolution equation
\qq
\partial_t\theta\,+\,\bar v\,\partial_x\theta\ =\ \varphi\,,
\label{insc}
\qqq where $\,\varphi\,$ represents an external source. As usual, the
solution of the initial value problem is given in terms of the PDF
$\,p(x,t;y,0\,|\,v)\,$ to find the backward Lagrangian trajectory at
$y$ at time $0$, given that at later time $t$ it passed by $x$, see
(\ref{trsol}). Except for the discrete set of time $t$ shock
locations, the backward trajectories are uniquely determined by
$x$. As a result, a smooth initial scalar will develop discontinuities
at shock locations but no stronger singularities. Since a given set of
points $\,(x_{_1},\dots,x_{_N})\equiv\underline{x}\,$ avoids the
shocks with probability 1, the joint backward PDF's of $N$
trajectories $\,{\cal P}_{_N}(\underline{x} ;\underline{y};-t)$, \,see
(\ref{pn}), should be regular for distinct $x_n$ and should possess
the collapse property (\ref{colla}). This leads to the conservation of
$\,\langle\theta^2\rangle\,$ in the absence of scalar sources
and to the linear pumping of the scalar variance into a soft mode when
a stationary source is present. Such behavior corresponds to an
inverse cascade of the passive scalar, as in the strongly compressible
phase of the Kraichnan model discussed in Section
(\ref{sec:cascades}).

The Burgers velocity itself and all its powers constitute an example
of advected scalars. Indeed, the equation of motion (\ref{sheq}) may
be also rewritten as \qq (\partial_t+\bar
v\,\partial_x-\lambda\,f)\,{\rm e}^{\,\lambda\,v}\ =\ 0 \qqq from
which the relation (\ref{insc}) for $\theta=v^n$ and $\varphi
=nfv^{n-1}$ follows. Of course, $v^n$ are active scalars so that in
the random case their initial data, the source terms, and the
Lagrangian trajectories are not independent, contrary to the case of
passive scalars. That correlation makes the unlimited growth of
$\langle v^2\rangle$ impossible: the larger the value of local
velocity, the faster it creates a shock and dissipates the energy.
The difference between active and passive tracers is thus substantial
enough to switch the direction of the energy cascade from inverse for
the passive scalar to direct for the velocity.

As usual in compressible flows, the advected density $\,n\,$ satisfies 
the continuity equation \qq \partial_t
n\,+\,\partial_x(\bar vn)\ =\ \varphi
\label{inden}
\qqq
different from (\ref{insc}) for the tracer. 
The solution of the initial value problem is given by the forward
Lagrangian PDF $\,p(y,0;x,t\,|\,v)$, \,see (\ref{trsol}).  Since the
trajectories collapse, a smooth initial density will become singular
under the evolution, with $\delta$-function contributions
concentrating all the mass from the regions compressed to shocks by
the Lagrangian flow. Since the trajectories are determined by the
initial point $y$, the joint forward PDF's $\,{\cal
P}_{_N}(\underline{y};\underline{x};t)\,$ should have the collapse
property (\ref{colla}) but they will also have contact terms in
$x_n$'s when the initial points $y_n$ are distinct.  Such terms signal
a finite probability of the trajectories to aggregate in the forward
evolution, the phenomenon that we have already met in the strongly
compressible Kraichnan model discussed in
Sect.\,\,(\ref{sec:compres}). The velocity gradient $\partial_x v$ is
an example of an (active) density satisfying equation (\ref{inden})
with $\varphi=\partial_xf$.

In (Bernard and Bauer, 1999), the behavior of the Lagrangian PDF's and
the advected scalars summarized above have been established by a
direct calculation in freely decaying Burgers velocities with random
Gaussian finitely-correlated initial potentials $\phi$.
\vskip 0.5cm

\subsection{Incompressible turbulence from a Lagrangian viewpoint}
\label{sec:NS}

As we learned from the study of passive fields, treating the
dissipation is rather easy as it is a linear mechanism. The main 
difficulty resides in proper understanding advection. 
For incompressible turbulence, the problem is even more
complicated than for the Burgers equation due to spatial
nonlocality of the pressure term.  The Euler equation may indeed
be written as the equation  
\begin{equation}
\ddot{\bm R}={\bm f}({\bm R},t)-{\bm\nabla} P\
,\label{Euler}
\end{equation}
for the Lagrangian trajectories ${\bf R}(t;{\bf r})$
where ${\bm f}$ is the external force and the pressure field is
determined by the incompressibility condition ${\bm\nabla}^2 P=-{\bm
\nabla}\cdot\left[{\bm v}\cdot{\bm \nabla}{\bm v}\right]$ with ${\bm
v}=\dot{\bm R}$ and the spatial derivatives taken with respect
to ${\bm R}$. The inversion of the Laplace operator in the previous
relation brings in nonlocality via the kernel decaying as a power
law.  We thus have a system of infinitely many particles interacting
strongly and nonlocally.  In such a situation, any attempt at an
analytic description looks unavoidably dependent on possible
simplifications in limiting cases.  The natural parameter to exploit
for the incompressible Euler equation is the space dimensionality,
varying between two and infinity. The two-dimensional case indeed
presents important simplifications since the vorticity is a scalar
Lagrangian invariant of the inviscid dynamics, as we shall discuss
hereafter. The opposite limit of the infinite-dimensional Euler
equation is very tempting for some kind of mean-field approximation to
the interaction among the fluid particles but nobody has derived it yet.
The level of discussion in this section is thus naturally different
from the rest of the review: as a consistent theory is absent, we present
a set of particular arguments and remarks that, on one hand, make contact
with the previously discussed subjects, and, on other hand, may
inspire further progress.

\subsubsection{Enstrophy cascade in two dimensions}
\label{sec:vort}

The Euler equation in any even-dimensional space has an infinite set
of integrals of motion besides energy. One may indeed show that
the determinant of the matrix $\omega_{ij}=\nabla_j v^i-\nabla_i
v^j$ is the nonnegative density of an integral of motion, i.e.  $\int
F(\det{\omega})d{\bm r}$ is conserved for any function $F$. The
quadratic invariant $\int (\det{\omega})^2 d{\bm r}$ is
called enstrophy. In the presence of an external pumping $\phi$
injecting energy and enstrophy, it is clear that both quantities may
flow throughout the scales. If both cascades are present, they cannot
go in the same direction: the different dependence of energy and
enstrophy on the scale prevents their fluxes to be both constant
in the same interval. A finite energy dissipation would imply an
infinite enstrophy dissipation in the limit $\nu\to0$.  The natural
conclusion is that, given a single pumping at some intermediate scale
and assuming the presence of two cascades, the energy and the
enstrophy flow toward the large and the small scales, respectively
(Kraichnan, 1967; Batchelor, 1969). This is indeed the case for the 
two-dimensional case.

In this Section, we shall focus on the $2d$ direct enstrophy cascade.
The basic knowledge of the Lagrangian dynamics presented in Sections
\ref{sec:general} and \ref{sec:batch} is essentially everything
needed. Vorticity in $2d$ is a scalar and the analogy between
vorticity and passive scalar was noticed by Batchelor and Kraichnan
already in the sixties. Vorticity is not passive though and such
analogies may be very misleading, as it is the case for vorticity
and  magnetic field in $3d$ and for velocity and passive scalar in 1d Burgers. 
The basic flux relation for the
enstrophy cascade is analogous to (\ref{Yaglom}):
\begin{equation}\langle({\bbox v}_1\cdot{\bm\nabla}_1 
+{\bbox v}_2\cdot{\bm\nabla}_2)\omega_1\omega_2
\rangle=\langle\varphi_1\omega_2+\varphi_2\omega_1\rangle=P_2\
.\label{vflux} \end{equation} The subscripts indicate the spatial
points ${\bm r}_1$ and ${\bm r}_2$ and the pumping is assumed to be
Gaussian with $\langle\varphi({\bm r},t)\phi({\bm
0},0)\rangle=\delta(t)\Phi(r/L)$ decaying rapidly for $r>L$.  The
constant $P_2\equiv\Phi(0)$, having dimensionality $time^{-3}$, is the
input rate of the enstrophy $\omega^2$. Equation (\ref{vflux}) states
that the enstrophy flux is constant in the inertial range, that is for
$r_{12}$ much smaller than $L$ and much larger than the viscous
scale. A simple power counting suggests that the velocity differences
and the vorticity scale as the first and the zeroth power of $r_{12}$,
respectively. That fits the idea of a scalar cascade in a spatially
smooth velocity: scalar correlation functions are indeed logarithmic
in that case, as it was discussed in Sect.~\ref{sec:batch}.

Even though one can imagine hypothetical power-law vorticity spectra
(Saffman, 1971; Moffatt, 1986; Polyakov, 1993), one can argue that
they are structurally unstable (Falkovich and Lebedev, 1994). Indeed,
imagine for a moment that the pumping at $L$ produces the spectrum
$\langle(\omega_1-\omega_2)^p\rangle\propto r_{12}^{\zeta_p}$ at
$r_{12}\ll L$. Regularity of the Euler equation in $2d$ requires
$\zeta_p>0$, see, e.g., (Eyink, 1995) and references therein.  In the
spirit of the stability theory of Kolmogorov spectra (Zakharov {\em et
al.}, 1992), let us add an infinitesimal pumping at some $\ell$ in the
inertial interval producing a small yet nonzero flux of
enstrophy. Small perturbations $\delta\omega$ obey the equation
$\,\partial_t\delta\omega+(v\nabla)\delta\omega+ (\delta v\nabla)
\omega=\nu{\nabla}^2\delta\omega$. \,Here, $\delta v$ is the velocity
perturbation related to $\delta\omega$. The perturbation
$\delta\omega$ has the typical scale $\ell$ while vorticity $\omega$,
associated with the main spectrum, is concentrated at $L$ when
$\zeta_2>0$.  The third term can be neglected as it is $(\ell/L)^{2}$
times smaller than the second one. Therefore, $\delta\omega$ behaves
as a passive scalar convected by main turbulence, i.e. the Batchelor
regime from Section \ref{sec:batch} takes place.  The correlation
functions of the scalar are logarithmic in this case for any velocity
statistics.  The perturbation in any vorticity structure function thus
decreases downscales slower than the contribution of the main flow.
That means that any hypothetical power-law spectrum is structurally
unstable with respect to the pumping variation. Stability analysis
cannot of course describe the spectrum downscales where the
perturbations is getting comparable to the main flow. It is logical
though to assume that since only the logarithmic regime may be
neutrally stable, it represents the universal small-scale asymptotics
of steady forced turbulence. Experiments (Rutgers, 1998; Paret {\em et
al.}, 1999; Jullien {\em et al.}, 2000) and numerical simulations
(Borue, 1993; Gotoh, 1998; Bowman {\em et al.}, 1999) are compatible
with that conclusion.

The physics of the enstrophy cascade is thus basically the same as
that for a passive scalar: a fluid blob embedded into a larger-scale
velocity shear is extended along the direction of a positive strain
and compressed along its negative eigendirection; such stretching
provides for the vorticity flux toward the small scales, with the rate
of transfer proportional to the strain. The vorticity rotates the blob
decelerating the stretching due to the rotation of the axes of
positive and negative strain. One can show that the vorticity
correlators are indeed solely determined by the influence of larger
scales (that give exponential separation of the fluid particles)
rather than smaller scales (that would lead to a diffusive growth as
the square root of time). The subtle differences from the passive
scalar case come from the active nature of the vorticity. Consider for
example the relation (\ref{Gauss}) expressing the fact that the
correlation function of a passive scalar is essentially the time spent
by the particles at distances smaller than $L$.  The passive nature of
the scalar makes Lagrangian trajectories independent of scalar pumping
which is crucial in deriving (\ref{Gauss}) by two independent
averages. For an active scalar, the two averages are coupled since the
forcing affects the velocity and thus the Lagrangian trajectories.  In
particular, the statistics of the forcing along the Lagrangian
trajectories $\phi\left({\bm R}(t)\right)$ involves nonvanishing
multipoint correlations at different times.  Falkovich and Lebedev
(1994) argued that, as far as the dominant logarithmic scaling of the
correlation functions is concerned, the active nature of vorticity
simply amounts to the following: the field can be treated as a passive
scalar, but the strain acting on it must be
renormalized with the scale. Their arguments are based on the analysis
of the infinite system of equations for the variational derivatives of
the vorticity correlation functions with respect to the pumping and
the relations between the strain and the vorticity correlation
functions. The law of renormalization is then established as follows,
along the line suggested earlier by Kraichnan (1967, 1971, 1975). From
(\ref{disp2}), one has the dimensional relation that $times$ behave as
$\,\omega^{-1}\ln(L/r)$. Furthermore, by using the relation
(\ref{Gauss}) for the vorticity correlation function, one has
$\,\langle\omega\,\omega\rangle\propto P_2\times time$. Combining the
two previous relations, the scaling $\,\omega \sim[P_2\ln(L/r)]^{1/3}$
follows. The consequences are that the distance between two fluid
particles satisfies: $\ln ( R/r)\sim P_2^{1/2}t^{3/2}$, and that the
pair correlation function $\langle \omega_1\omega_2 \rangle \sim
(P_2\ln(L/r_{12}))^{2/3}$. Note that the fluxes of higher powers
$\omega^{2n}$ are not constant in the inertial range due to the same
phenomenon of ``distributed pumping'' discussed in (vi) of
Sect.\,\ref{sec:passKrai}.  The vorticity statistics is thus
determined by the enstrophy production rate alone.

It is worth stressing that the logarithmic regime described above is a
small-scale asymptotics of a steady turbulence. Depending on the
conditions of excitation and dissipation, different other regimes can
be observed either during an intermediate time or in an intermediate
interval of scales.  First, a constant friction that provides for 
the velocity decay rate $\alpha$ prevails (if present) over
viscosity at scales larger than $(\nu/\alpha)^{1/2}$. At such scales,
the vorticity correlation functions are expected to behave as power
laws rather than logarithmically, very much like for the passive scalar
as described in (Chertkov, 1999) and in Section~\ref{sec:cascades}.  
Indeed, the advective and the friction
terms ${\bm v}\cdot{\bm\nabla}{\bm v}$ and $-\alpha {\bm v}$ have
again the same dimension for a smooth velocity. Nontrivial scaling is
therefore expected, including for the second-order correlation
function (and hence for the energy spectrum).  The difficulty is of
course that the system is now nonlinear and exact closed equations,
such as those in Section~\ref{sec:cascades}, are not available. For
theoretical attempts to circumvent that problem by some
approximations, not quite controlled yet, see (Bernard, 2000; Nam {\em
et al.}, 2000). Second, strong large-scale vortices often exist with
their (steeper) spectrum masking the enstrophy cascade in some
intermediate interval of scales (Legras {\em et al.}, 1988).

\subsubsection{On the energy cascades in incompressible turbulence} 
\label{sec:2dcasc}

The phenomenology of the energy cascade suggests that the energy flux
$\bar\epsilon$ is a major quantity characterizing the velocity statistics.  
It is interesting to understand the difference between the direct and the
inverse energy cascades from the Lagrangian perspective.  The 
mean Lagrangian time derivative of the squared velocity difference 
is as follows
\begin{equation}
\label{lagrder}
\left\langle {d(\Delta {\bm v})^2\over dt}\right\rangle=2\Big\langle
\Delta {\bm v}\,\Delta {\bm f}+\nu(2{\bm v}\cdot{\nabla}^2{\bm v}-
{\bm v}_1\cdot{\nabla}^2{\bm v}_2-{\bm v}_2\cdot{\nabla}^2{\bm
v}_1)\Big\rangle\ .\end{equation} The right hand side coincides 
with minus twice the flux and this gives
the Lagrangian interpretation of the flux relations. In the $2d$ inverse
energy cascade, there is no energy dissipative anomaly and the right
hand side in the inertial range is determined by the injection term
$4\langle {\bm f}\cdot{\bm v} \rangle$.  The energy flux is negative
(directed upscale) and the mean Lagrangian derivative is positive.  On
the contrary, in the $3d$ direct cascade the injection terms cancel and
the right hand side becomes equal to $-4\nu\langle\left(\nabla{\bm
v}\right)^2\rangle$. The mean Lagrangian derivative is negative while
the flux is positive (directed downscale).  This is natural as a
small-scale stirring causes opposite effects with respect to a
small-scale viscous dissipation. The negative sign of the mean
Lagrangian time derivative in $3d$ does not contradict the fact that any
couple of Lagrangian trajectories eventually separates and their
velocity difference increases. It tells however that the squared
velocity difference between two trajectories generally behaves in a
nonmonotonic way: the transverse contraction of a fluid element makes
initially the difference between the two velocities decrease, while
eventually the stretching along the trajectories takes over (Pumir and
Shraiman, 2000).

The Eulerian form of (\ref{lagrder}) is the generalization of
$4/5$-law (\ref{45}) for the $d$-dimensional case: $\langle(\Delta_r {
v})^3\rangle=-12\bar\epsilon r/d(d+2)$ if $\Delta_r v$ is
the longitudinal velocity increment and $\bar\epsilon$ is
positive for the direct cascade and negative for the inverse one.
Since the average velocity difference vanishes, a negative
$\langle(\Delta_r v)^3\rangle$ means that small longitudinal velocity
differences are predominantly positive, while large ones are negative.
In other words, in $3d$ if the longitudinal velocities of two particles
differ strongly then the particles are likely to attract each other; if 
the velocities are close, then the particles preferentially repel each
other. The opposite behavior takes place in $2d$, where the third-order
moment of the longitudinal velocity difference is positive.  Another
Lagrangian meaning of the flux law in $3d$ can be appreciated by
extrapolating it down to the viscous interval. Here, $\Delta_r {\bm
v}\approx \sigma {\bm r}$ and the positivity of the flux is likely to
be related to the fact that the negative Lyapunov exponent is the
largest one (in absolute value) in $3d$ incompressible turbulence.

If one assumes (after Kolmogorov) that $\,\bar\epsilon\,$ is the only
pumping-related quantity that determines the statistics then the
separation between the particles ${\bm R}_{12}={\bm R}(t;{\bm
r}_1)-{\bm R}(t;{\bm r}_2)$ has to obey the already mentioned
Richardson law: $\langle R_{12}^2\rangle\propto \bar\epsilon t^3$. The
equation for the separation immediately follows from the Euler
equation (\ref{Euler}): $\partial_t^2{\bm R}_{12} ={\bm
f}_{12}-{\bm\nabla}P_{12}$.  The corresponding forcing term ${\bm
f}_{12}\equiv{\bm f}({\bm R}(t;{\bm r}_1)) -{\bm f}({\bm R}(t;{\bm
r}_2))$ has completely different properties for an inverse energy
cascade in $2d$ than for a direct energy cascade in $3d$.  For the
former, $R_{12}$ in the inertial range is much larger than the forcing
correlation length.  The forcing can therefore be considered
short-correlated both in time and in space.  Was the pressure term
absent, one would get the separation growth: $\langle
R_{12}^2\rangle/\bar\epsilon t^3 =4/3$. The experimental data by
Jullien {\em et al.}\,\,(1999) give a smaller numerical factor $\simeq
0.5$, which is quite natural since the incompressibility constrains
the motion.  What is however important to note is that already the
forcing term prescribes the law $\langle R_{12}^2\rangle \propto t^3$
consistent with the scaling of the energy cascade. Conversely, for the
direct cascade the forcing is concentrated at the large scales and
$f_{12}\propto R_{12}$ in the inertial range. The forcing term is thus
negligible and even the scaling behavior comes entirely from the
advective terms (the viscous term should be accounted as well).
Another amazing aspect of the $2d$ inverse energy cascade can be
inferred if one considers it from the viewpoint of vorticity.
Enstrophy is transferred toward the small scales and its flux at the
large scales (where the inverse energy cascade is taking place)
vanishes. By analogy with the passive scalar behavior at the large
scales discussed in Sect.\,\ref{sec:passKrai}, one may expect the
behavior $\langle\omega_1\omega_2\rangle\propto r_{12}^{1-\alpha-d}$,
where $\alpha$ is the scaling exponent of the velocity. The
self-consistency of the argument dictated by the relation
$\omega={\bm\nabla}\times{\bm v}$ requires $1-\alpha-d=2\alpha-2$
which indeed gives the Kolmogorov scaling $\alpha=1/3$ for
$d=2$. Experiments (Paret and Tabeling, 1997, 1998) as well as
numerical simulations (Smith and Yakhot, 193; Boffetta {\it et al.},
2000) indicate that the inverse energy cascade has a normal Kolmogorov
scaling for all measured correlation functions. No consistent theory
is available yet, but the previous arguments based on the enstrophy
equipartition might give an interesting clue.  To avoid
misunderstanding, note that in considering the inverse cascades one
ought to have some large-scale dissipation (like bottom and wall
friction in the experiments with a fluid layer) to avoid the growth of
condensate modes on the scale of the container. Another example of
inverse cascade is that of the magnetic vector potential in
two-dimensional magnetohydrodynamics, where the numerical simulations
also indicate that intermittency is suppressed (Biskamp and Bremer,
1994). The generality of the absence of intermittency for inverse
cascades and its physical reasons is still an open problem.  The only
inverse cascade fully understood is that of the passive scalar in
Section\,\ref{sec:cascades}, where the absence of anomalous scaling
was related to the uniqueness of the trajectories in strongly
compressible flow.  That explanation applies neither to $2d$
Navier-Stokes nor to magnetohydrodynamics since the scalar is active
in both cases.  Qualitatively, it is likely that the scale-invariance
of an inverse cascade is physically associated to the growth of the
typical times with the scale. As the cascade proceeds, the
fluctuations have indeed time to get smoothed out and not
multiplicatively transferred as in the direct cascades, where the
typical times decrease in the direction of the cascade.

An interesting phenomenological Lagrangian model of $3d$ turbulence
based on the consideration of four particles was introduced by
Chertkov, Pumir {\it et al.}, (1999). As far as an anomalous scaling
observed in the $3d$ energy cascade is concerned, the primary target is
to understand the nature of the statistical integrals of motion
responsible for it. Note that the velocity exponent $\sigma_3=1$ and 
experiments demonstrate that $\sigma_p \to ap$ as $p\to0$ with $a$ 
exceeding $1/3$ beyond the measurement error, see (Sreenivasan and Antonia, 
1997) and the references therein. The convexity of $\sigma_p$ means
then that $\sigma_2>2/3$. In other words, already the pair correlation 
function should be determined by some nontrivial conservation law 
(like for magnetic fields in Sect.\,\ref{sec:magnana}).

\section{Conclusions}
\label{sec:con}

This review is intended to bring home to the reader two main points:
the power of the Lagrangian approach to fluid turbulence and the
importance of statistical integrals of motion for systems far from
equilibrium.

As it was shown in Sects.\,\ref{sec:part} and \ref{sec:fields}, the
Lagrangian approach allows for a systematic description of most
important aspects of particle and field statistics.  In a spatially
smooth flow, Lagrangian chaos and exponentially separating
trajectories are generally present.  The associated statistics of
passive scalar and vector fields is related to the statistics of large
deviations of stretching and contraction rates in a way that is well
understood. The theory is quite general and it finds a natural domain
of application in the viscous range of turbulence. The most important
open problem here seems to be the understanding of the back-reaction
of the advected field on the velocity.  That would include an account
of the buoyancy force in inhomogeneously heated fluids, the saturation
of the small-scale magnetic dynamo and the polymer drag reduction. In
nonsmooth velocities, pertaining to the inertial interval of developed
turbulence, the main Lagrangian phenomenon is the intrinsic
stochasticity of the particle trajectories that accounts for the
energy depletion at short distances. This phenomenon is fully captured
in the Kraichnan model of nonsmooth time-decorrelated velocities.  To
exhibit it for more realistic nonsmooth velocities and to relate it to
the hydrodynamical evolution equations governing the velocity field
remains an open problem. The spontaneous stochasticity of Lagrangian
trajectories enhances the interaction between fluid particles leading
to intricate multi-particle stochastic conservation laws. Here, there
are open problems already in the framework of the Kraichnan
model. First, there is the issue of whether one can build an operator
product expansion, classifying the zero modes and revealing their
possible underlying algebraic structure, both at large and small
scales.  The second class of problems is related to a consistent
description of high-order moments of scalar, vector and tensor fields.
In the situations where the amplitudes of the fields are growing, this
would be an important step towards a description of feedback effects.

Our inability to derive the Lagrangian statistics directly from the
Navier-Stokes equations of motion for the fluid particles is related
to the fact that the particle coupling is strong and nonlocal due to
pressure effects. Some small parameter for perturbative approaches,
like those discussed for the Kraichnan model, has often been sought
for.  We would like to stress, however, that most strongly coupled
systems, even if local, are not analytically solvable and that not all
measurable quantities may be derived from first principles. In fluid
turbulence, it seems more important to reach basic understanding of
the underlying physical mechanisms, than it is to find out the
numerical values of the scaling exponents. Such an understanding has
been achieved in passive scalar and magnetic fields through the
statistical conservation laws. We consider the notion of statistical
integrals of motion to be of central importance for fluid turbulence
and general enough to apply to other systems in non-equilibrium
statistical physics. Indeed, non-dimensionally scaling correlation
functions appearing in such systems should generally be dominated by
terms that solve dynamical equations in the absence of forcing (zero
modes). As we explained throughout the review, for passively advected
fields, such terms describe conservation laws that are related to the
geometry either of the configuration of particles (for the scalar) or
of the particle-plus-field configurations (for the magnetic field). It
is a major open problem to identify the appropriate configurations for
active and nonlocal cases. New particle-tracking methods (La Porta
{\it et al.}, 2001) open promising experimental possibilities in this
direction.  An investigation of geometrical statistics of fluid
turbulence by combined analytical, experimental and numerical methods
aimed at identifying the underlying conservation laws is a challenge
for future research.
\vskip 0.3cm

\noindent {\bf Acknowledgements}

We are grateful for hospitality to the Institute of Theoretical
Physics in Santa Barbara and the Institut des Hautes Etudes
Scientifiques in Bures-sur-Yvette where part of this work was
done. The project was supported by the National Science Foundation
under Grant No. PHY94-07194, the Israel Science Foundation, the
French-Israeli Cooperation Program Arc-en-Ciel, and the European Union
under Contract HPRN-CT-2000-00162.

\begin{figure} \caption{An illustration of the breakdown of the
Lagrangian flow in spatially non-smooth flows: infinitesimally close
particles reach a finite separation in a finite time.  The consequence
is the cloud observed in the figure.  The particles evolve in a fixed
realization of the velocity field and in the absence of any molecular
noise.}  \label{fig:panache}\end{figure}

\begin{figure} \caption{The implosion of Lagrangian
trajectories in a strongly compressible flow. Particles that are
initially released uniformly across a sizable span of the interval
are compressed and tend to produce a singular density field. }
\label{fig:compress}\end{figure}

\begin{figure} \caption{An example of Lagrangian
trajectories of three particles. The probability density of the
positions $\underline{{\bm R}}$, conditional to the $\underline{{\bm
r}}$'s, is described by the PDF $\,p({\bm r},s;{\bm R},t\,\vert\,{\bm
v})\,$ (in a fixed realization of the velocity). Its average over the
statistics of the velocity field gives the Green functions ${\cal P}
(\underline{\bm r};\,\underline{\bm R};\,t-s)$.}
\label{fig:particelle} \end{figure}

\begin{figure} \caption{The contour lines of a three particle
zero mode as a function of the shape of the triangle defined by the
particles.}  \label{fig:threepartcls} \end{figure}

\begin{figure} \caption{The fourth-order anomalous exponent
$2\zeta_2-\zeta_4$ of the scalar field {\it vs} the roughness
parameter $\xi$ of the velocity field in the Kraichnan model.  The
circles and the stars refer to the three-dimensional and the
two-dimensional cases, respectively.
The dashed lines are the perturbative predictions for small $\xi$ and
$2-\xi$ in 3d.}  \label{fig:anomaly} \end{figure}

\begin{figure} \caption{A typical snapshot of a scalar field
transported by a turbulent flow.}  \label{fig:snap} \end{figure}

\begin{figure} \caption{The PDF's ${\cal P}(\Delta_r \theta)$
of the scalar increments $\Delta_r\theta=\theta({\bm r})-\theta({\bm 0})$ 
for three values of $r$ inside the inertial range of scales, 
multiplied by the factor $r^{-\zeta_\infty}$.  The observed 
collapse of the curves implies the saturation of the scaling
exponents of the scalar structure functions.}  \label{fig:pdfcollapse}
\end{figure}

\begin{figure} \caption{Geometric construction of the Hopf-Cole
inviscid solution of the 1D Burgers equation. The inverted parabola
$C-{1\over 2t}(r-r')^2$ is moved upwards until the first contact point
with the profile of the initial potential $\phi(r',0)$. The
corresponding height $C$ gives the potential $\phi(r,t)$ at time
$t$. Shocks correspond to positions $r$ where there are several
contact points $r'$, as for the first parabola on the
left.}\label{fig:burgers} \end{figure}

\newpage
\appendix
\section{Regularization of stochastic integrals}

For the $\delta$-correlated strain, the equation (\ref{disp2}) becomes
a stochastic differential equation.  Let us present here a few
elementary facts about such equations for that simple case. The differential
equation (\ref{disp2}) is equivalent to the integral equation \qq {\bm
R}(t)={\bm R}(0)\,+\,\int\limits_0^t\sigma(s) \,ds\,{\bm
R}(s)\,,
\label{intsto}
\qqq 
where ${\bm R}\equiv{\bm R}_{12}$.
\,The right hand side involves a stochastic integral whose
distinctive feature is that $\,\sigma(t)\,dt\,$ is of order
$\,(dt)^{1/2}$, \,as indicated by the relation $\,\Big\langle\Big (
\int_0^{t}\sigma_{ij}(s)\,ds\Big)^2\Big\rangle\propto t$. \,Such
integrals require second-order manipulations of differentials and,
in general, are not unambiguously defined without the choice of a 
defining convention.
The most popular are the It\^{o}, the Stratonovich and the
anti-It\^{o} ones. Physically, different choices reflect finer details
of the strain correlations wiped out in the white-noise scaling limit,
like the presence or the absence of time-reversibility of the velocity
distribution. The It\^{o}, Stratonovich and anti-It\^{o} versions of
the stochastic integral in (\ref{intsto}) are given by the limits over
partitions of the time interval of different Riemann sums: 
\qq \int\limits_0^t\sigma(s)\,ds\,{\bm R}(s)\
=\ \cases{ \lim\,\sum\limits_{n}\int\limits_{t_{n}}
^{t_{n+1}}\sigma(s)\,ds\,\,{\bm R}(t_{n})\,,\cr
\lim\,\sum\limits_n\int\limits_{t_{n}}^{t_{n+1}}\sigma(s)\,ds
\,\,{1\over2}[{\bm R}(t_{n})+{\bm R}(t_{n+1})]\,,\cr
\lim\,\sum\limits_{n}\int\limits_{t_{n}}^{t_{n+1}}
\sigma(s)\,ds\,\,{\bm R}(t_{n+1})\,,}
\label{proced}
\qqq respectively, where $\,0=t_0<t_1<\cdots<t_N=t$. \,It is not difficult 
to compare the different choices.  For example, the difference between 
the second and the first one is \qq
{_1\over^2}\,\lim\,\sum\limits_n\smallint\limits_{t_{n}}^{t_{n+1}}\sigma(s)
\,ds\,\,[{\bm R}(t_{n+1})-{\bm R}(t_{n})]
&=&{_1\over^2}\,\lim\,\sum\limits_n\Big(\smallint\limits_{t_{n}}^{t_{n+1}}
\sigma(s)\,ds\Big)^2\,{\bm R}(t_n)\cr &=&\smallint\limits_0^t
\tilde C\,{\bm R}(s)\,\,ds, \qqq where $\,\tilde
C_{i\ell}=C_{ijj\ell}\,$ (sum over $j$) and the last equality is a
consequence of the Central Limit Theorem that suppresses the
fluctuations in
$\,\lim\,\sum\limits_n\Big(\int\limits_{t_{n}}^{t_{n+1}}
\sigma(s)\,ds\Big)^2$. \,Similarly, the difference between the
anti-It\^{o} and the It\^{o} procedure is twice the latter expression.  In
other words, the Stratonovich and anti-It\^{o} versions of
(\ref{disp2}) are equivalent, respectively, to the It\^{o} stochastic
equations \qq d{\bm R}\ =\ \cases{[\sigma(t)\,dt\,+\,\tilde
C\,dt]\,{\bm R} \,,\cr [\sigma(t)\,dt\,+\,2\tilde C\,dt]\,{\bm
R}\,.}
\label{difsto12}
\qqq Given the stochastic equation (\ref{disp2}) with a fixed
convention, its solution can be obtained by iteration from
(\ref{intsto}) and has the form (\ref{batch0}) with $W(t)$ given by
(\ref{disp6}) and the integrals interpreted with the same convention.
Note that the value of $\,\Big\langle\int\limits_{0}^t\sigma(s)
\,ds\int\limits_0^s\sigma(s')\,ds'\Big\rangle\,$ depends on the choice
of the convention: it vanishes for the It\^{o} one and is equal to
$\tilde C\,t\,$ for the Stratonovich and twice that for the
anti-It\^{o} ones.  The conventions are clearly related to the
time-reversibility of the finite-correlated strain before the
white-noise scaling limit is taken. For example, time-reversible
strains have even 2-point correlation functions and produce the
Stratonovich value for the above integrals in the white-noise scaling
limit. Most of these stochastic subtleties may be forgotten for the
incompressible strain where $\,\tilde C=0\,$ as follows from
(\ref{shortstrain3}) so that the difference between different
conventions for (\ref{disp2}) disappears.

We shall often have to consider functions of solutions of
stochastic differential equations so one should be aware that the
latter behave under such operation in a somewhat peculiar way.  
For the It\^{o} convention, this is the content of the so-called 
It\^{o} formula which results from straightforward second-order 
manipulations of the stochastic differentials and takes in the case 
of (\ref{disp2}) the form: \qq
d\,f({\bm R})\ =\ \Big(\sigma(t)\,dt\,{\bm R}\Big)\cdot{\bm
\nabla}f({\bm R})\ +\ C_{ijk\ell}\,R^j\,R^\ell
\,\nabla_i\nabla_kf({\bm R})\,dt\,.
\label{Itof}
\qqq Note the extra second-order term absent in the normal rules of
differential calculus. The latter are, however, preserved in the
Stratonovich convention. Note that the latter difference may appear
even for the incompressible strain.

More general stochastic equations may be treated similarly.
For example, (\ref{sde}) for Kraichnan velocities may be rewritten
as an integral equation involving the stochastic integral
$\,\int\limits_0^t{\bm v}({\bm R}(s),s)\,\,ds$.
\,The latter is defined as 
\qq
\lim\,\sum\limits_n\int\limits_{t_{n}}^{t_{n+1}}
\,\,{1\over2}[{\bm v}({\bm R}(t_{n}),s)+{\bm v}({\bm R}(t_{n+1}),s)]
\,\,ds
\qqq
in the Stratonovich convention, with the last $t_{n+1}$ ($t_n$) replaced
by $t_n$ ($t_{n+1}$) in the (anti-)It\^{o} one. The difference is 
\qq
\pm{_1\over^2}\,\lim\,\sum\limits_n\smallint\limits_{t_{n}}^{t_{n+1}}
\smallint\limits_{t_{n}}^{t_{n+1}}({\bm v}({\bm R}(t_n),s')\cdot{\bm\nabla})
\,{\bm v}({\bm R}(t_n),s)\ ds'\,ds\,=\pm\nabla_jD^{ij}({\bm 0})\,t\,,
\qqq 
where the left hand side was replaced by its mean by virtue of the
Central Limit Theorem. The last term vanishes if the velocity 2-point 
function is isotropic and parity invariant. Hence the choice 
of the convention is unimportant here even for the compressible velocities.

It may seem bizarre that the choice 
of convention in compressible velocities does not matter for individual 
trajectories but it does for the  equation (\ref{disp2}) which describes 
the evolution of small trajectory differences. It is not difficult to explain
this discrepancy (Horvai, 2000). The stochastic equation for ${\bm R}_{12}
\equiv{\bm R}$ leads (in the absence of noise) to the integral equation 
\qq
{\bm R}(t)\ =\ {\bm R}(0)\,+\,\int\limits_0^t\left[\,{\bm v}({\bm R}(s)+
{\bm R}_2(s),s)\,-\,{\bm v}({\bm R}_2(s),s)\,\right]\,ds\,.
\label{A8}
\qqq 
The difference between the Stratonovich and the It\^{o} conventions 
for the latter integral is
\qq
&&{_1\over^2}\,\lim\,\sum\limits_n\smallint\limits_{t_{n}}^{t_{n+1}}
\smallint\limits_{t_{n}}^{t_{n+1}} 
\left\{[{\bm v}({\bm R}(t_n)+{\bm R}_2(t_n),s')-{\bm v}({\bm R}_2(t_n),s')]
\cdot{\bm\nabla}\right\}{\bm v}({\bm R}(t_n)+{\bm R}_2(t_n),s)\cr
+&&{_1\over^2}\,\lim\,\sum\limits_n\smallint\limits_{t_{n}}^{t_{n+1}}
\smallint\limits_{t_{n}}^{t_{n+1}}{\bm v}({\bm R}_2(t_n),s')
\cdot{\bm\nabla})[{\bm v}({\bm R}(t_n)+{\bm R}_2(t_n),s)-{\bm v}
({\bm R}_2(t_n),s)]\,\,ds\cr
=&&\ \int\limits_0^t [\nabla_jD^{ij}({\bm R}(s))\,-\,\nabla_jD^{ij}
({\bm 0})]\,\,ds\ +\ \int\limits_0^t [\nabla_jD^{ij}({\bm 0})\,
-\,\nabla_jD^{ij}({\bm R}(s))]\,\,ds\ =\ 0\,,
\label{A9}
\qqq
where the two first lines that cancel each other are due to the time 
dependence of ${\bm R}$ and ${\bm R}_2$, respectively. If we replace
${\bm R}$ by $\epsilon{\bm R}$ then, when $\epsilon\to0$, the right
hand side of (\ref{A8}) is replaced by the right hand side of 
(\ref{intsto}) if we use the It\^{o} convention for the stochastic
integrals. The similar limiting procedure applied to the (vanishing) 
difference (\ref{A9}) does not reproduce the difference between the 
values of the integral (\ref{intsto}) for different conventions.
The latter corresponds to the limit of the first line of (\ref{A9}) 
only and does not reproduce the limit of the term of (\ref{A9}) due 
to the ${\bm R}_2$-dependence. The approximation (\ref{disp2}) is then 
valid only within the It\^{o} convention.

As for the PDF (\ref{dtp}), it may be viewed as a (generalized) function
of the trajectory ${\bm R}(t)$. The advection equation (\ref{ade0}) results
then from the equation for ${\bm R}(t)$ by applying the standard rules 
of differential calculus which hold when the Stratonovich convention
is used. The It\^{o} rules produce the equivalent It\^{o} form 
of the equation with an additional second order term containing 
the eddy diffusion generator $\,D_0\nabla_{_{\bm R}}^2$.

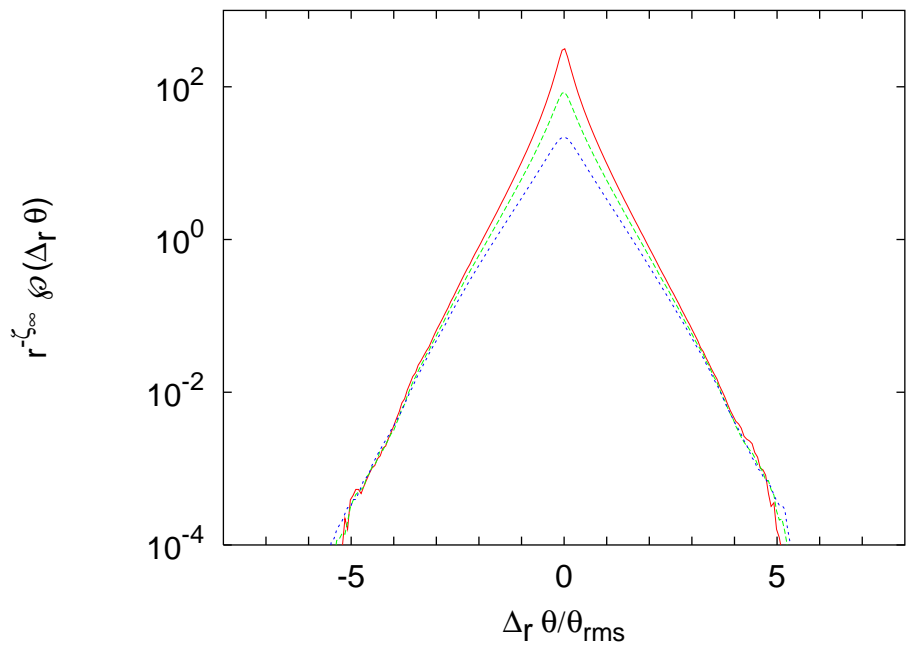
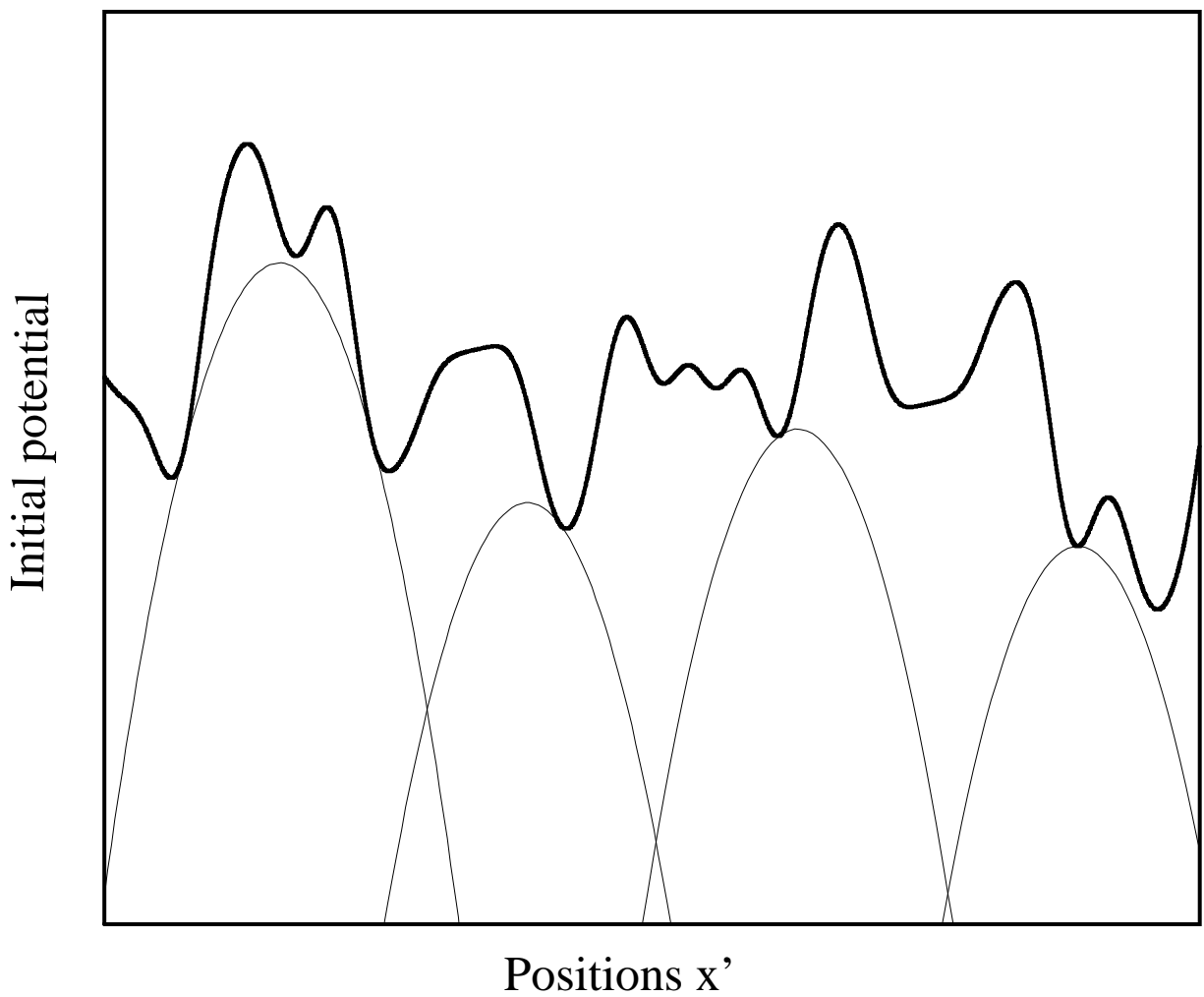
\end{document}